\documentclass[11pt,a4paper]{article}
\pdfoutput=1
\usepackage{jheppub}
\usepackage{graphicx}
\usepackage{lipsum}
\usepackage{cancel}
\usepackage{slashed}
\usepackage{soul}
\usepackage[capitalise, english]{cleveref}
\usepackage{mathtools}
\usepackage{bm}
\usepackage{float}
\usepackage{booktabs}
\usepackage{caption}
\usepackage{amsmath, amssymb}
\usepackage{bbold}
\usepackage[all]{hypcap}
\usepackage{makecell}
\usepackage{tikz-feynman}
\usepackage{pgfplots}
\usepackage{bbm}
\usepackage{xcolor}
\definecolor{nicered}{rgb}{0.7,0.1,0.1}
\newcommand{\ja}[1]{[{\textcolor{nicered}{JA: #1}}]}
\usepackage[font=small,labelfont=bf]{caption}
\usepackage{bbm}
\usepackage{color, colortbl}
\usepackage{subcaption}
\usepackage[font=footnotesize,labelfont=bf]{caption}
\usepackage{array}
\usepackage[export]{adjustbox}
\allowdisplaybreaks

\usepackage{graphicx}
\newcommand{\rank}{\ensuremath{\mathrm{Rank}}}

\newcommand{\sig}{\sigma}
\newcommand{\WFR}[1]{\big(#1\otimes 1\big)}
\newcommand{\WF}[2]{p_{#1}\,\big(#2\otimes 1\big)}
\newcommand{\lzm}{\left(}
\newcommand{\dzm}{\right)}
\newcommand{\lzs}{\left[}
\newcommand{\dzs}{\right]}
\newcommand{\lzv}{\left\{}
\newcommand{\dzv}{\right\}}
\newcommand{\lzu}{\left|}
\newcommand{\dzu}{\right|}
\newcommand{\VLL}{\mathrm{VLL}}
\newcommand{\VLR}{\mathrm{VLR}}
\newcommand{\SLL}{\mathrm{SLL}}
\newcommand{\TLL}{\mathrm{TLL}}
\newcommand{\SRL}{\mathrm{SRL}}
\newcommand{\cL}{\mathcal{L}}
\newcommand{\cO}{\mathcal{O}}
\newcommand{\cR}{\mathcal{R}}
\newcommand{\cM}{{\mathcal M}}
\newcommand{\cW}{{\mathcal W}}
\newcommand{\cB}{{\mathcal B}}
\newcommand{\cD}{{\mathcal D}}
\newcommand{\cQ}{{\mathcal Q}}
\newcommand{\cZ}{{\mathcal Z}}
\newcommand{\cI}{{\mathcal I}}
\newcommand{\cC}{{\mathcal C}}
\newcommand{\cS}{{\mathcal S}}
\newcommand{\cG}{{\mathcal G}}
\newcommand{\cU}{{\mathcal U}}
\newcommand{\cX}{{\mathcal X}}
\newcommand{\cY}{{\mathcal Y}}
\newcommand{\cH}{{\mathcal H}}
\newcommand{\gs}{{\Gamma_\phi}}
\newcommand{\gst}{{\tilde{\Gamma}_\phi}}
\newcommand{\eps}{{\epsilon}}
\newcommand{\kev}{\mathrm{keV}}
\newcommand{\mev}{\mathrm{MeV}}
\newcommand{\gev}{\mathrm{GeV}}
\newcommand{\tev}{\mathrm{TeV}}
\newcommand{\re}{{\mathrm{Re}} \,}
\newcommand{\im}{{\mathrm{Im}} \,}
\newcommand{\hermc}{\text{h.c.}}
\newcommand{\ud}[2]{\phantom{}^{#1}\phantom{}_{#2}}
\newcommand{\du}[2]{\phantom{}_{#1}\phantom{}^{#2}} 
\newcommand{\U}{\mathrm{U}}
\newcommand{\SU}{\mathrm{SU}}
\newcommand{\eminus}{\vcenter{\hbox{\scalebox{0.6}[1]{$ - $}}}}	
\newcommand{\rep}[1]{\mathbf{#1}}
\newcommand{\repbar}[1]{\overline{\mathbf{#1}}}
\newcommand{\sscript}[1]{{\scriptscriptstyle \mathrm{#1}}}
\newcommand{\UV}{\sscript{UV}}
\newcommand{\EFT}{\sscript{EFT}}
\newcommand{\SMEFT}{\sscript{SMEFT}}
\newcommand{\fdev}[1]{\dfrac{\delta}{\delta #1}}
\newcommand{\sdet}{\mathop{\mathrm{sDet}}}
\newcommand{\str}{\mathop{\mathrm{sTr}}}

\definecolor{deepskyblue}{rgb}{0.0, 0.75, 1.0}
\definecolor{aqua}{rgb}{0.0, 1.0, 1.0}
\definecolor{bronze}{rgb}{0.8, 0.5, 0.2}
\definecolor{electricyellow}{rgb}{1.0, 1.0, 0.0}
\definecolor{goldenyellow}{rgb}{1.0, 0.87, 0.0}
\definecolor{glaucous}{rgb}{0.38, 0.51, 0.71}
\newcommand{\soft}[1]{{\color{glaucous} #1}}
\newcommand{\hard}[1]{{\color{red} #1}}
\newcommand{\tealemp}[1]{{\color{teal} #1}}
\newcommand{\redemp}[1]{{\color{red} #1}}
\newcommand{\redf}{{\color{red}{f}}}
\newcommand{\gugu}[1]{{\color{orange!80!black} [GG: {shape #1}]}}
\newcommand{\ap}[1]{{\color{red!80!black} [AP: {shape #1}]}}
\newcommand{\mar}[1]{{\color{teal} [MP: { #1}]}}

\usepackage{soul} 
\colorlet{blueRef}{blue!80!black}
\colorlet{CodeColor}{red!70!black} 
\newcommand{\coloredbox}[3]{\colorbox{#1}{\texttt{{\color{#2}#3}}}}
\newcommand{\X}{\mathcal{X}}

\usepackage{xspace}

\newcolumntype{P}[1]{>{\centering\arraybackslash}p{#1}}
\allowdisplaybreaks
\newcommand{\be}{\begin{equation}}
\newcommand{\ee}{\end{equation}}
\newcommand{\bea}{\begin{eqnarray}}
\newcommand{\eea}{\end{eqnarray}}
\newcommand{\beq}{\begin{equation}}
\newcommand{\eeq}{\end{equation}}
\newcommand{\no}{\nonumber}
\newcommand{\comment}[1]{{\color{red}{#1}}}
\newcommand{\mdfd}[1]{{\color{magenta}{#1}}}

\usepackage{tikz}
\usetikzlibrary{decorations.pathmorphing,decorations.markings,arrows.meta,patterns}

\tikzset{
  ph/.style={decorate, decoration={snake, amplitude=1.1pt,
             segment length=5pt, pre length=1pt, post length=1pt}},
  phfit/.style={ph, decoration={segment length=#1}},
  phend/.style={ph, decoration={segment length=#1, post length=0pt}},
  scline/.style={line width=0.55pt, dash pattern=on 3pt off 2pt,
             line cap=round, line join=round},
  arrow at/.style={postaction={decorate, decoration={markings,
               mark=at position #1 with
                 {\arrow{Stealth[length=5.5pt,width=4.5pt]}}}}},
  sc/.style={scline, arrow at=0.5},
  blob/.style={draw, preaction={fill=white}, pattern=north east lines},
}

\newcommand{\VV}{%
  \draw (-1,1.12) -- (0,0.12) -- (1,1.12);
  \draw (-1,-1.12) -- (0,-0.12) -- (1,-1.12);}
\newcommand{\num}[1]{\node at (0,-1.6) {(#1)};}
\newcommand{\panel}[1]{\begin{tikzpicture}[x=1cm,y=1cm,baseline=0pt]#1\end{tikzpicture}}

\begin{document}

\title{Universal Two-Loop Current--Current Renormalization of Four-Fermion Operators}
\author{Jason Aebischer{$^{\, a}$},
Sudeepan Datta{$^{\, a}$},
Pol Morell{$^{\, b,c}$},
Marko Pesut{$^{\, a}$}
and
Javier Virto{$^{\, b,c}$}}

\affiliation{{$^{\, a}$} 
PSI Center for Neutron and Muon Sciences, 5232 Villigen PSI, Switzerland}

\affiliation{{$^{\, b}$}
Departament de Física Quàntica i Astrofísica, Universitat de Barcelona,\\
Martí Franquès 1, E08028 Barcelona, Catalunya}

\affiliation{
{$^{\, c}$}
Institut de Ciències del Cosmos (ICCUB), Universitat de Barcelona,\\
Martí Franquès 1, E08028 Barcelona, Catalunya
}

\emailAdd{jason.aebischer@psi.ch}
\emailAdd{sudeepan.datta@psi.ch}
\emailAdd{pmorell@icc.ub.edu}
\emailAdd{marko.pesut@psi.ch}
\emailAdd{jvirto@icc.ub.edu}

\abstract{The complete one- and two-loop gauge and scalar current--current contributions to the running of arbitrary dimension-six four-fermion operators are reported. The results are obtained in a basis-independent and renormalization-scheme-agnostic fashion and are valid for arbitrary gauge groups. The complete gauge and scalar contributions, including the genuine one- and two-loop poles, counterterm insertions, wave-function renormalization, finite evanescent insertions, diagram multiplicities, and the associated color, charge and Yukawa factors are provided for the most general renormalizable Lagrangian with gauge and scalar interactions. The results enable the derivation of leading- and next-to-leading-order current--current anomalous-dimension matrices for arbitrary Effective Field Theories (EFTs) and constitute a significant step toward the complete two-loop renormalization of general EFTs.}

\maketitle
\newpage
\section{Introduction}
Effective Field Theories (EFTs) provide a coherent formalism to describe the dynamics and degrees of freedom of a Quantum Field Theory at a given energy scale. The scale dependence is encoded in the coupling constants and Wilson coefficients of the EFT and is governed by the Renormalization Group Equations (RGEs). They describe the dependence of theory parameters on the unphysical scale $\mu$ introduced when employing dimensional regularization. Knowledge of these RGEs is crucial for any EFT, and significant efforts have focused on developing new computational techniques and applying them to a wide variety of effective theories. In particular, extensive work has been devoted in recent years to the derivation of the RGEs of the Standard Model Effective Field Theory (SMEFT) \cite{Buchmuller:1985jz,Grzadkowski:2010es,Brivio:2017vri,Isidori:2023pyp,Aebischer:2025qhh} and the Weak Effective Theory (WET)~\cite{Jenkins:2017jig}. The one-loop SMEFT RGEs were derived for dimension-five \cite{Chankowski:1993tx,Babu:1993qv,Antusch:2001ck}, dimension-six~\cite{Jenkins:2013zja,Jenkins:2013wua,Alonso:2013hga,Alonso:2014zka}, dimension-seven \cite{Zhang:2023ndw,Zhang:2024clp} and dimension-eight operators~\cite{Chala:2021pll,DasBakshi:2022mwk,DasBakshi:2023htx,Chala:2023xjy,Bakshi:2024wzz,Liao:2024xel,Wu:2025qto,DasBakshi:2026ief} and the one-loop WET running is known as well~\cite{Jenkins:2017dyc,Aebischer:2017gaw,Naterop:2023dek}. Two-loop RGEs for the WET \cite{Aebischer:2022anv,Aebischer:2025hsx,Naterop:2024cfx,Naterop:2025cwg,Jenkins:2023bls,Jenkins:2023rtg,Buras:1992zv,Buras:2000if,Buras:1991jm,Buras:1992tc,Morell:2024aml,Aebischer:2021raf,Naterop:2025lzc} and for the SMEFT \cite{Ibarra:2024tpt,Zhang:2025ywe,Born:2026xkr,Born:2024mgz,DiNoi:2024ajj,Duhr:2025zqw,DiNoi:2025arz,Haisch:2025vqj,Duhr:2025yor,DiNoi:2025tka,Banik:2025wpi} have also been computed recently, enabling phenomenological analyses at next-to-leading-logarithmic accuracy.

In the diagrammatic approach, RGEs can be derived from the $1/\epsilon$ ultraviolet divergences of Feynman diagrams regularized in $d = 4 - 2\epsilon$ dimensions. Since RGEs depend on the field content and gauge symmetries of the theory, these divergences must, in principle, be computed anew for each EFT. However, by adopting a generic EFT framework with arbitrary field content and gauge symmetries, it is possible to compute the ultraviolet divergences in a model-independent way. These UV poles can then be provided independently of the remaining steps of the calculation, which amount to Dirac algebra simplifications and group-theoretical calculations. Within this approach, the RGE coefficients for a wide class of EFTs can be directly obtained by combining the reported UV poles with the appropriate Dirac-algebra and group-theoretical factors.

For dimension-four interactions, several generic RGE results are available in the literature. Early two-loop results were obtained in~\cite{Jack:1984vj,Machacek:1983tz,Machacek:1983fi,Machacek:1984zw,Luo:2002ti,Schienbein:2018fsw}, while the two-loop renormalization of vacuum expectation values was derived in~\cite{Sperling:2013eva,Sperling:2013xqa}. The three-loop beta functions of gauge and Yukawa couplings were computed respectively in~\cite{Pickering:2001aq,Poole:2019kcm,Mihaila:2012pz} and~\cite{Davies:2021mnc,Poole:2019txl}. For general scalar–fermion theories the three-loop gaugeless RGEs were obtained in~\cite{Jack:2023zjt,Steudtner:2021fzs} and the quartic scalar coupling was renormalized at the three-loop level~\cite{Steudtner:2024teg}. The four-loop gauge beta function was determined in~\cite{Bednyakov:2021qxa}, while the running of renormalizable scalar couplings is known up to six loops~\cite{Bednyakov:2021ojn,Bednyakov:2025sri}, and even higher-order results for scalar $\phi^4$ theories are known \cite{Schnetz:2022nsc}.

In this article, following the spirit of the original classification of UV poles \cite{Buras:1992zv,Buras:2000if,Buras:1991jm,Buras:1992tc,Ciuchini:1993vr,Ciuchini:1993ks} and the recent computations of RGEs in generic EFT frameworks \cite{Aebischer:2025zxg,Misiak:2025xzq,Fonseca:2025zjb,Guedes:2025sax,Fonseca:2025cls,Aebischer:2025ddl} using functional \cite{Henning:2014wua,Drozd:2015rsp,delAguila:2016zcb,Henning:2016lyp,Fuentes-Martin:2016uol,Zhang:2016pja,Cohen:2020fcu,Cohen:2020qvb,Fuentes-Martin:2020udw,Fuentes-Martin:2024agf,Buchalla:2019wsc,Born:2024mgz,Fuentes-Martin:2023ljp} and on-shell methods\footnote{RGEs can also be computed using geometric techniques, see for instance \cite{Helset:2022pde,Assi:2023zid,Jenkins:2023rtg,Jenkins:2023bls}.} \cite{Chala:2024llp,LopezMiras:2025gar,Baratella:2020lzz,Machado:2022ozb,Caron-Huot:2016cwu,Panico:2018hal,Ma:2019gtx,Jiang:2020mhe,AccettulliHuber:2021uoa,EliasMiro:2020tdv,Bern:2020ikv,Bern:2019wie,Cheung:2015aba,Bresciani:2024shu,Bresciani:2023jsu}, we provide a complete list of UV divergences at the one- and two-loop orders from current--current contributions to the renormalization of generic dimension-six four-fermion operators. Adopting the philosophy of general EFTs, we assume the most general gauge and scalar sectors invariant under a generic gauge group for the renormalizable dimension-four Lagrangian. In this setup we report the one-loop and two-loop pole structures of all current--current contributions to four-fermion operators, as well as the necessary counterterm insertions and wave-function renormalization constants. The results are provided in a generic form, without performing Dirac simplifications, and can therefore be used in conjunction with arbitrary renormalization schemes. Furthermore, the pole structures are reported for generic four-fermion operators, which also makes them independent of the operator basis considered. 

Our results therefore allow to obtain all Leading-Order (LO) and Next-to-Leading-Order (NLO) current--current contributions to anomalous dimension matrices (ADMs) for arbitrary four-fermion operators. All necessary manipulations reduce to straightforward Dirac-algebra simplifications. Furthermore, the formalism is compatible with any choice of Dirac and evanescent prescription and is valid for arbitrary gauge groups and operator bases. To demonstrate the usefulness of our results, we compute the two-loop gauge contributions to the JMS basis \cite{Jenkins:2017jig} in Naive Dimensional Regularization (NDR), thereby reproducing the results of \cite{Aebischer:2025hsx}.  Crucially, NLO running is needed to consistently account for the scheme dependence introduced in the matching procedure \cite{Dekens:2019ept,Carmona:2021xtq,Fuentes-Martin:2022jrf}. By combining the NDR results in our evanescent basis with the \emph{shift} approach previously developed in \cite{Aebischer:2022aze,Aebischer:2022rxf,Aebischer:2023djt,Aebischer:2024xnf,Buras:1991jm, Herrlich:1994kh, Chetyrkin:1997gb,Gorbahn:2004my}, it is possible to facilitate changes of both the physical and evanescent operator bases as well as of the renormalization scheme. 

The rest of the article is organized as follows. In~\cref{sec:methodology}, we describe the notation and procedure used to
derive the one- and two-loop ultraviolet divergences. Furthermore, the construction of the corresponding current--current contributions to the RGEs, including the relevant Dirac, evanescent, and group-theoretical ingredients is outlined. The structure and application of the results are discussed in Section~\ref{sec:results}. In Section~\ref{sec:example}, we present a detailed example illustrating how our results can be used to compute NLO anomalous-dimension matrices of four-fermion operators. We conclude in Section~\ref{sec:conclusion}.
The generic pole structures generated by gauge and scalar interactions are collected in Appendices~\ref{app:genericGauge} and \ref{app:genericscalar}, respectively, while the diagram multiplicities and the corresponding color, charge, and Yukawa factors are given in Appendices~\ref{app:multiplicity} and \ref{app:colorfactors}. The projected gauge results in the NDR scheme are collected in Appendix~\ref{app:NDR}.

\section{Methodology}
\label{sec:methodology}
The main result of this work is a complete list of two-loop ultraviolet (UV) divergences generated by arbitrary gauge and scalar interactions in current-current four-fermion amplitudes. To this end the most general interaction Lagrangian governing fermions and scalars is adopted. For fermions this amounts to the regular Dirac Lagrangian invariant under a generic gauge theory (see~\cref{app:gencol}), whereas the scalar interaction Lagrangian is specified in \cref{eq:scallag}.  Furthermore, in order to obtain the relevant UV divergences in a basis-independent fashion we assume a generic four-fermion operator of the form
\begin{equation}\label{eq:genop}
    \mathcal{Q}_{\Gamma_1\Gamma_2}^{ijkl}
    =
    \big(\bar\psi_i \Gamma_1 \psi_j\big)
    \big(\bar\psi_k \Gamma_2 \psi_l\big)\,,
\end{equation}
where $\Gamma_1\otimes\Gamma_2$ denotes a generic Dirac structure. Typical examples of physical Dirac structures include
\begin{equation}
    \Gamma_1\otimes \Gamma_2
    \in
    \left\{
    \mathbb{1}\otimes \mathbb{1}\,,
    \gamma^\mu\otimes \gamma_\mu\,,
    \sigma^{\mu\nu}\otimes\sigma_{\mu\nu}
    \right\}\,, \quad \sigma^{\mu\nu} \equiv \frac{i}{2}\left[\gamma^\mu,\gamma^\nu\right]\,,
\end{equation}
or combinations including insertions of projection operators $P_{L,R} \equiv \frac{1 \mp \gamma_5}{2}$ or $\gamma_5$ matrices. Using two independent private implementations of the algorithmic procedure to extract UV poles as detailed in  \cite{Chetyrkin:1997fm} and the classification of relevant current-current Feynman diagrams in \cite{Buras:1992zv,Buras:2000if,Buras:1991jm,Buras:1992tc}, all one-loop and two-loop divergences are computed for the general operator in \cref{eq:genop}. The calculations were also cross-checked against an independent workflow based on \textsc{SmeftFR} \cite{Dedes:2023zws}, \textsc{FeynArts} \cite{Hahn:2000kx}, \textsc{FeynCalc} \cite{Shtabovenko:2020gxv,Shtabovenko:2023xyz,Shtabovenko:2025lxq}, and \textsc{FeynHelpers} \cite{Shtabovenko:2016whf}, the latter providing a straightforward interface to the program \textsc{Kira} \cite{Lange:2025fba} for the reduction of all two-loop scalar Feynman integrals to a smaller set of master integrals, whose analytic results are known in the literature. The results are reported without performing Dirac-algebra simplifications and can therefore be used in arbitrary renormalization schemes. The $1/\epsilon^2$ and $1/\epsilon$ poles are listed separately for the individual current--current diagrams and are collected for gauge interactions in \cref{app:genericGauge} and for scalar interactions in \cref{app:genericscalar}. 

In order to extract the ADM of a given theory the field content, gauge group, operator basis and the renormalization scheme need to be specified. After performing group-theory and Dirac simplifications the results of this article can be used to compute the ADM of the underlying theory. In the remainder of this section, we briefly review the procedure for obtaining the one- and two-loop ADM of four-fermi operators:

Consider the following renormalized Lagrangian given by\footnote{Here multiple operator insertions are neglected. For a general discussion we refer to \cite{Aebischer:2025hsx}.}
\begin{equation}
\mathcal{L}=\mathcal{L}_{d \leq4}+\sum_{k, l} \mathcal{C}_l Z_{l k} Z_{\mathcal{O}_k} \mathcal{O}_k \,,
\end{equation}
where the renormalized operators $\mathcal{O}_k=\left\{\mathcal{Q}_k, \mathcal{E}_k\right\}$ include physical $\left(\mathcal{Q}_k\right)$ and evanescent $\left(\mathcal{E}_k\right)$ operators. The $\mathcal{C}_l$ denote renormalized Wilson coefficients, while $Z_{lk}$ are the corresponding renormalization constants that induce operator mixing. The renormalization factor $Z_{\mathcal{O}_k}$ accounts for the wave-function renormalization of the fields appearing in the operator $\mathcal{O}_k$. The $\mu$-dependence of the
renormalized Wilson coefficients is given by the RGEs
\begin{equation}
\frac{d \mathcal{C}_i}{d \log \mu}=\gamma_{j i} \mathcal{C}_j \equiv \frac{1}{16 \pi^2} \dot{\mathcal{C}}_i \,,
\end{equation}
where $\gamma_{ij}$ denote the components of the ADM $\hat{\gamma}$. It governs the operator running and can be expanded perturbatively as follows
\begin{equation}
\hat{\gamma}=\tilde{\alpha}_{\mathcal{G}_1} \hat{\gamma}^{(1,0)}+\tilde{\alpha}_{\mathcal{G}_2}  \hat{\gamma}^{(0,1)}+\tilde{\alpha}_{\mathcal{G}_1}  \tilde{\alpha}_{\mathcal{G}_2} \hat{\gamma}^{(1,1)}+\tilde{\alpha}_{\mathcal{G}_1}^2 \hat{\gamma}^{(2,0)}+\tilde{\alpha}_{\mathcal{G}_2}^2 \hat{\gamma}^{(0,2)}+\cdots,
\end{equation}
where $\hat{\gamma}^{(n,m)}$ are the expansion coefficients of the ADM in powers of $\tilde{\alpha}_{\mathcal{G}_i} \equiv \alpha_{\mathcal{G}_i}/(4\pi) = g_i^2/(16\pi^2)$, where $\mathcal{G}_i$ labels different gauge groups. Furthermore, the renormalization matrix $\hat{Z}$ is related to the ADM as follows:
\begin{equation}
\hat{\gamma}=\hat{Z} \frac{d \hat{Z}^{-1}}{d \log \mu} \,,
\end{equation}
where $\hat{Z}$ depends on the renormalization scale via the gauge factors $\tilde{\alpha}_{\mathcal{G}_i}(\mu)$,
\begin{equation}
\hat{Z}=\sum_{n, m=0}^{\infty} \sum_{\ell=0}^{n+m} \frac{\tilde{\alpha}_{\mathcal{G}_1}^n \tilde{\alpha}_{\mathcal{G}_2}^m}{\epsilon^{\ell}} \hat{Z}^{(n, m ; \ell)},
\end{equation}
with $Z_{ij}^{(0,0 ; 0)}=\delta_{i j}$. In this scheme, $Z_{\mathcal{Q}_i \mathcal{E}_j}^{(n, m ; 0)}=0$, which leads to
\begin{align}
& \hat{\gamma}^{(1,0)}=2 \hat{Z}^{(1,0 ; 1)}\,, \\ %
& \hat{\gamma}^{(0,1)}=2 \hat{Z}^{(0,1 ; 1)}\,, \\ 
& \hat{\gamma}^{(2,0)}=4 \hat{Z}^{(2,0 ; 1)}-2 \hat{Z}^{(1,0 ; 1)} \hat{Z}^{(1,0 ; 0)}\,, \\ %
& \hat{\gamma}^{(0,2)}=4 \hat{Z}^{(0,2 ; 1)}-2 \hat{Z}^{(0,1 ; 1)} \hat{Z}^{(0,1 ; 0)}\,, \\ 
& \hat{\gamma}^{(1,1)}=4 \hat{Z}^{(1,1 ; 1)}-2 \hat{Z}^{(1,0 ; 1)} \hat{Z}^{(0,1 ; 0)}-2 \hat{Z}^{(0,1 ; 1)} \hat{Z}^{(1,0 ; 0)}\,.
\end{align}
In the $\overline{\mathrm{MS}}$ scheme the renormalization constants can be calculated in terms of the amputated renormalized matrix elements of the physical operators $\mathcal{Q}_i$. To arbitrary loop order, they are given by
\begin{align}
\left\langle \mathcal{Q}_i\right\rangle & =\sum_{n, m=0}^{\infty} \tilde{\mu}^{2 \epsilon(n+m)} \tilde{\alpha}_{\mathcal{G}_1}^n \tilde{\alpha}_{\mathcal{G}_2}^m\left\langle \mathcal{Q}_i\right\rangle^{(n, m)}, \\
\left\langle \mathcal{Q}_i\right\rangle^{(n, m)} & =\sum_{k=0}^{n+m} \frac{1}{\epsilon^k}\left[a_{\mathcal{Q}_i \mathcal{Q}_j}^{(n, m ; k)}\left\langle \mathcal{Q}_j\right\rangle^{(0)}+a_{\mathcal{Q}_i \mathcal{E}_j}^{(n, m ; k)}\left\langle \mathcal{E}_j\right\rangle^{(0)}\right],
\end{align}
where $\tilde{\mu}^2 \equiv \mu^2 e^{\gamma_E} / 4 \pi$ and $a_{\mathcal{Q}_i \mathcal{O}_j}^{(n, m ; k)}$ denote the finite coefficient matrices. The latter result from the $1 / \epsilon^k$ poles of the $(n+m)$-loop corrections to the operator $\mathcal{Q}_i$. In this notation one finds the following renormalization constants: 
\begin{align}
\hat{Z}^{(1,0 ; 1)} = & -\hat{a}^{(1,0 ; 1)}-\hat{Z}_{\mathcal{Q}}^{(1,0 ; 1)}\,, \\ %
\hat{Z}^{(0,1 ; 1)} = & -\hat{a}^{(0,1 ; 1)}-\hat{Z}_{\mathcal{Q}}^{(0,1 ; 1)}\,, \\ 
\hat{Z}^{(2,0 ; 1)} = & -\hat{a}^{(2,0 ; 1)}+\hat{a}^{(1,0 ; 1)} \hat{a}^{(1,0 ; 0)}-\hat{Z}^{(1,0 ; 0)} \hat{a}^{(1,0 ; 1)}-\hat{Z}_{\mathcal{Q}}^{(2,0 ; 1)}\,, \\ %
\hat{Z}^{(0,2 ; 1)} = & -\hat{a}^{(0,2 ; 1)}+\hat{a}^{(0,1 ; 1)} \hat{a}^{(0,1 ; 0)}-\hat{Z}^{(0,1 ; 0)} \hat{a}^{(0,1 ; 1)}-\hat{Z}_{\mathcal{Q}}^{(0,2 ; 1)}\,, \\ 
\hat{Z}^{(1,1 ; 1)} = & -\hat{a}^{(1,1 ; 1)}+\hat{a}^{(1,0 ; 1)} \hat{a}^{(0,1 ; 0)}+\hat{a}^{(0,1 ; 1)} \hat{a}^{(1,0 ; 0)}-\hat{Z}^{(1,0 ; 0)} \hat{a}^{(0,1 ; 1)} \notag\\
& -\hat{Z}^{(0,1 ; 0)} \hat{a}^{(1,0 ; 1)}-\hat{Z}_{\mathcal{Q}}^{(1,1 ; 1)}\,,
\end{align}
up to two loops in $\tilde{\alpha}_{\mathcal{G}_1}$ and $\tilde{\alpha}_{\mathcal{G}_2}$. The wave-function renormalization constant for an operator $\mathcal{Q}$ is defined by $\left(\hat{Z}_{\mathcal{Q}}\right)_{i j}=Z_{\mathcal{Q}_i} \delta_{\mathcal{Q}_i \mathcal{Q}_j}$, which in the case of four-fermion operators is given by
\begin{equation}
    Z_{\mathcal{Q}_i}^{(n, m ; 1)}=\frac{1}{2} Z_{\psi_1}^{(n, m ; 1)}+\frac{1}{2} Z_{\psi_2}^{(n, m ; 1)}+\frac{1}{2} Z_{\psi_3}^{(n, m ; 1)}+\frac{1}{2} Z_{\psi_4}^{(n, m ; 1)}\,. %
\end{equation}
The wave-function renormalization depends on group-theoretical contractions, which are defined as follows. For a fermion $\psi$ transforming in a representation $R_\psi$ of a generic gauge group $\mathcal{G}$, we use the following notation for the generators and Casimirs
\begin{equation}\label{eq:genneralgen}
T_{R_\psi}^a T_{R_\psi}^a
=
C_2(R_\psi)\,\mathbbm{1}\,,
\qquad
\mathrm{Tr}\!\left(T_{R_\psi}^aT_{R_\psi}^b\right)
=
T(R_\psi)\,\delta^{ab}\,,
\qquad
f^{acd}f^{bcd}
=
C_A\,\delta^{ab}\,.
\end{equation}
The abelian generators for $U(1)$ groups are simply denoted by $Q_\psi$. We denote different gauge factors by superscripts $[i]$, where $i$ labels the corresponding gauge group. Thus $Q_\psi^{[1]}$ is the charge of the fermion $\psi$ under $U(1)_1$, while $C_2^{[1]}(R_\psi)$ is the quadratic Casimir of its representation $R_\psi$ under a non-abelian group $\mathcal{G}_1$.

The fermion-loop factors entering the two-loop wave-function renormalization are defined as
\begin{equation}
N_T^{[i]}
\equiv
\sum_f d_f\,T^{[i]}(R_f)\,,
\qquad
N_Q^{[i]}
\equiv
\sum_f d_f\,\left(Q_f^{[i]}\right)^2\,.
\label{eq:ntnq}
\end{equation}
Here the sum runs over Dirac fermions charged under the corresponding gauge group, and $d_f$ denotes the multiplicity of the fermion $f$ under spectator gauge groups. For instance, for $N_F^{[i]}$ Dirac fermions in the fundamental representation of $SU(N_i)$ with no additional spectator multiplicity, one has $N_T^{[i]}=T (F) N_F^{[i]}=N_F^{[i]}/2$.

Finally, a simplification for the derivation of the ADMs can be achieved by adopting the Buras-Weisz scheme in which $Z_{\mathcal{E}_i \mathcal{Q}_j}^{(1,0;0)}=-a_{\mathcal{E}_i \mathcal{Q}_j}^{(1,0)}$ for $(i, j)$ being (evanescent, physical), and zero otherwise. This leads to the following master formulae
\begin{align}
\gamma_{i j}^{(1,0)}= & -2 a_{\mathcal{Q}_i \mathcal{Q}_j}^{(1,0 ; 1)}-2 Z_{\mathcal{Q}_i}^{(1,0 ; 1)} \delta_{\mathcal{Q}_i \mathcal{Q}_j} \,,\\
\gamma_{i j}^{(0,1)}= & -2 a_{\mathcal{Q}_i \mathcal{Q}_j}^{(0,1 ; 1)}-2 Z_{\mathcal{Q}_i}^{(0,1 ; 1)} \delta_{\mathcal{Q}_i \mathcal{Q}_j} \,,\\
\gamma_{i j}^{(2,0)}= & -4 a_{\mathcal{Q}_i \mathcal{Q}_j}^{(2,0 ; 1)}+4 a_{\mathcal{Q}_i \mathcal{Q}_k}^{(1,0 ; 1)} a_{\mathcal{Q}_k \mathcal{Q}_j}^{(1,0 ; 0)}+2 a_{\mathcal{Q}_i \mathcal{E}_k}^{(1,0 ; 1)} a_{\mathcal{E}_k \mathcal{Q}_j}^{(1,0 ; 0)}-4 Z_{\mathcal{Q}_i}^{(2,0 ; 1)} \delta_{\mathcal{Q}_i \mathcal{Q}_j} \label{eq:QED2ADM} \,,\\
\gamma_{i j}^{(0,2)}= & -4 a_{\mathcal{Q}_i \mathcal{Q}_j}^{(0,2 ; 1)}+4 a_{\mathcal{Q}_i \mathcal{Q}_k}^{(0,1 ; 1)} a_{\mathcal{Q}_k \mathcal{Q}_j}^{(0,1 ; 0)}+2 a_{\mathcal{Q}_i \mathcal{E}_k}^{(0,1 ; 1)} a_{\mathcal{E}_k \mathcal{Q}_j}^{(0,1 ; 0)}-4 Z_{\mathcal{Q}_i}^{(0,2 ; 1)} \delta_{\mathcal{Q}_i \mathcal{Q}_j} \,,\\
\gamma_{i j}^{(1,1)}= & -4 a_{\mathcal{Q}_i \mathcal{Q}_j}^{(1,1 ; 1)}+4 a_{\mathcal{Q}_i \mathcal{Q}_k}^{(1,0 ; 1)} a_{\mathcal{Q}_k \mathcal{Q}_j}^{(0,1 ; 0)}+4 a_{\mathcal{Q}_i \mathcal{Q}_k}^{(0,1 ; 1)} a_{\mathcal{Q}_k \mathcal{Q}_j}^{(1,0 ; 0)}+2 a_{\mathcal{Q}_i \mathcal{E}_k}^{(1,0 ; 1)} a_{\mathcal{E}_k \mathcal{Q}_j}^{(0,1 ; 0)} \notag\\
& +2 a_{\mathcal{Q}_i \mathcal{E}_k}^{(0,1 ; 1)} a_{\mathcal{E}_k \mathcal{Q}_j}^{(1,0 ; 0)}-4 Z_{\mathcal{Q}_i}^{(1,1 ; 1)} \delta_{\mathcal{Q}_i \mathcal{Q}_j}\,,
\end{align}
which allow to compute the one- and two-loop ADMs in this particular scheme.

\section{Overview of results}\label{sec:results}
In this section we give a brief overview of the results obtained in this article. They consist of generic wave-function renormalization contributions, one-loop corrections, counterterm insertions and two-loop corrections related to gauge and scalar interactions, which are reported in \cref{app:genericGauge} and \ref{app:genericscalar}, respectively. The multiplicity factors as well as the gauge and Yukawa factors corresponding to these diagrams are collected in \cref{app:multiplicity} and \ref{app:colorfactors}. Based on these general results an explicit example is worked out for the case of the JMS basis in the WET at two-loop order in the NDR scheme. For this scheme, we use the notation $\SLL$, $\TLL$, $\VLL$, $\VLR$, and $\SRL$ for the Dirac structures
\begin{equation}
\begin{aligned}
\SLL &\equiv
P_L\otimes P_L\,,
&
\TLL &\equiv
\sigma^{\mu\nu}P_L\otimes\sigma_{\mu\nu}P_L\,, &
\VLL &\equiv
\gamma^\mu P_L\otimes\gamma_\mu P_L\,,
\\
\VLR &\equiv
\gamma^\mu P_L\otimes\gamma_\mu P_R\,, &
\SRL &\equiv
P_R\otimes P_L\,.
\end{aligned}
\label{eq:projected-Dirac-labels}
\end{equation}
The results of this computation are collected in \cref{app:NDR} and fully agree with the findings in \cite{Aebischer:2025hsx}. As a further cross-check, we reconstructed the diagram-by-diagram current--current pole tables of \cite{Buras:1989xd,Buras:2000if} from the results in \cref{app:NDR}. For each diagram, the corresponding entry is obtained by combining the genuine two-loop pole, the physical one-loop counterterm insertion, the finite evanescent insertion with the BMU normalization, the dimension-four counterterms, the wave-function contribution, and the appropriate diagram multiplicity and color factors.

A compact overview over all the obtained results in this article is provided in~\cref{tab:results-roadmap}. The second and third columns are related to the generic gauge and scalar case, respectively while the fourth column contains references to the results of the explicit computation in the WET.

\begin{table}[t!]
\centering
\small
\setlength{\tabcolsep}{2pt}
\renewcommand{\arraystretch}{1.18}
\resizebox{\linewidth}{!}{
\begin{tabular}{@{}|
>{\raggedright\arraybackslash}p{4.1cm} |
>{\raggedright\arraybackslash}p{4.15cm}|
>{\raggedright\arraybackslash}p{3.9cm}|
>{\raggedright\arraybackslash}p{4.1cm}|
@{}}
\Xhline{0.7mm}
\textbf{Building block}
&
\textbf{Generic gauge}
&
\textbf{Generic scalar}
&
\textbf{WET example}
\\
\Xhline{0.7mm}

One-loop operator insertions
(diagrams 1--3)
&
\cref{tab:generic-1L-gluon}
&
\cref{tab:generic-1L-scalar}
&
\crefrange{tab:1L_SLL}{tab:1L_SRL}
\\
\hline

One-loop counterterm insertions
(diagrams 1--3)
&
Definition:
\cref{tab:dim4-CT-decomposition,tab:dim4-CTs}, Results:
\crefrange{tab:generic-fermion-CT-D1D3}{tab:1L_OPCT};

&
\crefrange{tab:generic-scalar-fermion-CT-D1D3}
{tab:generic-scalar-propagator-CT-D1D3}
&
\crefrange{tab:CT_SLL}{tab:CT_SRL}
\\
\hline

Finite one-loop evanescent insertions
&
Definition:
\cref{sec:1Lpolesev}
&
\hspace{1.3cm} ---
&
\crefrange{tab:E4}{tab:E2RL}
\\
\hline

Fermion wave-function renormalization
&
Results: \cref{tab:wfr-open-ct,tab:wfr-open-twogauge},
Color factors:
\cref{tab:wfr-bare-color}
&
\crefrange{tab:wfr-open-scalar-ct}{tab:wfr-open-scalar-selfenergies}
&
\cref{tab:WFR}
\\
\hline

Two-loop exchange topologies
(diagrams 4--24)
&
\crefrange{tab:genericpart1}{tab:genericpart3}
&
\crefrange{tab:generic-scalar-D14-D10}
{tab:generic-scalar-D17-D24}
&
\crefrange{tab:2L_SLL}{tab:2L_SRL}
\\
\hline

Three-boson and mixed topologies
(diagrams 25--28)
&
\cref{tab:genericpart4}
&
\cref{tab:generic_SSG_D25_D28}
&
\crefrange{tab:trig_SLL}{tab:trig_SRL}
\\
\hline

Self-energy topologies
(diagrams 29--31)
&
\crefrange{tab:generic29}{tab:generic31}
&
\crefrange{tab:genericScalar29}{tab:genericScalar31}
&
Abelian contributions:
\crefrange{tab:2L_SLL}{tab:2L_SRL},

Non-abelian contributions:
\crefrange{tab:se29_SLL}{tab:se31_SRL}
\\
\hline

Diagram multiplicities and coupling / color factors
&
Multiplicity:
\cref{tab_multSUN};

Generic gauge contractions:
\cref{tab:generalcolorfac} and \ref{tab:generalcolorfacSE};

$SU(N)^2$:
\cref{tab:color-mixing};

$SU(N_1)\times SU(N_2)$:
\crefrange{tab:color-mixing-suN2-simple-slash}
{tab:color-mixing812firstgluon22};

BNV $SU(3)$:
\cref{tab:color-mixingBviol};

$SU(N)\times U(1)$: \cref{tab:color-mixingmixSUNxU1};

$U(1)_1\times U(1)_2$: \cref{tab:chargefactorsU1xU1};

$U(1)^2$: \cref{tab:chargefactorsU1};
&
Yukawa and flavor contractions:
\cref{tab:scalar-yukawa-contractions,tab:scalar-yukawa-SEs}
&
\hspace{1.7cm}---
\\
\Xhline{0.7mm}
\end{tabular}
}
\caption{Road map to the tabulated ingredients provided in this article. The entries for diagrams 4--24 are common to abelian and non-abelian interactions up to their diagram-dependent factors, while the non-abelian diagrams 25--28 and the self-energy diagrams 29--31 are reported separately.}
\label{tab:results-roadmap}
\end{table}

\section{Example: NLO ADM for a semileptonic tensor operator}
\label{sec:example}
In this section, we provide an explicit example illustrating how to use our results to construct a two-loop current--current ADM. As a concrete case, we consider the following four-fermion semileptonic tensor operator:
\begin{equation}\label{eq:exampleADMop}
    \mathcal{Q} \equiv T^{LL}_{\nu e du}= \left(\bar{\nu} \sigma^{\mu\nu}P_L e\right)\left(\bar{d} \sigma_{\mu \nu} P_L u\right) \,,
\end{equation}
which, up to hermitian conjugation and flavor labels, belongs to the charged-current WET description of non-standard semileptonic $d\to u\, e\,\bar\nu$ transitions. Such tensor interactions are probed, for instance, in neutron and nuclear beta decays and in radiative pion decays and, at the level of UV completions, can arise after integrating out e.g. a scalar leptoquark. The renormalization-group evolution of the corresponding Wilson coefficient is therefore an essential ingredient in precision constraints. For the present illustrative purpose, we derive the two-loop running induced by QED in the WET$(5)$, i.e. with fixed numbers of up-type, down-type and lepton flavors $n_u=2$ and $n_d=n_e=3$.

Although QED is vector-like and therefore does not flip chiralities, the operator in Eq.~\eqref{eq:exampleADMop} mixes with the Fierz-conjugated scalar left--left operator, since scalar and tensor structures are related by four-dimensional Fierz identities. In what follows we focus only on the running of the operator in Eq.~\eqref{eq:exampleADMop} into itself. The running into the Fierz-conjugated scalar operator is obtained in the same way, by extracting instead the coefficient multiplying the scalar Dirac structure.

First, we collect from \cref{tab:WFR} the two-loop wave-function renormalization for each fermion field $\psi$ entering the operator. In the present case, three of the four external fermions carry a non-vanishing electromagnetic charge, while the two quark fields carry the usual $N=3$ color multiplicity in the QED fermion sums. Setting
\begin{equation}
    Q_\nu=0\,,\qquad Q_e=-1\,,\qquad Q_d=-\frac{1}{3}\,,\qquad Q_u=\frac{2}{3}\,,
\end{equation}
and using Eq.~\eqref{eq:QED2ADM}, the wave-function contribution to the two-loop ADM is
\begin{equation}
   \gamma^{(2,0)}_{\mathcal{Q} \to \mathcal{Q}}\supset -4Z_{\mathcal{Q}}^{(2,0;1)}
   = - \frac{203}{9}\,.
   \label{eq:expiece1}
\end{equation}
Next, we consider the contribution involving the product of the
one-loop physical pole and finite part. Both quantities are obtained
from \cref{tab:1L_TLL}, after multiplication by the corresponding
$U(1)$ charge factors in \cref{tab:chargefactorsU1}. For the
$\mathcal Q\to\mathcal Q$ entry, we find
\begin{equation}
   \gamma^{(2,0)}_{\mathcal Q\to\mathcal Q}
   \supset
   4
   a_{\mathcal Q\mathcal Q_k}^{(1,0;1)}
   a_{\mathcal Q_k\mathcal Q}^{(1,0;0)}
   =
   4 \left(0+2+1\right)\left(0-1-\frac{7}{6}\right)
   =-26\,.
   \label{eq:expiece2}
\end{equation}
Here, the three entries in the first parenthesis are the one-loop pole coefficients from diagrams 1, 2, and 3, while the three entries in the second parenthesis are the corresponding finite coefficients, after multiplication by the electromagnetic charge factors.

We also need the evanescent-to-physical contribution. The one-loop
physical-to-evanescent mixing,
$a_{\mathcal{Q}_i \mathcal{E}_k}^{(1,0;1)}$, is obtained from
\cref{tab:1L_TLL}, while the finite evanescent-to-physical mixing,
$a_{\mathcal{E}_k \mathcal{Q}_i}^{(1,0;0)}$, is extracted from
\cref{tab:E4}. Including again the relevant $U(1)$ charge factors,
we find
\begin{equation}
   \gamma^{(2,0)}_{\mathcal Q\to\mathcal Q}
   \supset
   2 a_{\mathcal Q\mathcal E_k}^{(1,0;1)}
     a_{\mathcal E_k\mathcal Q}^{(1,0;0)}
   =
   2\left(0-\frac16+\frac1{12}\right)
    \left(\frac{32}{9}+\frac{32}{3}+0\right)
   =
   -\frac{64}{27}\,.
   \label{eq:expiece3}
\end{equation}

Finally, we determine
$a_{\mathcal Q\mathcal Q}^{(2,0;1)}$, which contains both the genuine
two-loop poles and the one-loop dimension-four counterterm
insertions. The former are obtained from \cref{tab:2L_TLL}, with the
corresponding $U(1)$ charge factors from
\cref{tab:chargefactorsU1}, while the latter are obtained from
\cref{tab:CT_TLL}. The multiplicity of each genuine two-loop diagram
is already included in the $U(1)$ charge factor, which contains the
sum over the inequivalent permutations for a given topology. We find
\begin{align}
   \gamma^{(2,0)}_{\mathcal Q\to\mathcal Q}
   \supset
   &-4a_{\mathcal Q\mathcal Q}^{(2,0;1)}
   \nonumber \\
   &=-4\Bigg(
   0-\frac{4}{3}-\frac{7}{9}+\frac{8}{81}
   -\frac{4}{9}-\frac{1}{9}-\frac{10}{81}
   -\frac{91}{27}-\frac{5}{3}
   +0
   +\frac{65}{27}
   +\frac{35}{27}
   -\frac{8}{27}
   +0
   \nonumber\\
   &
   +\frac{4}{27}
   +0
   -\frac{2}{3}
   -\frac{10}{9}
   +0+0+0+0+0+0+0
   +\frac{80}{81}
   -\frac{320}{27}
   -\frac{80}{27}
   +\frac{716}{243}
   \Bigg)
   \nonumber\\
   &=
   \frac{16360}{243}\,.
   \label{eq:expiece4}
\end{align}
The first 28 terms in brackets are the contributions from the genuine
two-loop diagrams 4--31, while the last term is the contribution from the dimension-four counterterm
insertions. Combining \crefrange{eq:expiece1}{eq:expiece4}, we obtain
\begin{equation}
   \gamma^{(2,0)}_{\mathcal Q\to\mathcal Q}
   =
   \frac{3985}{243}\,,
\end{equation}
in agreement with the known result \cite{Aebischer:2025hsx}.

\section{Conclusions}
\label{sec:conclusion}

In this work, we have presented a complete diagram-by-diagram determination of the one- and two-loop ultraviolet poles generated by current--current insertions of
dimension-six four-fermion operators. In addition to the generic gauge-group results, we have derived the analogous pole structures for scalar exchange with generic renormalizable Yukawa and scalar interactions, including mixed gauge–scalar contributions. Besides the genuine loop divergences, our results include the required dimension-four and operator counterterms, wave-function contributions, finite evanescent insertions, diagram multiplicities, and the corresponding color, charge, and Yukawa factors.

The main result of this work is therefore a universal set of building blocks from which the LO and NLO current--current contributions to four-fermion anomalous dimension matrices can be determined. We provide the results both with the Dirac strings left unreduced and for the special case of adopting a specific operator basis and the NDR scheme with anticommuting $\gamma_5$, using a specified $\overline{\mathrm{MS}}$ evanescent scheme. Once the field content, operator basis, and gauge and flavor quantum numbers of a theory have been specified, the remaining steps consist only of Dirac projections and group-theoretical and flavor contractions, no additional two-loop integration is required. The unreduced formulation, together with the explicit evanescent contributions also permits translations to other physical and evanescent operator bases, or to other renormalization schemes, through the appropriate finite NLO transformations. This is particularly important for combining one-loop matching with two-loop running in a scheme-consistent manner and for resumming log-enhanced corrections.

Our results generalize the classic diagrammatic classification developed for weak effective Hamiltonians \cite{Buras:1989xd,Buras:1991jm,Buras:1992zv,Buras:1992tc,Buras:2000if} to generic Dirac structures, renormalization schemes, product gauge groups, and scalar interactions. As non-trivial checks, we have reconstructed the corresponding diagram-by-diagram current--current pole tables of the literature and reproduced the known two-loop current-current QCD and QED WET running of four-fermion operators \cite{Aebischer:2025hsx}.

Natural extensions of this work involve the inclusion of penguin topologies and of insertions of operator classes beyond dimension-six four-fermion interactions, in particular dimension-five dipole operators. Furthermore, the factorized organization developed here is well suited to automation, allowing complete anomalous-dimension matrices to be generated directly from the operator content and quantum numbers of a given theory. Finally, the one- and two-loop results collected in this work provide the necessary lower-loop input for future three-loop calculations, which can be computed according to similar algorithmic procedures as described in \cite{Chetyrkin:1997fm}, and may become relevant whenever high perturbative precision is required for instance in flavor- and CP-violating observables.


\acknowledgments

We would like to thank Luis Hourtz for several useful cross-checks of the results. 

\noindent
The work of J.A., S.D. and M.P. is supported by the Swiss National Science Foundation (SNSF) through grant TMSGI2-225951.

\noindent
P.M. acknowledges funding from the Spanish MCIN/AEI/10.13039/501100011033: grant PRE2022-103999 funded by MCIN/AEI/10.13039/501100011033 and by "ESF Investing in your future", grant CEX2019-000918-M through the “Unit of Excellence Mar\'ia de Maeztu 2020-2023” award to the Institute of Cosmos Sciences.

\noindent
J.V. acknowledges funding from grant 2021-SGR-249 (Generalitat de Catalunya), and from the Spanish MCIN/AEI/10.13039/501100011033 through the following grants: grant CNS2022-135262 funded by the “European Union NextGenerationEU/PRTR”, grant CEX2019-000918-M through the “Unit of Excellence Mar\'ia de Maeztu 2020-2023” award to the Institute of Cosmos Sciences, and grant PID2022-136224NB-C21.

\newpage
\appendix

\section{Generic pole structures from gauge interactions}\label{app:genericGauge}

In this appendix general four-fermion pole structures resulting from gauge interactions at the one and two loops are reported. The corresponding one-loop and two-loop diagrams are depicted in \cref{fig:oneloopgauge,fig:twoloopgauge}, respectively. Poles of mirrored diagrams are equal to the ones shown and are not reported explicitly.

\subsection{General one-loop structures}
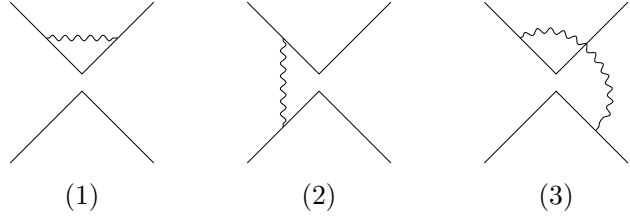
\begin{figure}[t]
\centering
\scalebox{0.95}{%
\begin{tikzpicture}[x=1cm,y=1cm]
\begin{scope}[shift={(0,0)}]      
  \VV
  \draw[ph] (-0.5,0.62) -- (0.5,0.62);
  \num{1}
\end{scope}
\begin{scope}[shift={(3.3,0)}]    
  \VV
  \draw[ph] (-0.5,0.62) -- (-0.5,-0.62);
  \num{2}
\end{scope}
\begin{scope}[shift={(6.6,0)}]    
  \VV
  \draw[ph] (-0.5,0.62) to[bend left=80,looseness=1.25] (0.55,-0.67);
  \num{3}
\end{scope}
\end{tikzpicture}%
}
\caption{One-loop diagrams for gauge interactions.}
\label{fig:oneloopgauge}
\end{figure}

The Dirac structures resulting from one-loop gauge interactions, depicted in \cref{fig:oneloopgauge}, are reported in \cref{tab:generic-1L-gluon}.

\begin{table}[!b]
\centering
\setlength{\tabcolsep}{7pt}
\renewcommand{\arraystretch}{1.05}
\begin{tabular}{@{}c c@{}}
\toprule[0.7mm]
Diagram & $1/\epsilon$ \\
\midrule[0.7mm]

1 &
$\displaystyle
\frac{1}{4}\,
(\gamma^\mu\gamma^\nu\Gamma_1\gamma_\nu\gamma_\mu\otimes\Gamma_2)$
\\ \midrule

2 &
$\displaystyle
-\frac{1}{4}\,
(\Gamma_1\gamma^\mu\gamma^\nu\otimes\Gamma_2\gamma_\mu\gamma_\nu)$
\\ \midrule

3 &
$\displaystyle
\frac{1}{4}\,
(\Gamma_1\gamma^\mu\gamma^\nu\otimes\gamma_\nu\gamma_\mu\Gamma_2)$
\\

\bottomrule[0.7mm]
\end{tabular}
\caption{One-loop divergences from gauge interactions for generic four-fermion insertions with Dirac structure $\Gamma_1\otimes\Gamma_2$.}
\label{tab:generic-1L-gluon}
\end{table}

\subsection{Counterterm insertions}
\label{sec:1LpolesCT}
In order to construct the two-loop ADM, the divergent parts of the one-loop counterterm insertions are required. They include contributions from fermion and gauge-boson wave-function counterterms, as well as from vertex and operator counterterms. In this subsection the resulting pole structures from these contributions are reported. When constructing the ADM they need to be multiplied by the appropriate color or charge factors, which coincide with the corresponding one-loop diagrams.

For generic renormalization constants $Z_X=1+\tilde \alpha_{\mathcal G}\delta Z_X$, gauge coupling constants $\tilde \alpha_{\mathcal G} = \frac{\alpha_{\mathcal G}}{4\pi}$ and gauge interaction generators $T_{R_\psi}^a$, one finds the following counterterm Feynman rules:
\begin{align}\label{eq:fermioCT}
\delta\Gamma_{\psi\psi}(p)
&=
i\tilde{\alpha}_{\mathcal G}\,
\delta Z_\psi\,\slashed p ,
\\
\delta\Gamma_{\bar\psi\psi A}^{a\mu}
&=
-i\tilde{\alpha}_{\mathcal G}g_{\mathcal G}\,
T_{R_\psi}^a\gamma^\mu\,\delta Z_V ,
\\
\delta\Gamma_{AA}^{ab,\mu\nu}(q)
&=
-i\tilde{\alpha}_{\mathcal G}\delta^{ab}
\left(q^2g^{\mu\nu}-q^\mu q^\nu\right)\delta Z_A
+i\tilde{\alpha}_{\mathcal G}\delta^{ab}
m^2g^{\mu\nu}\delta Z_A^{\rm mass}.
\end{align}
The last counterterm $\delta Z_A^{\rm mass}$ is an auxiliary mass counterterm associated with the infrared rearrangement used to extract UV poles \cite{Chetyrkin:1997fm}. At one loop, it is useful to decompose each renormalization constant as
\begin{equation}
\delta Z_X
=
\delta Z_X|_{\rm ab}
+
\delta Z_X|_{\rm n.ab}\,,
\end{equation}
where $\delta Z_X|_{\rm ab}$ denotes the abelian-like contribution, which is common to both abelian and non-abelian gauge interactions, while $\delta Z_X|_{\rm n.ab}$ denotes the non-abelian part. In Feynman gauge, the corresponding coefficients are summarized in \cref{tab:dim4-CT-decomposition}, while the group-theoretical factors ${\cal C}_\psi$ and ${\cal T}_{\mathcal G}$ are given in \cref{tab:dim4-CTs}.

\begin{table}[t]
\centering
\setlength{\tabcolsep}{5.5pt}
\renewcommand{\arraystretch}{1.25}
\resizebox{\linewidth}{!}{
\begin{tabular}{@{}l c c c@{}}
\toprule[0.7mm]
Counterterm
& Abelian-like part
& Non-abelian part
& Full non-abelian result
\\
\midrule[0.7mm]

Fermion kinetic term
&
$\displaystyle
\delta Z_\psi|_{\rm ab}
=
-\frac{{\cal C}_\psi}{\epsilon}$
&
$\displaystyle
\delta Z_\psi|_{\rm n.ab}=0$
&
$\displaystyle
\delta Z_\psi
=
-\frac{C_2(R_\psi)}{\epsilon}$
\\
\midrule

Gauge--fermion vertex
&
$\displaystyle
\delta Z_V|_{\rm ab}
=
-\frac{{\cal C}_\psi}{\epsilon}$
&
$\displaystyle
\delta Z_V|_{\rm n.ab}
=
-\frac{C_A}{\epsilon}$
&
$\displaystyle
\delta Z_V
=
-\frac{C_2(R_\psi)+C_A}{\epsilon}$
\\
\midrule

Gauge kinetic term
&
$\displaystyle
\delta Z_A|_{\rm ab}
=
-\frac{4{\cal T}_{\mathcal G}}{3\epsilon}$
&
$\displaystyle
\delta Z_A|_{\rm n.ab}
=
\frac{5C_A}{3\epsilon}$
&
$\displaystyle
\delta Z_A
=
\frac{5C_A-4N_T}{3\epsilon}$
\\
\midrule

Auxiliary gauge-boson mass
&
$\displaystyle
\delta Z_A^{\rm mass}|_{\rm ab}
=
-\frac{4{\cal T}_{\mathcal G}}{\epsilon}$
&
$\displaystyle
\delta Z_A^{\rm mass}|_{\rm n.ab}
=
-\frac{C_A}{\epsilon}$
&
$\displaystyle
\delta Z_A^{\rm mass}
=
-\frac{C_A+4N_T}{\epsilon}$
\\
\bottomrule[0.7mm]
\end{tabular}
}
\caption{One-loop dimension-four counterterms related to gauge interactions. The abelian-like terms are common to abelian and non-abelian gauge interactions, while the non-abelian terms vanish for abelian gauge groups. The $SU(N)$ specific factors are defined in \cref{eq:colorfactorsdef}.}
\label{tab:dim4-CT-decomposition}
\end{table}

\begin{table}[!b]
\centering
\setlength{\tabcolsep}{10pt}
\renewcommand{\arraystretch}{1.2}
\begin{tabular}{@{}l c c c@{}}
\toprule[0.7mm]
Gauge group
& Generator
& ${\cal C}_\psi$
& ${\cal T}_{\mathcal G}$
\\
\midrule[0.7mm]
Abelian
& $Q_\psi$
& $Q_\psi^2$
& $\displaystyle
   N_Q\equiv\sum_f d_f Q_f^2$
\\
\midrule
Non-abelian
& $T_{R_\psi}^a$
& $C_2(R_\psi)$
& $\displaystyle
   N_T\equiv\sum_f d_f T(R_f)$
\\
\bottomrule[0.7mm]
\end{tabular}
\caption{Group-theoretical factors relevant for the one-loop counterterms in \cref{tab:dim4-CT-decomposition}. The sum runs over dynamical fermions $f$ transforming in representations $R_f$ under the generic gauge group $\mathcal{G}$, while $d_f$ denotes their multiplicity factors under spectator gauge groups.}
\label{tab:dim4-CTs}
\end{table}

\subsubsection{Generic one-loop counterterm insertions }
In this subsection the counterterm insertions are reported for fermion contributions (\cref{tab:generic-fermion-CT-D1D3}), vertex contributions (\cref{tab:generic-vertex-CT-D1D3}), gauge boson contributions (\cref{tab:generic-photon-CT-D1D3}) and operator contributions (\cref{tab:1L_OPCT}).

\begin{table}[H]
\centering
\setlength{\tabcolsep}{6pt}
\renewcommand{\arraystretch}{0.45}
\begin{tabular}{@{}c c@{}}
\toprule[0.7mm]
Diagram
& $1/\epsilon^2$
\\
\midrule[0.7mm]

1 &
\makecell[l]{$\displaystyle
\begin{aligned}[t]
&-\frac{\delta Z_\psi}{24\epsilon}\,
(\gamma^\mu\gamma^\nu\Gamma_1
\gamma^\rho\gamma_\rho\gamma_\nu\gamma_\mu
\otimes\Gamma_2)
\\
&-\frac{\delta Z_\psi}{24\epsilon}\,
(\gamma^\mu\gamma^\nu\Gamma_1
\gamma^\rho\gamma_\nu\gamma_\rho\gamma_\mu
\otimes\Gamma_2)
\\
&-\frac{\delta Z_\psi}{24\epsilon}\,
(\gamma^\mu\gamma^\nu\Gamma_1
\gamma_\nu\gamma^\rho\gamma_\rho\gamma_\mu
\otimes\Gamma_2)
\end{aligned}$}

\\
\midrule

2 &
\makecell[l]{$\displaystyle
\begin{aligned}[t]
&\frac{\delta Z_\psi}{24\epsilon}\,
(\Gamma_1\gamma^\mu\gamma_\mu\gamma^\nu\gamma^\rho
\otimes
\Gamma_2\gamma_\nu\gamma_\rho)
\\
&+\frac{\delta Z_\psi}{24\epsilon}\,
(\Gamma_1\gamma^\mu\gamma^\nu\gamma_\mu\gamma^\rho
\otimes
\Gamma_2\gamma_\nu\gamma_\rho)
\\
&+\frac{\delta Z_\psi}{24\epsilon}\,
(\Gamma_1\gamma^\mu\gamma^\nu\gamma_\nu\gamma^\rho
\otimes
\Gamma_2\gamma_\mu\gamma_\rho)
\end{aligned}$}
\\
\midrule

3 &
\makecell[l]{$\displaystyle
\begin{aligned}[t]
&-\frac{\delta Z_\psi}{24\epsilon}\,
(\Gamma_1\gamma^\mu\gamma_\mu\gamma^\nu\gamma^\rho
\otimes
\gamma_\rho\gamma_\nu\Gamma_2)
\\
&-\frac{\delta Z_\psi}{24\epsilon}\,
(\Gamma_1\gamma^\mu\gamma^\nu\gamma_\mu\gamma^\rho
\otimes
\gamma_\rho\gamma_\nu\Gamma_2)
\\
&-\frac{\delta Z_\psi}{24\epsilon}\,
(\Gamma_1\gamma^\mu\gamma^\nu\gamma_\nu\gamma^\rho
\otimes
\gamma_\rho\gamma_\mu\Gamma_2)
\end{aligned}$}

\\

\bottomrule[0.7mm]
\end{tabular}
\caption{Generic Dirac structures for a fermion counterterm insertion in the one-loop topologies 1--3. The counterterm coefficient $\delta Z_\psi$ is defined in \cref{tab:dim4-CT-decomposition}.}
\label{tab:generic-fermion-CT-D1D3}
\end{table}

\begin{table}[H]
\centering
\setlength{\tabcolsep}{6pt}
\renewcommand{\arraystretch}{0.45}
\begin{tabular}{@{}c c@{}}
\toprule[0.7mm]
Diagram
& $1/\epsilon^2$
\\
\midrule[0.7mm]

1 &
\makecell[l]{$\displaystyle
\frac{\delta Z_V}{4\epsilon}\,
(\gamma^\mu\gamma^\nu\Gamma_1
\gamma_\nu\gamma_\mu
\otimes\Gamma_2)$}
\\
\midrule

2 &
\makecell[l]{$\displaystyle
-\frac{\delta Z_V}{4\epsilon}\,
(\Gamma_1\gamma^\mu\gamma^\nu
\otimes
\Gamma_2\gamma_\mu\gamma_\nu)$}
\\
\midrule

3 &
\makecell[l]{$\displaystyle
\frac{\delta Z_V}{4\epsilon}\,
(\Gamma_1\gamma^\mu\gamma^\nu
\otimes
\gamma_\nu\gamma_\mu\Gamma_2)$}
\\

\bottomrule[0.7mm]
\end{tabular}
\caption{Generic Dirac structures for a gauge--fermion vertex counterterm insertion in the one-loop topologies 1--3. The counterterm coefficient $\delta Z_V$ is defined in \cref{tab:dim4-CT-decomposition}.}
\label{tab:generic-vertex-CT-D1D3}
\end{table}

\begin{table}[H]
\centering
\setlength{\tabcolsep}{5pt}
\renewcommand{\arraystretch}{0.45}
\begin{tabular}{@{}c c c@{}}
\toprule[0.7mm]
Diagram
& $1/\epsilon^2$
& $1/\epsilon$
\\
\midrule[0.7mm]

1 &
\makecell[l]{$\displaystyle
\begin{aligned}[t]
&-\frac{5\,\delta Z_A}{24\epsilon}\,
(\gamma^\mu\gamma^\nu\Gamma_1
 \gamma_\nu\gamma_\mu\otimes\Gamma_2)
\\
&+\frac{\delta Z_A}{24\epsilon}\,
(\gamma^\mu\gamma_\mu\Gamma_1
 \gamma^\nu\gamma_\nu\otimes\Gamma_2)
\\
&+\frac{\delta Z_A}{24\epsilon}\,
(\gamma^\mu\gamma^\nu\Gamma_1
 \gamma_\mu\gamma_\nu\otimes\Gamma_2)
\end{aligned}$}
&
\makecell[l]{$\displaystyle
\begin{aligned}[t]
&\frac{\delta Z_A}{12}\,
(\gamma^\mu\gamma^\nu\Gamma_1
 \gamma_\nu\gamma_\mu\otimes\Gamma_2)
\\
&-\frac{\delta Z_{A}^{\rm mass}}{12}\,
(\gamma^\mu\gamma^\nu\Gamma_1
 \gamma_\nu\gamma_\mu\otimes\Gamma_2)
\end{aligned}$}
\\
\midrule

2 &
\makecell[l]{$\displaystyle
\begin{aligned}[t]
&\frac{5\,\delta Z_A}{24\epsilon}\,
(\Gamma_1\gamma^\mu\gamma^\nu
 \otimes\Gamma_2\gamma_\mu\gamma_\nu)
\\
&-\frac{\delta Z_A}{24\epsilon}\,
(\Gamma_1\gamma^\mu\gamma_\mu
 \otimes\Gamma_2\gamma^\nu\gamma_\nu)
\\
&-\frac{\delta Z_A}{24\epsilon}\,
(\Gamma_1\gamma^\mu\gamma^\nu
 \otimes\Gamma_2\gamma_\nu\gamma_\mu)
\end{aligned}$}
&
\makecell[l]{$\displaystyle
\begin{aligned}[t]
&-\frac{\delta Z_A}{12}\,
(\Gamma_1\gamma^\mu\gamma^\nu
 \otimes\Gamma_2\gamma_\mu\gamma_\nu)
\\
&+\frac{\delta Z_{A}^{\rm mass}}{12}\,
(\Gamma_1\gamma^\mu\gamma^\nu
 \otimes\Gamma_2\gamma_\mu\gamma_\nu)
\end{aligned}$}
\\
\midrule

3 &
\makecell[l]{$\displaystyle
\begin{aligned}[t]
&-\frac{5\,\delta Z_A}{24\epsilon}\,
(\Gamma_1\gamma^\mu\gamma^\nu
 \otimes\gamma_\nu\gamma_\mu\Gamma_2)
\\
&+\frac{\delta Z_A}{24\epsilon}\,
(\Gamma_1\gamma^\mu\gamma_\mu
 \otimes\gamma^\nu\gamma_\nu\Gamma_2)
\\
&+\frac{\delta Z_A}{24\epsilon}\,
(\Gamma_1\gamma^\mu\gamma^\nu
 \otimes\gamma_\mu\gamma_\nu\Gamma_2)
\end{aligned}$}
&
\makecell[l]{$\displaystyle
\begin{aligned}[t]
&\frac{\delta Z_A}{12}\,
(\Gamma_1\gamma^\mu\gamma^\nu
 \otimes\gamma_\nu\gamma_\mu\Gamma_2)
\\
&-\frac{\delta Z_{A}^{\rm mass}}{12}\,
(\Gamma_1\gamma^\mu\gamma^\nu
 \otimes\gamma_\nu\gamma_\mu\Gamma_2)
\end{aligned}$}
\\

\bottomrule[0.7mm]
\end{tabular}
\caption{Generic Dirac structures for the full gauge-boson
propagator counterterm insertion in the one-loop topologies 1--3. The counterterm coefficients $\delta Z_{A}$ and $\delta Z_{A}^{\rm mass}$ are defined in \cref{tab:dim4-CT-decomposition}.}
\label{tab:generic-photon-CT-D1D3}
\end{table}

\begin{table}[H]
\centering
\setlength{\tabcolsep}{5pt}
\renewcommand{\arraystretch}{0.45}
\begin{tabular}{@{}c c @{}}
\toprule[0.7mm]
Diagram & $1/\epsilon^2$   \\
\midrule[0.7mm] 
1 & $\frac{\delta Z_\mathrm{Op}}{4\epsilon}(\gamma^\mu \gamma^\nu \Gamma_1 \gamma_\nu \gamma_\mu \otimes \Gamma_2)$ 
\\ 
\midrule 
2 & $-\frac{\delta Z_\mathrm{Op}}{4\epsilon}( \Gamma_1 \gamma^\mu \gamma^\nu \otimes \Gamma_2 \gamma_\mu \gamma_\nu)$  
\\ 
\midrule 
3 & $\frac{\delta Z_\mathrm{Op}}{4\epsilon}(\Gamma_1 \gamma^\mu \gamma^\nu \otimes \gamma_\nu \gamma_\mu \Gamma_2)$
\\ 
\bottomrule[0.7mm]
\end{tabular}
\caption{General operator counterterm insertion for a four-fermion operator of the form
$\Gamma_1 \otimes \Gamma_2$. For a given operator, $\delta Z_\mathrm{Op}$ denotes the corresponding one-loop operator counterterm.}
\label{tab:1L_OPCT}
\end{table}

\subsubsection{Counterterm insertions for fermion wave-function renormalization}

In order to obtain the two-loop wave-function renormalization constant of the fermion, the one-loop counterterm insertions of the fermion two-point function are needed. They are given in \cref{tab:wfr-open-ct}.

\begin{table}[H]
\centering
\setlength{\tabcolsep}{5pt}
\renewcommand{\arraystretch}{0.72}
\begin{tabular}{@{}c c c@{}}
\toprule[0.7mm]
Counterterm insertion
& $1/\epsilon^2$
& $1/\epsilon$
\\
\midrule[0.7mm]

Fermion-propagator CT
&
\makecell[l]{$\displaystyle
\begin{aligned}[t]
&\frac{\delta Z_\psi}{12\epsilon}\,
\gamma^\mu\gamma^\nu\gamma_\nu
\gamma^\lambda\gamma_\mu
\\
&+\frac{\delta Z_\psi}{12\epsilon}\,
\gamma^\mu\gamma^\nu\gamma^\lambda
\gamma_\nu\gamma_\mu
\\
&+\frac{\delta Z_\psi}{12\epsilon}\,
\gamma^\mu\gamma^\lambda\gamma^\nu
\gamma_\nu\gamma_\mu
\end{aligned}$}
&
\makecell[l]{$\displaystyle 0$}
\\
\midrule

Gauge--fermion vertex CT
&
\makecell[l]{$\displaystyle
-\frac{\delta Z_V}{2\epsilon}\,
\gamma^\mu\gamma^\lambda\gamma_\mu$}
&
\makecell[l]{$\displaystyle 0$}
\\
\midrule

Gauge kinetic CT
&
\makecell[l]{$\displaystyle
\begin{aligned}[t]
&\frac{\delta Z_A}{12\epsilon}\,
\gamma^\mu\gamma_\mu\gamma^\lambda
\\
&+\frac{\delta Z_A}{12\epsilon}\,
\gamma^\lambda\gamma^\mu\gamma_\mu
\\
&+\frac{\delta Z_A}{3\epsilon}\,
\gamma^\mu\gamma^\lambda\gamma_\mu
\end{aligned}$}
&
\makecell[l]{$\displaystyle
-\frac{\delta Z_A}{3}\,
\gamma^\mu\gamma^\lambda\gamma_\mu$}
\\
\midrule

Auxiliary gauge-mass CT
&
\makecell[l]{$\displaystyle 0$}
&
\makecell[l]{$\displaystyle
\frac{\delta Z_A^{\rm mass}}{3}\,
\gamma^\mu\gamma^\lambda\gamma_\mu$}
\\

\bottomrule[0.7mm]
\end{tabular}
\caption{Generic Dirac structures of the one-loop fermion self-energy diagrams with dimension-four counterterm insertions.
The counterterm coefficients $\delta Z_X$ are defined in \cref{tab:dim4-CT-decomposition}. A common
overall factor $C_2(R_\psi)$ is removed. The index $\lambda$ denotes the Lorentz index carried by the external momentum.}
\label{tab:wfr-open-ct}
\end{table}

\subsection{Two-loop fermion wave-function renormalization}

The wave-function renormalization of fermion fields provides a contribution to the renormalization of four-fermion operators. We report in \cref{tab:wfr-open-twogauge} the generic Dirac structures of fermion two-point functions at the one- and two-loop level.

\begin{table}[H]
\centering
\setlength{\tabcolsep}{5pt}
\renewcommand{\arraystretch}{0.55}
\begin{tabular}{@{}c c c@{}}
\toprule[0.7mm]
Diagram & $1/\epsilon^2$ & $1/\epsilon$ \\
\midrule[0.7mm]

One-loop
&
\makecell[l]{$\displaystyle \hspace{1cm}-$}
&
\makecell[l]{$\displaystyle
-\frac{1}{2}\,
\gamma^\mu\gamma^\lambda\gamma_\mu$}
\\
\midrule

Nested
&
\makecell[l]{$\displaystyle\small
\begin{aligned}[t]
&-\frac{1}{48}\,
\gamma^\mu\gamma^\nu\gamma^\rho\gamma^\lambda
\gamma_\rho\gamma_\nu\gamma_\mu
\\
&-\frac{1}{48}\,
\gamma^\mu\gamma^\lambda\gamma^\nu\gamma^\rho
\gamma_\nu\gamma_\rho\gamma_\mu
\\
&-\frac{1}{48}\,
\gamma^\mu\gamma^\nu\gamma^\rho
\gamma_\nu\gamma_\rho\gamma^\lambda\gamma_\mu
\end{aligned}$}
&
\makecell[l]{$\displaystyle\small
\begin{aligned}[t]
&-\frac{1}{288}\,
\gamma^\mu\gamma^\nu\gamma^\rho\gamma^\lambda
\gamma_\rho\gamma_\nu\gamma_\mu
\\
&-\frac{1}{288}\,
\gamma^\mu\gamma^\lambda\gamma^\nu\gamma^\rho
\gamma_\nu\gamma_\rho\gamma_\mu
\\
&-\frac{1}{288}\,
\gamma^\mu\gamma^\nu\gamma^\rho
\gamma_\nu\gamma_\rho\gamma^\lambda\gamma_\mu
\end{aligned}$}
\\
\midrule

Crossed
&
\makecell[l]{$\displaystyle\small
\begin{aligned}[t]
&-\frac{1}{16}\,
\gamma^\mu\gamma^\lambda\gamma^\nu\gamma^\rho
\gamma_\mu\gamma_\rho\gamma_\nu
\\
&-\frac{1}{16}\,
\gamma^\mu\gamma^\nu\gamma^\rho
\gamma_\nu\gamma_\mu\gamma^\lambda\gamma_\rho
\end{aligned}$}
&
\makecell[l]{$\displaystyle\small
\begin{aligned}[t]
&-\frac{1}{32}\,
\gamma^\mu\gamma^\lambda\gamma^\nu\gamma^\rho
\gamma_\mu\gamma_\rho\gamma_\nu
\\
&-\frac{1}{32}\,
\gamma^\mu\gamma^\nu\gamma^\rho
\gamma_\nu\gamma_\mu\gamma^\lambda\gamma_\rho
\end{aligned}$}
\\
\midrule

Triple-gluon
&
\makecell[l]{$\displaystyle
\begin{aligned}[t]
&-\frac{1}{16}\,
\gamma^\mu\gamma^\lambda\gamma_\mu\gamma^\nu\gamma_\nu
\\
&+\frac{1}{8}\,
\gamma^\mu\gamma^\nu\gamma_\mu\gamma^\lambda\gamma_\nu
\\
&+\frac{1}{8}\,
\gamma^\mu\gamma^\lambda\gamma^\nu\gamma_\mu\gamma_\nu
\\
&-\frac{1}{16}\,
\gamma^\mu\gamma_\mu\gamma^\nu\gamma^\lambda\gamma_\nu
\\
&-\frac{1}{16}\,
\gamma^\mu\gamma^\lambda\gamma^\nu\gamma_\nu\gamma_\mu
\\
&-\frac{1}{16}\,
\gamma^\mu\gamma^\nu\gamma_\nu\gamma^\lambda\gamma_\mu
\end{aligned}$}
&
\makecell[l]{$\displaystyle
\begin{aligned}[t]
&\frac{1}{32}\,
\gamma^\mu\gamma^\lambda\gamma_\mu\gamma^\nu\gamma_\nu
\\
&+\frac{1}{8}\,
\gamma^\mu\gamma^\nu\gamma_\mu\gamma^\lambda\gamma_\nu
\\
&-\frac{1}{16}\,
\gamma^\mu\gamma^\nu\gamma_\mu\gamma_\nu\gamma^\lambda
\\
&+\frac{1}{8}\,
\gamma^\mu\gamma^\lambda\gamma^\nu\gamma_\mu\gamma_\nu
\\
&+\frac{1}{32}\,
\gamma^\mu\gamma_\mu\gamma^\nu\gamma^\lambda\gamma_\nu
\\
&-\frac{1}{16}\,
\gamma^\lambda\gamma^\mu\gamma^\nu\gamma_\mu\gamma_\nu
\\
&-\frac{5}{32}\,
\gamma^\mu\gamma^\lambda\gamma^\nu\gamma_\nu\gamma_\mu
\\
&-\frac{5}{32}\,
\gamma^\mu\gamma^\nu\gamma_\nu\gamma^\lambda\gamma_\mu
\\
&+\frac{1}{8}\,
\gamma^\mu\gamma^\nu\gamma^\lambda\gamma_\nu\gamma_\mu
\end{aligned}$}
\\ \midrule
Fermion bubble
&
\makecell[l]{$\displaystyle\small
\begin{aligned}[t]
&\frac{2}{9}N_T\,
\gamma^\mu\gamma^\lambda\gamma_\mu
\\
&+\frac{1}{18}N_T\,
\gamma^\lambda\gamma^\mu\gamma_\mu
\\
&+\frac{1}{18}N_T\,
\gamma^\mu\gamma_\mu\gamma^\lambda
\end{aligned}$}
&
\makecell[l]{$\displaystyle\small
\begin{aligned}[t]
&\frac{67}{54}N_T\,
\gamma^\mu\gamma^\lambda\gamma_\mu
\\
&-\frac{1}{108}N_T\,
\gamma^\lambda\gamma^\mu\gamma_\mu
\\
&-\frac{1}{108}N_T\,
\gamma^\mu\gamma_\mu\gamma^\lambda
\end{aligned}$}
\\
\midrule

Ghost bubble
&
\makecell[l]{$\displaystyle\small
\begin{aligned}[t]
&\frac{5}{144}\,
\gamma^\mu\gamma^\lambda\gamma_\mu
\\
&-\frac{1}{144}\,
\gamma^\lambda\gamma^\mu\gamma_\mu
\\
&-\frac{1}{144}\,
\gamma^\mu\gamma_\mu\gamma^\lambda
\end{aligned}$}
&
\makecell[l]{$\displaystyle\small
\begin{aligned}[t]
&\frac{163}{864}\,
\gamma^\mu\gamma^\lambda\gamma_\mu
\\
&+\frac{1}{864}\,
\gamma^\lambda\gamma^\mu\gamma_\mu
\\
&+\frac{1}{864}\,
\gamma^\mu\gamma_\mu\gamma^\lambda
\end{aligned}$}
\\
\midrule

Gauge-boson bubble
&
\makecell[l]{$\displaystyle\small
\begin{aligned}[t]
&-\frac{35}{144}\,
\gamma^\mu\gamma^\lambda\gamma_\mu
\\
&-\frac{11}{144}\,
\gamma^\lambda\gamma^\mu\gamma_\mu
\\
&-\frac{11}{144}\,
\gamma^\mu\gamma_\mu\gamma^\lambda
\end{aligned}$}
&
\makecell[l]{$\displaystyle\small
\begin{aligned}[t]
&\frac{1319}{864}\,
\gamma^\mu\gamma^\lambda\gamma_\mu
\\
&-\frac{13}{864}\,
\gamma^\lambda\gamma^\mu\gamma_\mu
\\
&-\frac{13}{864}\,
\gamma^\mu\gamma_\mu\gamma^\lambda
\end{aligned}$}
\\
\midrule

Gauge tadpole
&
\makecell[l]{$\displaystyle 0$}
&
\makecell[l]{$\displaystyle
\,-\gamma^\mu\gamma^\lambda\gamma_\mu$}
\\
\bottomrule[0.7mm]
\end{tabular}
\caption{Generic Dirac structures of the one-loop and two-loop wave-function renormalization diagrams. The index $\lambda$ denotes the external-momentum index.}
\label{tab:wfr-open-twogauge}
\end{table}

Combining the bare contributions in
\cref{tab:wfr-open-twogauge} with the corresponding color factors and with the dimension-four counterterm insertions given in \cref{tab:wfr-open-ct}, and subsequently reducing the Dirac strings using the NDR scheme, one obtains the fermion wave-function renormalization constants reported in \cref{tab:WFR}. For a single $SU(N)$ gauge factor and a fermion in the
fundamental representation, the simple-pole coefficients are
\begin{equation}
Z_\psi^{(1,0;1)}
=
-C_F,
\qquad
Z_\psi^{(2,0;1)}
=
C_F
\left[
\frac{3}{4}C_F
-\frac{17}{4}C_A
+N_T
\right]\,,
\label{eq:wfr-buras-check}
\end{equation}
which agree with the result of Buras and Weisz~\cite{Buras:1989xd}. The generalization of these results involving combinations of different $SU(N)$ and $U(1)$ factors is reported in \cref{tab:WFR} and agrees with the literature results \cite{Machacek:1983tz}.

\begin{table}[t]
\centering
\small
\setlength{\tabcolsep}{5.2pt}
\renewcommand{\arraystretch}{1.15}
\begin{tabular}{l c c c}
\toprule[0.7mm]
Gauge group 
& $Z_\psi^{(1,0 ; 1)}$ 
& $Z_\psi^{(1,1 ; 1)}$ 
& $Z_\psi^{(2,0 ; 1)}$ \\[1mm]
\toprule[0.7mm]
$SU(N_1) \times SU(N_2)$  
&
$-C_2^{[1]}(R_\psi)$
&
$\displaystyle \frac{3}{2}\,
C_2^{[1]}(R_\psi)\,C_2^{[2]}(R_\psi)$
&
$\displaystyle
C_2^{[1]}(R_\psi)
\left[
\frac{3}{4}C_2^{[1]}(R_\psi)
-\frac{17}{4}C_A^{[1]}
+N_T^{[1]}
\right]$
\\[2mm]
\midrule
$U(1)_1 \times U(1)_2$  
&
$-\left(Q_\psi^{[1]}\right)^2$
&
$\displaystyle
\frac{3}{2}
\left(Q_\psi^{[1]}\right)^2
\left(Q_\psi^{[2]}\right)^2$
&
$\displaystyle
\left(Q_\psi^{[1]}\right)^2
\left[
\frac{3}{4}\left(Q_\psi^{[1]}\right)^2
+N_Q^{[1]}
\right]$
\\[2mm]
\midrule
$SU(N_1) \times U(1)_2$  
&
$*$
&
$\displaystyle
\frac{3}{2}\,
C_2(R_\psi)\left(Q_\psi^{[2]}\right)^2$
&
$\displaystyle
*$
\\[1.5mm]
\bottomrule[0.7mm]
\end{tabular}
\caption{Wave-function renormalization constants for different gauge groups. For the first two rows, the corresponding pure-gauge coefficients for the second gauge factor are obtained by the replacement $[1]\leftrightarrow[2]$. The asterisk ($*$) entries follow from the corresponding $SU(N)$ and $U(1)$ expressions in the previous rows, depending on the gauge factor considered.}
\label{tab:WFR}
\end{table}

\subsection{General two-loop structures}
\label{sec:2Lpolesgeneric}
\begin{figure}[H]
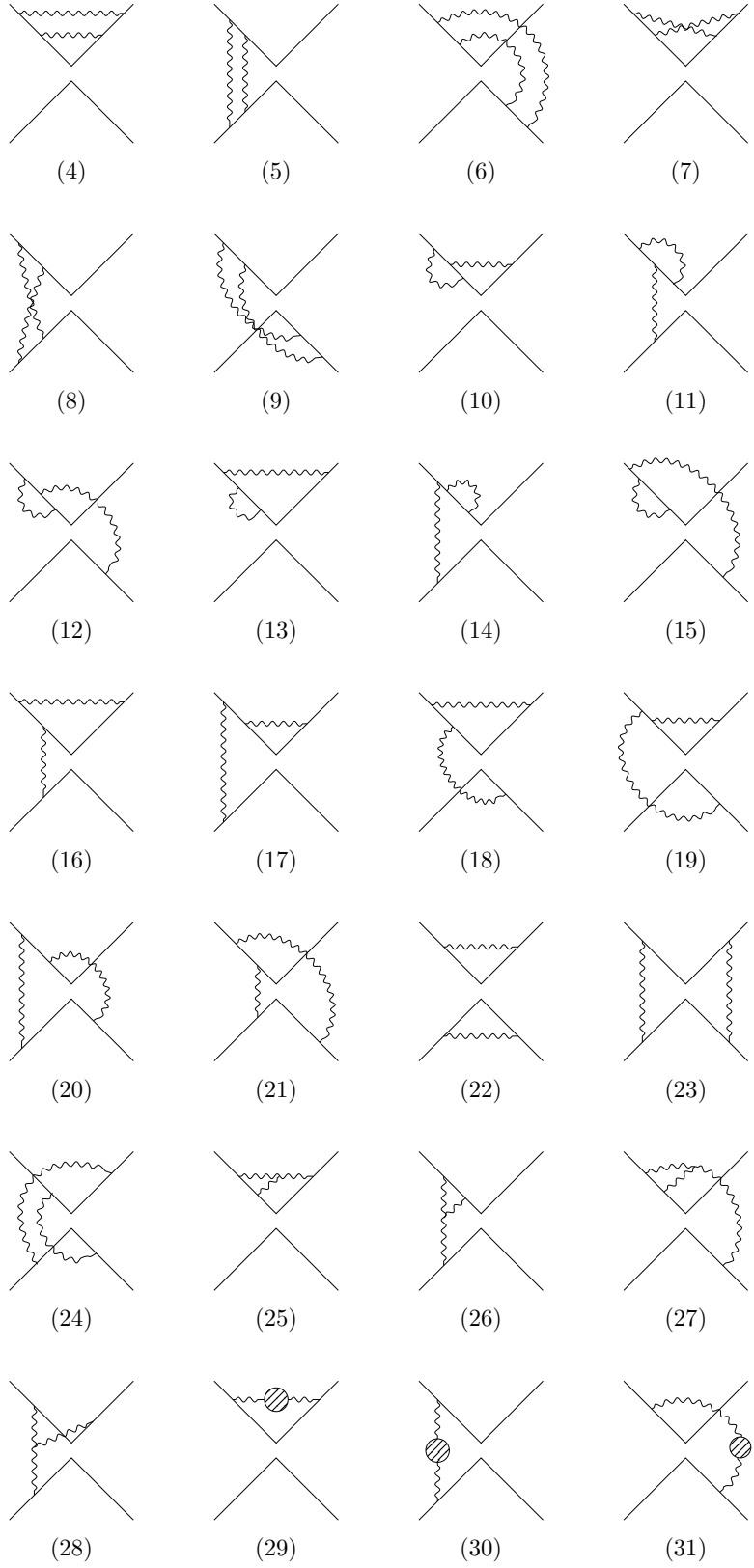

\centering
\scalebox{0.85}{%
%
}
\caption{Two-loop diagrams from gauge interactions.}
\label{fig:twoloopgauge}
\end{figure}

In this subsection we report the divergent parts of the two-loop diagrams in \cref{fig:twoloopgauge} for generic insertions of four-fermion operators with Dirac structure $\Gamma_1\otimes\Gamma_2$. 
Both the double and single poles, proportional to $1/\epsilon^2$ and $1/\epsilon$, are given explicitly. At this stage no evanescent-operator prescription is assumed for the Dirac structures. 
We therefore keep the Dirac strings in the form in which they arise from the loop calculation, without performing any projection, contraction, or reduction onto a physical operator basis. The results below can then be combined with any chosen evanescent prescription to obtain the corresponding two-loop anomalous-dimension matrix.

\crefrange{tab:genericpart1}{tab:genericpart3} contain the poles of diagrams 4--24. The divergences of diagrams corresponding to non-abelian interactions are collected in~\crefrange{tab:genericpart4}{tab:generic31}.

\subsubsection{Genuine two-loop abelian structures}

\begin{table}[H]
\centering
\small
\setlength{\tabcolsep}{6pt}
\renewcommand{\arraystretch}{0.3} 
\begin{tabular}{@{}c c c@{}}
\toprule[0.7mm]
Diagram & $1/\epsilon^2$ & $1/\epsilon$ \\
\midrule[0.7mm]

4 &
\makecell[l]{$\displaystyle\small
\begin{aligned}[t]
&\frac{1}{32}\,
(\gamma^\mu\gamma^\nu\gamma^\rho\gamma^\sigma\Gamma_1
\gamma_\sigma\gamma_\rho\gamma_\nu\gamma_\mu\otimes\Gamma_2)
\end{aligned}$}
&
\makecell[l]{$\displaystyle\small
\begin{aligned}[t]
&\frac{5}{192}\,
(\gamma^\mu\gamma^\nu\gamma^\rho\gamma^\sigma\Gamma_1
\gamma_\sigma\gamma_\rho\gamma_\nu\gamma_\mu\otimes\Gamma_2)\\
&+\frac{1}{96}\,
(\gamma^\mu\gamma^\nu\gamma^\rho\gamma^\sigma\Gamma_1
\gamma_\nu\gamma_\rho\gamma_\sigma\gamma_\mu\otimes\Gamma_2)\\
&+\frac{1}{96}\,
(\gamma^\mu\gamma^\nu\gamma^\rho\gamma_\nu\Gamma_1
\gamma^\sigma\gamma_\rho\gamma_\sigma\gamma_\mu\otimes\Gamma_2)
\end{aligned}$}
\\ \midrule

5 &
\makecell[l]{$\displaystyle\small
\begin{aligned}[t]
&\frac{1}{32}\,
(\Gamma_1\gamma^\mu\gamma^\nu\gamma^\rho\gamma^\sigma
\otimes\Gamma_2\gamma_\mu\gamma_\nu\gamma_\rho\gamma_\sigma)
\end{aligned}$}
&
\makecell[l]{$\displaystyle\small
\begin{aligned}[t]
&\frac{5}{192}\,
(\Gamma_1\gamma^\mu\gamma^\nu\gamma^\rho\gamma^\sigma
\otimes\Gamma_2\gamma_\mu\gamma_\nu\gamma_\rho\gamma_\sigma)\\
&+\frac{1}{96}\,
(\Gamma_1\gamma^\mu\gamma^\nu\gamma^\rho\gamma^\sigma
\otimes\Gamma_2\gamma_\rho\gamma_\nu\gamma_\mu\gamma_\sigma)\\
&+\frac{1}{96}\,
(\Gamma_1\gamma^\mu\gamma^\nu\gamma_\mu\gamma^\rho
\otimes\Gamma_2\gamma^\sigma\gamma_\nu\gamma_\sigma\gamma_\rho)
\end{aligned}$}
\\
\end{tabular}
\caption{Divergences of the two-loop diagrams 4--10 for a generic Dirac structure $\Gamma_1\otimes\Gamma_2$.}
\label{tab:genericpart1}
\end{table}

\begin{table}[H]
\centering
\small
\setlength{\tabcolsep}{8pt}
\renewcommand{\arraystretch}{0.9} 
\begin{tabular}{@{}c c c@{}}
\toprule[0.7mm]
Diagram & $1/\epsilon^2$ & $1/\epsilon$ \\
\midrule[0.7mm]
11 &
\makecell[l]{$\displaystyle\small
\begin{aligned}[t]
&-\frac{1}{32}\,
(\Gamma_1\gamma^\mu\gamma^\nu\gamma^\rho\gamma^\sigma\gamma_\rho\gamma_\nu\otimes\Gamma_2\gamma_\mu\gamma_\sigma)
\end{aligned}$}
&
\makecell[l]{$\displaystyle\small
\begin{aligned}[t]
&-\frac{1}{192}\,
(\Gamma_1\gamma^\mu\gamma^\nu\gamma^\rho\gamma^\sigma\gamma_\rho\gamma_\nu\otimes\Gamma_2\gamma_\mu\gamma_\sigma)\\
&+\frac{1}{96}\,
(\Gamma_1\gamma^\mu\gamma^\nu\gamma^\rho\gamma^\sigma\gamma_\mu\gamma_\nu\otimes\Gamma_2\gamma_\rho\gamma_\sigma)\\
&+\frac{1}{96}\,
(\Gamma_1\gamma^\mu\gamma^\nu\gamma_\mu\gamma^\rho\gamma^\sigma\gamma_\nu\otimes\Gamma_2\gamma_\sigma\gamma_\rho)
\end{aligned}$}
\\ \midrule
12 &
\makecell[l]{$\displaystyle\small
\begin{aligned}[t]
&\frac{1}{32}\,
(\Gamma_1\gamma^\mu\gamma^\nu\gamma^\rho\gamma^\sigma\gamma_\rho\gamma_\nu\otimes\gamma_\sigma\gamma_\mu\Gamma_2)
\end{aligned}$}
&
\makecell[l]{$\displaystyle\small
\begin{aligned}[t]
&\frac{1}{192}\,
(\Gamma_1\gamma^\mu\gamma^\nu\gamma^\rho\gamma^\sigma\gamma_\rho\gamma_\nu\otimes\gamma_\sigma\gamma_\mu\Gamma_2)\\
&-\frac{1}{96}\,
(\Gamma_1\gamma^\mu\gamma^\nu\gamma^\rho\gamma^\sigma\gamma_\mu\gamma_\nu\otimes\gamma_\sigma\gamma_\rho\Gamma_2)\\
&-\frac{1}{96}\,
(\Gamma_1\gamma^\mu\gamma^\nu\gamma_\mu\gamma^\rho\gamma^\sigma\gamma_\nu\otimes\gamma_\rho\gamma_\sigma\Gamma_2)
\end{aligned}$}
\\
\end{tabular}
\caption{Divergences of the two-loop diagrams 11--18 for a generic Dirac structure $\Gamma_1\otimes\Gamma_2$.}
\label{tab:genericpart2}
\end{table}

\begin{table}[H]
\centering
\setlength{\tabcolsep}{2pt}
\renewcommand{\arraystretch}{0.6} 
\begin{tabular}{@{}c c c@{}}
\toprule[0.7mm]
Diagram & $1/\epsilon^2$ & $1/\epsilon$ \\
\midrule[0.7mm]
19 &
\makecell[l]{$\displaystyle
\begin{aligned}[t]
&\frac{1}{32}\,
(\gamma^\mu\gamma^\nu\Gamma_1\gamma_\nu\gamma_\mu\gamma^\rho\gamma^\sigma\otimes\gamma_\sigma\gamma_\rho\Gamma_2)
\end{aligned}$}
&
\makecell[l]{$\displaystyle
\begin{aligned}[t]
&\frac{1}{192}\,
(\gamma^\mu\gamma^\nu\Gamma_1\gamma_\nu\gamma_\mu\gamma^\rho\gamma^\sigma\otimes\gamma_\sigma\gamma_\rho\Gamma_2)\\
&-\frac{1}{96}\,
(\gamma^\mu\gamma^\nu\Gamma_1\gamma^\rho\gamma_\mu\gamma_\nu\gamma^\sigma\otimes\gamma_\sigma\gamma_\rho\Gamma_2)\\
&-\frac{1}{96}\,
(\gamma^\mu\gamma^\nu\Gamma_1\gamma^\rho\gamma_\mu\gamma_\rho\gamma^\sigma\otimes\gamma_\sigma\gamma_\nu\Gamma_2)
\end{aligned}$}
\\ \midrule
20 &
\makecell[l]{$\displaystyle
\begin{aligned}[t]
&-\frac{1}{32}\,
(\Gamma_1\gamma^\mu\gamma^\nu\gamma^\rho\gamma^\sigma\otimes\gamma_\nu\gamma_\mu\Gamma_2\gamma_\rho\gamma_\sigma)
\end{aligned}$}
&
\makecell[l]{$\displaystyle
\begin{aligned}[t]
&-\frac{1}{192}\,
(\Gamma_1\gamma^\mu\gamma^\nu\gamma^\rho\gamma^\sigma\otimes\gamma_\nu\gamma_\mu\Gamma_2\gamma_\rho\gamma_\sigma)\\
&+\frac{1}{96}\,
(\Gamma_1\gamma^\mu\gamma^\nu\gamma^\rho\gamma^\sigma\otimes\gamma_\nu\gamma_\rho\Gamma_2\gamma_\mu\gamma_\sigma)\\
&+\frac{1}{96}\,
(\Gamma_1\gamma^\mu\gamma^\nu\gamma_\mu\gamma^\rho\otimes\gamma_\nu\gamma^\sigma\Gamma_2\gamma_\sigma\gamma_\rho)
\end{aligned}$}
\\
\end{tabular}
\caption{Divergences of the two-loop diagrams 19--24 for a generic Dirac structure $\Gamma_1\otimes\Gamma_2$.}
\label{tab:genericpart3}
\end{table}

\subsubsection{Two-loop three-gauge-boson contributions}
\hspace{-1cm}
\begin{table}[H]
\centering
\small
\setlength{\tabcolsep}{8pt}
\renewcommand{\arraystretch}{0.6} 
\resizebox{0.75\linewidth}{!}{
\begin{tabular}{@{}c c c@{}}
\toprule[0.7mm]
Diagram & $1/\epsilon^2$ & $1/\epsilon$ \\
\midrule[0.7mm]
25 &
\makecell[l]{$\displaystyle\small
\begin{aligned}[t]
&\frac{1}{32}\,
(\gamma^\mu\gamma^\nu\Gamma_1\gamma_\nu\gamma^\rho\gamma_\rho\gamma_\mu\otimes\Gamma_2)\\
&-\frac{1}{16}\,
(\gamma^\mu\gamma^\nu\Gamma_1\gamma_\nu\gamma^\rho\gamma_\mu\gamma_\rho\otimes\Gamma_2)\\
&+\frac{1}{32}\,
(\gamma^\mu\gamma^\nu\Gamma_1\gamma_\nu\gamma_\mu\gamma^\rho\gamma_\rho\otimes\Gamma_2)
\end{aligned}$}
&
\makecell[l]{$\displaystyle\small
\begin{aligned}[t]
&\frac{3}{64}\,
(\gamma^\mu\gamma^\nu\Gamma_1\gamma_\nu\gamma^\rho\gamma_\rho\gamma_\mu\otimes\Gamma_2)\\
&+\frac{1}{32}\,
(\gamma^\mu\gamma^\nu\Gamma_1\gamma^\rho\gamma_\nu\gamma_\rho\gamma_\mu\otimes\Gamma_2)\\
&+\frac{1}{32}\,
(\gamma^\mu\gamma^\nu\Gamma_1\gamma^\rho\gamma_\rho\gamma_\nu\gamma_\mu\otimes\Gamma_2)\\
&-\frac{1}{32}\,
(\gamma^\mu\gamma^\nu\Gamma_1\gamma_\nu\gamma^\rho\gamma_\mu\gamma_\rho\otimes\Gamma_2)\\
&-\frac{1}{64}\,
(\gamma^\mu\gamma^\nu\Gamma_1\gamma_\nu\gamma_\mu\gamma^\rho\gamma_\rho\otimes\Gamma_2)\\
&-\frac{1}{32}\,
(\gamma^\mu\gamma^\nu\Gamma_1\gamma^\rho\gamma_\mu\gamma_\rho\gamma_\nu\otimes\Gamma_2)\\
&-\frac{1}{32}\,
(\gamma^\mu\gamma^\nu\Gamma_1\gamma^\rho\gamma_\mu\gamma_\nu\gamma_\rho\otimes\Gamma_2)
\end{aligned}$}
\\
\end{tabular}
}
\caption{Divergences of the two-loop diagrams 25--28 for a generic Dirac structure $\Gamma_1\otimes\Gamma_2$.}
\label{tab:genericpart4}
\end{table}

\subsubsection{Two-loop self-energy contributions}

\begin{figure}[t]
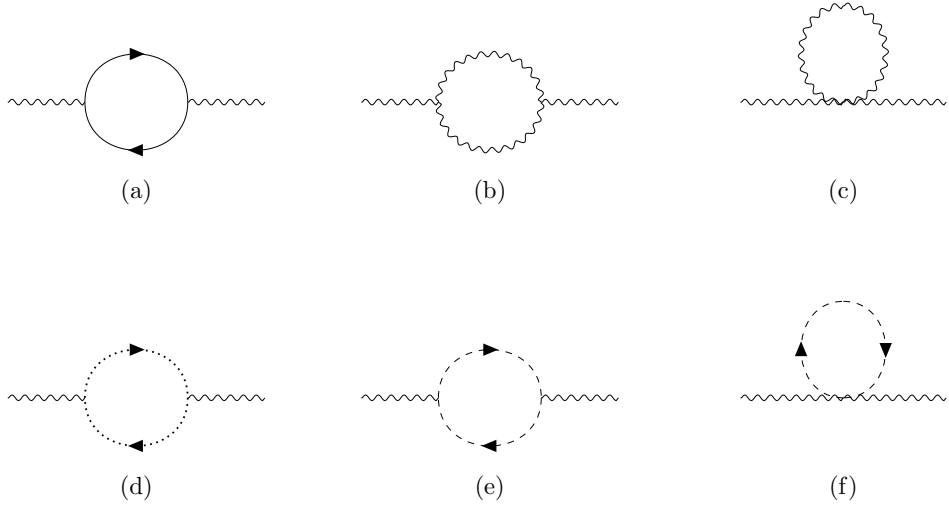

\centering
\scalebox{0.85}{%
%
%
}
\caption{One-loop contributions to the gauge-boson self-energy: (a) fermion
bubble, (b) gauge-boson bubble, (c) gauge-boson tadpole, (d) ghost bubble,
(e) scalar bubble and (f) scalar seagull.}
\label{fig:gluonSE}
\end{figure}

The two-loop self-energy diagrams contain six contributions related to the one-loop topologies shown in \cref{fig:gluonSE}. The divergences for the diagrams 29-31 are reported in~\crefrange{tab:generic29}{tab:generic31}.

\begin{table}[H]
\centering
\setlength{\tabcolsep}{8pt}
\renewcommand{\arraystretch}{0.9}
\begin{tabular}{@{}c c c@{}}
\toprule[0.7mm]
Diagram & $1/\epsilon^2$ & $1/\epsilon$ \\
\midrule[0.7mm]

29a &
\makecell[l]{$\displaystyle
\begin{aligned}[t]
&-N_T\frac{5}{36}\,
(\gamma^\mu\gamma^\nu\Gamma_1\gamma_\nu\gamma_\mu\otimes\Gamma_2)\\
&+N_T\frac{1}{36}\,
(\gamma^\mu\gamma^\nu\Gamma_1\gamma_\mu\gamma_\nu\otimes\Gamma_2)\\
&+N_T\frac{1}{36}\,
(\gamma^\mu\gamma_\mu\Gamma_1\gamma^\nu\gamma_\nu\otimes\Gamma_2)
\end{aligned}$}
&
\makecell[l]{$\displaystyle
\begin{aligned}[t]
&-N_T\frac{55}{216}\,
(\gamma^\mu\gamma^\nu\Gamma_1\gamma_\nu\gamma_\mu\otimes\Gamma_2)\\
&-N_T\frac{1}{216}\,
(\gamma^\mu\gamma^\nu\Gamma_1\gamma_\mu\gamma_\nu\otimes\Gamma_2)\\
&-N_T\frac{1}{216}\,
(\gamma^\mu\gamma_\mu\Gamma_1\gamma^\nu\gamma_\nu\otimes\Gamma_2)
\end{aligned}$}
\\ \midrule

29b &
\makecell[l]{$\displaystyle
\begin{aligned}[t]
&\frac{23}{144}\,
(\gamma^\mu\gamma^\nu\Gamma_1\gamma_\nu\gamma_\mu\otimes\Gamma_2)\\
&-\frac{11}{288}\,
(\gamma^\mu\gamma^\nu\Gamma_1\gamma_\mu\gamma_\nu\otimes\Gamma_2)\\
&-\frac{11}{288}\,
(\gamma^\mu\gamma_\mu\Gamma_1\gamma^\nu\gamma_\nu\otimes\Gamma_2)
\end{aligned}$}
&
\makecell[l]{$\displaystyle
\begin{aligned}[t]
&-\frac{353}{864}\,
(\gamma^\mu\gamma^\nu\Gamma_1\gamma_\nu\gamma_\mu\otimes\Gamma_2)\\
&-\frac{13}{1728}\,
(\gamma^\mu\gamma^\nu\Gamma_1\gamma_\mu\gamma_\nu\otimes\Gamma_2)\\
&-\frac{13}{1728}\,
(\gamma^\mu\gamma_\mu\Gamma_1\gamma^\nu\gamma_\nu\otimes\Gamma_2)
\end{aligned}$}
\\ \midrule

29c &
\makecell[l]{$\displaystyle
\begin{aligned}[t]
&\hspace{1.6cm} 0
\end{aligned}$}
&
\makecell[l]{$\displaystyle
\begin{aligned}[t]
&\frac{1}{4}\,
(\gamma^\mu\gamma^\nu\Gamma_1\gamma_\nu\gamma_\mu\otimes\Gamma_2)
\end{aligned}$}
\\
\end{tabular}
\caption{Divergences of the two-loop gauge-boson self-energy diagram 29 for a generic Dirac structure $\Gamma_1\otimes\Gamma_2$. The individual contributions from the fermion bubble, gauge-boson bubble, gauge-boson tadpole, ghost bubble, scalar bubble and scalar seagull are denoted by a, b, c, d, e and f, respectively.}
\label{tab:generic29}
\end{table}

\begin{table}[H]
\centering
\setlength{\tabcolsep}{8pt}
\renewcommand{\arraystretch}{0.9}
\begin{tabular}{@{}c c c@{}}
\toprule[0.7mm]
Diagram & $1/\epsilon^2$ & $1/\epsilon$ \\
\midrule[0.7mm]

30a &
\makecell[l]{$\displaystyle
\begin{aligned}[t]
&N_T\frac{5}{36}\,
(\Gamma_1\gamma^\mu\gamma^\nu\otimes\Gamma_2\gamma_\mu\gamma_\nu)\\
&-N_T\frac{1}{36}\,
(\Gamma_1\gamma^\mu\gamma^\nu\otimes\Gamma_2\gamma_\nu\gamma_\mu)\\
&-N_T\frac{1}{36}\,
(\Gamma_1\gamma^\mu\gamma_\mu\otimes\Gamma_2\gamma^\nu\gamma_\nu)
\end{aligned}$}
&
\makecell[l]{$\displaystyle
\begin{aligned}[t]
&N_T\frac{55}{216}\,
(\Gamma_1\gamma^\mu\gamma^\nu\otimes\Gamma_2\gamma_\mu\gamma_\nu)\\
&+N_T\frac{1}{216}\,
(\Gamma_1\gamma^\mu\gamma^\nu\otimes\Gamma_2\gamma_\nu\gamma_\mu)\\
&+N_T\frac{1}{216}\,
(\Gamma_1\gamma^\mu\gamma_\mu\otimes\Gamma_2\gamma^\nu\gamma_\nu)
\end{aligned}$}
\\ \midrule

30b &
\makecell[l]{$\displaystyle
\begin{aligned}[t]
&-\frac{23}{144}\,
(\Gamma_1\gamma^\mu\gamma^\nu\otimes\Gamma_2\gamma_\mu\gamma_\nu)\\
&+\frac{11}{288}\,
(\Gamma_1\gamma^\mu\gamma^\nu\otimes\Gamma_2\gamma_\nu\gamma_\mu)\\
&+\frac{11}{288}\,
(\Gamma_1\gamma^\mu\gamma_\mu\otimes\Gamma_2\gamma^\nu\gamma_\nu)
\end{aligned}$}
&
\makecell[l]{$\displaystyle
\begin{aligned}[t]
&+\frac{353}{864}\,
(\Gamma_1\gamma^\mu\gamma^\nu\otimes\Gamma_2\gamma_\mu\gamma_\nu)\\
&+\frac{13}{1728}\,
(\Gamma_1\gamma^\mu\gamma^\nu\otimes\Gamma_2\gamma_\nu\gamma_\mu)\\
&+\frac{13}{1728}\,
(\Gamma_1\gamma^\mu\gamma_\mu\otimes\Gamma_2\gamma^\nu\gamma_\nu)
\end{aligned}$}
\\ \midrule

30c &
\makecell[l]{$\displaystyle
\begin{aligned}[t]
&\hspace{1.6cm} 0
\end{aligned}$}
&
\makecell[l]{$\displaystyle
\begin{aligned}[t]
&-\frac{1}{4}\,
(\Gamma_1\gamma^\mu\gamma^\nu\otimes\Gamma_2\gamma_\mu\gamma_\nu)
\end{aligned}$}
\\
\end{tabular}
\caption{Divergences of the two-loop gauge-boson self-energy diagram 30 for a generic Dirac structure $\Gamma_1\otimes\Gamma_2$. The individual contributions from the fermion bubble, gauge-boson bubble, gauge-boson tadpole, ghost bubble, scalar bubble and scalar seagull are denoted by a, b, c, d, e and f, respectively.}
\label{tab:generic30}
\end{table}

\begin{table}[H]
\centering
\setlength{\tabcolsep}{8pt}
\renewcommand{\arraystretch}{0.9}
\begin{tabular}{@{}c c c@{}}
\toprule[0.7mm]
Diagram & $1/\epsilon^2$ & $1/\epsilon$ \\
\midrule[0.7mm]

31a &
\makecell[l]{$\displaystyle
\begin{aligned}[t]
&-N_T\frac{5}{36}\,
(\Gamma_1\gamma^\mu\gamma^\nu\otimes\gamma_\nu\gamma_\mu\Gamma_2)\\
&+N_T\frac{1}{36}\,
(\Gamma_1\gamma^\mu\gamma^\nu\otimes\gamma_\mu\gamma_\nu\Gamma_2)\\
&+N_T\frac{1}{36}\,
(\Gamma_1\gamma^\mu\gamma_\mu\otimes\gamma^\nu\gamma_\nu\Gamma_2)
\end{aligned}$}
&
\makecell[l]{$\displaystyle
\begin{aligned}[t]
&-N_T\frac{55}{216}\,
(\Gamma_1\gamma^\mu\gamma^\nu\otimes\gamma_\nu\gamma_\mu\Gamma_2)\\
&-N_T\frac{1}{216}\,
(\Gamma_1\gamma^\mu\gamma^\nu\otimes\gamma_\mu\gamma_\nu\Gamma_2)\\
&-N_T\frac{1}{216}\,
(\Gamma_1\gamma^\mu\gamma_\mu\otimes\gamma^\nu\gamma_\nu\Gamma_2)
\end{aligned}$}
\\ \midrule

31b &
\makecell[l]{$\displaystyle
\begin{aligned}[t]
&\frac{23}{144}\,
(\Gamma_1\gamma^\mu\gamma^\nu\otimes\gamma_\nu\gamma_\mu\Gamma_2)\\
&-\frac{11}{288}\,
(\Gamma_1\gamma^\mu\gamma^\nu\otimes\gamma_\mu\gamma_\nu\Gamma_2)\\
&-\frac{11}{288}\,
(\Gamma_1\gamma^\mu\gamma_\mu\otimes\gamma^\nu\gamma_\nu\Gamma_2)
\end{aligned}$}
&
\makecell[l]{$\displaystyle
\begin{aligned}[t]
&-\frac{353}{864}\,
(\Gamma_1\gamma^\mu\gamma^\nu\otimes\gamma_\nu\gamma_\mu\Gamma_2)\\
&-\frac{13}{1728}\,
(\Gamma_1\gamma^\mu\gamma^\nu\otimes\gamma_\mu\gamma_\nu\Gamma_2)\\
&-\frac{13}{1728}\,
(\Gamma_1\gamma^\mu\gamma_\mu\otimes\gamma^\nu\gamma_\nu\Gamma_2)
\end{aligned}$}
\\ \midrule

31c &
\makecell[l]{$\displaystyle
\begin{aligned}[t]
&\hspace{1.6cm} 0
\end{aligned}$}
&
\makecell[l]{$\displaystyle
\begin{aligned}[t]
&\frac{1}{4}\,
(\Gamma_1\gamma^\mu\gamma^\nu\otimes\gamma_\nu\gamma_\mu\Gamma_2)
\end{aligned}$}
\\
\end{tabular}
\caption{Divergences of the two-loop gauge-boson self-energy diagram 31 for a generic Dirac structure $\Gamma_1\otimes\Gamma_2$. The individual contributions from the fermion bubble, gauge-boson bubble, gauge-boson tadpole, ghost bubble, scalar bubble and scalar seagull are denoted by a, b, c, d, e and f, respectively.}
\label{tab:generic31}
\end{table}

\newpage
\section{Generic pole structures from scalar interactions}\label{app:genericscalar}

In this appendix we collect the generic one- and two-loop pole structures generated by
scalar exchange, keeping the scalar vertices and the Dirac structure
$\Gamma_1\otimes\Gamma_2$ generic. We denote the exchanged scalars generically by
$\phi_i$, where the index $i$ labels possible scalar flavors or gauge components. The special case of a real gauge-singlet scalar is recovered by taking the scalar index to be trivial, imposing $\phi^\dagger=\phi$, and setting all gauge generators to zero.

The relevant renormalizable interactions are written as
\begin{equation}\label{eq:scallag}
\begin{aligned}
\mathcal{L}_{\phi}
=&\,
(D_\mu\phi)_i^\dagger(D^\mu\phi)_i
-
m_\phi^2\,\phi_i^\dagger\phi_i
-\left(y^i_{kl}\,\bar f_k\gs_i f_l\,\phi_i+\mathrm{h.c.}\right)
\\
&\,
-\frac{1}{3!}\,\kappa_{ijk}\,\phi_i\phi_j\phi_k
-\frac{1}{4!}\,\lambda_{ijkl}\,\phi_i\phi_j\phi_k\phi_l\, ,
\end{aligned}
\end{equation}
and the covariant derivative is given by
\begin{equation}
(D_\mu\phi)_i
=
\left(\partial_\mu\delta_{ij}
+i g_A\,\theta^a_{ij} A_\mu^a\right)\phi_j \,,
\end{equation}
where $\theta^a$ denotes the generators acting on the scalar multiplet and $g_A$ is the generic gauge coupling constant. In our convention $\gs_i$ denotes the Dirac structure associated with
the scalar--fermion interaction, whereas the conjugate version is denoted by
\begin{equation}
\tilde{\Gamma}_{\phi_i}
\equiv
\gamma^0\gs_i^\dagger\gamma^0 \,.
\end{equation}
The resulting scalar--fermion Feynman rules read
\begin{equation}
\phi_i\,\bar f_k f_l:
\qquad
-i\,y^i_{kl}\gs_i\,,
\qquad
\phi_i^\dagger\,\bar f_l f_k:
\qquad
-i\,y^{i*}_{kl}\tilde{\Gamma}_{\phi_i} \,.
\end{equation}
Similarly, in the single-real-scalar limit,
\begin{equation}
\phi\phi\phi:\qquad -i\kappa\,,
\qquad
\phi\phi\phi\phi:\qquad -i\lambda \,.
\end{equation}

In the following subsections the pole tables corresponding to scalar interactions are reported. To increase readability all coupling constants, group-theory factors and flavor contractions associated with these vertices are not shown explicitly. Our convention for the displayed Dirac strings is that the first scalar vertex encountered when following a fermion line is denoted by $\gs$, corresponding to scalar flow entering the fermion line, while the opposite orientation is denoted by $\gst$, see \cref{fig:oneloopscalar,fig:twoloopscalar}. For the scalar-exchange topologies 4--24 this is only a relabelling convention. It becomes relevant for the mixed scalar--scalar--gauge topologies 25--28, where the relative scalar orientation fixes the sign and ordering of the Dirac strings displayed in the tables.

\subsection{General one-loop structures}
\begin{figure}[H]
\centering
\scalebox{0.85}{%
\begin{tikzpicture}[x=1cm,y=1cm]
\begin{scope}[shift={(0,0)}]      
  \VV
  \draw[scline, arrow at=0.6] (-0.5,0.62) -- (0.5,0.62);
  \num{1}
\end{scope}
\begin{scope}[shift={(3.3,0)}]    
  \VV                             
  \draw[scline, arrow at=0.58] (-0.5,-0.62) -- (-0.5,0.62);
  \num{2}
\end{scope}
\begin{scope}[shift={(6.6,0)}]    
  \VV                             
  \draw[scline, arrow at=0.38] (0.55,-0.67) to[bend right=80,looseness=1.25] (-0.5,0.62);
  \num{3}
\end{scope}
\end{tikzpicture}%
}
\caption{One-loop scalar corrections to four-fermi interactions.}
\label{fig:oneloopscalar}
\end{figure}
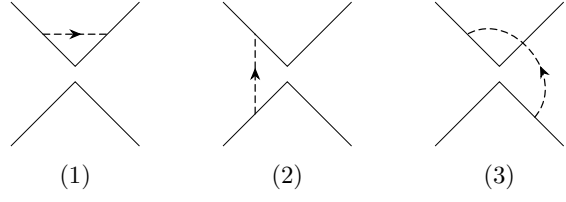

The one-loop scalar corrections to four-fermion interactions are depicted in \cref{fig:oneloopscalar} and the corresponding poles are reported in \cref{tab:generic-1L-scalar}. 

\begin{table}[b]
\centering
\setlength{\tabcolsep}{7pt}
\renewcommand{\arraystretch}{1.05}
\begin{tabular}{@{}c c@{}}
\toprule[0.7mm]
Diagram & $1/\epsilon$ \\
\midrule[0.7mm]

1 &
$\displaystyle
-\frac{1}{4}\,
(\gs\gamma^\mu\Gamma_1\gamma_\mu\gst\otimes\Gamma_2)$
\\ \midrule
2 &
$\displaystyle
\frac{1}{4}\,
(\Gamma_1\gamma^\mu\gs\otimes\Gamma_2\gamma_\mu\gst)$
\\ \midrule
3 &
$\displaystyle
-\frac{1}{4}\,
(\Gamma_1\gamma^\mu\gs\otimes\gst\gamma_\mu\Gamma_2)$
\\
\bottomrule[0.7mm]
\end{tabular}
\caption{One-loop divergences from scalar interactions for generic four-fermion insertions with Dirac structure $\Gamma_1\otimes\Gamma_2$.}
\label{tab:generic-1L-scalar}
\end{table}

\subsection{One-loop scalar renormalization}
For the renormalization of the scalar sector we introduce the following renormalization constants for the field, Yukawa coupling and scalar mass: 

\begin{equation}\label{eq:scalren}
\begin{aligned}
\phi^{(0)} &= \phi\,\sqrt{1+\Delta_\phi}\,, \\
y^{(0)}&= y \mu^\epsilon(1+\Delta_{\text{yuk}})\,, \\
(m_\phi^{(0)})^2 &= (m_\phi^2+\Delta m_\phi^2)\,.
\end{aligned}
\end{equation}
The mass counterterm $\Delta m_\phi^2$ contains contributions from all massive parameters of the theory like the scalar mass $m_\phi$, the trilinear coupling $\kappa$, fermion masses $m_\psi$ and the infrared regulator mass $m$. For the computation of the ADM only the term proportional to $m^2$ is relevant, which we will denote as follows:
\begin{equation}
    \Delta m_\phi^2 \supset m^2 \Delta_{\phi,m}\,.
\end{equation}
The assignments in \cref{eq:scalren} lead to the following counterterm Feynman rules for the scalar propagator and the Yukawa vertex:
\begin{align}
\delta\Gamma_{\phi\phi}(p)&=i \left[(p^2-m_\phi^2)\,\Delta_\phi-\Delta m_\phi^2\right]\,, \\
\delta\Gamma_{\psi\psi\phi}&=-i y\gs \Delta_Y\,.
\end{align}
Here we used the abbreviation
\begin{equation}
\Delta_Y = \Delta_{\text{yuk}}+\Delta_{\psi}+\frac{1}{2}\Delta_{\phi}\,,
\end{equation}
where $\Delta_\psi$ denotes the scalar part of the fermion wave-function renormalization constant and which should be understood, for a generic Yukawa interaction $y^i_{kl}$, as $\frac{1}{2}(\Delta_{\psi_k}+\Delta_{\psi_l})$.

\begin{figure}[t]
\centering
\scalebox{0.8}{
\tikzfeynmanset{compat=1.1.0}

\begin{tikzpicture}
\begin{scope}[shift={(0,0)}]
\begin{feynman}
\vertex (a1) at (-2.0,0);
\vertex (b1) at (-0.8,0);
\vertex (c1) at ( 0.8,0);
\vertex (d1) at ( 2.0,0);
\diagram*{
(a1) -- [charged scalar] (b1),
(b1) -- [fermion, half left, looseness=1.6] (c1),
(c1) -- [fermion, half left, looseness=1.6] (b1),
(c1) -- [charged scalar] (d1),
};
\end{feynman}
\node at (0,-1.4) {(a)};
\end{scope}

\begin{scope}[shift={(5.5,0)}]
\begin{feynman}
\vertex (a2) at (-2.0,0);
\vertex (b2) at (-0.8,0);
\vertex (c2) at ( 0.8,0);
\vertex (d2) at ( 2.0,0);
\diagram*{
(a2) -- [charged scalar] (b2),
(b2) -- [charged scalar, half left, looseness=1.6] (c2),
(c2) -- [charged scalar, half left, looseness=1.6] (b2),
(c2) -- [charged scalar] (d2),
};
\end{feynman}
\node at (0,-1.4) {(b)};
\end{scope}

\begin{scope}[shift={(11,0)}]
\begin{feynman}
\vertex (a3) at (-2.0,0);
\vertex (b3) at (-0.8,0);
\vertex (c3) at ( 0.8,0);
\vertex (d3) at ( 2.0,0);
\diagram*{
(a3) -- [charged scalar] (b3),
(c3) -- [charged boson, half left, looseness=1.6] (b3),
(b3) -- [charged scalar, half left, looseness=1.6] (c3),
(c3) -- [charged scalar] (d3),
};
\end{feynman}
\node at (0,-1.4) {(c)};
\end{scope}

\begin{scope}[shift={(2.75,-4.2)}]
\begin{feynman}
\vertex (a4) at (-1.6,0);
\vertex (b4) at ( 0.0,0);
\vertex (c4) at ( 1.6,0);
\vertex (t4) at ( 0.0,1.5);
\diagram*{
(a4) -- [charged scalar] (b4) -- [charged scalar] (c4),
(b4) -- [charged scalar, half left, looseness=1.8] (t4),
(t4) -- [charged scalar, half left, looseness=1.8] (b4),
};
\end{feynman}
\node at (0,-1.0) {(d)};
\end{scope}

\begin{scope}[shift={(8.25,-4.2)}]
\begin{feynman}
\vertex (a5) at (-1.6,0);
\vertex (b5) at ( 0.0,0);
\vertex (c5) at ( 1.6,0);
\vertex (t5) at ( 0.0,1.5);
\diagram*{
(a5) -- [charged scalar] (b5) -- [charged scalar] (c5),
(b5) -- [charged boson, half left, looseness=1.8] (t5),
(t5) -- [charged boson, half left, looseness=1.8] (b5),
};
\end{feynman}
\node at (0,-1.0) {(e)};
\end{scope}
\end{tikzpicture}
}
\caption{One-loop contributions to the scalar self-energy: (a) fermion
bubble, (b) scalar bubble, (c) scalar--gauge bubble, (d) scalar tadpole
and (e) gauge tadpole.}
\label{fig:scalar-self-en}
\end{figure}
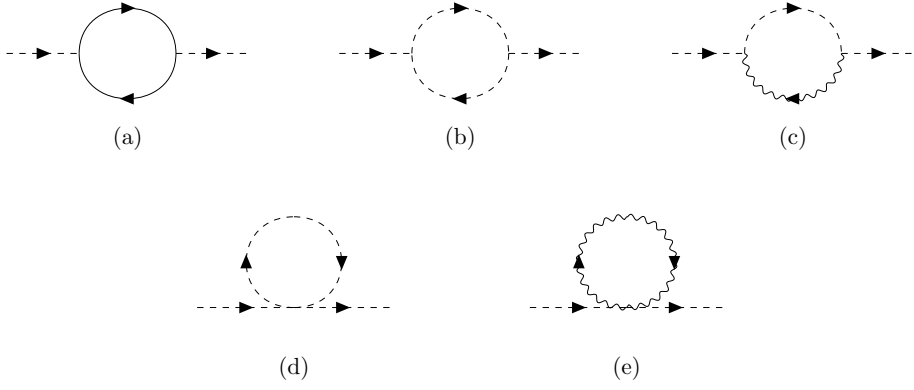

The wave-function and mass renormalization constants $\Delta_\phi$ and $\Delta m_\phi^2$ of the scalar field are derived from the one-loop scalar self-energy depicted in \cref{fig:scalar-self-en}, corresponding to the fermion, scalar and scalar-gauge bubble as well as the scalar and gauge tadpole, respectively. The resulting renormalization constants $\Delta_\phi$ and $\Delta m_\phi^2$ can be derived from the contributions in \cref{tab:scalar-propagator-CTsP} and \ref{tab:scalar-propagator-CTsM}, respectively.

\begin{figure}[t]
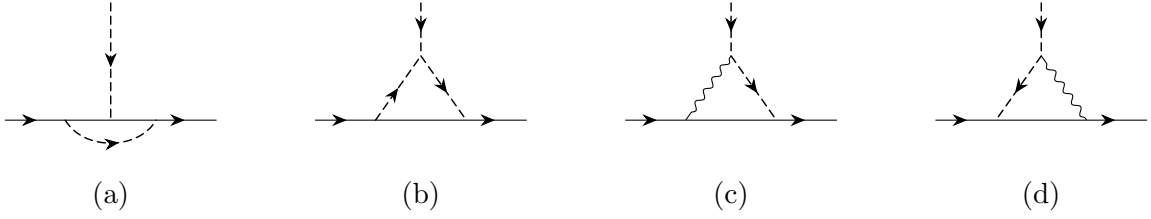

\centering
\panel{%
  \draw[arrow at=0.15, arrow at=0.85] (-1.4,0) -- (1.4,0);
  \draw[scline, arrow at=0.55] (0,1.55) -- (0,0);
  \draw[scline, arrow at=0.58] (-0.6,0) to[bend right=60] (0.6,0);
  \node at (0,-1.0) {(a)};}
\hfill
\panel{%
  \draw[arrow at=0.15, arrow at=0.85] (-1.4,0) -- (1.4,0);
  \draw[scline, arrow at=0.55] (0,1.55) -- (0,0.85);
  \draw[scline, arrow at=0.58] (0,0.85) -- (0.6,0);
  \draw[scline, arrow at=0.5] (-0.6,0) -- (0,0.85);
  \node at (0,-1.0) {(b)};}
\hfill
\panel{%
  \draw[arrow at=0.15, arrow at=0.85] (-1.4,0) -- (1.4,0);
  \draw[scline, arrow at=0.55] (0,1.55) -- (0,0.85);
  \draw[scline, arrow at=0.58] (0,0.85) -- (0.6,0);
  \draw[phfit=5.45pt] (0,0.85) -- (-0.6,0);
  \node at (0,-1.0) {(c)};}
\hfill
\panel{%
  \draw[arrow at=0.15, arrow at=0.85] (-1.4,0) -- (1.4,0);
  \draw[scline, arrow at=0.55] (0,1.55) -- (0,0.85);
  \draw[phfit=5.45pt] (0.6,0) -- (0,0.85);
  \draw[scline, arrow at=0.58] (0,0.85) -- (-0.6,0);
  \node at (0,-1.0) {(d)};}
\caption{One-loop corrections to the fermion--fermion--scalar vertex: (a) genuine scalar vertex correction, (b) three-scalar correction, (c) three-point correction with gauge boson attached to incoming fermion, (d) three-point correction with gauge boson attached to outgoing fermion.}
\label{fig:ffsvertex}
\end{figure}

Finally, the Yukawa vertex counterterm $\Delta_Y$ is computed from the diagrams depicted in \cref{fig:ffsvertex}, for which the corresponding results are collected in \cref{tab:scalar-propagator-CTsVERT}. 

\begin{table}[H]
\centering
\setlength{\tabcolsep}{6pt}
\renewcommand{\arraystretch}{1.15}
\begin{tabular}{@{}l c@{}}
\toprule[0.7mm]
Diagram
& $1/\epsilon$
\\
\midrule[0.7mm]

Fermion bubble
&
$\frac{p^2}{12}\operatorname{Tr}\,[\gamma_\mu \gs\gamma^\mu \gst]+\frac{1}{6}\operatorname{Tr}\,[\slashed{p} \,\gs\,\slashed{p}\,\gst]$
\\
\midrule

Scalar bubble
&
$0$
\\
\midrule

Scalar--gauge bubble
&
$-2p^2$
\\
\midrule

Scalar tadpole
&
$0$
\\
\midrule

Gauge tadpole
&
$0$
\\
\bottomrule[0.7mm]
\end{tabular}
\caption{One-loop contributions to the scalar wave-function renormalization $\Delta_\phi$.}
\label{tab:scalar-propagator-CTsP}
\end{table}

\begin{table}[H]
\centering
\setlength{\tabcolsep}{6pt}
\renewcommand{\arraystretch}{1.15}
\begin{tabular}{@{}l c@{}}
\toprule[0.7mm]
Diagram
& $1/\epsilon$
\\
\midrule[0.7mm]

Fermion bubble
&
$-m_\psi^2\operatorname{Tr}\,[\gs \gst]-\frac{m_\psi^2+m^2}{2}\operatorname{Tr}\,[\gamma_\mu \,\gs\,\gamma^\mu\,\gst]$
\\
\midrule

Scalar bubble
&
$\frac12 \kappa^2$
\\
\midrule

Scalar--gauge bubble
&
$-m_\phi^2-2m^2$
\\
\midrule

Scalar tadpole
&
$\frac12 \lambda(m_\phi^2+m^2)$
\\
\midrule

Gauge tadpole
&
$2m^2$
\\
\bottomrule[0.7mm]
\end{tabular}
\caption{One-loop contributions to the scalar mass counterterm $\Delta m_\phi^2$.}
\label{tab:scalar-propagator-CTsM}
\end{table}

\begin{table}[H]
\centering
\setlength{\tabcolsep}{6pt}
\renewcommand{\arraystretch}{1.15}
\begin{tabular}{@{}l c@{}}
\toprule[0.7mm]
Diagram
& $1/\epsilon$
\\
\midrule[0.7mm]

Genuine
&
$\frac{1}{4}\gs \gamma_\mu \,\gs\,\gamma^\mu\,\gst$
\\
\midrule

Trilinear
&
0
\\
\midrule

Gauge boson-left
&
$\frac14 \gs \gamma_\mu \gamma^\mu$
\\
\midrule

Gauge boson-right
&
$-\frac14 \gamma_\mu \gamma^\mu\gs$
\\
\bottomrule[0.7mm]
\end{tabular}
\caption{One-loop contributions to the vertex counterterm $\Delta_Y$.}
\label{tab:scalar-propagator-CTsVERT}
\end{table}

\subsection{One-loop counterterm insertions }

\subsubsection{Generic one-loop counterterm insertions}

In this subsection the generic one-loop counterterm insertions in the topologies 1--3 corresponding to the fermion (\cref{tab:generic-scalar-fermion-CT-D1D3}), vertex (\cref{tab:generic-scalar-vertex-CT-D1D3}) and scalar (\cref{tab:generic-scalar-propagator-CT-D1D3}) contributions are reported. The expressions for the renormalization constants $\Delta_\psi$, $\Delta_Y$, $\Delta_\phi$ and $\Delta_{\phi,m}$ are discussed in the previous subsection. The counterterm contributions proportional to $\gst$ are omitted.

\begin{table}[H]
\centering
\small
\setlength{\tabcolsep}{6pt}
\renewcommand{\arraystretch}{0.55}
\begin{tabular}{@{}c c@{}}
\toprule[0.7mm]
Diagram & $1/\epsilon^2$ \\
\midrule[0.7mm]

1 
&
\makecell[l]{$\displaystyle\small
\begin{aligned}[t]
&\frac{\Delta_\psi}{24}\,
(\gs\gamma^\mu\Gamma_1
 \gamma^\nu\gamma_\nu\gamma_\mu\gst
 \otimes\Gamma_2)
\\
&+\frac{\Delta_\psi}{24}\,
(\gs\gamma^\mu\Gamma_1
 \gamma^\nu\gamma_\mu\gamma_\nu\gst
 \otimes\Gamma_2)
\\
&+\frac{\Delta_\psi}{24}\,
(\gs\gamma^\mu\Gamma_1
 \gamma_\mu\gamma^\nu\gamma_\nu\gst
 \otimes\Gamma_2)
\end{aligned}$}
\\
\midrule

2 
&
\makecell[l]{$\displaystyle\small
\begin{aligned}[t]
&-\frac{\Delta_\psi}{24}\,
(\Gamma_1\gamma^\mu\gamma_\mu\gamma^\nu\gs
 \otimes\Gamma_2\gamma_\nu\gst)
\\
&-\frac{\Delta_\psi}{24}\,
(\Gamma_1\gamma^\mu\gamma^\nu\gamma_\mu\gs
 \otimes\Gamma_2\gamma_\nu\gst)
\\
&-\frac{\Delta_\psi}{24}\,
(\Gamma_1\gamma^\mu\gamma^\nu\gamma_\nu\gs
 \otimes\Gamma_2\gamma_\mu\gst)
\end{aligned}$}
\\
\midrule

3 
&
\makecell[l]{$\displaystyle\small
\begin{aligned}[t]
&\frac{\Delta_\psi}{24}\,
(\Gamma_1\gamma^\mu\gamma_\mu\gamma^\nu\gs
 \otimes\gst\gamma_\nu\Gamma_2)
\\
&+\frac{\Delta_\psi}{24}\,
(\Gamma_1\gamma^\mu\gamma^\nu\gamma_\mu\gs
 \otimes\gst\gamma_\nu\Gamma_2)
\\
&+\frac{\Delta_\psi}{24}\,
(\Gamma_1\gamma^\mu\gamma^\nu\gamma_\nu\gs
 \otimes\gst\gamma_\mu\Gamma_2)
\end{aligned}$}
\\

\bottomrule[0.7mm]
\end{tabular}
\caption{Generic Dirac structures for a fermion counterterm insertion in the one-loop topologies 1--3.}
\label{tab:generic-scalar-fermion-CT-D1D3}
\end{table}

\begin{table}[H]
\centering
\small
\setlength{\tabcolsep}{7pt}
\renewcommand{\arraystretch}{0.85}
\begin{tabular}{@{}c c@{}}
\toprule[0.7mm]
Diagram & $1/\epsilon^2$ \\
\midrule[0.7mm]

1
&
\makecell[l]{$\displaystyle
-\frac{\Delta_Y}{4}\,
(\gs\gamma^\mu\Gamma_1\gamma_\mu\gst
 \otimes\Gamma_2)$}
\\
\midrule

2 
&
\makecell[l]{$\displaystyle
\frac{\Delta_Y}{4}\,
(\Gamma_1\gamma^\mu\gs
 \otimes\Gamma_2\gamma_\mu\gst)$}
\\
\midrule

3 
&
\makecell[l]{$\displaystyle
-\frac{\Delta_Y}{4}\,
(\Gamma_1\gamma^\mu\gs
 \otimes\gst\gamma_\mu\Gamma_2)$}
\\

\bottomrule[0.7mm]
\end{tabular}
\caption{Generic Dirac structures for a scalar--fermion vertex counterterm insertion in the one-loop topologies 1--3.}
\label{tab:generic-scalar-vertex-CT-D1D3}
\end{table}

\begin{table}[H]
\centering
\small
\setlength{\tabcolsep}{7pt}
\renewcommand{\arraystretch}{0.85}
\begin{tabular}{@{}c c c@{}}
\toprule[0.7mm]
Diagram & $1/\epsilon^2$ & $1/\epsilon$ \\
\midrule[0.7mm]

1
&
\makecell[l]{$\displaystyle
\frac{\Delta_\phi}{4}\,
(\gs\gamma^\mu\Gamma_1\gamma_\mu\gst
 \otimes\Gamma_2)$}
&
\makecell[l]{$\displaystyle
\frac{\Delta_{\phi,m}-\Delta_\phi}{12}\,
(\gs\gamma^\mu\Gamma_1\gamma_\mu\gst
 \otimes\Gamma_2)$}
\\
\midrule

2
&
\makecell[l]{$\displaystyle
-\frac{\Delta_\phi}{4}\,
(\Gamma_1\gamma^\mu\gs
 \otimes\Gamma_2\gamma_\mu\gst)$}
&
\makecell[l]{$\displaystyle
-\frac{\Delta_{\phi,m}-\Delta_\phi}{12}\,
(\Gamma_1\gamma^\mu\gs
 \otimes\Gamma_2\gamma_\mu\gst)$}
\\
\midrule

3
&
\makecell[l]{$\displaystyle
\frac{\Delta_\phi}{4}\,
(\Gamma_1\gamma^\mu\gs
 \otimes\gst\gamma_\mu\Gamma_2)$}
&
\makecell[l]{$\displaystyle
\frac{\Delta_{\phi,m}-\Delta_\phi}{12}\,
(\Gamma_1\gamma^\mu\gs
 \otimes\gst\gamma_\mu\Gamma_2)$}
\\

\bottomrule[0.7mm]
\end{tabular}
\caption{Generic Dirac structures for the full scalar propagator counterterm insertion in the one-loop topologies 1--3.}
\label{tab:generic-scalar-propagator-CT-D1D3}
\end{table}

\subsubsection{Counterterm insertions for fermion wave-function renormalization}

\begin{table}[H]
\centering
\setlength{\tabcolsep}{5pt}
\renewcommand{\arraystretch}{0.72}
\begin{tabular}{@{}c c c@{}}
\toprule[0.7mm]
Diagram & $1/\epsilon^2$ & $1/\epsilon$ \\
\midrule[0.7mm]

Fermion-propagator CT
&
\makecell[l]{$\displaystyle
\begin{aligned}[t]
&-\frac{\Delta_\psi}{12}\,
 \gs\gamma^\nu\gamma_\nu\gamma^\lambda\gst
\\
&-\frac{\Delta_\psi}{12}\,
 \gs\gamma^\nu\gamma^\lambda\gamma_\nu\gst
\\
&-\frac{\Delta_\psi}{12}\,
 \gs\gamma^\lambda\gamma^\nu\gamma_\nu\gst
\end{aligned}$}
&
\makecell[l]{$\displaystyle 0$}
\\
\midrule

Scalar--fermion vertex CT
&
\makecell[l]{$\displaystyle
\frac{\Delta_Y}{2}\,
\gs\gamma^\lambda\gst$}
&
\makecell[l]{$\displaystyle 0$}
\\
\midrule

Scalar kinetic CT
&
\makecell[l]{$\displaystyle
-\frac{\Delta_\phi}{2}\,
\gs\gamma^\lambda\gst$}
&
\makecell[l]{$\displaystyle
\frac{\Delta_\phi}{3}\,
\gs\gamma^\lambda\gst$}
\\
\midrule

Auxiliary scalar-mass CT
&
\makecell[l]{$\displaystyle 0$}
&
\makecell[l]{$\displaystyle
-\frac{\Delta_{\phi,m}}{3}\,
\gs\gamma^\lambda\gst$}
\\

\bottomrule[0.7mm]
\end{tabular}
\caption{Generic Dirac structures of the one-loop fermion self-energy diagrams with dimension-four counterterm insertions. The index $\lambda$ denotes the external-momentum index.}
\label{tab:wfr-open-scalar-ct}
\end{table}

\subsection{Two-loop fermion wave-function renormalization}

\begin{table}[H]
\centering
\setlength{\tabcolsep}{5pt}
\renewcommand{\arraystretch}{0.70}
\begin{tabular}{@{}c c c@{}}
\toprule[0.7mm]
Diagram & $1/\epsilon^2$ & $1/\epsilon$ \\
\midrule[0.7mm]

One-loop
&
\makecell[l]{\hspace{1cm} --}
&
\makecell[l]{\hspace{0.8cm}$\displaystyle
\frac{1}{2}\,
\gs\gamma^\lambda\gst$}
\\
\midrule

Nested
&
\makecell[l]{$\displaystyle
\begin{aligned}[t]
&-\frac{1}{48}\,
 \gs\gamma^\nu\gs\gamma^\lambda
 \gst\gamma_\nu\gst
\\
&-\frac{1}{48}\,
 \gs\gamma^\lambda\gs\gamma^\nu
 \gst\gamma_\nu\gst
\\
&-\frac{1}{48}\,
 \gs\gamma^\nu\gs\gamma_\nu
 \gst\gamma^\lambda\gst
\end{aligned}$}
&
\makecell[l]{$\displaystyle
\begin{aligned}[t]
&-\frac{1}{288}\,
 \gs\gamma^\nu\gs\gamma^\lambda
 \gst\gamma_\nu\gst
\\
&-\frac{1}{288}\,
 \gs\gamma^\lambda\gs\gamma^\nu
 \gst\gamma_\nu\gst
\\
&-\frac{1}{288}\,
 \gs\gamma^\nu\gs\gamma_\nu
 \gst\gamma^\lambda\gst
\end{aligned}$}
\\
\midrule

Crossed
&
\makecell[l]{$\displaystyle
\begin{aligned}[t]
&-\frac{1}{16}\,
 \gs\gamma^\lambda\gs\gamma^\nu
 \gst\gamma_\nu\gst
\\
&-\frac{1}{16}\,
 \gs\gamma^\nu\gs\gamma_\nu
 \gst\gamma^\lambda\gst
\end{aligned}$}
&
\makecell[l]{$\displaystyle
\begin{aligned}[t]
&-\frac{1}{32}\,
 \gs\gamma^\lambda\gs\gamma^\nu
 \gst\gamma_\nu\gst
\\
&-\frac{1}{32}\,
 \gs\gamma^\nu\gs\gamma_\nu
 \gst\gamma^\lambda\gst
\end{aligned}$}
\\

\bottomrule[0.7mm]
\end{tabular}
\caption{Generic Dirac structure of the one-loop and two-loop wave-function renormalization diagrams. The index $\lambda$ denotes the external-momentum index.}
\label{tab:wfr-open-purescalar}
\end{table}

\begin{table}[H]
\centering
\setlength{\tabcolsep}{4pt}
\renewcommand{\arraystretch}{0.62}
\begin{tabular}{@{}c c c@{}}
\toprule[0.7mm]
Orientation & $1/\epsilon^2$ & $1/\epsilon$ \\
\midrule[0.7mm]

$G_1S_{23}$
&
\makecell[l]{$\displaystyle
\begin{aligned}[t]
&-\frac{1}{8}\,
 \gamma^\nu\gamma^\lambda\gs\gamma_\nu\gst
\\
&+\frac{1}{16}\,
 \gamma^\nu\gamma_\nu\gs\gamma^\lambda\gst
\end{aligned}$}
&
\makecell[l]{$\displaystyle
\begin{aligned}[t]
&-\frac{1}{8}\,
 \gamma^\nu\gamma^\lambda\gs\gamma_\nu\gst
\\
&-\frac{1}{32}\,
 \gamma^\nu\gamma_\nu\gs\gamma^\lambda\gst
\\
&+\frac{1}{16}\,
 \gamma^\lambda\gamma^\nu\gs\gamma_\nu\gst
\end{aligned}$}
\\
\midrule

$G_2S_{13}$
&
\makecell[l]{$\displaystyle
\begin{aligned}[t]
&-\frac{1}{16}\,
 \gs\gamma^\lambda\gamma^\nu\gamma_\nu\gst
\\
&-\frac{1}{16}\,
 \gs\gamma^\nu\gamma_\nu\gamma^\lambda\gst
\end{aligned}$}
&
\makecell[l]{$\displaystyle
\begin{aligned}[t]
&-\frac{5}{32}\,
 \gs\gamma^\lambda\gamma^\nu\gamma_\nu\gst
\\
&-\frac{5}{32}\,
 \gs\gamma^\nu\gamma_\nu\gamma^\lambda\gst
\\
&+\frac{1}{8}\,
 \gs\gamma^\nu\gamma^\lambda\gamma_\nu\gst
\end{aligned}$}
\\
\midrule

$G_3S_{12}$
&
\makecell[l]{$\displaystyle
\begin{aligned}[t]
&\frac{1}{16}\,
 \gs\gamma^\lambda\gst\gamma^\nu\gamma_\nu
\\
&-\frac{1}{8}\,
 \gs\gamma^\nu\gst\gamma^\lambda\gamma_\nu
\end{aligned}$}
&
\makecell[l]{$\displaystyle
\begin{aligned}[t]
&-\frac{1}{32}\,
 \gs\gamma^\lambda\gst\gamma^\nu\gamma_\nu
\\
&-\frac{1}{8}\,
 \gs\gamma^\nu\gst\gamma^\lambda\gamma_\nu
\\
&+\frac{1}{16}\,
 \gs\gamma^\nu\gst\gamma_\nu\gamma^\lambda
\end{aligned}$}
\\

\bottomrule[0.7mm]
\end{tabular}
\caption{Generic Dirac structure of the two-loop scalar-gauge boson contributions to the fermion wave-function renormalization. The index $\lambda$ denotes the external-momentum index. The notation $G_iS_{jk}$ specifies the ordering of the gauge boson and the two scalars along the fermion line.}
\label{tab:wfr-open-scalar-ssg}
\end{table}

\begin{table}[H]
\centering
\setlength{\tabcolsep}{4pt}
\renewcommand{\arraystretch}{0.68}
\begin{tabular}{@{}c c c@{}}
\toprule[0.7mm]
Diagram & $1/\epsilon^2$ & $1/\epsilon$ \\
\midrule[0.7mm]

Fermion bubble
&
\makecell[l]{$\displaystyle
\begin{aligned}[t]
&-\frac{5}{144}\,
 \operatorname{Tr}\!\left[
  \gs\gamma^\nu\gst\gamma_\nu
 \right]
 \gs\gamma^\lambda\gst
\\
&+\frac{1}{144}\,
 \operatorname{Tr}\!\left[
  \gs\gamma^\lambda\gst\gamma^\nu
 \right]
 \gs\gamma_\nu\gst
\\
&+\frac{1}{144}\,
 \operatorname{Tr}\!\left[
  \gs\gamma^\nu\gst\gamma^\lambda
 \right]
 \gs\gamma_\nu\gst
\end{aligned}$}
&
\makecell[l]{$\displaystyle
\begin{aligned}[t]
&-\frac{163}{864}\,
 \operatorname{Tr}\!\left[
  \gs\gamma^\nu\gst\gamma_\nu
 \right]
 \gs\gamma^\lambda\gst
\\
&-\frac{1}{864}\,
 \operatorname{Tr}\!\left[  \gs\gamma^\lambda\gst\gamma^\nu
 \right]
 \gs\gamma_\nu\gst
\\
&-\frac{1}{864}\,
 \operatorname{Tr}\!\left[
  \gs\gamma^\nu\gst\gamma^\lambda
 \right]
 \gs\gamma_\nu\gst
\end{aligned}$}
\\
\midrule

Scalar bubble
in scalar line
&
\makecell[l]{$\displaystyle 0$}
&
\makecell[l]{$\displaystyle
-\frac{1}{6}\,
\gs\gamma^\lambda\gst$}
\\
\midrule

Scalar--gauge bubble
&
\makecell[l]{$\displaystyle
\frac{1}{2}\,
\gs\gamma^\lambda\gst$}
&
\makecell[l]{$\displaystyle
-\frac{7}{12}\,
\gs\gamma^\lambda\gst$}
\\
\midrule

Scalar tadpole
in scalar line
&
\makecell[l]{$\displaystyle 0$}
&
\makecell[l]{$\displaystyle
\frac{1}{6}\,
\gs\gamma^\lambda\gst$}
\\
\midrule

Gauge tadpole
&
\makecell[l]{$\displaystyle 0$}
&
\makecell[l]{$\displaystyle
\frac{2}{3}\,
\gs\gamma^\lambda\gst$}
\\
\midrule

Scalar bubble
in gauge line
&
\makecell[l]{$\displaystyle
\begin{aligned}[t]
&\frac{1}{18}\,
 \gamma^\nu\gamma^\lambda\gamma_\nu
\\
&+\frac{1}{72}\,
 \gamma^\lambda\gamma^\nu\gamma_\nu
\\
&+\frac{1}{72}\,
 \gamma^\nu\gamma_\nu\gamma^\lambda
\end{aligned}$}
&
\makecell[l]{$\displaystyle
\begin{aligned}[t]
&\frac{151}{216}\,
 \gamma^\nu\gamma^\lambda\gamma_\nu
\\
&+\frac{5}{432}\,
 \gamma^\lambda\gamma^\nu\gamma_\nu
\\
&+\frac{5}{432}\,
 \gamma^\nu\gamma_\nu\gamma^\lambda
\end{aligned}$}
\\
\midrule

Scalar seagull
in gauge line
&
\makecell[l]{$\displaystyle 0$}
&
\makecell[l]{$\displaystyle
-\frac{1}{3}\,
\gamma^\nu\gamma^\lambda\gamma_\nu$}
\\

\bottomrule[0.7mm]
\end{tabular}
\caption{Two-loop self-energy contributions to the fermion wave-function renormalization. The index $\lambda$ denotes the external-momentum index. The first five entries correspond to diagrams containing those shown in \cref{fig:scalar-self-en} as subdiagrams. The last two entries contain the last two contributions in \cref{fig:gluonSE} as subdiagrams.}
\label{tab:wfr-open-scalar-selfenergies}
\end{table}

\subsection{General two-loop structures}
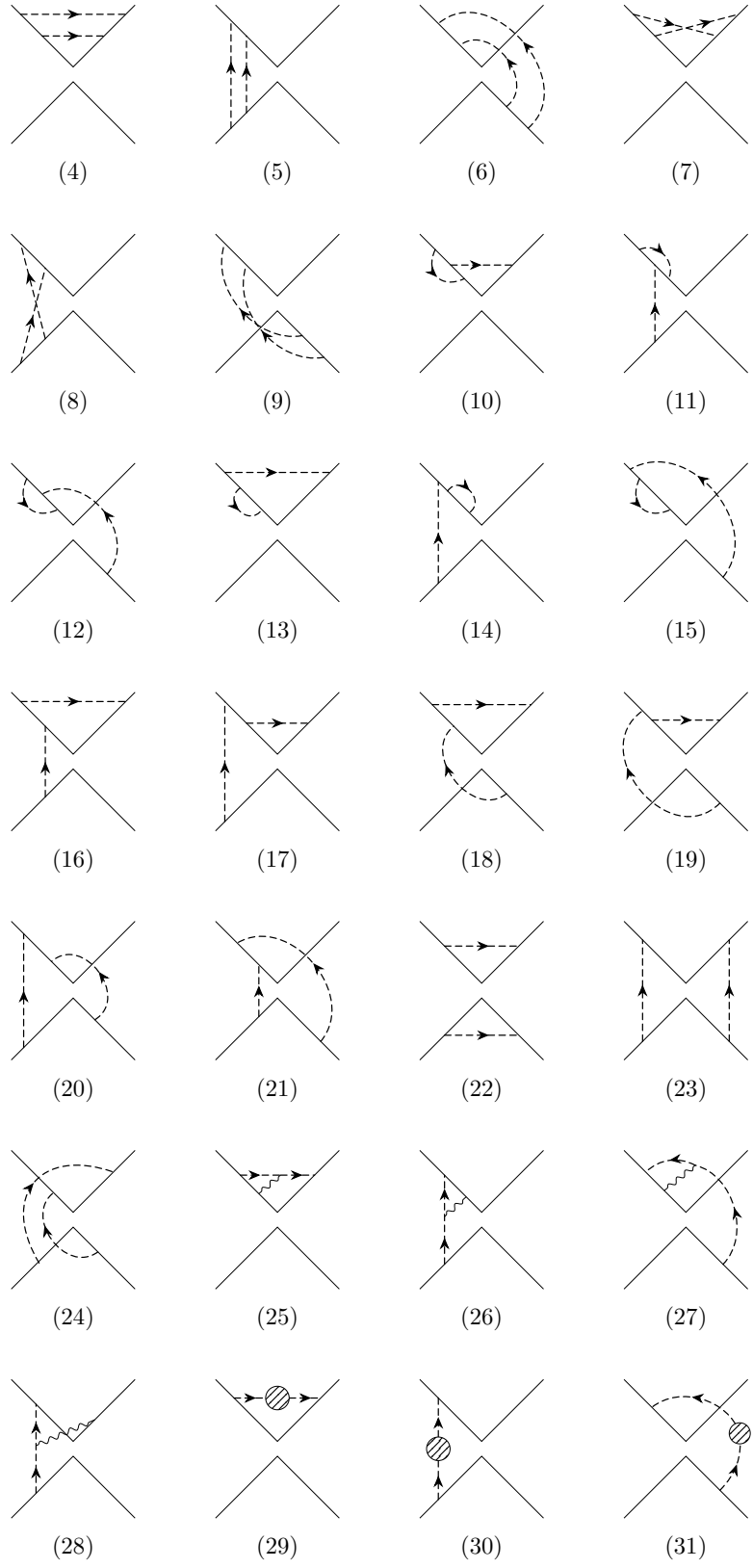
\begin{figure}[H]
\centering
\scalebox{0.85}{%
\begin{tikzpicture}[x=1cm,y=1cm]
\begin{scope}[shift={(0,0)}]      
  \VV                             
  \draw[scline, arrow at=0.56] (-0.85,0.97) -- (0.85,0.97);
  \draw[scline, arrow at=0.6] (-0.5,0.62)  -- (0.5,0.62);
  \num{4}
\end{scope}
\begin{scope}[shift={(3.3,0)}]    
  \VV                             
  \draw[scline, arrow at=0.62] (-0.75,-0.87) -- (-0.75,0.87);
  \draw[scline, arrow at=0.62] (-0.5,-0.62)  -- (-0.5,0.62);
  \num{5}
\end{scope}
\begin{scope}[shift={(6.6,0)}]    
  \VV                             
  \draw[sc] (0.75,-0.87) to[bend right=90,looseness=1.3] (-0.7,0.82);
  \draw[sc] (0.4,-0.52)  to[bend right=90,looseness=1.3] (-0.35,0.47);
  \num{6}
\end{scope}
\begin{scope}[shift={(9.9,0)}]    
  \VV                             
  \draw[sc] (-0.85,0.97) -- (0.5,0.62);
  \draw[scline, arrow at=0.68] (-0.5,0.62)  -- (0.85,0.97);
  \num{7}
\end{scope}
\begin{scope}[shift={(0,-3.7)}]   
  \VV                             
  \draw[scline, arrow at=0.7] (-0.45,-0.57) -- (-0.85,0.97);
  \draw[sc] (-0.85,-0.97) -- (-0.45,0.57);
  \num{8}
\end{scope}
\begin{scope}[shift={(3.3,-3.7)}] 
  \VV                             
  \draw[sc] (0.4,-0.52)  to[bend left=60] (-0.85,0.97);
  \draw[sc] (0.75,-0.87) to[bend left=60] (-0.5,0.62);
  \num{9}
\end{scope}
\begin{scope}[shift={(6.6,-3.7)}] 
  \VV                             
  \draw[sc] (-0.5,0.62) -- (0.5,0.62);
  \draw[sc] (-0.75,0.87) to[bend right=70,looseness=1.6] (-0.28,0.4);
  \num{10}
\end{scope}
\begin{scope}[shift={(9.9,-3.7)}] 
  \VV                             
  \draw[sc] (-0.5,-0.62) -- (-0.5,0.62);
  \draw[sc] (-0.75,0.87) to[bend left=70,looseness=1.2] (-0.28,0.4);
  \num{11}
\end{scope}
\begin{scope}[shift={(0,-7.4)}]   
  \VV                             
  \draw[sc] (0.55,-0.67) to[bend right=80,looseness=1.25] (-0.5,0.62);
  \draw[sc] (-0.75,0.87) to[bend right=70,looseness=1.6] (-0.25,0.37);
  \num{12}
\end{scope}
\begin{scope}[shift={(3.3,-7.4)}] 
  \VV                             
  \draw[sc] (-0.85,0.97) -- (0.85,0.97);
  \draw[sc] (-0.6,0.72) to[bend right=100,looseness=1.9] (-0.25,0.37);
  \num{13}
\end{scope}
\begin{scope}[shift={(6.6,-7.4)}] 
  \VV                             
  \draw[sc] (-0.7,-0.82) -- (-0.7,0.82);
  \draw[sc] (-0.55,0.67) to[bend left=100,looseness=1.8] (-0.2,0.32);
  \num{14}
\end{scope}
\begin{scope}[shift={(9.9,-7.4)}] 
  \VV                             
  \draw[sc] (-0.75,0.87) to[bend right=70,looseness=1.6] (-0.25,0.37);
  \draw[scline, arrow at=0.62] (0.6,-0.72) to[bend right=80,looseness=1.2] (-0.9,1.02);
  \num{15}
\end{scope}
\begin{scope}[shift={(0,-11.1)}]  
  \VV                             
  \draw[scline, arrow at=0.56] (-0.85,0.97) -- (0.85,0.97);
  \draw[sc] (-0.45,-0.57) -- (-0.45,0.57);
  \num{16}
\end{scope}
\begin{scope}[shift={(3.3,-11.1)}]
  \VV                             
  \draw[scline, arrow at=0.6] (-0.5,0.62) -- (0.5,0.62);
  \draw[sc] (-0.85,-0.97) -- (-0.85,0.97);
  \num{17}
\end{scope}
\begin{scope}[shift={(6.6,-11.1)}]
  \VV                             
  \draw[scline, arrow at=0.56] (-0.8,0.92) -- (0.8,0.92);
  \draw[scline, arrow at=0.65] (0.4,-0.52) to[bend left=90,looseness=1.3] (-0.45,0.57);
  \num{18}
\end{scope}
\begin{scope}[shift={(9.9,-11.1)}]
  \VV                             
  \draw[scline, arrow at=0.59] (-0.55,0.67) -- (0.55,0.67);
  \draw[scline, arrow at=0.65] (0.55,-0.67) to[bend left=95,looseness=1.5] (-0.7,0.82);
  \num{19}
\end{scope}
\begin{scope}[shift={(0,-14.8)}]  
  \VV                             
  \draw[sc] (-0.8,-0.92) -- (-0.8,0.92);
  \draw[sc] (0.35,-0.47) to[bend right=95,looseness=1.5] (-0.35,0.47);
  \num{20}
\end{scope}
\begin{scope}[shift={(3.3,-14.8)}]
  \VV                             
  \draw[sc] (0.7,-0.82) to[bend right=80,looseness=1.2] (-0.65,0.77);
  \draw[scline, arrow at=0.6] (-0.3,-0.42) -- (-0.3,0.42);
  \num{21}
\end{scope}
\begin{scope}[shift={(6.6,-14.8)}]
  \VV                             
  \draw[scline, arrow at=0.6] (-0.6,0.72)  -- (0.6,0.72);
  \draw[scline, arrow at=0.6] (-0.6,-0.72) -- (0.6,-0.72);
  \num{22}
\end{scope}
\begin{scope}[shift={(9.9,-14.8)}]
  \VV                             
  \draw[scline, arrow at=0.56] (-0.7,-0.82) -- (-0.7,0.82);
  \draw[scline, arrow at=0.56] (0.7,-0.82)  -- (0.7,0.82);
  \num{23}
\end{scope}
\begin{scope}[shift={(0,-18.5)}]  
  \VV                              
  \draw[scline, arrow at=0.65] (0.4,-0.52)  to[bend left=90,looseness=1.4] (-0.30,0.45);
  \draw[sc] (-0.55,-0.7) to[bend left=70,looseness=1.70] (0.65,0.77);
  \num{24}
\end{scope}
\begin{scope}[shift={(3.3,-18.5)}]
  \VV                             
  \draw[sc] (-0.6,0.72) -- (0.05,0.72);
  \draw[scline, arrow at=0.65] (0.05,0.72) -- (0.6,0.72);
  \draw[ph] (0.05,0.72) -- (-0.3,0.42);
  \num{25}
\end{scope}
\begin{scope}[shift={(6.6,-18.5)}]
  \VV                             
  \draw[scline, arrow at=0.3, arrow at=0.85] (-0.6,-0.72) -- (-0.6,0.72);
  \draw[ph] (-0.6,0.05)  -- (-0.25,0.37);
  \num{26}
\end{scope}
\begin{scope}[shift={(9.9,-18.5)}]
  \VV                             
  \draw[sc] (0.6,-0.72) to[bend right=60] (0.15,0.88);
  \draw[sc] (0.15,0.88) to[bend right=35] (-0.65,0.77);
  \draw[ph] (0.15,0.88) -- (-0.35,0.47);
  \num{27}
\end{scope}
\begin{scope}[shift={(0,-22.2)}]  
  \VV                             
  \draw[scline, arrow at=0.3, arrow at=0.8] (-0.6,-0.72) -- (-0.6,0.72);
  \draw[ph] (-0.6,0.05)  -- (0.35,0.47);
  \num{28}
\end{scope}
\begin{scope}[shift={(3.3,-22.2)}]
  \VV                             
  \draw[scline, arrow at=0.7] (-0.7,0.82) -- (-0.19,0.82);
  \draw[scline, arrow at=0.7] (0.19,0.82) -- (0.7,0.82);
  \filldraw[blob] (0,0.82) circle (0.19);
  \num{29}
\end{scope}
\begin{scope}[shift={(6.6,-22.2)}]
  \VV                             
  \draw[scline, arrow at=0.6] (-0.7,-0.82) -- (-0.7,-0.17);
  \draw[sc] (-0.7,0.19)  -- (-0.7,0.84);
  \filldraw[blob] (-0.7,0) circle (0.19);
  \num{30}
\end{scope}
\begin{scope}[shift={(9.9,-22.2)}]
  \VV                             
  \draw[sc] (0.55,-0.67) to[bend right=25] (0.87,0.25);
  \draw[scline, arrow at=0.6] (0.87,0.25)  to[bend right=55] (-0.55,0.67);
  \filldraw[blob] (0.87,0.25) circle (0.17);
  \num{31}
\end{scope}
\end{tikzpicture}%
}
\caption{Two-loop diagrams from scalar interactions.}
\label{fig:twoloopscalar}
\end{figure}

In this subsection we report the divergent parts of the two-loop diagrams in \cref{fig:twoloopscalar} for generic insertions of four-fermion operators with Dirac structure $\Gamma_1\otimes\Gamma_2$. 

Tables~\ref{tab:generic-scalar-D14-D10}--\ref{tab:generic-scalar-D17-D24} contain the poles of diagrams 4--24. The divergences of diagrams corresponding to scalar-gauge boson interactions are collected in \cref{tab:generic_SSG_D25_D28} and the self-energy contributions are given in Tables \ref{tab:genericScalar29}--\ref{tab:genericScalar31}.

\subsubsection{Genuine two-loop abelian-like structures}

\begin{table}[H]
\centering
\setlength{\tabcolsep}{6pt}
\renewcommand{\arraystretch}{0.65}
\begin{tabular}{@{}c c c@{}}
\toprule[0.7mm]
Diagram & $1/\epsilon^2$ & $1/\epsilon$ \\
\midrule[0.7mm]
4 &
\makecell[l]{$\displaystyle\small
\begin{aligned}[t]
&\frac{1}{32}\,
(\gs \gamma^\mu\gs \gamma^\nu\Gamma_1\gamma_\nu\gst\gamma_\mu\gst\otimes\Gamma_2)
\end{aligned}$}
&
\makecell[l]{$\displaystyle\small
\begin{aligned}[t]
&\frac{5}{192}\,
(\gs \gamma^\mu\gs \gamma^\nu\Gamma_1\gamma_\nu\gst\gamma_\mu\gst\otimes\Gamma_2)\\
&+\frac{1}{96}\,
(\gs \gamma^\mu\gs \gamma^\nu\Gamma_1\gamma_\mu\gst\gamma_\nu\gst\otimes\Gamma_2)\\
&+\frac{1}{96}\,
(\gs \gamma^\mu\gs \gamma_\mu\Gamma_1\gamma^\nu\gst\gamma_\nu\gst\otimes\Gamma_2)
\end{aligned}$}
\\ \midrule

5 &
\makecell[l]{$\displaystyle\small
\begin{aligned}[t]
&\frac{1}{32}\,
(\Gamma_1\gamma^\mu\gs \gamma^\nu\gs \otimes\Gamma_2\gamma_\mu\gst\gamma_\nu\gst)
\end{aligned}$}
&
\makecell[l]{$\displaystyle\small
\begin{aligned}[t]
&\frac{5}{192}\,
(\Gamma_1\gamma^\mu\gs \gamma^\nu\gs \otimes\Gamma_2\gamma_\mu\gst\gamma_\nu\gst)\\
&+\frac{1}{96}\,
(\Gamma_1\gamma^\mu\gs \gamma^\nu\gs \otimes\Gamma_2\gamma_\nu\gst\gamma_\mu\gst)\\
&+\frac{1}{96}\,
(\Gamma_1\gamma^\mu\gs \gamma_\mu\gs \otimes\Gamma_2\gamma^\nu\gst\gamma_\nu\gst)
\end{aligned}$}
\\ \midrule

6 &
\makecell[l]{$\displaystyle\small
\begin{aligned}[t]
&\frac{1}{32}\,
(\Gamma_1\gamma^\mu\gs \gamma^\nu\gs \otimes\gst\gamma_\nu\gst\gamma_\mu\Gamma_2)
\end{aligned}$}
&
\makecell[l]{$\displaystyle\small
\begin{aligned}[t]
&\frac{5}{192}\,
(\Gamma_1\gamma^\mu\gs \gamma^\nu\gs \otimes\gst\gamma_\nu\gst\gamma_\mu\Gamma_2)\\
&+\frac{1}{96}\,
(\Gamma_1\gamma^\mu\gs \gamma^\nu\gs \otimes\gst\gamma_\mu\gst\gamma_\nu\Gamma_2)\\
&+\frac{1}{96}\,
(\Gamma_1\gamma^\mu\gs \gamma_\mu\gs \otimes\gst\gamma^\nu\gst\gamma_\nu\Gamma_2)
\end{aligned}$}
\\ \midrule

7 &
\hspace{-1cm}$0$
&
\makecell[l]{$\displaystyle\small
\begin{aligned}[t]
&\frac{1}{48}\,
(\gs \gamma^\mu\gs \gamma_\mu\Gamma_1\gamma^\nu\gst\gamma_\nu\gst\otimes\Gamma_2)\\
&+\frac{1}{48}\,
(\gs \gamma^\mu\gs \gamma^\nu\Gamma_1\gamma_\mu\gst\gamma_\nu\gst\otimes\Gamma_2)\\
&-\frac{1}{24}\,
(\gs \gamma^\mu\gs \gamma^\nu\Gamma_1\gamma_\nu\gst\gamma_\mu\gst\otimes\Gamma_2)
\end{aligned}$}
\\ \midrule

8 &
\hspace{-1cm}$0$
&
\makecell[l]{$\displaystyle\small
\begin{aligned}[t]
&\frac{1}{48}\,
(\Gamma_1\gamma^\mu\gs \gamma^\nu\gs \otimes\Gamma_2\gamma_\nu\gst\gamma_\mu\gst)\\
&-\frac{1}{24}\,
(\Gamma_1\gamma^\mu\gs \gamma^\nu\gs \otimes\Gamma_2\gamma_\mu\gst\gamma_\nu\gst)\\
&+\frac{1}{48}\,
(\Gamma_1\gamma^\mu\gs \gamma_\mu\gs \otimes\Gamma_2\gamma^\nu\gst\gamma_\nu\gst)
\end{aligned}$}
\\ \midrule

9 &
\hspace{-1cm}$0$
&
\makecell[l]{$\displaystyle\small
\begin{aligned}[t]
&\frac{1}{48}\,
(\Gamma_1\gamma^\mu\gs \gamma_\mu\gs \otimes\gst \gamma^\nu\gst\gamma_\nu\Gamma_2)\\
&-\frac{1}{24}\,
(\Gamma_1\gamma^\mu\gs \gamma^\nu\gs \otimes\gst\gamma_\nu\gst\gamma_\mu\Gamma_2)\\
&+\frac{1}{48}\,
(\Gamma_1\gamma^\mu\gs \gamma^\nu\gs \otimes\gst\gamma_\mu\gst\gamma_\nu\Gamma_2)
\end{aligned}$}
\\ \midrule

10 &
\makecell[l]{$\displaystyle\small
\begin{aligned}[t]
&\frac{1}{32}\,
(\gs \gamma^\mu\Gamma_1\gamma_\mu\gs \gamma^\nu\gst\gamma_\nu\gst\otimes\Gamma_2)
\end{aligned}$}
&
\makecell[l]{$\displaystyle\small
\begin{aligned}[t]
&\frac{1}{192}\,
(\gs \gamma^\mu\Gamma_1\gamma_\mu\gs \gamma^\nu\gst\gamma_\nu\gst\otimes\Gamma_2)\\
&-\frac{1}{96}\,
(\gs \gamma^\mu\Gamma_1\gamma^\nu\gs \gamma_\nu\gst\gamma_\mu\gst\otimes\Gamma_2)\\
&-\frac{1}{96}\,
(\gs \gamma^\mu\Gamma_1\gamma^\nu\gs \gamma_\mu\gst\gamma_\nu\gst\otimes\Gamma_2)
\end{aligned}$}
\\ 
\bottomrule[0.7mm]
\end{tabular}
\caption{Divergences of the two-loop diagrams 4--10 for a generic Dirac structure $\Gamma_1\otimes\Gamma_2$.\label{tab:generic-scalar-D14-D10}}
\end{table}

\begin{table}[H]
\centering
\setlength{\tabcolsep}{6pt}
\renewcommand{\arraystretch}{0.65}
\begin{tabular}{@{}c c c@{}}
\toprule[0.7mm]
Diagram & $1/\epsilon^2$ & $1/\epsilon$ \\
\midrule[0.7mm]
11 &
\makecell[l]{$\displaystyle\small
\begin{aligned}[t]
&-\frac{1}{32}\,
(\Gamma_1\gamma^\mu\gs \gamma^\nu\gs \gamma_\nu\gst\otimes\Gamma_2\gamma_\mu \gst)
\end{aligned}$}
&
\makecell[l]{$\displaystyle\small
\begin{aligned}[t]
&-\frac{1}{192}\,
(\Gamma_1\gamma^\mu\gs \gamma^\nu\gs \gamma_\nu\gst\otimes\Gamma_2\gamma_\mu\gst)\\
&+\frac{1}{96}\,
(\Gamma_1\gamma^\mu\gs \gamma^\nu\gs \gamma_\mu\gst\otimes\Gamma_2\gamma_\nu\gst)\\
&+\frac{1}{96}\,
(\Gamma_1\gamma^\mu\gs \gamma_\mu\gs \gamma^\nu\gst\otimes\Gamma_2\gamma_\nu\gst)
\end{aligned}$}
\\ \midrule

12 &
\makecell[l]{$\displaystyle\small
\begin{aligned}[t]
&\frac{1}{32}\,
(\Gamma_1\gamma^\mu\gs\gamma^\nu\gs\gamma_\nu\gst\otimes\gst\gamma_\mu\Gamma_2)
\end{aligned}$}
&
\makecell[l]{$\displaystyle\small
\begin{aligned}[t]
&\frac{1}{192}\,
(\Gamma_1\gamma^\mu\gs\gamma^\nu\gs\gamma_\nu\gst\otimes\gst\gamma_\mu\Gamma_2)\\
&-\frac{1}{96}\,
(\Gamma_1\gamma^\mu\gs\gamma^\nu\gs\gamma_\mu\gst\otimes\gst\gamma_\nu\Gamma_2)\\
&-\frac{1}{96}\,
(\Gamma_1\gamma^\mu\gs\gamma_\mu\gs\gamma^\nu\gst\otimes\gst\gamma_\nu\Gamma_2)
\end{aligned}$}
\\ \midrule

13 &
\makecell[l]{$\displaystyle\small
\begin{aligned}[t]
&\frac{1}{96}\,
(\gs\gamma^\mu\Gamma_1\gamma_\mu\gs\gamma^\nu\gst\gamma_\nu\gst\otimes\Gamma_2)\\
&+\frac{1}{96}\,
(\gs\gamma^\mu\Gamma_1\gamma^\nu\gs\gamma_\nu\gst\gamma_\mu\gst\otimes\Gamma_2)\\
&+\frac{1}{96}\,
(\gs\gamma^\mu\Gamma_1\gamma^\nu\gs\gamma_\mu\gst\gamma_\nu\gst\otimes\Gamma_2)
\end{aligned}$}
&
\makecell[l]{$\displaystyle\small
\begin{aligned}[t]
&\frac{1}{576}\,
(\gs\gamma^\mu\Gamma_1\gamma_\mu\gs\gamma^\nu\gst\gamma_\nu\gst\otimes\Gamma_2)\\
&+\frac{1}{576}\,
(\gs\gamma^\mu\Gamma_1\gamma^\nu\gs\gamma_\nu\gst\gamma_\mu\gst\otimes\Gamma_2)\\
&+\frac{1}{576}\,
(\gs\gamma^\mu\Gamma_1\gamma^\nu\gs\gamma_\mu\gst\gamma_\nu\gst\otimes\Gamma_2)
\end{aligned}$}
\\ \midrule

14 &
\makecell[l]{$\displaystyle\small
\begin{aligned}[t]
&-\frac{1}{96}\,
(\Gamma_1\gamma^\mu\gs\gamma^\nu\gst\gamma_\mu\gs\otimes\Gamma_2\gamma_\nu\gst)\\
&-\frac{1}{96}\,
(\Gamma_1\gamma^\mu\gs\gamma^\nu\gst\gamma_\nu\gs\otimes\Gamma_2\gamma_\mu\gst)\\
&-\frac{1}{96}\,
(\Gamma_1\gamma^\mu\gs\gamma_\mu\gst\gamma^\nu\gs\otimes\Gamma_2\gamma_\nu\gst)
\end{aligned}$}
&
\makecell[l]{$\displaystyle\small
\begin{aligned}[t]
&-\frac{1}{576}\,
(\Gamma_1\gamma^\mu\gs\gamma^\nu\gst\gamma_\mu\gs\otimes\Gamma_2\gamma_\nu\gst)\\
&-\frac{1}{576}\,
(\Gamma_1\gamma^\mu\gs\gamma^\nu\gst\gamma_\nu\gs\otimes\Gamma_2\gamma_\mu\gst)\\
&-\frac{1}{576}\,
(\Gamma_1\gamma^\mu\gs\gamma_\mu\gst\gamma^\nu\gs\otimes\Gamma_2\gamma_\nu\gst)
\end{aligned}$}
\\ \midrule

15 &
\makecell[l]{$\displaystyle\small
\begin{aligned}[t]
&\frac{1}{96}\,
(\Gamma_1\gamma^\mu\gs\gamma^\nu\gst\gamma_\mu\gs\otimes\gst\gamma_\nu\Gamma_2)\\
&+\frac{1}{96}\,
(\Gamma_1\gamma^\mu\gs\gamma^\nu\gst\gamma_\nu\gs\otimes\gst\gamma_\mu\Gamma_2)\\
&+\frac{1}{96}\,
(\Gamma_1\gamma^\mu\gs\gamma_\mu\gst\gamma^\nu\gs\otimes\gst\gamma_\nu\Gamma_2)
\end{aligned}$}
&
\makecell[l]{$\displaystyle\small
\begin{aligned}[t]
&\frac{1}{576}\,
(\Gamma_1\gamma^\mu\gs\gamma^\nu\gst\gamma_\mu\gs\otimes\gst\gamma_\nu\Gamma_2)\\
&+\frac{1}{576}\,
(\Gamma_1\gamma^\mu\gs\gamma^\nu\gst\gamma_\nu\gs\otimes\gst\gamma_\mu\Gamma_2)\\
&+\frac{1}{576}\,
(\Gamma_1\gamma^\mu\gs\gamma_\mu\gst\gamma^\nu\gs\otimes\gst\gamma_\nu\Gamma_2)
\end{aligned}$}
\\ \midrule

16 &
\makecell[l]{$\displaystyle\small
\begin{aligned}[t]
&-\frac{1}{32}\,
(\gs\gamma^\mu\Gamma_1\gamma^\nu\gs\gamma_\mu\gst\otimes\Gamma_2\gamma_\nu\gst)
\end{aligned}$}
&
\makecell[l]{$\displaystyle\small
\begin{aligned}[t]
&-\frac{1}{192}\,
(\gs\gamma^\mu\Gamma_1\gamma^\nu\gs\gamma_\mu\gst\otimes\Gamma_2\gamma_\nu\gst)\\
&+\frac{1}{96}\,
(\gs\gamma^\mu\Gamma_1\gamma^\nu\gs\gamma_\nu\gst\otimes\Gamma_2\gamma_\mu\gst)\\
&+\frac{1}{96}\,
(\gs\gamma^\mu\Gamma_1\gamma_\mu\gs\gamma^\nu\gst\otimes\Gamma_2\gamma_\nu\gst)
\end{aligned}$}
\\
\bottomrule[0.7mm]
\end{tabular}
\caption{Divergences of the two-loop diagrams 11--16 for a generic Dirac structure $\Gamma_1\otimes\Gamma_2$.}\label{tab:generic-scalar-D11-D16}
\end{table}

\begin{table}[H]
\centering
\setlength{\tabcolsep}{6pt}
\renewcommand{\arraystretch}{0.65}
\begin{tabular}{@{}c c c@{}}
\toprule[0.7mm]
Diagram & $1/\epsilon^2$ & $1/\epsilon$ \\
\midrule[0.7mm]
17 &
\makecell[l]{$\displaystyle\small
\begin{aligned}[t]
&-\frac{1}{32}\,
(\gs\gamma^\mu\Gamma_1\gamma_\mu\gst\gamma^\nu\gs\otimes\Gamma_2\gamma_\nu\gst)
\end{aligned}$}
&
\makecell[l]{$\displaystyle\small
\begin{aligned}[t]
&-\frac{1}{192}\,
(\gs\gamma^\mu\Gamma_1\gamma_\mu\gst\gamma^\nu\gs\otimes\Gamma_2\gamma_\nu\gst)\\
&+\frac{1}{96}\,
(\gs\gamma^\mu\Gamma_1\gamma^\nu\gst\gamma_\nu\gs\otimes\Gamma_2\gamma_\mu\gst)\\
&+\frac{1}{96}\,
(\gs\gamma^\mu\Gamma_1\gamma^\nu\gst\gamma_\mu\gs\otimes\Gamma_2\gamma_\nu\gst)
\end{aligned}$}
\\ \midrule

18 &
\makecell[l]{$\displaystyle\small
\begin{aligned}[t]
&\frac{1}{32}\,
(\gs\gamma^\mu\Gamma_1\gamma^\nu\gs\gamma_\mu\gst\otimes\gst\gamma_\nu\Gamma_2)
\end{aligned}$}
&
\makecell[l]{$\displaystyle\small
\begin{aligned}[t]
&\frac{1}{192}\,
(\gs\gamma^\mu\Gamma_1\gamma^\nu\gs\gamma_\mu\gst\otimes\gst\gamma_\nu\Gamma_2)\\
&-\frac{1}{96}\,
(\gs\gamma^\mu\Gamma_1\gamma^\nu\gs\gamma_\nu\gst\otimes\gst\gamma_\mu\Gamma_2)\\
&-\frac{1}{96}\,
(\gs\gamma^\mu\Gamma_1\gamma_\mu\gs\gamma^\nu\gst\otimes\gst\gamma_\nu\Gamma_2)
\end{aligned}$}
\\ \midrule

19 &
\makecell[l]{$\displaystyle\small
\begin{aligned}[t]
&\frac{1}{32}\,
(\gs\gamma^\mu\Gamma_1\gamma_\mu\gst\gamma^\nu\gs\otimes\gst\gamma_\nu\Gamma_2)
\end{aligned}$}
&
\makecell[l]{$\displaystyle\small
\begin{aligned}[t]
&\frac{1}{192}\,
(\gs\gamma^\mu\Gamma_1\gamma_\mu\gst\gamma^\nu\gs\otimes\gst\gamma_\nu\Gamma_2)\\
&-\frac{1}{96}\,
(\gs\gamma^\mu\Gamma_1\gamma^\nu\gst\gamma_\mu\gs\otimes\gst\gamma_\nu\Gamma_2)\\
&-\frac{1}{96}\,
(\gs\gamma^\mu\Gamma_1\gamma^\nu\gst\gamma_\nu\gs\otimes\gst\gamma_\mu\Gamma_2)
\end{aligned}$}
\\ \midrule

20 &
\makecell[l]{$\displaystyle\small
\begin{aligned}[t]
&-\frac{1}{32}\,
(\Gamma_1\gamma^\mu\gs\gamma^\nu\gs\otimes\gst\gamma_\mu\Gamma_2\gamma_\nu\gst)
\end{aligned}$}
&
\makecell[l]{$\displaystyle\small
\begin{aligned}[t]
&-\frac{1}{192}\,
(\Gamma_1\gamma^\mu\gs\gamma^\nu\gs\otimes\gst\gamma_\mu\Gamma_2\gamma_\nu\gst)\\
&+\frac{1}{96}\,
(\Gamma_1\gamma^\mu\gs\gamma^\nu\gs\otimes\gst\gamma_\nu\Gamma_2\gamma_\mu\gst)\\
&+\frac{1}{96}\,
(\Gamma_1\gamma^\mu\gs\gamma_\mu\gs\otimes\gst\gamma^\nu\Gamma_2\gamma_\nu\gst)
\end{aligned}$}
\\ \midrule

21 &
\makecell[l]{$\displaystyle\small
\begin{aligned}[t]
&-\frac{1}{32}\,
(\Gamma_1\gamma^\mu\gs \gamma^\nu\gs\otimes\gst\gamma_\nu\Gamma_2\gamma_\mu\gst)
\end{aligned}$}
&
\makecell[l]{$\displaystyle\small
\begin{aligned}[t]
&-\frac{1}{192}\,
(\Gamma_1\gamma^\mu\gs\gamma^\nu\gs\otimes\gst\gamma_\nu\Gamma_2\gamma_\mu\gst)\\
&+\frac{1}{96}\,
(\Gamma_1\gamma^\mu\gs\gamma^\nu\gs\otimes\gst\gamma_\mu\Gamma_2\gamma_\nu\gst)\\
&+\frac{1}{96}\,
(\Gamma_1\gamma^\mu\gs\gamma_\mu\gs\otimes\gst\gamma^\nu\Gamma_2\gamma_\nu\gst)
\end{aligned}$}
\\ \midrule

22 &
\makecell[l]{$\displaystyle\small
\begin{aligned}[t]
&\frac{1}{16}\,
(\gs\gamma^\mu\Gamma_1\gamma_\mu\gst\otimes\gs\gamma^\nu\Gamma_2\gamma_\nu\gst)
\end{aligned}$}
&
\makecell[l]{$\displaystyle\small
\begin{aligned}[t]
&\hspace{1cm}0
\end{aligned}$}
\\ \midrule

23 &
\makecell[l]{$\displaystyle\small
\begin{aligned}[t]
&\frac{1}{16}\,
(\gs\gamma^\mu\Gamma_1\gamma^\nu\gs\otimes\gst\gamma_\mu\Gamma_2\gamma_\nu\gst)
\end{aligned}$}
&
\makecell[l]{$\displaystyle\small
\begin{aligned}[t]
&\hspace{1cm}0
\end{aligned}$}
\\ \midrule

24 &
\makecell[l]{$\displaystyle\small
\begin{aligned}[t]
&\frac{1}{16}\,
(\gs\gamma^\mu\Gamma_1\gamma^\nu\gs\otimes\gst\gamma_\nu\Gamma_2\gamma_\mu\gst)
\end{aligned}$}
&
\makecell[l]{$\displaystyle\small
\begin{aligned}[t]
&\hspace{1cm}0
\end{aligned}$}
\\

\bottomrule[0.7mm]
\end{tabular}
\caption{Divergences of the two-loop diagrams 17--24 for a generic Dirac structure $\Gamma_1\otimes\Gamma_2$.}\label{tab:generic-scalar-D17-D24}
\end{table}

\subsubsection{Two-loop scalar-gauge-boson contributions}
\hspace{-1cm}
\begin{table}[H]
\centering
\setlength{\tabcolsep}{6pt}
\renewcommand{\arraystretch}{0.65}
\resizebox{0.7\linewidth}{!}{
\begin{tabular}{@{}c c c@{}}
\toprule[0.7mm]
Diagram & $1/\epsilon^2$ & $1/\epsilon$ \\
\midrule[0.7mm]

25 &
\makecell[l]{$\displaystyle\small
\begin{aligned}[t]
&\frac{1}{16}
(\gamma^\mu\gamma^\nu\Gamma_1\gamma_\nu\gs
 \gamma_\mu\gst\otimes\Gamma_2)\\[1mm]
&+\frac{1}{32}
(\gs\gamma^\mu\Gamma_1\gamma_\mu
 \gamma^\nu\gamma_\nu\gst\otimes\Gamma_2)\\[1mm]
&-\frac{1}{32}
(\gs\gamma^\mu\Gamma_1\gamma_\mu
 \gst\gamma^\nu\gamma_\nu\otimes\Gamma_2)
\end{aligned}$}
&
\makecell[l]{$\displaystyle\small
\begin{aligned}[t]
&\frac{1}{32}
(\gamma^\mu\gamma^\nu\Gamma_1\gamma_\nu\gs
 \gamma_\mu\gst\otimes\Gamma_2)\\[1mm]
&+\frac{3}{64}
(\gs\gamma^\mu\Gamma_1\gamma_\mu
 \gamma^\nu\gamma_\nu\gst\otimes\Gamma_2)\\
&+\frac{1}{32}
(\gs\gamma^\mu\Gamma_1\gamma^\nu
 \gamma_\mu\gamma_\nu\gst\otimes\Gamma_2)\\
&+\frac{1}{32}
(\gs\gamma^\mu\Gamma_1\gamma^\nu
 \gamma_\nu\gamma_\mu\gst\otimes\Gamma_2)\\[1mm]
&+\frac{1}{64}
(\gs\gamma^\mu\Gamma_1\gamma_\mu
 \gst\gamma^\nu\gamma_\nu\otimes\Gamma_2)\\
&+\frac{1}{32}
(\gs\gamma^\mu\Gamma_1\gamma^\nu
 \gst\gamma_\nu\gamma_\mu\otimes\Gamma_2)\\
&+\frac{1}{32}
(\gs\gamma^\mu\Gamma_1\gamma^\nu
 \gst\gamma_\mu\gamma_\nu\otimes\Gamma_2)
\end{aligned}$}
\\ \midrule

26 &
\makecell[l]{$\displaystyle\small
\begin{aligned}[t]
&\frac{1}{32}
(\Gamma_1\gamma^\mu\gamma^\nu\gamma_\nu\gs
 \otimes\Gamma_2\gamma_\mu\gst)\\[1mm]
&-\frac{1}{32}
(\Gamma_1\gamma^\mu\gs\gamma^\nu\gamma_\nu
 \otimes\Gamma_2\gamma_\mu\gst)\\[1mm]
&-\frac{1}{16}
(\Gamma_1\gamma^\mu\gs\gamma^\nu\gst
 \otimes\Gamma_2\gamma_\mu\gamma_\nu)
\end{aligned}$}
&
\makecell[l]{$\displaystyle\small
\begin{aligned}[t]
&\frac{3}{64}
(\Gamma_1\gamma^\mu\gamma^\nu\gamma_\nu\gs
 \otimes\Gamma_2\gamma_\mu\gst)\\
&+\frac{1}{32}
(\Gamma_1\gamma^\mu\gamma_\mu\gamma^\nu\gs
 \otimes\Gamma_2\gamma_\nu\gst)\\
&+\frac{1}{32}
(\Gamma_1\gamma^\mu\gamma^\nu\gamma_\mu\gs
 \otimes\Gamma_2\gamma_\nu\gst)\\[1mm]
&+\frac{1}{64}
(\Gamma_1\gamma^\mu\gs\gamma^\nu\gamma_\nu
 \otimes\Gamma_2\gamma_\mu\gst)\\
&+\frac{1}{32}
(\Gamma_1\gamma^\mu\gs\gamma^\nu\gamma_\mu
 \otimes\Gamma_2\gamma_\nu\gst)\\
&+\frac{1}{32}
(\Gamma_1\gamma^\mu\gs\gamma_\mu\gamma^\nu
 \otimes\Gamma_2\gamma_\nu\gst)\\[1mm]
&-\frac{1}{32}
(\Gamma_1\gamma^\mu\gs\gamma^\nu\gst
 \otimes\Gamma_2\gamma_\mu\gamma_\nu)
\end{aligned}$}
\\ \midrule

27 &
\makecell[l]{$\displaystyle\small
\begin{aligned}[t]
&-\frac{1}{32}
(\Gamma_1\gamma^\mu\gamma^\nu\gamma_\nu\gs
 \otimes\gst\gamma_\mu\Gamma_2)\\[1mm]
&+\frac{1}{32}
(\Gamma_1\gamma^\mu\gs\gamma^\nu\gamma_\nu
 \otimes\gst\gamma_\mu\Gamma_2)\\[1mm]
&+\frac{1}{16}
(\Gamma_1\gamma^\mu\gs\gamma^\nu\gst
 \otimes\gamma_\nu\gamma_\mu\Gamma_2)
\end{aligned}$}
&
\makecell[l]{$\displaystyle\small
\begin{aligned}[t]
&-\frac{3}{64}
(\Gamma_1\gamma^\mu\gamma^\nu\gamma_\nu\gs
 \otimes\gst\gamma_\mu\Gamma_2)\\
&-\frac{1}{32}
(\Gamma_1\gamma^\mu\gamma_\mu\gamma^\nu\gs
 \otimes\gst\gamma_\nu\Gamma_2)\\
&-\frac{1}{32}
(\Gamma_1\gamma^\mu\gamma^\nu\gamma_\mu\gs
 \otimes\gst\gamma_\nu\Gamma_2)\\[1mm]
&-\frac{1}{64}
(\Gamma_1\gamma^\mu\gs\gamma^\nu\gamma_\nu
 \otimes\gst\gamma_\mu\Gamma_2)\\
&-\frac{1}{32}
(\Gamma_1\gamma^\mu\gs\gamma^\nu\gamma_\mu
 \otimes\gst\gamma_\nu\Gamma_2)\\
&-\frac{1}{32}
(\Gamma_1\gamma^\mu\gs\gamma_\mu\gamma^\nu
 \otimes\gst\gamma_\nu\Gamma_2)\\[1mm]
&+\frac{1}{32}
(\Gamma_1\gamma^\mu\gs\gamma^\nu\gst
 \otimes\gamma_\nu\gamma_\mu\Gamma_2)
\end{aligned}$}
\\ \midrule

28 &
$0$
&
\makecell[l]{$\displaystyle\small
\begin{aligned}[t]
&\frac{1}{16}
(\gamma^\mu\gamma^\nu\Gamma_1\gamma_\nu\gs
 \otimes\Gamma_2\gamma_\mu\gst)\\
&-\frac{1}{16}
(\gamma^\mu\gamma^\nu\Gamma_1\gamma_\mu\gs
 \otimes\Gamma_2\gamma_\nu\gst)\\[1mm]
&+\frac{1}{16}
(\gs\gamma^\mu\Gamma_1\gamma_\mu\gamma^\nu
 \otimes\Gamma_2\gamma_\nu\gst)\\
&-\frac{1}{16}
(\gs\gamma^\mu\Gamma_1\gamma^\nu\gamma_\mu
 \otimes\Gamma_2\gamma_\nu\gst)\\[1mm]
&+\frac{1}{16}
(\gs\gamma^\mu\Gamma_1\gamma^\nu\gst
 \otimes\Gamma_2\gamma_\mu\gamma_\nu)\\
&-\frac{1}{16}
(\gs\gamma^\mu\Gamma_1\gamma^\nu\gst
 \otimes\Gamma_2\gamma_\nu\gamma_\mu)
\end{aligned}$}
\\

\bottomrule[0.7mm]
\end{tabular}
}
\caption{Divergences of the two-loop diagrams 25--28 for a generic Dirac structure $\Gamma_1\otimes\Gamma_2$.}
\label{tab:generic_SSG_D25_D28}
\end{table}

\subsubsection{Two-loop self-energy contributions}

The two-loop self-energy diagrams contain five contributions related to the one-loop topologies shown in \cref{fig:scalar-self-en}. The divergences for the diagrams 29-31 are reported in the Tables.~\ref{tab:genericScalar29}--\ref{tab:genericScalar31}.

\begin{table}[H]
\centering
\setlength{\tabcolsep}{8pt}
\renewcommand{\arraystretch}{0.9}
\begin{tabular}{@{}c c c@{}}
\toprule[0.7mm]
Diagram & $1/\epsilon^2$ & $1/\epsilon$ \\
\midrule[0.7mm]

29a &
\makecell[l]{$\displaystyle\small
\begin{aligned}[t]
&\frac{1}{72}\,
\operatorname{Tr}\!\left[\gs\gamma^\mu\gst\gamma_\mu\right]\,
(\gs\gamma^\nu\Gamma_1\gamma_\nu\gst\otimes\Gamma_2)\\
&+\frac{1}{288}\,
\operatorname{Tr}\!\left[\gs\gamma^\mu\gst\gamma^\nu\right]\,
(\gs\gamma_\mu\Gamma_1\gamma_\nu\gst\otimes\Gamma_2)\\
&+\frac{1}{288}\,
\operatorname{Tr}\!\left[\gs\gamma^\mu\gst\gamma^\nu\right]\,
(\gs\gamma_\nu\Gamma_1\gamma_\mu\gst\otimes\Gamma_2)
\end{aligned}$}
&
\makecell[l]{$\displaystyle\small
\begin{aligned}[t]
&\frac{5}{108}\,
\operatorname{Tr}\!\left[\gs\gamma^\mu\gst\gamma_\mu\right]\,
(\gs\gamma^\nu\Gamma_1\gamma_\nu\gst\otimes\Gamma_2)\\
&-\frac{1}{1728}\,
\operatorname{Tr}\!\left[\gs\gamma^\mu\gst\gamma^\nu\right]\,
(\gs\gamma_\mu\Gamma_1\gamma_\nu\gst\otimes\Gamma_2)\\
&-\frac{1}{1728}\,
\operatorname{Tr}\!\left[\gs\gamma^\mu\gst\gamma^\nu\right]\,
(\gs\gamma_\nu\Gamma_1\gamma_\mu\gst\otimes\Gamma_2)
\end{aligned}$}
\\ \midrule

29b &
\makecell[l]{$\displaystyle\small
\begin{aligned}[t]
&\hspace{1.6cm}0
\end{aligned}$}
&
\makecell[l]{$\displaystyle\small
\begin{aligned}[t]
&-\frac{1}{24}\,
(\gs\gamma^\mu\Gamma_1\gamma_\mu\gst\otimes\Gamma_2)
\end{aligned}$}
\\ \midrule

29c &
\makecell[l]{$\displaystyle\small
\begin{aligned}[t]
&-\frac{1}{4}\,
(\gs\gamma^\mu\Gamma_1\gamma_\mu\gst\otimes\Gamma_2)
\end{aligned}$}
&
\makecell[l]{$\displaystyle\small
\begin{aligned}[t]
&\frac{5}{24}\,
(\gs\gamma^\mu\Gamma_1\gamma_\mu\gst\otimes\Gamma_2)
\end{aligned}$}
\\ \midrule

29d &
\makecell[l]{$\displaystyle\small
\begin{aligned}[t]
&\hspace{1.6cm}0
\end{aligned}$}
&
\makecell[l]{$\displaystyle\small
\begin{aligned}[t]
&-\frac{1}{24}\,
(\gs\gamma^\mu\Gamma_1\gamma_\mu\gst\otimes\Gamma_2)
\end{aligned}$}
\\ \midrule

29e &
\makecell[l]{$\displaystyle\small
\begin{aligned}[t]
&\hspace{1.6cm}0
\end{aligned}$}
&
\makecell[l]{$\displaystyle\small
\begin{aligned}[t]
&-\frac{1}{6}\,
(\gs\gamma^\mu\Gamma_1\gamma_\mu\gst\otimes\Gamma_2)
\end{aligned}$}
\\

\bottomrule[0.7mm]
\end{tabular}
\caption{Divergences of the two-loop scalar self-energy diagram 29 for a generic Dirac structure $\Gamma_1\otimes\Gamma_2$. The individual self-energy contributions from the fermion bubble, scalar bubble, scalar--gauge bubble, scalar tadpole, and gauge tadpole are denoted by a, b, c, d and e, respectively.}
\label{tab:genericScalar29}
\end{table}

\begin{table}[H]
\centering
\setlength{\tabcolsep}{8pt}
\renewcommand{\arraystretch}{0.9}
\begin{tabular}{@{}c c c@{}}
\toprule[0.7mm]
Diagram & $1/\epsilon^2$ & $1/\epsilon$ \\
\midrule[0.7mm]

30a &
\makecell[l]{$\displaystyle\small
\begin{aligned}[t]
&-\frac{1}{72}\,
\operatorname{Tr}\!\left[\gs\gamma^\mu\gst\gamma_\mu\right]\,
(\Gamma_1\gamma^\nu\gs\otimes\Gamma_2\gamma_\nu\gst)\\
&-\frac{1}{288}\,
\operatorname{Tr}\!\left[\gs\gamma^\mu\gst\gamma^\nu\right]\,
(\Gamma_1\gamma_\mu\gs\otimes\Gamma_2\gamma_\nu\gst)\\
&-\frac{1}{288}\,
\operatorname{Tr}\!\left[\gs\gamma^\mu\gst\gamma^\nu\right]\,
(\Gamma_1\gamma_\nu\gs\otimes\Gamma_2\gamma_\mu\gst)
\end{aligned}$}
&
\makecell[l]{$\displaystyle\small
\begin{aligned}[t]
&-\frac{5}{108}\,
\operatorname{Tr}\!\left[\gs\gamma^\mu\gst\gamma_\mu\right]\,
(\Gamma_1\gamma^\nu\gs\otimes\Gamma_2\gamma_\nu\gst)\\
&+\frac{1}{1728}\,
\operatorname{Tr}\!\left[\gs\gamma^\mu\gst\gamma^\nu\right]\,
(\Gamma_1\gamma_\mu\gs\otimes\Gamma_2\gamma_\nu\gst)\\
&+\frac{1}{1728}\,
\operatorname{Tr}\!\left[\gs\gamma^\mu\gst\gamma^\nu\right]\,
(\Gamma_1\gamma_\nu\gs\otimes\Gamma_2\gamma_\mu\gst)
\end{aligned}$}
\\ \midrule

30b &
\makecell[l]{$\displaystyle\small
\begin{aligned}[t]
&\hspace{1.6cm}0
\end{aligned}$}
&
\makecell[l]{$\displaystyle\small
\begin{aligned}[t]
&\frac{1}{24}\,
(\Gamma_1\gamma^\mu\gs\otimes\Gamma_2\gamma_\mu\gst)
\end{aligned}$}
\\ \midrule

30c &
\makecell[l]{$\displaystyle\small
\begin{aligned}[t]
&\frac{1}{4}\,
(\Gamma_1\gamma^\mu\gs\otimes\Gamma_2\gamma_\mu\gst)
\end{aligned}$}
&
\makecell[l]{$\displaystyle\small
\begin{aligned}[t]
&-\frac{5}{24}\,
(\Gamma_1\gamma^\mu\gs\otimes\Gamma_2\gamma_\mu\gst)
\end{aligned}$}
\\ \midrule

30d &
\makecell[l]{$\displaystyle\small
\begin{aligned}[t]
&\hspace{1.6cm}0
\end{aligned}$}
&
\makecell[l]{$\displaystyle\small
\begin{aligned}[t]
&\frac{1}{24}\,
(\Gamma_1\gamma^\mu\gs\otimes\Gamma_2\gamma_\mu\gst)
\end{aligned}$}
\\ \midrule

30e &
\makecell[l]{$\displaystyle\small
\begin{aligned}[t]
&\hspace{1.6cm}0
\end{aligned}$}
&
\makecell[l]{$\displaystyle\small
\begin{aligned}[t]
&\frac{1}{6}\,
(\Gamma_1\gamma^\mu\gs\otimes\Gamma_2\gamma_\mu\gst)
\end{aligned}$}
\\

\bottomrule[0.7mm]
\end{tabular}
\caption{Divergences of the two-loop scalar self-energy diagram 30 for a generic Dirac structure $\Gamma_1\otimes\Gamma_2$. The individual self-energy contributions from the fermion bubble, scalar bubble, scalar--gauge bubble, scalar tadpole, and gauge tadpole are denoted by a, b, c, d and e, respectively.}
\label{tab:genericScalar30}
\end{table}

\begin{table}[H]
\centering
\setlength{\tabcolsep}{8pt}
\renewcommand{\arraystretch}{0.9}
\begin{tabular}{@{}c c c@{}}
\toprule[0.7mm]
Diagram & $1/\epsilon^2$ & $1/\epsilon$ \\
\midrule[0.7mm]

31a &
\makecell[l]{$\displaystyle\small
\begin{aligned}[t]
&\frac{1}{72}\,
\operatorname{Tr}\!\left[\gs\gamma^\mu\gst\gamma_\mu\right]\,
(\Gamma_1\gamma^\nu\gs\otimes\gst\gamma_\nu\Gamma_2)\\
&+\frac{1}{288}\,
\operatorname{Tr}\!\left[\gs\gamma^\mu\gst\gamma^\nu\right]\,
(\Gamma_1\gamma_\nu\gs\otimes\gst\gamma_\mu\Gamma_2)\\
&+\frac{1}{288}\,
\operatorname{Tr}\!\left[\gs\gamma^\mu\gst\gamma^\nu\right]\,
(\Gamma_1\gamma_\mu\gs\otimes\gst\gamma_\nu\Gamma_2)
\end{aligned}$}
&
\makecell[l]{$\displaystyle\small
\begin{aligned}[t]
&\frac{5}{108}\,
\operatorname{Tr}\!\left[\gs\gamma^\mu\gst\gamma_\mu\right]\,
(\Gamma_1\gamma^\nu\gs\otimes\gst\gamma_\nu\Gamma_2)\\
&-\frac{1}{1728}\,
\operatorname{Tr}\!\left[\gs\gamma^\mu\gst\gamma^\nu\right]\,
(\Gamma_1\gamma_\nu\gs\otimes\gst\gamma_\mu\Gamma_2)\\
&-\frac{1}{1728}\,
\operatorname{Tr}\!\left[\gs\gamma^\mu\gst\gamma^\nu\right]\,
(\Gamma_1\gamma_\mu\gs\otimes\gst\gamma_\nu\Gamma_2)
\end{aligned}$}
\\ \midrule

31b &
\makecell[l]{$\displaystyle\small
\begin{aligned}[t]
&\hspace{1.6cm}0
\end{aligned}$}
&
\makecell[l]{$\displaystyle\small
\begin{aligned}[t]
&-\frac{1}{24}\,
(\Gamma_1\gamma^\mu\gs\otimes\gst\gamma_\mu\Gamma_2)
\end{aligned}$}
\\ \midrule

31c &
\makecell[l]{$\displaystyle\small
\begin{aligned}[t]
&-\frac{1}{4}\,
(\Gamma_1\gamma^\mu\gs\otimes\gst\gamma_\mu\Gamma_2)
\end{aligned}$}
&
\makecell[l]{$\displaystyle\small
\begin{aligned}[t]
&\frac{5}{24}\,
(\Gamma_1\gamma^\mu\gs\otimes\gst\gamma_\mu\Gamma_2)
\end{aligned}$}
\\ \midrule

31d &
\makecell[l]{$\displaystyle\small
\begin{aligned}[t]
&\hspace{1.6cm}0
\end{aligned}$}
&
\makecell[l]{$\displaystyle\small
\begin{aligned}[t]
&-\frac{1}{24}\,
(\Gamma_1\gamma^\mu\gs\otimes\gst\gamma_\mu\Gamma_2)
\end{aligned}$}
\\ \midrule

31e &
\makecell[l]{$\displaystyle\small
\begin{aligned}[t]
&\hspace{1.6cm}0
\end{aligned}$}
&
\makecell[l]{$\displaystyle\small
\begin{aligned}[t]
&-\frac{1}{6}\,
(\Gamma_1\gamma^\mu\gs\otimes\gst\gamma_\mu\Gamma_2)
\end{aligned}$}
\\

\bottomrule[0.7mm]
\end{tabular}
\caption{Divergences of the two-loop scalar self-energy diagram 31 for a generic Dirac structure $\Gamma_1\otimes\Gamma_2$. The individual self-energy contributions from the fermion bubble, scalar bubble, scalar--gauge bubble, scalar tadpole, and gauge tadpole are denoted by a, b, c, d and e, respectively.}
\label{tab:genericScalar31}
\end{table}

\newpage
\section{Multiplicity for $SU(N_1)\times SU(N_2)$}\label{app:multiplicity}

In this section the multiplicity factors for the case of two different $SU(N)$ groups are reported. They are given for the one-loop diagrams (\cref{fig:oneloopgauge}) and the two-loop diagrams (\cref{fig:twoloopgauge}) in \cref{tab_multSUN}.
In our notation the external fermion lines are labelled according to 
\begin{equation}
    \mathcal{O}_{1234} = \left(\bar{\psi}_1 \Gamma_1 \psi_2\right)\left(\bar{\psi}_3 \Gamma_2 \psi_4\right) \,.
\end{equation}
Furthermore, the factor $\delta_{ij}^{N_k}$ indicates whether the fermion lines $i$ and
$j$ are both charged under the gauge group $SU(N_k)$. 
\begin{equation}
\delta_{ij}^{N_k}\, \equiv \,\,
\begin{cases}
1\,, & \text{fermions } i \text{ and } j \text{ transform under } SU(N_k),\\
0\,, & \text{otherwise}.
\end{cases}
\end{equation}
The symbol ``perm.'' denotes the additional terms obtained by exchanging the two gauge groups, $N_1\leftrightarrow N_2$.  The results are also valid where there is a single $SU(N)$ gauge group, in which case the permuted contributions need to be dropped.

\begin{table}[H]
\centering
\setlength{\tabcolsep}{8pt}
\renewcommand{\arraystretch}{0.6} 
\begin{tabular}{@{}c c@{}}
\toprule[0.7mm]
Diagram & Multiplicity \\
\midrule[0.7mm]
1 & $\delta_{12}^{N_1}+\delta_{34}^{N_1}$  \\ \midrule
2 & $\delta_{24}^{N_1}+\delta_{13}^{N_1}$  \\ \midrule
3 & $\delta_{14}^{N_1}+\delta_{23}^{N_1}$ \\ \midrule
4  & $\delta_{12}^{N_1}\delta_{12}^{N_2}+\delta_{34}^{N_1}\delta_{34}^{N_2}+\text{perm.}$ \\ \midrule
5  & $\delta_{24}^{N_1}\delta_{24}^{N_2}+\delta_{13}^{N_1}\delta_{13}^{N_2}+\text{perm.}$  \\ \midrule
6  & $\delta_{23}^{N_1}\delta_{23}^{N_2}+\delta_{14}^{N_1}\delta_{14}^{N_2}+\text{perm.}$ \\ \midrule
7  & $\delta_{12}^{N_1}\delta_{12}^{N_2}+\delta_{34}^{N_1}\delta_{34}^{N_2}+\text{perm.}$ \\ \midrule
8  & $\delta_{24}^{N_1}\delta_{24}^{N_2}+\delta_{13}^{N_1}\delta_{13}^{N_2}+\text{perm.}$ \\ \midrule
9  & $\delta_{23}^{N_1}\delta_{23}^{N_2}+\delta_{14}^{N_1}\delta_{14}^{N_2}+\text{perm.}$ \\ \midrule
10 & $\delta_{12}^{N_1}\delta_{22}^{N_2}+\delta_{12}^{N_1}\delta_{11}^{N_2}+\delta_{34}^{N_1}\delta_{44}^{N_2}+\delta_{34}^{N_1}\delta_{33}^{N_2}+\text{perm.}$ \\ \midrule
11 & $\delta_{24}^{N_1}\delta_{22}^{N_2}+\delta_{24}^{N_1}\delta_{44}^{N_2}+\delta_{13}^{N_1}\delta_{11}^{N_2}+\delta_{13}^{N_1}\delta_{33}^{N_2}+\text{perm.}$ \\ \midrule
12 & $\delta_{23}^{N_1}\delta_{22}^{N_2}+\delta_{23}^{N_1}\delta_{33}^{N_2}+\delta_{14}^{N_1}\delta_{11}^{N_2}+\delta_{14}^{N_1}\delta_{44}^{N_2}+\text{perm.}$ \\ \midrule
13 & $\delta_{12}^{N_1}\delta_{22}^{N_2}+\delta_{12}^{N_1}\delta_{11}^{N_2}+\delta_{34}^{N_1}\delta_{33}^{N_2}+\delta_{34}^{N_1}\delta_{44}^{N_2}+\text{perm.}$ \\ \midrule
14 & $\delta_{24}^{N_1}\delta_{22}^{N_2}+\delta_{24}^{N_1}\delta_{44}^{N_2}+\delta_{13}^{N_1}\delta_{11}^{N_2}+\delta_{13}^{N_1}\delta_{33}^{N_2}+\text{perm.}$ \\ \midrule
15 & $\delta_{23}^{N_1}\delta_{22}^{N_2}+\delta_{23}^{N_1}\delta_{33}^{N_2}+\delta_{14}^{N_1}\delta_{44}^{N_2}+\delta_{14}^{N_1}\delta_{11}^{N_2}+\text{perm.}$ \\ \midrule
16 & $\delta_{12}^{N_1}\delta_{24}^{N_2}+\delta_{12}^{N_1}\delta_{13}^{N_2}+\delta_{34}^{N_1}\delta_{24}^{N_2}+\delta_{34}^{N_1}\delta_{13}^{N_2}+\text{perm.}$ \\ \midrule
17 & $\delta_{12}^{N_1}\delta_{24}^{N_2}+\delta_{12}^{N_1}\delta_{13}^{N_2}+\delta_{34}^{N_1}\delta_{24}^{N_2}+\delta_{34}^{N_1}\delta_{13}^{N_2}+\text{perm.}$ \\ \midrule
18 & $\delta_{12}^{N_1}\delta_{23}^{N_2}+\delta_{12}^{N_1}\delta_{14}^{N_2}+\delta_{34}^{N_1}\delta_{23}^{N_2}+\delta_{34}^{N_1}\delta_{14}^{N_2}+\text{perm.}$ \\ \midrule
19 & $\delta_{12}^{N_1}\delta_{23}^{N_2}+\delta_{12}^{N_1}\delta_{14}^{N_2}+\delta_{34}^{N_1}\delta_{23}^{N_2}+\delta_{34}^{N_1}\delta_{14}^{N_2}+\text{perm.}$ \\ \midrule
20 & $\delta_{23}^{N_1}\delta_{24}^{N_2}+\delta_{24}^{N_1}\delta_{14}^{N_2}+\delta_{13}^{N_1}\delta_{14}^{N_2}+\delta_{13}^{N_1}\delta_{23}^{N_2}+\text{perm.}$ \\ \midrule
21 & $\delta_{23}^{N_1}\delta_{24}^{N_2}+\delta_{24}^{N_1}\delta_{14}^{N_2}+\delta_{13}^{N_1}\delta_{14}^{N_2}+\delta_{13}^{N_1}\delta_{23}^{N_2}+\text{perm.}$ \\ \midrule
22 & $\delta_{12}^{N_1}\delta_{34}^{N_2}+\text{perm.}$ \\ \midrule
23 & $\delta_{13}^{N_1}\delta_{24}^{N_2}+\text{perm.}$ \\ \midrule
24 & $\delta_{14}^{N_1}\delta_{23}^{N_2}+\text{perm.}$ \\ \midrule
25 & $2\delta_{12}^{N_1}+2\delta_{34}^{N_1}$ \\ \midrule
26 & $2\delta_{24}^{N_1}+2\delta_{13}^{N_1}$ \\ \midrule
27 & $2\delta_{14}^{N_1}+2\delta_{23}^{N_1}$ \\ \midrule
28 & $\delta_{24}^{N_1}\delta_{11}^{N_1}+\delta_{24}^{N_1}\delta_{33}^{N_1}+\delta_{13}^{N_1}\delta_{22}^{N_1}+\delta_{13}^{N_1}\delta_{44}^{N_1}$ \\ \midrule
29 & $\delta_{12}^{N_1}+\delta_{34}^{N_1}$ \\ \midrule
30 & $\delta_{24}^{N_1}+\delta_{13}^{N_1}$ \\ \midrule
31 & $\delta_{14}^{N_1}+\delta_{23}^{N_1}$ \\
\bottomrule[0.7mm]
\end{tabular}
\caption{Multiplicity factors for $SU(N_1)\times SU(N_2)$ one-loop and two-loop corrections. Permutations are understood as swapping only $N_1 \leftrightarrow N_2$. If there is a single $SU(N)$ gauge group, these terms are absent. Entries without ``$+ \text{ perm.}$'' are only relevant for diagrams that involve a single $SU(N)$ factor.}
\label{tab_multSUN}
\end{table}

\section{Color, charge and Yukawa factors}\label{app:colorfactors}
In this section the color and charge factors that result from the gauge interactions discussed in \cref{app:genericGauge} as well as the Yukawa factors resulting from scalar interactions in \cref{app:genericscalar} are reported. 

\subsection{General gauge factors}\label{app:gencol}
In this subsection we report the color factors resulting from a generic gauge group $\mathcal G$. The covariant derivative of a fermion $\psi$ is assumed to have the following form:

\begin{equation}
    D_\mu \psi = (\partial_\mu +i g_{\mathcal G} G_\mu^aT^a)\psi\,,
\end{equation}
where the general gauge field is denoted by $G_\mu^a$ and the generators satisfy the commutation relation:

\begin{equation}
[T^a,T^b] = i f^{abc}T^c\,,
\end{equation}
with the structure constants $f^{abc}$. The gauge contractions are reported for a generic four-fermion operator of the form
\begin{equation}
    (\bar\psi_1 \Gamma_1 C_1 \psi_2)(\bar\psi_3 \Gamma_2 C_2 \psi_4)\,,
\end{equation} 
where $C_1$ and $C_2$ are the color structures of the two bilinears. The gauge contractions are given for all one-loop and two-loop diagrams in \cref{tab:generalcolorfac} and \cref{tab:generalcolorfacSE}. We denote by $T_i^a$ the generator associated with the fermion on the $i$th external leg. Furthermore, we stress that, for fermion-loop self-energy diagrams (29a, 30a and 31a), the factor $\mathrm{Tr}(T^aT^b)$ is not included, since it is already contained in the $N_T$ factor multiplying the corresponding pole. For the tri-gauge diagrams (25--28), contractions with $f^{abc}$ generate an explicit factor of $i$. This factor has already been taken into account in the pole tables, such that the tabulated entries are real. Finally, diagrams 29d, 30d and 31d carry an additional minus sign from the closed ghost loop, which is not included in the gauge contraction and must be supplied separately to agree with the pole convention reported in the tables.

\begin{table}[H]
\centering
\setlength{\tabcolsep}{8pt}
\renewcommand{\arraystretch}{0.6}
\begin{tabular}{@{}c c@{}}
\toprule[0.7mm]
Diagram & Contraction \\ 
\midrule[0.7mm]
1   & $(T^a_1 C_1 T^a_2)_{\alpha\beta}\,(C_2)_{\gamma\delta}$ \\ \midrule
2   & $(C_1 T^a_2)_{\alpha\beta}\,(C_2 T^a_4)_{\gamma\delta}$ \\ \midrule
3   & $(C_1 T^a_2)_{\alpha\beta}\,(T^a_3 C_2)_{\gamma\delta}$ \\ \midrule
4   & $(T^a_1 T^b_1 C_1 T^b_2 T^a_2)_{\alpha\beta}\,(C_2)_{\gamma\delta}$ \\ \midrule
5   & $(C_1 T^a_2 T^b_2)_{\alpha\beta}\,(C_2 T^a_4 T^b_4)_{\gamma\delta}$ \\ \midrule
6   & $(C_1 T^b_2 T^a_2)_{\alpha\beta}\,(T^a_3 T^b_3 C_2)_{\gamma\delta}$ \\ \midrule
7   & $(T^a_1 T^b_1 C_1 T^a_2 T^b_2)_{\alpha\beta}\,(C_2)_{\gamma\delta}$ \\ \midrule
8   & $(C_1 T^a_2 T^b_2)_{\alpha\beta}\,(C_2 T^b_4 T^a_4)_{\gamma\delta}$ \\ \midrule
9   & $(C_1 T^a_2 T^b_2)_{\alpha\beta}\,(T^a_3 T^b_3 C_2)_{\gamma\delta}$ \\ \midrule
10  & $(T^a_1 C_1 T^b_2 T^a_2 T^b_2)_{\alpha\beta}\,(C_2)_{\gamma\delta}$ \\ \midrule
11  & $(C_1 T^b_2 T^a_2 T^b_2)_{\alpha\beta}\,(C_2 T^a_4)_{\gamma\delta}$ \\ \midrule
12  & $(C_1 T^b_2 T^a_2 T^b_2)_{\alpha\beta}\,(T^a_3 C_2)_{\gamma\delta}$ \\ \midrule
13  & $(T^a_1 C_1 T^b_2 T^b_2 T^a_2)_{\alpha\beta}\,(C_2)_{\gamma\delta}$ \\ \midrule
14  & $(C_1 T^b_2 T^b_2 T^a_2)_{\alpha\beta}\,(C_2 T^a_4)_{\gamma\delta}$ \\ \midrule
15  & $(C_1 T^b_2 T^b_2 T^a_2)_{\alpha\beta}\,(T^a_3 C_2)_{\gamma\delta}$ \\ \midrule
16  & $(T^a_1 C_1 T^b_2 T^a_2)_{\alpha\beta}\,(C_2 T^b_4)_{\gamma\delta}$ \\ \midrule
17  & $(T^a_1 C_1 T^a_2 T^b_2)_{\alpha\beta}\,(C_2 T^b_4)_{\gamma\delta}$ \\ \midrule
18  & $(T^a_1 C_1 T^b_2 T^a_2)_{\alpha\beta}\,(T^b_3 C_2)_{\gamma\delta}$ \\ \midrule
19  & $(T^a_1 C_1 T^a_2 T^b_2)_{\alpha\beta}\,(T^b_3 C_2)_{\gamma\delta}$ \\ \midrule
20  & $(C_1 T^a_2 T^b_2)_{\alpha\beta}\,(T^a_3 C_2 T^b_4)_{\gamma\delta}$ \\ \midrule
21  & $(C_1 T^b_2 T^a_2)_{\alpha\beta}\,(T^a_3 C_2 T^b_4)_{\gamma\delta}$ \\ \midrule
22  & $(T^a_1 C_1 T^a_2)_{\alpha\beta}\,(T^b_3 C_2 T^b_4)_{\gamma\delta}$ \\ \midrule
23  & $(T^a_1 C_1 T^b_2)_{\alpha\beta}\,(T^a_3 C_2 T^b_4)_{\gamma\delta}$ \\ \midrule
24  & $(T^a_1 C_1 T^b_2)_{\alpha\beta}\,(T^b_3 C_2 T^a_4)_{\gamma\delta}$ \\ \midrule
25  & $f^{abc}\,(T^a_1 C_1 T^b_2 T^c_2)_{\alpha\beta}\,(C_2)_{\gamma\delta}$ \\ \midrule
26  & $ f^{abc}\,(C_1 T^a_2 T^b_2)_{\alpha\beta}\,(C_2 T^c_4)_{\gamma\delta}$ \\ \midrule
27  & $f^{abc}\,(C_1 T^a_2 T^b_2)_{\alpha\beta}\,(T^c_3 C_2)_{\gamma\delta}$ \\ \midrule
28  & $f^{abc}\,(T^a_1 C_1 T^b_2)_{\alpha\beta}\,(C_2 T^c_4)_{\gamma\delta}$ \\ 
\bottomrule[0.7mm]
\end{tabular}
\caption{General gauge contractions for the one- and two-loop diagrams 1--28. The indices $\alpha\beta$ and $\gamma\delta$ denote the color indices of the first and second bilinear, respectively.\label{tab:generalcolorfac}}
\end{table}

\begin{table}[H]
\centering
\setlength{\tabcolsep}{8pt}
\renewcommand{\arraystretch}{0.6}
\begin{tabular}{@{}c c@{}}
\toprule[0.7mm]
Diagram & Contraction \\ 
\midrule[0.7mm]
29a & $(T^a_1 C_1 T^b_2)_{\alpha\beta}\,(C_2)_{\gamma\delta}$ \\ \midrule
29b & $f^{acd} f^{bcd}\,(T^a_1 C_1 T^b_2)_{\alpha\beta}\,(C_2)_{\gamma\delta}$ \\ \midrule
29c & $f^{acd} f^{bcd}\,(T^a_1 C_1 T^b_2)_{\alpha\beta}\,(C_2)_{\gamma\delta}$ \\ \midrule
29d & $f^{acd} f^{bcd}\,(T^a_1 C_1 T^b_2)_{\alpha\beta}\,(C_2)_{\gamma\delta}$ \\ \midrule
29e & ${\cal T}_{\phi}^{ab}\,
(T^a_1 C_1 T^b_2)_{\alpha\beta}\,(C_2)_{\gamma\delta}$ \\ \midrule
29f & $2{\cal T}_{\phi}^{ab}\,
(T^a_1 C_1 T^b_2)_{\alpha\beta}\,(C_2)_{\gamma\delta}$ \\ \midrule
30a & $(C_1 T^a_2)_{\alpha\beta}\,(C_2T^b_4)_{\gamma\delta}$ \\ \midrule
30b & $f^{acd} f^{bcd}\,(C_1 T^a_2)_{\alpha\beta}\,(C_2T^b_4)_{\gamma\delta}$ \\ \midrule
30c & $f^{acd} f^{bcd}\,(C_1 T^a_2)_{\alpha\beta}\,(C_2T^b_4)_{\gamma\delta}$ \\ \midrule
30d & $f^{acd} f^{bcd}\,(C_1 T^a_2)_{\alpha\beta}\,(C_2T^b_4)_{\gamma\delta}$ \\ \midrule
30e & ${\cal T}_{\phi}^{ab}\,
(C_1 T^a_2)_{\alpha\beta}\,(C_2T^b_4)_{\gamma\delta}$ \\ \midrule
30f & $2{\cal T}_{\phi}^{ab}\,
(C_1 T^a_2)_{\alpha\beta}\,(C_2T^b_4)_{\gamma\delta}$ \\ \midrule
31a & $(C_1 T^a_2)_{\alpha\beta}\,(T^b_3C_2)_{\gamma\delta}$ \\ \midrule
31b & $f^{acd} f^{bcd}\,(C_1 T^a_2)_{\alpha\beta}\,(T^b_3C_2)_{\gamma\delta}$ \\ \midrule
31c & $f^{acd} f^{bcd}\,(C_1 T^a_2)_{\alpha\beta}\,(T^b_3C_2)_{\gamma\delta}$ \\ \midrule
31d & $f^{acd} f^{bcd}\,(C_1 T^a_2)_{\alpha\beta}\,(T^b_3C_2)_{\gamma\delta}$ \\ \midrule
31e & ${\cal T}_{\phi}^{ab}\,
(C_1 T^a_2)_{\alpha\beta}\,(T^b_3C_2)_{\gamma\delta}$ \\ \midrule
31f & $2{\cal T}_{\phi}^{ab}\,
(C_1 T^a_2)_{\alpha\beta}\,(T^b_3C_2)_{\gamma\delta}$ \\
\bottomrule[0.7mm]
\end{tabular}
\caption{General gauge contractions for the two-loop gauge-boson self-energy diagrams 29--31. The indices $\alpha\beta$ and $\gamma\delta$ denote the color indices of the first and second bilinear, respectively. \label{tab:generalcolorfacSE}}
\end{table}

For scalar multiplets, we define
\begin{equation}
{\cal T}_{\phi}^{ab}
\equiv
\sum_s d_s\,\operatorname{Tr}\!\left(\theta_s^a\theta_s^b\right)
=
N_{T,\phi}\,\delta^{ab},
\qquad
N_{T,\phi}\equiv\sum_s d_s\,T(R_s)\,,
\end{equation}
where $d_s$ denotes the multiplicity under spectator gauge groups.

\subsection{$SU(N)$ color factors for fermion wave-function renormalization}
The color factors of the one-loop and two-loop diagrams of \cref{tab:wfr-open-twogauge} relevant for the fermion  wave-function renormalization are given in \cref{tab:wfr-bare-color}. 

\begin{table}[b]
\centering
\setlength{\tabcolsep}{10pt}
\renewcommand{\arraystretch}{1.05}
\begin{tabular}{@{}c c@{}}
\toprule[0.7mm]
Diagram & Color factor \\
\midrule[0.7mm] 

One-loop&
$C_F$
\\
\midrule 

Nested &
$C_F^2$
\\
\midrule 

Crossed &
$C_F\left(C_F-C_A/2\right)$
\\
\midrule 

Triple-gluon &
$\dfrac{1}{2}C_AC_F$
\\
\midrule 

Fermion bubble &
$C_F$
\\
\midrule 

Ghost bubble &
$-C_A C_F$
\\
\midrule 

Gauge-boson bubble &
$C_A C_F$
\\
\midrule 

Gauge tadpole &
$C_A C_F$
\\

\bottomrule[0.7mm]
\end{tabular}
\caption{$SU(N)$ factors for the diagrams contributing to the fermion wave-function renormalization.}
\label{tab:wfr-bare-color}
\end{table}

\subsection{$SU(N)^2$ color factors}
\label{app:colorfactorsSUN}
In this subsection the color factors for $SU(N)^2$ are reported in the following color basis:
\begin{equation}
    \mathcal{O}_1 = (\bar \psi^\alpha \Gamma_1 \psi^\alpha)(\bar \psi^\beta \Gamma_2 \psi^\beta)\,,\quad  \mathcal{O}_8 = (T^A)_{\alpha \beta}(T^A)_{\rho \sigma}(\bar \psi^\alpha \Gamma_1 \psi^\beta)(\bar \psi^\rho \Gamma_2 \psi^\sigma)\,.
\end{equation}
Other types of bases are easily obtained using the $SU(N)$ completeness relation. The results for all one-loop and two-loop diagrams are reported in \cref{tab:color-mixing}, where the notation
\begin{equation}
\mathcal{O}_i\to \mathcal{O}_j\,,
\end{equation}
denotes the insertion of the operator $\mathcal{O}_i$ and reading off the color coefficient of operator $\mathcal{O}_j$.

Concerning the $SU(N)$ conventions, in terms of the general notation introduced in \cref{eq:genneralgen} and assuming that the fermions transform in the fundamental representation $F$ one finds:

\begin{equation}
C_2(F)\equiv C_F=\frac{N^2-1}{2N},
\qquad
T(F)=\frac12,
\qquad
C_A=N.
\label{eq:colorfactorsdef}
\end{equation}
This leads to:

\begin{equation}
\begin{aligned}
\operatorname{Tr}\!\left(T_F^aT_F^b\right)
&=
\frac{1}{2}\,\delta^{ab}\,,
&
\left(T_F^aT_F^a\right)_{ij}
&=
C_F\,\delta_{ij}\,,
\\
f^{acd}f^{bcd}
&=
C_A\,\delta^{ab}\,,
&
f^{abc}T_F^bT_F^c
&=
\frac{i}{2}\,C_A T_F^a\,,
\\
(T_F^a)_{ij}(T_F^a)_{kl}
&=
\frac{1}{2}
\left(
\delta_{il}\delta_{kj}
-\frac{1}{N}\delta_{ij}\delta_{kl}
\right)\,,
&
f^{abc}T_F^aT_F^bT_F^c
&=
\frac{i}{2}\,C_A C_F\,\mathbbm{1}\,.
\end{aligned}
\label{eq:SUNcolordef}
\end{equation}

\begin{table}[H]
\centering
\small
\setlength{\tabcolsep}{3pt}
\renewcommand{\arraystretch}{0.6}
\hspace*{-1cm}
\resizebox{1.1\linewidth}{!}{
\begin{tabular}{@{}c c c c c@{}}
\toprule[0.7mm]
Diagram & $\mathcal{O}_1\to \mathcal{O}_1$ & $\mathcal{O}_1\to \mathcal{O}_8$ & $\mathcal{O}_8\to \mathcal{O}_1$ & $\mathcal{O}_8\to \mathcal{O}_8$ \\
\midrule[0.7mm]
1  & $C_F$ & $0$ & $0$ & $-(C_A/2)+C_F$ \\ \midrule
2  & $0$ & $1$ & $1/4-1/(4N^2)$ & $-1/N$ \\ \midrule
3  & $0$ & $1$ & $C_F/(2N)$ & $C_F-1/(2N)$ \\ \midrule
4  & $C_F^2$ & $0$ & $0$ & $1/4(C_A-2C_F)^2$ \\ \midrule
5  & $1/4-1/(4N^2)$ & $-1/N$ & $-(-1+N^2)/(4N^3)$ & $1/4+3/(4N^2)$ \\ \midrule
6  & $C_F/(2N)$ & $C_F-1/(2N)$ & $(C_F(-1+2C_FN))/(4N^2)$ & $(1+2C_FN(-1+2C_FN))/(4N^2)$ \\ \midrule
7  & $-(C_A C_F)/2+C_F^2$ & $0$ & $0$ & $(C_A-2C_F+N)/(4N)$ \\ \midrule
8  & $C_F/(2N)$ & $C_F-1/(2N)$ & $((-1+2C_FN)(-1+N^2))/(8N^3)$ & $(1-C_FN)/(2N^2)$ \\ \midrule
9  & $1/4-1/(4N^2)$ & $-1/N$ & $(1+(-1-2C_A C_F+4C_F^2)N^2)/(8N^3)$ & $-(C_A C_F)/2+C_F^2+1/(2N^2)$ \\ \midrule
10 & $-(C_A C_F)/2+C_F^2$ & $0$ & $0$ & $1/4(C_A-2C_F)^2$ \\ \midrule
11 & $0$ & $-(C_A/2)+C_F$ & $-((C_A-2C_F)(-1+N^2))/(8N^2)$ & $(C_A-2C_F)/(2N)$ \\ \midrule
12 & $0$ & $-(C_A/2)+C_F$ & $-((C_A-2C_F)C_F)/(4N)$ & $-((C_A-2C_F)(-1+2C_FN))/(4N)$ \\ \midrule
13 & $C_F^2$ & $0$ & $0$ & $-(C_A C_F)/2+C_F^2$ \\ \midrule
14 & $0$ & $C_F$ & $(C_F(-1+N^2))/(4N^2)$ & $-C_F/N$ \\ \midrule
15 & $0$ & $C_F$ & $C_F^2/(2N)$ & $C_F^2-C_F/(2N)$ \\ \midrule
16 & $0$ & $-(C_A/2)+C_F$ & $(1+N^2(-1+2C_FN))/(8N^3)$ & $1/(2N^2)$ \\ \midrule
17 & $0$ & $C_F$ & $-((C_A-2C_F)(-1+N^2))/(8N^2)$ & $(C_A-2C_F)/(2N)$ \\ \midrule
18 & $0$ & $-(C_A/2)+C_F$ & $(C_F(-1+N^2))/(4N^2)$ & $(1-2C_FN)/(4N^2)$ \\ \midrule
19 & $0$ & $C_F$ & $-((C_A-2C_F)C_F)/(4N)$ & $-((C_A-2C_F)(-1+2C_FN))/(4N)$ \\ \midrule
20 & $1/4-1/(4N^2)$ & $-1/N$ & $((-1+2C_FN)(-1+N^2))/(8N^3)$ & $(1-C_FN)/(2N^2)$ \\ \midrule
21 & $C_F/(2N)$ & $C_F-1/(2N)$ & $(C_A-C_A N^2+2C_F(-2+N^2))/(8N^2)$ & $(1+C_A N-4C_FN)/(4N^2)$ \\ \midrule
22 & $C_F^2$ & $0$ & $0$ & $1/4(C_A-2C_F)^2$ \\ \midrule
23 & $1/4-1/(4N^2)$ & $-1/N$ & $-(-1+N^2)/(4N^3)$ & $1/4+3/(4N^2)$ \\ \midrule
24 & $C_F/(2N)$ & $C_F-1/(2N)$ & $(C_F(-1+2C_FN))/(4N^2)$ & $(1+2C_FN(-1+2C_FN))/(4N^2)$ \\ \midrule
25 & $C_A C_F/2$ & $0$ & $0$ & $(2C_AC_F-C_A^2)/4$ \\ \midrule
26 & $0$ & $C_A/2$ & $(C_AN^2-C_A)/8N^2$ & $-C_A/2N$ \\ \midrule
27 & $0$ & $C_A/2$ & $C_AC_F/4N$ & $(2C_AC_FN-C_A)/4N$ \\ \midrule
28 & $0$ & $-C_A/2$ & $(C_AN^2-C_A)/8N^2$ & $0$ \\ \midrule
29a & $C_F$ & $0$ & $0$ & $(C_F-C_A/2)$ \\ \midrule
29b & $C_A C_F$ & $0$ & $0$ & $C_A(C_F-C_A/2)$ \\ \midrule
29c & $C_A C_F$ & $0$ & $0$ & $C_A(C_F-C_A/2)$ \\ \midrule
29d & $-C_AC_F$ & $0$ & $0$ & $C_A(C_A/2-C_F)$ \\ \midrule
30a & $0$ & $1$ & $-(1-N^2)/(4N^2)$ & $-1/N$ \\ \midrule
30b & $0$ & $C_A$ & $-(1-N^2)C_A/(4N^2)$ & $-C_A/N$ \\ \midrule
30c & $0$ & $C_A $ & $-(1-N^2)C_A/(4N^2)$ & $-C_A/N$ \\ \midrule
30d & $0$ & $-C_A$ & $(1-N^2)C_A/(4N^2)$ & $C_A/N$ \\ \midrule
31a & $0$ & $1$ & $C_F/(2N)$ & $(-1+2C_FN)/(2N)$ \\ \midrule
31b & $0$ & $C_A$ & $C_AC_F/(2N)$ & $(-1+2C_FN)C_A/(2N)$ \\ \midrule
31c & $0$ & $C_A $ & $C_AC_F/(2N)$ & $(-1+2C_FN)C_A/(2N)$ \\ \midrule
31d & $0$ & $-C_A$ & $-C_AC_F/(2N)$ & $(1-2C_FN)C_A/(2N)$ \\ 
\bottomrule[0.7mm]
\end{tabular}
}
\caption{One- and two-loop color mixing coefficients for a single $SU(N)$ gauge group in the singlet/adjoint basis $\{\mathcal{O}_1,\mathcal{O}_8\}$. Entries give the color factor multiplying each diagram after insertion of the operator indicated by the column.}
\label{tab:color-mixing}
\end{table}

\subsection{$SU(N_1) \times SU(N_2)$ color factors}
\label{app:colorfactorsSUNSUN}
In this subsection we report the color factors for diagrams involving two gauge bosons belonging to the product group $SU(N_1)\times SU(N_2)$. We work in the basis 
\begin{equation}\label{eq:SUN1N2basis}
\{\mathcal{O}_1,\mathcal{O}_8^{[1]},\mathcal{O}_8^{[2]},\mathcal{O}_8^{[12]}\}\,,    
\end{equation}
where $\mathcal{O}_1$ is a singlet under both groups, $\mathcal{O}_8^{[i]}$ denotes an adjoint contraction with generators of $SU(N_i)$, and $\mathcal{O}_8^{[12]}$ contains one generator of each gauge group. When the tables distinguish whether the first gauge boson belongs to $SU(N_1)$ or $SU(N_2)$, ``first'' refers to the gauge boson encountered first when traversing the corresponding fermion line in the direction used to define the amplitude. The results of all possible insertions and mixings are collected in Tables.~\ref{tab:color-mixing-suN2-simple-slash}--\ref{tab:color-mixing812firstgluon22}.

\begin{table}[H]
\centering
\small
\setlength{\tabcolsep}{8pt}
\renewcommand{\arraystretch}{0.6}
\begin{tabular}{@{}c c c c c@{}}
\toprule[0.7mm]
Diagram & $\mathcal{O}_1\to \mathcal{O}_1$ & $\mathcal{O}_1\to \mathcal{O}_8^{[1]}$ & $\mathcal{O}_1\to \mathcal{O}_8^{[2]}$ & $\mathcal{O}_1\to \mathcal{O}_8^{[12]}$ \\
\midrule[0.7mm]
4  & $C_F^{[1]}C_F^{[2]}$ & $0$ & $0$ & $0$ \\ \midrule
5  & $0$ & $0$ & $0$ & $1$ \\ \midrule
6  & $0$ & $0$ & $0$ & $1$ \\ \midrule
7  & $C_F^{[1]}C_F^{[2]}$ & $0$ & $0$ & $0$ \\ \midrule
8  & $0$ & $0$ & $0$ & $1$ \\ \midrule
9  & $0$ & $0$ & $0$ & $1$ \\ \midrule
10 & $C_F^{[1]}C_F^{[2]}$ & $0$ & $0$ & $0$ \\ \midrule
11 & $0$ & $0$ & $C_F^{[1]}$ & $0$ \\ \midrule
12 & $0$ & $0$ & $C_F^{[1]}$ & $0$ \\ \midrule
13 & $C_F^{[1]}C_F^{[2]}$ & $0$ & $0$ & $0$ \\ \midrule
14 & $0$ & $0$ & $C_F^{[1]}$ & $0$ \\ \midrule
15 & $0$ & $0$ & $C_F^{[1]}$ & $0$ \\ \midrule
16 & $0$ & $0$ & $C_F^{[1]}$ & $0$ \\ \midrule
17 & $0$ & $0$ & $C_F^{[1]}$ & $0$ \\ \midrule
18 & $0$ & $0$ & $C_F^{[1]}$ & $0$ \\ \midrule
19 & $0$ & $0$ & $C_F^{[1]}$ & $0$ \\ \midrule
20 & $0$ & $0$ & $0$ & $1$ \\ \midrule
21 & $0$ & $0$ & $0$ & $1$ \\ \midrule
22 & $C_F^{[1]}C_F^{[2]}$ & $0$ & $0$ & $0$ \\ \midrule
23 & $0$ & $0$ & $0$ & $1$ \\ \midrule
24 & $0$ & $0$ & $0$ & $1$ \\
\bottomrule[0.7mm]
\end{tabular}
\caption{Color mixing coefficients for $SU(N_1)\times SU(N_2)$ in the basis $\{\mathcal{O}_1,\mathcal{O}_8^{[1]},\mathcal{O}_8^{[2]},\mathcal{O}_8^{[12]}\}$, for an insertion of the singlet operator $\mathcal{O}_1$. The first exchanged gauge boson is associated with the $SU(N_1)$ factor.}
\label{tab:color-mixing-suN2-simple-slash}
\end{table}

\begin{table}[H]
\centering
\small
\setlength{\tabcolsep}{8pt}
\renewcommand{\arraystretch}{0.6}
\begin{tabular}{@{}c c c c c@{}}
\toprule[0.7mm]
Diagram & $\mathcal{O}_1\to \mathcal{O}_1$ & $\mathcal{O}_1\to \mathcal{O}_8^{[1]}$ & $\mathcal{O}_1\to \mathcal{O}_8^{[2]}$ & $\mathcal{O}_1\to \mathcal{O}_8^{[12]}$ \\
\midrule[0.7mm]
4  & $C_F^{[1]}C_F^{[2]}$ & $0$ & $0$ & $0$ \\ \midrule
5  & $0$ & $0$ & $0$ & $1$ \\ \midrule
6  & $0$ & $0$ & $0$ & $1$ \\ \midrule
7  & $C_F^{[1]}C_F^{[2]}$ & $0$ & $0$ & $0$ \\ \midrule
8  & $0$ & $0$ & $0$ & $1$ \\ \midrule
9  & $0$ & $0$ & $0$ & $1$ \\ \midrule
10 & $C_F^{[1]}C_F^{[2]}$ & $0$ & $0$ & $0$ \\ \midrule
11 & $0$ & $C_F^{[2]}$ & $0$ & $0$ \\ \midrule
12 & $0$ & $C_F^{[2]}$ & $0$ & $0$ \\ \midrule
13 & $C_F^{[1]}C_F^{[2]}$ & $0$ & $0$ & $0$ \\ \midrule
14 & $0$ & $C_F^{[2]}$ & $0$ & $0$ \\ \midrule
15 & $0$ & $C_F^{[2]}$ & $0$ & $0$ \\ \midrule
16 & $0$ & $C_F^{[2]}$ & $0$ & $0$ \\ \midrule
17 & $0$ & $C_F^{[2]}$ & $0$ & $0$ \\ \midrule
18 & $0$ & $C_F^{[2]}$ & $0$ & $0$ \\ \midrule
19 & $0$ & $C_F^{[2]}$ & $0$ & $0$ \\ \midrule
20 & $0$ & $0$ & $0$ & $1$ \\ \midrule
21 & $0$ & $0$ & $0$ & $1$ \\ \midrule
22 & $C_F^{[1]}C_F^{[2]}$ & $0$ & $0$ & $0$ \\ \midrule
23 & $0$ & $0$ & $0$ & $1$ \\ \midrule
24 & $0$ & $0$ & $0$ & $1$ \\
\bottomrule[0.7mm]
\end{tabular}
\caption{Color mixing coefficients for $SU(N_1)\times SU(N_2)$ in the basis $\{\mathcal{O}_1,\mathcal{O}_8^{[1]},\mathcal{O}_8^{[2]},\mathcal{O}_8^{[12]}\}$, for an insertion of the singlet operator $\mathcal{O}_1$. The first exchanged gauge boson is associated with the $SU(N_2)$ factor.}
\label{tab:color-mixing-suN2-no-section}
\end{table}

\begin{table}[H]
\centering
\small
\setlength{\tabcolsep}{8pt}
\renewcommand{\arraystretch}{0.6}
\begin{tabular}{@{}c c c c c@{}}
\toprule[0.7mm]
Diagram & $\mathcal{O}_8^{[1]}\to \mathcal{O}_1$ & $\mathcal{O}_8^{[1]}\to \mathcal{O}_8^{[1]}$ & $\mathcal{O}_8^{[1]}\to \mathcal{O}_8^{[2]}$ & $\mathcal{O}_8^{[1]}\to \mathcal{O}_8^{[12]}$ \\
\midrule[0.7mm]
4  & $0$ & $-C_F^{[2]}/(2N_1)$ & $0$ & $0$ \\ \midrule
5  & $0$ & $0$ & $1/4-1/(2N_1)^2$ & $-1/N_1$ \\ \midrule
6  & $0$ & $0$ & $C_F^{[1]}/(2N_1)$ & $C_F^{[1]}-1/(2N_1)$ \\ \midrule
7  & $0$ & $-C_F^{[2]}/(2N_1)$ & $0$ & $0$ \\ \midrule
8  & $0$ & $0$ & $1/4-1/(2N_1)^2$ & $-1/N_1$ \\ \midrule
9  & $0$ & $0$ & $C_F^{[1]}/(2N_1)$ & $C_F^{[1]}-1/(2N_1)$ \\ \midrule
10 & $0$ & $-C_F^{[2]}/(2N_1)$ & $0$ & $0$ \\ \midrule
11 & $0$ & $0$ & $0$ & $C_F^{[1]}$ \\ \midrule
12 & $0$ & $0$ & $0$ & $C_F^{[1]}$ \\ \midrule
13 & $0$ & $-C_F^{[2]}/(2N_1)$ & $0$ & $0$ \\ \midrule
14 & $0$ & $0$ & $0$ & $C_F^{[1]}$ \\ \midrule
15 & $0$ & $0$ & $0$ & $C_F^{[1]}$ \\ \midrule
16 & $0$ & $0$ & $0$ & $-1/(2N_1)$ \\ \midrule
17 & $0$ & $0$ & $0$ & $-1/(2N_1)$ \\ \midrule
18 & $0$ & $0$ & $0$ & $-1/(2N_1)$ \\ \midrule
19 & $0$ & $0$ & $0$ & $-1/(2N_1)$ \\ \midrule
20 & $0$ & $0$ & $C_F^{[1]}/(2N_1)$ & $C_F^{[1]}-1/(2N_1)$ \\ \midrule
21 & $0$ & $0$ & $1/4-1/(2N_1)^2$ & $-1/N_1$ \\ \midrule
22 & $0$ & $-C_F^{[2]}/(2N_1)$ & $0$ & $0$ \\ \midrule
23 & $0$ & $0$ & $1/4-1/(2N_1)^2$ & $-1/N_1$ \\ \midrule
24 & $0$ & $0$ & $C_F^{[1]}/(2N_1)$ & $C_F^{[1]}-1/(2N_1)$ \\ 
\bottomrule[0.7mm]
\end{tabular}
\caption{Color mixing coefficients for $SU(N_1)\times SU(N_2)$ in the basis $\{\mathcal{O}_1,\mathcal{O}_8^{[1]},\mathcal{O}_8^{[2]},\mathcal{O}_8^{[12]}\}$, for an insertion of the adjoint operator $\mathcal{O}_8^{[1]}$. The first exchanged gauge boson is associated with the $SU(N_1)$ factor.}
\label{tab:color-mixingfirstgluon1}
\end{table}

\begin{table}[H]
\centering
\small
\setlength{\tabcolsep}{8pt}
\renewcommand{\arraystretch}{0.6}
\begin{tabular}{@{}c c c c c@{}}
\toprule[0.7mm]
Diagram & $\mathcal{O}_8^{[1]}\to \mathcal{O}_1$ & $\mathcal{O}_8^{[1]}\to \mathcal{O}_8^{[1]}$ & $\mathcal{O}_8^{[1]}\to \mathcal{O}_8^{[2]}$ & $\mathcal{O}_8^{[1]}\to \mathcal{O}_8^{[12]}$ \\
\midrule[0.7mm]
4  & $0$ & $-C_F^{[2]}/(2N_1)$ & $0$ & $0$ \\ \midrule
5  & $0$ & $0$ & $1/4-1/(2N_1)^2$ & $-1/N_1$ \\ \midrule
6  & $0$ & $0$ & $C_F^{[1]}/(2N_1)$ & $C_F^{[1]}-1/(2N_1)$ \\ \midrule
7  & $0$ & $-C_F^{[2]}/(2N_1)$ & $0$ & $0$ \\ \midrule
8  & $0$ & $0$ & $1/4-1/(2N_1)^2$ & $-1/N_1$ \\ \midrule
9  & $0$ & $0$ & $C_F^{[1]}/(2N_1)$ & $C_F^{[1]}-1/(2N_1)$ \\ \midrule
10 & $0$ & $C_F^{[1]}C_F^{[2]}$ & $0$ & $0$ \\ \midrule
11 & $C_F^{[2]}( -1 + (N_1)^2 )/(2N_1)^2$ & $-C_F^{[2]}/N_1$ & $0$ & $0$ \\ \midrule
12 & $C_F^{[1]}C_F^{[2]}/(2N_1)$ & $C_F^{[1]}C_F^{[2]}-C_F^{[2]}/(2N_1)$ & $0$ & $0$ \\ \midrule
13 & $0$ & $C_F^{[1]}C_F^{[2]}$ & $0$ & $0$ \\ \midrule
14 & $C_F^{[2]}( -1 + (N_1)^2 )/(2N_1)^2$ & $-C_F^{[2]}/N_1$ & $0$ & $0$ \\ \midrule
15 & $C_F^{[1]}C_F^{[2]}/(2N_1)$ & $C_F^{[1]}C_F^{[2]}-C_F^{[2]}/(2N_1)$ & $0$ & $0$ \\ \midrule
16 & $C_F^{[2]}( -1 + (N_1)^2 )/(2N_1)^2$ & $-C_F^{[2]}/N_1$ & $0$ & $0$ \\ \midrule
17 & $C_F^{[2]}( -1 + (N_1)^2 )/(2N_1)^2$ & $-C_F^{[2]}/N_1$ & $0$ & $0$ \\ \midrule
18 & $C_F^{[1]}C_F^{[2]}/(2N_1)$ & $C_F^{[1]}C_F^{[2]}-C_F^{[2]}/(2N_1)$ & $0$ & $0$ \\ \midrule
19 & $C_F^{[1]}C_F^{[2]}/(2N_1)$ & $C_F^{[1]}C_F^{[2]}-C_F^{[2]}/(2N_1)$ & $0$ & $0$ \\ \midrule
20 & $0$ & $0$ & $1/4-1/(2N_1)^2$ & $-1/N_1$ \\ \midrule
21 & $0$ & $0$ & $C_F^{[1]}/(2N_1)$ & $C_F^{[1]}-1/(2N_1)$ \\ \midrule
22 & $0$ & $-C_F^{[2]}/(2N_1)$ & $0$ & $0$ \\ \midrule
23 & $0$ & $0$ & $1/4-1/(2N_1)^2$ & $-1/N_1$ \\ \midrule
24 & $0$ & $0$ & $C_F^{[1]}/(2N_1)$ & $C_F^{[1]}-1/(2N_1)$ \\ 
\bottomrule[0.7mm]
\end{tabular}
\caption{Color mixing coefficients for $SU(N_1)\times SU(N_2)$ in the basis $\{\mathcal{O}_1,\mathcal{O}_8^{[1]},\mathcal{O}_8^{[2]},\mathcal{O}_8^{[12]}\}$, for an insertion of the adjoint operator $\mathcal{O}_8^{[1]}$. The first exchanged gauge boson is associated with the $SU(N_2)$ factor.}
\label{tab:color-mixingfirstgluon2}
\end{table}

\begin{table}[H]
\centering
\small
\setlength{\tabcolsep}{8pt}
\renewcommand{\arraystretch}{0.6}
\begin{tabular}{@{}c c c c c@{}}
\toprule[0.7mm]
Diagram & $\mathcal{O}_8^{[2]}\to \mathcal{O}_1$ & $\mathcal{O}_8^{[2]}\to \mathcal{O}_8^{[1]}$ & $\mathcal{O}_8^{[2]}\to \mathcal{O}_8^{[2]}$ & $\mathcal{O}_8^{[2]}\to \mathcal{O}_8^{[12]}$ \\
\midrule[0.7mm]
4  & $0$ & $0$ & $-C_F^{[1]}/(2N_2)$ & $0$ \\ \midrule
5  & $0$ & $1/4-1/(2N_2)^2$ & $0$ & $-1/N_2$ \\ \midrule
6  & $0$ & $C_F^{[2]}/(2N_2)$ & $0$ & $(C_F^{[2]}-1/(2N_2))$ \\ \midrule
7  & $0$ & $0$ & $-C_F^{[1]}/(2N_2)$ & $0$ \\ \midrule
8  & $0$ & $1/4-1/(2N_2)^2$ & $0$ & $-1/N_2$ \\ \midrule
9  & $0$ & $C_F^{[2]}/(2N_2)$ & $0$ & $(C_F^{[2]}-1/(2N_2))$ \\ \midrule
10 & $0$ & $0$ & $C_F^{[1]}C_F^{[2]}$ & $0$ \\ \midrule
11 & $C_F^{[1]}(-1+(N_2)^2)/(2N_2)^2$ & $0$ & $-C_F^{[1]}/N_2$ & $0$ \\ \midrule
12 & $C_F^{[1]}C_F^{[2]}/(2N_2)$ & $0$ & $C_F^{[1]}(C_F^{[2]}-1/(2N_2))$ & $0$ \\ \midrule
13 & $0$ & $0$ & $C_F^{[1]}C_F^{[2]}$ & $0$ \\ \midrule
14 & $C_F^{[1]}(-1+(N_2)^2)/(2N_2)^2$ & $0$ & $-C_F^{[1]}/N_2$ & $0$ \\ \midrule
15 & $C_F^{[1]}C_F^{[2]}/(2N_2)$ & $0$ & $C_F^{[1]}(C_F^{[2]}-1/(2N_2))$ & $0$ \\ \midrule
16 & $C_F^{[1]}(-1+(N_2)^2)/(2N_2)^2$ & $0$ & $-C_F^{[1]}/N_2$ & $0$ \\ \midrule
17 & $C_F^{[1]}(-1+(N_2)^2)/(2N_2)^2$ & $0$ & $-C_F^{[1]}/N_2$ & $0$ \\ \midrule
18 & $C_F^{[1]}C_F^{[2]}/(2N_2)$ & $0$ & $C_F^{[1]}(C_F^{[2]}-1/(2N_2))$ & $0$ \\ \midrule
19 & $C_F^{[1]}C_F^{[2]}/(2N_2)$ & $0$ & $C_F^{[1]}(C_F^{[2]}-1/(2N_2))$ & $0$ \\ \midrule
20 & $0$ & $1/4-1/(2N_2)^2$ & $0$ & $-1/N_2$ \\ \midrule
21 & $0$ & $C_F^{[2]}/(2N_2)$ & $0$ & $(C_F^{[2]}-1/(2N_2))$ \\ \midrule
22 & $0$ & $0$ & $-C_F^{[1]}/(2N_2)$ & $0$ \\ \midrule
23 & $0$ & $1/4-1/(2N_2)^2$ & $0$ & $-1/N_2$ \\ \midrule
24 & $0$ & $C_F^{[2]}/(2N_2)$ & $0$ & $(C_F^{[2]}-1/(2N_2))$ \\
\bottomrule[0.7mm]
\end{tabular}
\caption{Color mixing coefficients for $SU(N_1)\times SU(N_2)$ in the basis $\{\mathcal{O}_1,\mathcal{O}_8^{[1]},\mathcal{O}_8^{[2]},\mathcal{O}_8^{[12]}\}$, for an insertion of the adjoint operator $\mathcal{O}_8^{[2]}$. The first exchanged gauge boson is associated with the $SU(N_1)$ factor.}
\label{tab:color-mixing8firstgluon1}
\end{table}

\begin{table}[H]
\centering
\small
\setlength{\tabcolsep}{8pt}
\renewcommand{\arraystretch}{0.6}
\begin{tabular}{@{}c c c c c@{}}
\toprule[0.7mm]
Diagram & $\mathcal{O}_8^{[2]}\to \mathcal{O}_1$ & $\mathcal{O}_8^{[2]}\to \mathcal{O}_8^{[1]}$ & $\mathcal{O}_8^{[2]}\to \mathcal{O}_8^{[2]}$ & $\mathcal{O}_8^{[2]}\to \mathcal{O}_8^{[12]}$ \\
\midrule[0.7mm]
4  & $0$ & $0$ & $-C_F^{[1]}/(2N_2)$ & $0$ \\ \midrule
5  & $0$ & $1/4-1/(2N_2)^2$ & $0$ & $-1/N_2$ \\ \midrule
6  & $0$ & $C_F^{[2]}/(2N_2)$ & $0$ & $(C_F^{[2]}-1/(2N_2))$ \\ \midrule
7  & $0$ & $0$ & $-C_F^{[1]}/(2N_2)$ & $0$ \\ \midrule
8  & $0$ & $1/4-1/(2N_2)^2$ & $0$ & $-1/N_2$ \\ \midrule
9  & $0$ & $C_F^{[2]}/(2N_2)$ & $0$ & $(C_F^{[2]}-1/(2N_2))$ \\ \midrule
10 & $0$ & $0$ & $-C_F^{[1]}/(2N_2)$ & $0$ \\ \midrule
11 & $0$ & $0$ & $0$ & $C_F^{[2]}$ \\ \midrule
12 & $0$ & $0$ & $0$ & $C_F^{[2]}$ \\ \midrule
13 & $0$ & $0$ & $-C_F^{[1]}/(2N_2)$ & $0$ \\ \midrule
14 & $0$ & $0$ & $0$ & $C_F^{[2]}$ \\ \midrule
15 & $0$ & $0$ & $0$ & $C_F^{[2]}$ \\ \midrule
16 & $0$ & $0$ & $0$ & $-1/(2N_2)$ \\ \midrule
17 & $0$ & $0$ & $0$ & $-1/(2N_2)$ \\ \midrule
18 & $0$ & $0$ & $0$ & $-1/(2N_2)$ \\ \midrule
19 & $0$ & $0$ & $0$ & $-1/(2N_2)$ \\ \midrule
20 & $0$ & $C_F^{[2]}/(2N_2)$ & $0$ & $(C_F^{[2]}-1/(2N_2))$ \\ \midrule
21 & $0$ & $1/4-1/(2N_2)^2$ & $0$ & $-1/N_2$ \\ \midrule
22 & $0$ & $0$ & $-C_F^{[1]}/(2N_2)$ & $0$ \\ \midrule
23 & $0$ & $1/4-1/(2N_2)^2$ & $0$ & $-1/N_2$ \\ \midrule
24 & $0$ & $C_F^{[2]}/(2N_2)$ & $0$ & $(C_F^{[2]}-1/(2N_2))$ \\
\bottomrule[0.7mm]
\end{tabular}
\caption{Color mixing coefficients for $SU(N_1)\times SU(N_2)$ in the basis $\{\mathcal{O}_1,\mathcal{O}_8^{[1]},\mathcal{O}_8^{[2]},\mathcal{O}_8^{[12]}\}$, for an insertion of the adjoint operator $\mathcal{O}_8^{[2]}$. The first exchanged gauge boson is associated with the $SU(N_2)$ factor.}
\label{tab:color-mixing88firstgluon2}
\end{table}

\begin{table}[H]
\centering
\small
\setlength{\tabcolsep}{8pt}
\renewcommand{\arraystretch}{0.6}
\begin{tabular}{@{}c c c@{}}
\toprule[0.7mm]
Diagram & $\mathcal{O}_8^{[12]}\to \mathcal{O}_1$ & $\mathcal{O}_8^{[12]}\to \mathcal{O}_8^{[1]}$ \\
\midrule[0.7mm]
4  & $0$ & $0$ \\ \midrule
5  & $((-1+(N_1)^2)(-1+(N_2)^2))/(16(N_1)^2(N_2)^2)$ & $-(-1+(N_2)^2)/(4N_1(N_2)^2)$ \\ \midrule
6  & $C_F^{[1]}C_F^{[2]}/(4N_1N_2)$ & $C_F^{[2]}(-1+2C_F^{[1]}N_1)/(4N_1N_2)$ \\ \midrule
7  & $0$ & $0$ \\ \midrule
8  & $((-1+(N_1)^2)(-1+(N_2)^2))/(16(N_1)^2(N_2)^2)$ & $-(-1+(N_2)^2)/(4N_1(N_2)^2)$ \\ \midrule
9  & $C_F^{[1]}C_F^{[2]}/(4N_1N_2)$ & $C_F^{[2]}(-1+2C_F^{[1]}N_1)/(4N_1N_2)$ \\ \midrule
10 & $0$ & $0$ \\ \midrule
11 & $0$ & $C_F^{[1]}(-1+(N_2)^2)/(2N_2)^2$ \\ \midrule
12 & $0$ & $C_F^{[1]}C_F^{[2]}/(2N_2)$ \\ \midrule
13 & $0$ & $0$ \\ \midrule
14 & $0$ & $C_F^{[1]}(-1+(N_2)^2)/(2N_2)^2$ \\ \midrule
15 & $0$ & $C_F^{[1]}C_F^{[2]}/(2N_2)$ \\ \midrule
16 & $0$ & $-(-1+(N_2)^2)/(8N_1(N_2)^2)$ \\ \midrule
17 & $0$ & $-(-1+(N_2)^2)/(8N_1(N_2)^2)$ \\ \midrule
18 & $0$ & $-C_F^{[2]}/(4N_1N_2)$ \\ \midrule
19 & $0$ & $-C_F^{[2]}/(4N_1N_2)$ \\ \midrule
20 & $C_F^{[1]}(-1+(N_2)^2)/(8N_1(N_2)^2)$ & $(-1+2C_F^{[1]}N_1)(-1+(N_2)^2)/(8N_1(N_2)^2)$ \\ \midrule
21 & $C_F^{[2]}(-1+(N_1)^2)/(8(N_1)^2N_2)$ & $-C_F^{[2]}/(2N_1N_2)$ \\ \midrule
22 & $0$ & $0$ \\ \midrule
23 & $((-1+(N_1)^2)(-1+(N_2)^2))/(16(N_1)^2(N_2)^2)$ & $-(-1+(N_2)^2)/(4N_1(N_2)^2)$ \\ \midrule
24 & $C_F^{[1]}C_F^{[2]}/(4N_1N_2)$ & $C_F^{[2]}(-1+2C_F^{[1]}N_1)/(4N_1N_2)$ \\
\bottomrule[0.7mm]
\end{tabular}
\caption{Color mixing coefficients for $SU(N_1)\times SU(N_2)$ in the basis $\{\mathcal{O}_1,\mathcal{O}_8^{[1]},\mathcal{O}_8^{[2]},\mathcal{O}_8^{[12]}\}$, for an insertion of the mixed-adjoint operator $\mathcal{O}_8^{[12]}$. The first exchanged gauge boson is associated with the $SU(N_1)$ factor, and only the $\mathcal{O}_1$ and $\mathcal{O}_8^{[1]}$ output components are shown.}
\label{tab:color-mixing88firstgluon1}
\end{table}

\begin{table}[H]
\centering
\small
\setlength{\tabcolsep}{8pt}
\renewcommand{\arraystretch}{0.6}
\begin{tabular}{@{}c c c@{}}
\toprule[0.7mm]
Diagram & $\mathcal{O}_8^{[12]}\to \mathcal{O}_1$ & $\mathcal{O}_8^{[12]}\to \mathcal{O}_8^{[1]}$ \\
\midrule[0.7mm]
4  & $0$ & $0$ \\ \midrule
5  & $((-1+(N_1)^2)(-1+(N_2)^2))/(16 (N_1)^2 (N_2)^2)$ & $-(-1+(N_2)^2)/(4 N_1 (N_2)^2)$ \\ \midrule
6  & $C_F^{[1]}C_F^{[2]}/(4 N_1 N_2)$ & $C_F^{[2]}(-1+2C_F^{[1]}N_1)/(4 N_1 N_2)$ \\ \midrule
7  & $0$ & $0$ \\ \midrule
8  & $((-1+(N_1)^2)(-1+(N_2)^2))/(16 (N_1)^2 (N_2)^2)$ & $-(-1+(N_2)^2)/(4 N_1 (N_2)^2)$ \\ \midrule
9  & $C_F^{[1]}C_F^{[2]}/(4 N_1 N_2)$ & $C_F^{[2]}(-1+2C_F^{[1]}N_1)/(4 N_1 N_2)$ \\ \midrule
10 & $0$ & $0$ \\ \midrule
11 & $0$ & $0$ \\ \midrule
12 & $0$ & $0$ \\ \midrule
13 & $0$ & $0$ \\ \midrule
14 & $0$ & $0$ \\ \midrule
15 & $0$ & $0$ \\ \midrule
16 & $0$ & $0$ \\ \midrule
17 & $0$ & $0$ \\ \midrule
18 & $0$ & $0$ \\ \midrule
19 & $0$ & $0$ \\ \midrule
20 & $C_F^{[2]}(-1+(N_1)^2)/(8 (N_1)^2 N_2)$ & $-C_F^{[2]}/(2 N_1 N_2)$ \\ \midrule
21 & $C_F^{[1]}(-1+(N_2)^2)/(8 N_1 (N_2)^2)$ & $(-1+2C_F^{[1]}N_1)(-1+(N_2)^2)/(8 N_1 (N_2)^2)$ \\ \midrule
22 & $0$ & $0$ \\ \midrule
23 & $((-1+(N_1)^2)(-1+(N_2)^2))/(16 (N_1)^2 (N_2)^2)$ & $-(-1+(N_2)^2)/(4 N_1 (N_2)^2)$ \\ \midrule
24 & $C_F^{[1]}C_F^{[2]}/(4 N_1 N_2)$ & $C_F^{[2]}(-1+2C_F^{[1]}N_1)/(4 N_1 N_2)$ \\
\bottomrule[0.7mm]
\end{tabular}
\caption{Color mixing coefficients for $SU(N_1)\times SU(N_2)$ in the basis $\{\mathcal{O}_1,\mathcal{O}_8^{[1]},\mathcal{O}_8^{[2]},\mathcal{O}_8^{[12]}\}$, for an insertion of the mixed-adjoint operator $\mathcal{O}_8^{[12]}$. The first exchanged gauge boson is associated with the $SU(N_2)$ factor, and only the $\mathcal{O}_1$ and $\mathcal{O}_8^{[1]}$ output components are shown.}
\label{tab:color-mixing812firstgluon2}
\end{table}

\begin{table}[H]
\centering
\small
\setlength{\tabcolsep}{8pt}
\renewcommand{\arraystretch}{0.6}
\begin{tabular}{@{}c c c@{}}
\toprule[0.7mm]
Diagram & $\mathcal{O}_8^{[12]}\to \mathcal{O}_8^{[2]}$ & $\mathcal{O}_8^{[12]}\to \mathcal{O}_8^{[12]}$ \\
\midrule[0.7mm]
4  & $0$ & $1/(4N_1N_2)$ \\ \midrule
5  & $-(-1+(N_1)^2)/(4(N_1)^2N_2)$ & $1/(N_1N_2)$ \\ \midrule
6  & $C_F^{[1]}(-1+2C_F^{[2]}N_2)/(4N_1N_2)$ & $((-1+2C_F^{[1]}N_1)(-1+2C_F^{[2]}N_2))/(4N_1N_2)$ \\ \midrule
7  & $0$ & $1/(4N_1N_2)$ \\ \midrule
8  & $-(-1+(N_1)^2)/(4(N_1)^2N_2)$ & $1/(N_1N_2)$ \\ \midrule
9  & $C_F^{[1]}(-1+2C_F^{[2]}N_2)/(4N_1N_2)$ & $((-1+2C_F^{[1]}N_1)(-1+2C_F^{[2]}N_2))/(4N_1N_2)$ \\ \midrule
10 & $0$ & $-C_F^{[2]}/(2N_1)$ \\ \midrule
11 & $0$ & $-C_F^{[1]}/N_2$ \\ \midrule
12 & $0$ & $C_F^{[1]}(C_F^{[2]}-1/(2N_2))$ \\ \midrule
13 & $0$ & $-C_F^{[2]}/(2N_1)$ \\ \midrule
14 & $0$ & $-C_F^{[1]}/N_2$ \\ \midrule
15 & $0$ & $C_F^{[1]}(C_F^{[2]}-1/(2N_2))$ \\ \midrule
16 & $0$ & $1/(2N_1N_2)$ \\ \midrule
17 & $0$ & $1/(2N_1N_2)$ \\ \midrule
18 & $0$ & $(1-2C_F^{[2]}N_2)/(4N_1N_2)$ \\ \midrule
19 & $0$ & $(1-2C_F^{[2]}N_2)/(4N_1N_2)$ \\ \midrule
20 & $-C_F^{[1]}/(2N_1N_2)$ & $(1-2C_F^{[1]}N_1)/(2N_1N_2)$ \\ \midrule
21 & $(-1+(N_1)^2)(-1+2C_F^{[2]}N_2)/(8(N_1)^2N_2)$ & $(1-2C_F^{[2]}N_2)/(2N_1N_2)$ \\ \midrule
22 & $0$ & $1/(4N_1N_2)$ \\ \midrule
23 & $-(-1+(N_1)^2)/(4(N_1)^2N_2)$ & $1/(N_1N_2)$ \\ \midrule
24 & $C_F^{[1]}(-1+2C_F^{[2]}N_2)/(4N_1N_2)$ & $((-1+2C_F^{[1]}N_1)(-1+2C_F^{[2]}N_2))/(4N_1N_2)$ \\
\bottomrule[0.7mm]
\end{tabular}
\caption{Color mixing coefficients for $SU(N_1)\times SU(N_2)$ in the basis $\{\mathcal{O}_1,\mathcal{O}_8^{[1]},\mathcal{O}_8^{[2]},\mathcal{O}_8^{[12]}\}$, for an insertion of the mixed-adjoint operator $\mathcal{O}_8^{[12]}$. The first exchanged gauge boson is associated with the $SU(N_1)$ factor, and only the $\mathcal{O}_8^{[2]}$ and $\mathcal{O}_8^{[12]}$ output components are shown.}
\label{tab:color-mixing812firstgluon1}
\end{table}

\begin{table}[H]
\centering
\small
\setlength{\tabcolsep}{8pt}
\renewcommand{\arraystretch}{0.6}
\begin{tabular}{@{}c c c@{}}
\toprule[0.7mm]
Diagram & $\mathcal{O}_8^{[12]}\to \mathcal{O}_8^{[2]}$ & $\mathcal{O}_8^{[12]}\to \mathcal{O}_8^{[12]}$ \\
\midrule[0.7mm]
4  & $0$ & $1/(4 N_1 N_2)$ \\ \midrule
5  & $-(-1+(N_1)^2)/(4 (N_1)^2 N_2)$ & $1/(N_1 N_2)$ \\ \midrule
6  & $C_F^{[1]}(-1+2C_F^{[2]}N_2)/(4 N_1 N_2)$ & $((-1+2C_F^{[1]}N_1)(-1+2C_F^{[2]}N_2))/(4 N_1 N_2)$ \\ \midrule
7  & $0$ & $1/(4 N_1 N_2)$ \\ \midrule
8  & $-(-1+(N_1)^2)/(4 (N_1)^2 N_2)$ & $1/(N_1 N_2)$ \\ \midrule
9  & $C_F^{[1]}(-1+2C_F^{[2]}N_2)/(4 N_1 N_2)$ & $((-1+2C_F^{[1]}N_1)(-1+2C_F^{[2]}N_2))/(4 N_1 N_2)$ \\ \midrule
10 & $0$ & $-C_F^{[1]}/(2 N_2)$ \\ \midrule
11 & $C_F^{[2]}(-1+(N_1)^2)/(4(N_1)^2)$ & $-C_F^{[2]}/N_1$ \\ \midrule
12 & $C_F^{[1]}C_F^{[2]}/(2 N_1)$ & $C_F^{[1]}C_F^{[2]}-C_F^{[2]}/(2 N_1)$ \\ \midrule
13 & $0$ & $-C_F^{[1]}/(2 N_2)$ \\ \midrule
14 & $C_F^{[2]}(-1+(N_1)^2)/(4(N_1)^2)$ & $-C_F^{[2]}/N_1$ \\ \midrule
15 & $C_F^{[1]}C_F^{[2]}/(2 N_1)$ & $C_F^{[1]}C_F^{[2]}-C_F^{[2]}/(2 N_1)$ \\ \midrule
16 & $-(-1+(N_1)^2)/(8 (N_1)^2 N_2)$ & $1/(2 N_1 N_2)$ \\ \midrule
17 & $-(-1+(N_1)^2)/(8 (N_1)^2 N_2)$ & $1/(2 N_1 N_2)$ \\ \midrule
18 & $-C_F^{[1]}/(4 N_1 N_2)$ & $(1-2 C_F^{[1]} N_1)/(4 N_1 N_2)$ \\ \midrule
19 & $-C_F^{[1]}/(4 N_1 N_2)$ & $(1-2 C_F^{[1]} N_1)/(4 N_1 N_2)$ \\ \midrule
20 & $((-1+(N_1)^2)(-1+2 C_F^{[2]} N_2))/(8 (N_1)^2 N_2)$ & $(1-2 C_F^{[2]} N_2)/(2 N_1 N_2)$ \\ \midrule
21 & $-C_F^{[1]}/(2 N_1 N_2)$ & $(1-2 C_F^{[1]} N_1)/(2 N_1 N_2)$ \\ \midrule
22 & $0$ & $1/(4 N_1 N_2)$ \\ \midrule
23 & $-(-1+(N_1)^2)/(4 (N_1)^2 N_2)$ & $1/(N_1 N_2)$ \\ \midrule
24 & $C_F^{[1]}(-1+2C_F^{[2]}N_2)/(4 N_1 N_2)$ & $((-1+2C_F^{[1]}N_1)(-1+2C_F^{[2]}N_2))/(4 N_1 N_2)$ \\
\bottomrule[0.7mm]
\end{tabular}
\caption{Color mixing coefficients for $SU(N_1)\times SU(N_2)$ in the basis $\{\mathcal{O}_1,\mathcal{O}_8^{[1]},\mathcal{O}_8^{[2]},\mathcal{O}_8^{[12]}\}$, for an insertion of the mixed-adjoint operator $\mathcal{O}_8^{[12]}$. The first exchanged gauge boson is associated with the $SU(N_2)$ factor, and only the $\mathcal{O}_8^{[2]}$ and $\mathcal{O}_8^{[12]}$ output components are shown.}
\label{tab:color-mixing812firstgluon22}
\end{table}

\subsection{$SU(3)$ color factors for Baryon number violating operators}
\label{app:colorfactorsBviol}
We also consider $SU(3)$ corrections to Baryon number violating operators of the form
\begin{equation}
    \epsilon_{\alpha\beta\gamma}(q_1^{\alpha\,T}C\,\Gamma_1 q_2^\beta)(q_3^{\gamma\,T}C\,\Gamma_2\ell)\,.
\end{equation}
The corresponding color factors for all one- and two-loop diagrams are reported in \cref{tab:color-mixingBviol}.

\begin{table}[H]
\centering
\setlength{\tabcolsep}{8pt}
\renewcommand{\arraystretch}{0.6}
\begin{tabular}{@{}c c@{}}
\toprule[0.7mm]
Diagram & Color factor \\ 
\midrule[0.7mm]
1  & $(1+N)/(2N)$ \\ \midrule
2  & $-(1+N)/(2N)$ \\ \midrule
3  & $(1+N)/(2N)$ \\ \midrule
4  & $((1+N)^2)/(4N^2)$ \\ \midrule
5  & $((1+N)^2)/(4N^2)$ \\ \midrule
6  & $((1+N)^2)/(4N^2)$ \\ \midrule
7  & $(1+2N-N^3)/(4N^2)$ \\ \midrule
8  & $(1+2N-N^3)/(4N^2)$ \\ \midrule
9  & $(1+2N-N^3)/(4N^2)$ \\ \midrule
10 & $-(1+N)/(4N^2)$ \\ \midrule
11 & $(1+N)/(4N^2)$ \\ \midrule
12 & $-(1+N)/(4N^2)$ \\ \midrule
13 & $C_F(1+N)/(2N)$ \\ \midrule
14 & $-C_F(1+N)/(2N)$ \\ \midrule
15 & $C_F(1+N)/(2N)$ \\ \midrule
16 & $-(1+N)^2/(4N^2)$ \\ \midrule
17 & $-(1+N)^2/(4N^2)$ \\ \midrule
18 & $(1+N)^2/(4N^2)$ \\ \midrule
19 & $(1+N)^2/(4N^2)$ \\ \midrule
20 & $-(1+N)^2/(4N^2)$ \\ \midrule
21 & $-(1+N)^2/(4N^2)$ \\ \midrule
22 & $0$ \\ \midrule
23 & $0$ \\ \midrule
24 & $0$ \\ \midrule
25 & $(1+N)/4$ \\ \midrule
26 & $-(1+N)/4$ \\ \midrule
27 & $(1+N)/4$ \\ \midrule
28 & $0$ \\ \midrule
29 & $(1+N)/(2N)$$\times$ factors from loop \\ \midrule
30 & $-(1+N)/(2N)$$\times$ factors from loop \\ \midrule
31 & $(1+N)/(2N)$$\times$ factors from loop \\
\bottomrule[0.7mm]
\end{tabular}
\caption{$SU(3)$ color factors for Baryon number violating four-fermion operators with color contraction $\epsilon_{\alpha\beta\gamma}(q_1^{\alpha\,T}C\,\Gamma_1 q_2^\beta)(q_3^{\gamma\,T}C\,\Gamma_2\ell)$.}
\label{tab:color-mixingBviol}
\end{table}

\subsection{$SU(N) \times U(1)$ charge factors}
\label{app:colorfactorsSUNxU1}
The mixed $SU(N)\times U(1)$ color factors are reported in \cref{tab:color-mixingmixSUNxU1}, where the following abbreviations were used:

\begin{equation}
\begin{aligned}
    D_1 &= Q_1Q_2+Q_3Q_4\,, \\
    D_2 &= Q_1Q_3+Q_2Q_4\,, \\
    D_3 &= Q_1Q_4+Q_2Q_3\,, \\
    Q^2 &= Q_1^2+Q_2^2+Q_3^2+Q_4^2\,.
\end{aligned}
\label{eq:d1d2d3}
\end{equation}

\begin{table}[H]
\centering
\footnotesize
\setlength{\tabcolsep}{4pt}
\renewcommand{\arraystretch}{0.6}
\resizebox{\linewidth}{!}{
\begin{tabular}{@{}c c c c c@{}}
\toprule[0.7mm]
Diagram & $\mathcal{O}_1\to \mathcal{O}_1$ & $\mathcal{O}_1\to \mathcal{O}_8$ & $\mathcal{O}_8\to \mathcal{O}_1$ & $\mathcal{O}_8\to \mathcal{O}_8$ \\
\midrule[0.7mm]
1  & $C_F$ & $0$ & $0$ & $(-(C_A/2)+C_F)$ \\ \midrule
2  & $0$ & $1$ & $(1/4-1/(4N^2))$ & $(-1/N)$ \\ \midrule
3  & $0$ & $1$ & $C_F/(2N)$ & $(C_F-1/(2N))$ \\ \midrule
4  & $2D_1C_F$ & $0$ & $0$ & $2D_1(-(C_A/2)+C_F)$ \\ \midrule
5  & $0$ & $2D_2$ & $2D_2(1/4-1/(4N^2))$ & $-2D_2/N$ \\ \midrule
6  & $0$ & $2D_3$ & $2D_3C_F/(2N)$ & $2D_3(C_F-1/(2N))$ \\ \midrule
7  & $2D_1C_F$ & $0$ & $0$ & $2D_1(-(C_A/2)+C_F)$ \\ \midrule
8  & $0$ & $2D_2$ & $2D_2(1/4-1/(4N^2))$ & $-2D_2/N$ \\ \midrule
9  & $0$ & $2D_3$ & $2D_3C_F/(2N)$ & $2D_3(C_F-1/(2N))$ \\ \midrule
10  & $Q^2C_F+2D_1C_F$ & $0$ & $0$ & $Q^2(-(C_A/2)+C_F)+2D_1C_F$ \\ \midrule
11  & $2D_2C_F$ & $Q^2$ & $Q^2(1/4-1/(4N^2))$ & $Q^2(-1/N)+2D_2C_F$ \\ \midrule
12  & $2D_3C_F$ & $Q^2$ & $Q^2C_F/(2N)$ & $Q^2(C_F-1/(2N))+2D_3C_F$ \\ \midrule
13  & $Q^2C_F+2D_1C_F$ & $0$ & $0$ & $Q^2(-(C_A/2)+C_F)+2D_1C_F$ \\ \midrule
14  & $2D_2C_F$ & $Q^2$ & $Q^2(1/4-1/(4N^2))$ & $Q^2(-1/N)+2D_2C_F$ \\ \midrule
15  & $2D_3C_F$ & $Q^2$ & $Q^2C_F/(2N)$ & $Q^2(C_F-1/(2N))+2D_3C_F$ \\ \midrule
16  & $2D_2C_F$ & $2D_1$ & $2D_1(1/4-1/(4N^2))$ & $2D_1(-1/N)+2D_2(-(C_A/2)+C_F)$ \\ \midrule
17  & $2D_2C_F$ & $2D_1$ & $2D_1(1/4-1/(4N^2))$ & $2D_1(-1/N)+2D_2(-(C_A/2)+C_F)$ \\ \midrule
18  & $2D_3C_F$  & $2D_1$ & $2D_1C_F/(2N)$ & $2D_1(C_F-1/(2N))+2D_3(-(C_A/2)+C_F)$ \\ \midrule
19  & $2D_3C_F$  & $2D_1$ & $2D_1C_F/(2N)$ & $2D_1(C_F-1/(2N))+2D_3(-(C_A/2)+C_F)$ \\ \midrule
20  & $0$ & $2D_3+2D_2$ & $2D_3(1/4-1/(4N^2))+2D_2C_F/(2N)$ & $2D_3(-1/N)+2D_2(C_F-1/(2N))$ \\ \midrule
21  & $0$ & $2D_3+2D_2$ & $2D_3(1/4-1/(4N^2))+2D_2C_F/(2N)$ & $2D_3(-1/N)+2D_2(C_F-1/(2N))$ \\ \midrule
22  & $D_1C_F$ & $0$ & $0$ & $D_1(-(C_A/2)+C_F)$ \\ \midrule
23  & $0$ & $D_2$ & $D_2(1/4-1/(4N^2))$ & $D_2(-1/N)$ \\ \midrule
24  & $0$ & $D_3$ & $D_3C_F/(2N)$ & $D_3(C_F-1/(2N))$ \\
\bottomrule[0.7mm]
\end{tabular}
}
\caption{Color and charge factors for mixed $SU(N)\times U(1)$ corrections in the singlet/adjoint basis $\{\mathcal{O}_1,\mathcal{O}_8\}$. The entries include the relevant diagram multiplicities. The charge combinations $D_{1,2,3}$ and $Q^2$ are defined in \cref{eq:d1d2d3}.}
\label{tab:color-mixingmixSUNxU1}
\end{table}

\subsection{$U(1)_1 \times U(1)_2$ charge factors}
\label{app:colorfactorsU1xU1}
\hspace{-1cm}
\begin{table}[H]
\centering
\small
\setlength{\tabcolsep}{8pt}
\renewcommand{\arraystretch}{0.6}
\begin{tabular}{@{}c c@{}}
\toprule[0.7mm]
Diagram & Charge factor \\ 
\midrule[0.7mm]
1  & $Q_1^{[1]} Q_2^{[1]} + Q_3^{[1]} Q_4^{[1]}$ \\ \midrule
2  & $Q_2^{[1]} Q_4^{[1]} + Q_1^{[1]} Q_3^{[1]}$ \\ \midrule
3  & $Q_1^{[1]} Q_4^{[1]} + Q_2^{[1]} Q_3^{[1]}$ \\ \midrule
4  & $Q_1^{[1]} Q_2^{[1]}Q_1^{[2]} Q_2^{[2]} + Q_3^{[1]} Q_4^{[1]}Q_3^{[2]} Q_4^{[2]} + \mathrm{perm.}$ \\ \midrule
5  & $Q_2^{[1]} Q_4^{[1]}Q_2^{[2]} Q_4^{[2]} + Q_1^{[1]} Q_3^{[1]}Q_1^{[2]} Q_3^{[2]} + \mathrm{perm.}$ \\ \midrule
6  & $Q_2^{[1]} Q_3^{[1]}Q_2^{[2]} Q_3^{[2]} + Q_1^{[1]} Q_4^{[1]}Q_1^{[2]} Q_4^{[2]} + \mathrm{perm.}$ \\ \midrule
7  & $Q_1^{[1]} Q_2^{[1]}Q_1^{[2]} Q_2^{[2]} + Q_3^{[1]} Q_4^{[1]}Q_3^{[2]} Q_4^{[2]} + \mathrm{perm.}$ \\ \midrule
8  & $Q_2^{[1]} Q_4^{[1]}Q_2^{[2]} Q_4^{[2]} + Q_1^{[1]} Q_3^{[1]}Q_1^{[2]} Q_3^{[2]} + \mathrm{perm.}$ \\ \midrule
9  & $Q_2^{[1]} Q_3^{[1]}Q_2^{[2]} Q_3^{[2]} + Q_1^{[1]} Q_4^{[1]}Q_1^{[2]} Q_4^{[2]} + \mathrm{perm.}$ \\ \midrule
10 & $Q_1^{[1]} Q_2^{[1]}\big((Q_1^{[2]})^2+(Q_2^{[2]})^2\big) + Q_3^{[1]} Q_4^{[1]}\big((Q_3^{[2]})^2+(Q_4^{[2]})^2\big) + \mathrm{perm.}$ \\ \midrule
11 & $Q_2^{[1]} Q_4^{[1]}\big((Q_2^{[2]})^2+(Q_4^{[2]})^2\big) + Q_1^{[1]} Q_3^{[1]}\big((Q_1^{[2]})^2+(Q_3^{[2]})^2\big) + \mathrm{perm.}$ \\ \midrule
12 & $Q_2^{[1]} Q_3^{[1]}\big((Q_2^{[2]})^2+(Q_3^{[2]})^2\big) + Q_1^{[1]} Q_4^{[1]}\big((Q_1^{[2]})^2+(Q_4^{[2]})^2\big) + \mathrm{perm.}$ \\ \midrule
13 & $Q_1^{[1]} Q_2^{[1]}\big((Q_1^{[2]})^2+(Q_2^{[2]})^2\big) + Q_3^{[1]} Q_4^{[1]}\big((Q_3^{[2]})^2+(Q_4^{[2]})^2\big) + \mathrm{perm.}$ \\ \midrule
14 & $Q_2^{[1]} Q_4^{[1]}\big((Q_2^{[2]})^2+(Q_4^{[2]})^2\big) + Q_1^{[1]} Q_3^{[1]}\big((Q_1^{[2]})^2+(Q_3^{[2]})^2\big) + \mathrm{perm.}$ \\ \midrule
15 & $Q_2^{[1]} Q_3^{[1]}\big((Q_2^{[2]})^2+(Q_3^{[2]})^2\big) + Q_1^{[1]} Q_4^{[1]}\big((Q_1^{[2]})^2+(Q_4^{[2]})^2\big) + \mathrm{perm.}$ \\ \midrule
16 & $(Q_1^{[1]}Q_2^{[1]}+Q_3^{[1]}Q_4^{[1]})(Q_2^{[2]}Q_4^{[2]}+Q_1^{[2]}Q_3^{[2]}) + \mathrm{perm.}$ \\ \midrule
17 & $(Q_1^{[1]}Q_2^{[1]}+Q_3^{[1]}Q_4^{[1]})(Q_2^{[2]}Q_4^{[2]}+Q_1^{[2]}Q_3^{[2]}) + \mathrm{perm.}$ \\ \midrule
18 & $(Q_1^{[1]}Q_2^{[1]}+Q_3^{[1]}Q_4^{[1]})(Q_2^{[2]}Q_3^{[2]}+Q_1^{[2]}Q_4^{[2]}) + \mathrm{perm.}$ \\ \midrule
19 & $(Q_1^{[1]}Q_2^{[1]}+Q_3^{[1]}Q_4^{[1]})(Q_2^{[2]}Q_3^{[2]}+Q_1^{[2]}Q_4^{[2]}) + \mathrm{perm.}$ \\ \midrule
20 & $(Q_2^{[1]}Q_4^{[1]}+Q_1^{[1]}Q_3^{[1]})(Q_2^{[2]}Q_3^{[2]}+Q_1^{[2]}Q_4^{[2]}) + \mathrm{perm.}$ \\ \midrule
21 & $(Q_2^{[1]}Q_4^{[1]}+Q_1^{[1]}Q_3^{[1]})(Q_2^{[2]}Q_3^{[2]}+Q_1^{[2]}Q_4^{[2]}) + \mathrm{perm.}$ \\ \midrule
22 & $Q_1^{[1]} Q_2^{[1]} Q_3^{[2]} Q_4^{[2]} + \mathrm{perm.}$ \\ \midrule
23 & $Q_1^{[1]} Q_3^{[1]} Q_2^{[2]} Q_4^{[2]} + \mathrm{perm.}$ \\ \midrule
24 & $Q_1^{[1]} Q_4^{[1]} Q_2^{[2]} Q_3^{[2]} + \mathrm{perm.}$ \\ \midrule
25 & $0$ \\ \midrule
26 & $0$ \\ \midrule
27 & $0$ \\ \midrule
28 & $0$ \\ \midrule
29 & $(Q_1^{[1]} Q_2^{[1]} + Q_3^{[1]} Q_4^{[1]})\times\text{factors from loop}$ \\ \midrule
30 & $(Q_2^{[1]} Q_4^{[1]} + Q_1^{[1]} Q_3^{[1]})\times\text{factors from loop}$ \\ \midrule
31 & $(Q_2^{[1]} Q_3^{[1]} + Q_1^{[1]} Q_4^{[1]})\times\text{factors from loop}$ \\
\bottomrule[0.7mm]
\end{tabular}
\caption{Charge factors for $U(1)_1\times U(1)_2$ corrections to the operator $(\bar\psi_1\Gamma_1\psi_2)(\bar\psi_3\Gamma_2\psi_4)$. The symbol ``perm.'' denotes the additional contribution obtained by exchanging the two  factors, $Q_i^{[1]}\leftrightarrow Q_i^{[2]}$. For a single $U(1)$ one sets $Q_i^{[2]}\to Q_i^{[1]}$ and drops the permuted terms. For diagrams 29--31 the displayed factors multiply the corresponding self-energy loop factor.} \label{tab:chargefactorsU1xU1}
\end{table}

\subsection{$U(1)^2$ charge factors}
\label{app:colorfactorsU1same}

\begin{table}[H]
\centering
\small
\setlength{\tabcolsep}{8pt}
\renewcommand{\arraystretch}{0.6}
\begin{tabular}{@{}c c@{}}
\toprule[0.7mm]
Diagram & Charge factor \\ 
\midrule[0.7mm]
1  & $Q_1 Q_2 + Q_3 Q_4$ \\ \midrule
2  & $Q_2 Q_4 + Q_1 Q_3$ \\ \midrule
3  & $Q_1 Q_4 + Q_2 Q_3$ \\ \midrule
4  & $Q_1^2 Q_2^2 + Q_3^2 Q_4^2$ \\ \midrule
5  & $Q_2^2 Q_4^2 + Q_1^2 Q_3^2$ \\ \midrule
6  & $Q_2^2 Q_3^2 + Q_1^2 Q_4^2$ \\ \midrule
7  & $Q_1^2 Q_2^2 + Q_3^2 Q_4^2$ \\ \midrule
8  & $Q_2^2 Q_4^2 + Q_1^2 Q_3^2$ \\ \midrule
9  & $Q_2^2 Q_3^2 + Q_1^2 Q_4^2$ \\ \midrule
10 & $Q_1 Q_2^3 + Q_1^3 Q_2 + Q_3^3 Q_4 + Q_3Q_4^3$ \\ \midrule
11 & $Q_2^3 Q_4 + Q_1^3 Q_3 + Q_2 Q_4^3 + Q_1 Q_3^3$ \\ \midrule
12 & $Q_2^3 Q_3 + Q_2 Q_3^3 + Q_1^3 Q_4 + Q_1Q_4^3$ \\ \midrule
13 & $Q_1 Q_2^3 + Q_1^3 Q_2 + Q_3^3 Q_4 + Q_3 Q_4^3$ \\ \midrule
14 & $Q_2^3 Q_4 + Q_2 Q_4^3 + Q_1^3 Q_3 + Q_1 Q_3^3$ \\ \midrule
15 & $Q_2^3 Q_3 + Q_2 Q_3^3 + Q_1^3 Q_4 + Q_1 Q_4^3$ \\ \midrule
16 & $Q_1 Q_2^2 Q_4 + Q_1^2 Q_2 Q_3 + Q_2 Q_3 Q_4^2 + Q_1 Q_3^2 Q_4$ \\ \midrule
17 & $Q_1 Q_2^2 Q_4 + Q_1^2 Q_2 Q_3 + Q_2 Q_3 Q_4^2  + Q_1 Q_3^2 Q_4$ \\ \midrule
18 & $Q_1 Q_2^2 Q_3 + Q_1^2 Q_2 Q_4 + Q_2 Q_3^2 Q_4 + Q_1 Q_3 Q_4^2$ \\ \midrule
19 & $Q_1 Q_2^2 Q_3 + Q_1^2 Q_2 Q_4 + Q_2 Q_3^2  Q_4 + Q_1 Q_3 Q_4^2$ \\ \midrule
20 & $Q_2^2 Q_3 Q_4 + Q_1 Q_2 Q_4^2 + Q_1 Q_2 Q_3^2 + Q_1^2 Q_3 Q_4$ \\ \midrule
21 & $Q_2^2 Q_3 Q_4 + Q_1 Q_2 Q_4^2 + Q_1 Q_2 Q_3^2 + Q_1^2 Q_3 Q_4$ \\ \midrule
22 & $Q_1 Q_2 Q_3 Q_4$ \\ \midrule
23 & $Q_1 Q_2 Q_3 Q_4$ \\ \midrule
24 & $Q_1 Q_2 Q_3 Q_4$ \\ \midrule
25 & $0$ \\ \midrule
26 & $0$ \\ \midrule
27 & $0$ \\ \midrule
28 & $0$ \\ \midrule
29 & $(Q_1 Q_2 + Q_3 Q_4)$$\times$ factors from loop \\ \midrule
30 & $(Q_2 Q_4 + Q_1 Q_3)$$\times$ factors from loop \\ \midrule
31 & $(Q_2 Q_3 + Q_1 Q_4)$$\times$ factors from loop \\
\bottomrule[0.7mm]
\end{tabular}
\caption{$U(1)^2$ charge factors for the operator $(\bar\psi_1 \Gamma_1 \psi_2)(\bar\psi_3 \Gamma_2 \psi_4)$. For diagrams 29--31 the displayed factors multiply the corresponding self-energy loop factor.\label{tab:chargefactorsU1}}
\end{table}

\subsection{Yukawa factors}

In this subsection the Yukawa factors resulting from one- and two-loop scalar exchanges depicted in \cref{fig:oneloopscalar} and \cref{fig:twoloopscalar} are reported. We denote by $Y^a$ the Yukawa matrix with entries
$(Y^a)_{mn}=y^a_{mn}$ and write the four-fermion interaction as
\begin{equation}
\mathcal L_{\rm EFT}\supset
C_{ijkl}\,(\bar \psi_i\Gamma_1 \psi_j)(\bar \psi_k\Gamma_2 \psi_l)\,.
\end{equation}
The results are collected for diagrams 1--28 in \cref{tab:scalar-yukawa-contractions} and for the self-energy contributions in \cref{tab:scalar-yukawa-SEs}. Scalar-species labels are suppressed for diagrams 1--24 and are understood to be contracted along the scalar propagators of the corresponding topology. The symbol
``perm.'' denotes the remaining inequivalent placements on the external fermion
legs, including interchange of the two currents, with $Y$ versus $Y^\dagger$ fixed
by the fermion flow. For example, the factor of diagram 4 includes
\begin{equation}
(YY)_{ia}C_{abkl}(Y^\dagger Y^\dagger)_{bj}
+(YY)_{ka}C_{ijab}(Y^\dagger Y^\dagger)_{bl}\,.
\end{equation}
For diagrams 25--28, $G_iS_{jk}$ denotes the ordered assignment in which line $i$ is
the gauge line and lines $j,k$ are scalar lines.

\begin{table}[H]
\centering
\small
\setlength{\tabcolsep}{5pt}
\renewcommand{\arraystretch}{0.55}
\begin{tabular}{@{}c c@{}}
\toprule[0.7mm]
Diagram & Coupling contraction \\
\midrule[0.7mm]

1  & $(Y_1)_{ia}C_{abkl}(Y_2^\dagger)_{bj}+\mathrm{perm.}$ \\ \midrule
2  & $C_{iakb}(Y_1)_{aj}(Y_2^\dagger)_{bl}+\mathrm{perm.}$ \\ \midrule
3  & $(Y_1^\dagger)_{kb}C_{iabl}(Y_2)_{aj}+\mathrm{perm.}$ \\ \midrule

4  & $(Y_1Y_2)_{ia}C_{abkl}(Y_3^\dagger Y_4^\dagger)_{bj}+\mathrm{perm.}$ \\ \midrule
5  & $C_{iakb}(Y_1Y_2)_{aj}(Y_3^\dagger Y_4^\dagger)_{bl}+\mathrm{perm.}$ \\ \midrule
6  & $(Y_1^\dagger Y_2^\dagger)_{kb}C_{iabl}(Y_3Y_4)_{aj}+\mathrm{perm.}$ \\ \midrule
7  & $(Y_1Y_2)_{ia}C_{abkl}(Y_3^\dagger Y_4^\dagger)_{bj}+\mathrm{perm.}$ \\ \midrule
8  & $C_{iakb}(Y_1Y_2)_{aj}(Y_3^\dagger Y_4^\dagger)_{bl}+\mathrm{perm.}$ \\ \midrule
9  & $(Y_1^\dagger Y_2^\dagger)_{kb}C_{iabl}(Y_3Y_4)_{aj}+\mathrm{perm.}$ \\ \midrule

10 & $(Y_1)_{ia}C_{abkl}(Y_2Y_3^\dagger Y_4^\dagger)_{bj}+\mathrm{perm.}$ \\ \midrule
11 & $C_{iakb}(Y_1Y_2Y_3^\dagger)_{aj}(Y_4^\dagger)_{bl}+\mathrm{perm.}$ \\ \midrule
12 & $(Y_1^\dagger)_{kb}C_{iabl}(Y_2Y_3Y_4^\dagger)_{aj}+\mathrm{perm.}$ \\ \midrule
13 & $(Y_1)_{ia}C_{abkl}(Y_2Y_3^\dagger Y_4^\dagger)_{bj}+\mathrm{perm.}$ \\ \midrule
14 & $C_{iakb}(Y_1Y_2^\dagger Y_3)_{aj}(Y_4^\dagger)_{bl}+\mathrm{perm.}$ \\ \midrule
15 & $(Y_1^\dagger)_{kb}C_{iabl}(Y_2Y_3^\dagger Y_4)_{aj}+\mathrm{perm.}$ \\ \midrule

16 & $(Y_1)_{ia}C_{abkc}(Y_2Y_3^\dagger)_{bj}(Y_4^\dagger)_{cl}+\mathrm{perm.}$ \\ \midrule
17 & $(Y_1)_{ia}C_{abkc}(Y_2^\dagger Y_3)_{bj}(Y_4^\dagger)_{cl}+\mathrm{perm.}$ \\ \midrule
18 & $(Y_1)_{ia}(Y_2^\dagger)_{kc}C_{abcl}(Y_3Y_4^\dagger)_{bj}+\mathrm{perm.}$ \\ \midrule
19 & $(Y_1)_{ia}(Y_2^\dagger)_{kc}C_{abcl}(Y_3^\dagger Y_4)_{bj}+\mathrm{perm.}$ \\ \midrule
20 & $(Y_1^\dagger)_{kc}C_{iacd}(Y_2Y_3)_{aj}(Y_4^\dagger)_{dl}+\mathrm{perm.}$ \\ \midrule
21 & $(Y_1^\dagger)_{kc}C_{iacd}(Y_2Y_3)_{aj}(Y_4^\dagger)_{dl}+\mathrm{perm.}$ \\ \midrule

22 & $(Y_1)_{ia}(Y_2)_{kc}C_{abcd}(Y_3^\dagger)_{bj}(Y_4^\dagger)_{dl}+\mathrm{perm.}$ \\ \midrule
23 & $(Y_1)_{ia}(Y_2^\dagger)_{kc}C_{abcd}(Y_3)_{bj}(Y_4^\dagger)_{dl}+\mathrm{perm.}$ \\ \midrule
24 & $(Y_1)_{ia}(Y_2^\dagger)_{kc}C_{abcd}(Y_3)_{bj}(Y_4^\dagger)_{dl}+\mathrm{perm.}$ \\ \midrule

25 &
\makecell[l]{$\displaystyle
\begin{aligned}[t]
G_1S_{23}:&\quad
C_{iakl}(Y_2Y_3^\dagger)_{aj}+\mathrm{perm.}\\
G_2S_{13}:&\quad
(Y_1)_{ia}C_{abkl}(Y_3^\dagger)_{bj}+\mathrm{perm.}\\
G_3S_{12}:&\quad
(Y_1)_{ia}C_{abkl}(Y_2^\dagger)_{bj}+\mathrm{perm.}
\end{aligned}$}
\\ \midrule

26 &
\makecell[l]{$\displaystyle
\begin{aligned}[t]
G_1S_{23}:&\quad
C_{iakb}(Y_2)_{aj}(Y_3^\dagger)_{bl}+\mathrm{perm.}\\
G_2S_{13}:&\quad
C_{iakb}(Y_1)_{aj}(Y_3^\dagger)_{bl}+\mathrm{perm.}\\
G_3S_{12}:&\quad
C_{iakl}(Y_1Y_2^\dagger)_{aj}+\mathrm{perm.}
\end{aligned}$}
\\ \midrule

27 &
\makecell[l]{$\displaystyle
\begin{aligned}[t]
G_1S_{23}:&\quad
(Y_3^\dagger)_{kb}C_{iabl}(Y_2)_{aj}+\mathrm{perm.}\\
G_2S_{13}:&\quad
(Y_3^\dagger)_{kb}C_{iabl}(Y_1)_{aj}+\mathrm{perm.}\\
G_3S_{12}:&\quad
C_{iakl}(Y_1Y_2^\dagger)_{aj}+\mathrm{perm.}
\end{aligned}$}
\\ \midrule

28 &
\makecell[l]{$\displaystyle
\begin{aligned}[t]
G_1S_{23}:&\quad
C_{iakb}(Y_2)_{aj}(Y_3^\dagger)_{bl}+\mathrm{perm.}\\
G_2S_{13}:&\quad
(Y_1)_{ia}C_{ajkb}(Y_3^\dagger)_{bl}+\mathrm{perm.}\\
G_3S_{12}:&\quad
(Y_1)_{ia}C_{abkl}(Y_2^\dagger)_{bj}+\mathrm{perm.}
\end{aligned}$}
\\

\bottomrule[0.7mm]
\end{tabular}
\caption{Coupling factors for the one- and two-loop topologies 1--28 for operator insertions of the form
$C_{ijkl}(\bar\psi_i\Gamma_1\psi_j)(\bar\psi_k\Gamma_2\psi_l)$.}
\label{tab:scalar-yukawa-contractions}
\end{table}

\begin{table}[H]
\centering
\small
\setlength{\tabcolsep}{5pt}
\renewcommand{\arraystretch}{0.55}
\begin{tabular}{@{}c c@{}}
\toprule[0.7mm]
Diagram & Coupling contraction \\
\midrule[0.7mm]

29a &
$(Y_1)_{ia}C_{abkl}(Y_2^\dagger)_{bj}\,
\operatorname{Tr}[Y_3Y_4^\dagger]
+\mathrm{perm.}$
\\ \midrule

29b &
$(Y_1)_{ia}C_{abkl}(Y_2^\dagger)_{bj}\,
\kappa^2+\mathrm{perm.}$
\\ \midrule

29c &
$g_A^2\,
(Y_1)_{ia}C_{abkl}(Y_2^\dagger)_{bj}
(\theta^a\theta^a)_{12}
+\mathrm{perm.}$
\\ \midrule

29d &
$(Y_1)_{ia}C_{abkl}(Y_2^\dagger)_{bj}\,
\lambda+\mathrm{perm.}$
\\ \midrule

29e &
$\displaystyle
g_A^2\,
(Y_1)_{ia}C_{abkl}(Y_2^\dagger)_{bj}
\{\theta^a,\theta^a\}_{12}
+\mathrm{perm.}$
\\ \midrule

30a &
$C_{iakb}(Y_1)_{aj}(Y_2^\dagger)_{bl}\,
\operatorname{Tr}[Y_3Y_4^\dagger]
+\mathrm{perm.}$
\\ \midrule

30b &
$C_{iakb}(Y_1)_{aj}(Y_2^\dagger)_{bl}\,
\kappa^2+\mathrm{perm.}$
\\ \midrule

30c &
$g_A^2\,
C_{iakb}(Y_1)_{aj}(Y_2^\dagger)_{bl}
(\theta^a\theta^a)_{12}
+\mathrm{perm.}$
\\ \midrule

30d &
$C_{iakb}(Y_1)_{aj}(Y_2^\dagger)_{bl}\,
\lambda+\mathrm{perm.}$
\\ \midrule

30e &
$\displaystyle
g_A^2\,
C_{iakb}(Y_1)_{aj}(Y_2^\dagger)_{bl}
\{\theta^a,\theta^a\}_{12}
+\mathrm{perm.}$
\\ \midrule

31a &
$(Y_1^\dagger)_{kb}C_{iabl}(Y_2)_{aj}\,
\operatorname{Tr}[Y_3Y_4^\dagger]
+\mathrm{perm.}$
\\ \midrule

31b &
$(Y_1^\dagger)_{kb}C_{iabl}(Y_2)_{aj}\,
\kappa^2+\mathrm{perm.}$
\\ \midrule

31c &
$g_A^2\,
(Y_1^\dagger)_{kb}C_{iabl}(Y_2)_{aj}
(\theta^a\theta^a)_{21}
+\mathrm{perm.}$
\\ \midrule

31d &
$(Y_1^\dagger)_{kb}C_{iabl}(Y_2)_{aj}\,
\lambda+\mathrm{perm.}$
\\ \midrule

31e &
$\displaystyle
g_A^2\,
(Y_1^\dagger)_{kb}C_{iabl}(Y_2)_{aj}
\{\theta^a,\theta^a\}_{21}
+\mathrm{perm.}$
\\

\bottomrule[0.7mm]
\end{tabular}
\caption{Coupling factors for the two-loop self-energy topologies 29--31 for operator insertions of the form
$C_{ijkl}(\bar\psi_i\Gamma_1\psi_j)(\bar\psi_k\Gamma_2\psi_l)$. }
\label{tab:scalar-yukawa-SEs}
\end{table}

\newpage
\section{NDR results}\label{app:NDR}
As an illustration of the generic results obtained in \cref{app:genericGauge}, we present in this appendix the explicit expressions for gauge corrections to the JMS basis defined in \cref{eq:projected-Dirac-labels}. For simplicity we adopt the NDR scheme together with the evanescent scheme defined in \cite{Dekens:2019ept}. The one-loop finite terms are retained for general evanescent-scheme parameters, while in the two-loop tables all scheme constants have been set to one. The overall loop factor, color/charge factors, and diagram multiplicities are omitted. Diagrams 4--24 are common to $U(1)$ and $SU(N)$, whereas diagrams 25--28 and the non-abelian components of diagrams 29--31 are displayed separately. The results are in agreement with the findings in \cite{Aebischer:2025hsx}. 

\subsection{One-loop operator insertions}
\label{sec:1Lpolesphysical}

\begin{table}[h]
\centering
\setlength{\tabcolsep}{8pt}
\renewcommand{\arraystretch}{1}
\begin{tabular}{@{}c c c c@{}}
\toprule[0.7mm]
Diagram & $\SLL/\epsilon$ & $\TLL/\epsilon$ & $\SLL$ \\
\midrule[0.7mm] 
1 & $4$ & $0$ & $-4$
\\ 
\midrule 
2 & $-1$ & $1/4$ & $1/2$
\\ 
\midrule 
3 & $1$ & $1/4$ & $-1/2$
\\ 
\bottomrule[0.7mm]
\end{tabular}
\caption{ Poles and finite parts of the one-loop $\SLL$ operator insertion.}
\label{tab:1L_SLL}
\end{table}

\begin{table}[H]
\centering
\setlength{\tabcolsep}{8pt}
\renewcommand{\arraystretch}{1}
\begin{tabular}{@{}c c c c c c@{}}
\toprule[0.7mm]
Diagram & $\SLL/\epsilon$ & $\TLL/\epsilon$ & $E^{(4)}_{LL}/\epsilon$ & $\SLL$ & $\TLL$ \\
\midrule[0.7mm] 
1 & $0$ & $0$ & $0$ & $0$ & $0$
\\ 
\midrule 
2 & $12$ & $-3$ & $1/4$ & $4-24d_{\rm ev}$ & $2e_{\rm ev}-1/2$
\\ 
\midrule 
3 & $12$ & $3$ & $1/4$ & $4-24d_{\rm ev}$ & $2e_{\rm ev}-11/2$
\\ 
\bottomrule[0.7mm]
\end{tabular}
\caption{ Poles and finite parts of the one-loop $\TLL$ operator insertion.}
\label{tab:1L_TLL}
\end{table}

\begin{table}[H]
\centering
\setlength{\tabcolsep}{8pt}
\renewcommand{\arraystretch}{1}
\begin{tabular}{@{}c c c c@{}}
\toprule[0.7mm]
Diagram & $\VLL/\epsilon$ & $E^{(3)}_{LL}/\epsilon$ & $\VLL$ \\
\midrule[0.7mm] 
1 & $1$ & $0$ & $-2$
\\ 
\midrule 
2 & $-4$ & $-1/4$ & $b_{\rm ev}$
\\ 
\midrule 
3 & $1$ & $-1/4$ & $b_{\rm ev}-3$
\\ 
\bottomrule[0.7mm]
\end{tabular}
\caption{ Poles and finite parts of the one-loop $\VLL$ operator insertion.}
\label{tab:1L_VLL}
\end{table}

\begin{table}[H]
\centering
\setlength{\tabcolsep}{8pt}
\renewcommand{\arraystretch}{1}
\begin{tabular}{@{}c c c c@{}}
\toprule[0.7mm]
Diagram & $\VLR/\epsilon$ & $E^{(3)}_{LR}/\epsilon$ & $\VLR$ \\
\midrule[0.7mm] 
1 & $1$ & $0$ & $-2$
\\ 
\midrule 
2 & $-1$ & $-1/4$ & $-c_{\rm ev}$
\\ 
\midrule 
3 & $4$ & $-1/4$ & $-c_{\rm ev}-3$
\\ 
\bottomrule[0.7mm]
\end{tabular}
\caption{ Poles and finite parts of the one-loop $\VLR$ operator insertion.}
\label{tab:1L_VLR}
\end{table}

\begin{table}[H]
\centering
\setlength{\tabcolsep}{8pt}
\renewcommand{\arraystretch}{1}
\begin{tabular}{@{}c c c c@{}}
\toprule[0.7mm]
Diagram & $\SRL/\epsilon$ & $E^{(2)}_{RL}/\epsilon$ & $\SRL$ \\
\midrule[0.7mm] 
1 & $4$ & $0$ & $-4$
\\ 
\midrule 
2 & $-1$ & $-1/4$ & $-a_{\rm ev}$
\\ 
\midrule 
3 & $1$ & $-1/4$ & $-a_{\rm ev}-1$
\\ 
\bottomrule[0.7mm]
\end{tabular}
\caption{Poles and finite parts of the one-loop $\SRL$ operator insertion.}
\label{tab:1L_SRL}
\end{table}

\subsection{One-loop counterterm insertions}\label{sec:1LCT}

\begin{table}[H]
\centering
\setlength{\tabcolsep}{10pt}
\begin{tabular}{@{}c c c c c@{}}
\toprule[0.7mm]
Diagram & $\SLL/\epsilon$ & $\TLL/\epsilon$ \; & $\SLL$ & $\TLL$ \\
\midrule[0.7mm]
1 (Fermion)
& $-4\,\delta Z_\psi$
& $0$
\; & $\dfrac{16}{3}\,\delta Z_\psi$
& $0$
\\
\midrule
2 (Fermion)
& $\delta Z_\psi$
& $-\dfrac{1}{4}\,\delta Z_\psi$
\; & $-\dfrac{5}{6}\,\delta Z_\psi$
& $\dfrac{1}{12}\,\delta Z_\psi$
\\
\midrule
3 (Fermion)
& $-\delta Z_\psi$
& $-\dfrac{1}{4}\,\delta Z_\psi$
\; & $\dfrac{5}{6}\,\delta Z_\psi$
& $\dfrac{1}{12}\,\delta Z_\psi$
\\
\midrule
1 (Vertex)
& $4\,\delta Z_V$
& $0$
\; & $-4\,\delta Z_V$
& $0$
\\
\midrule
2 (Vertex)
& $-\delta Z_V$
& $\dfrac{1}{4}\,\delta Z_V$
\; & $\dfrac{1}{2}\,\delta Z_V$
& $0$
\\
\midrule
3 (Vertex)
& $\delta Z_V$
& $\dfrac{1}{4}\,\delta Z_V$
\; & $-\dfrac{1}{2}\,\delta Z_V$
& $0$
\\
\midrule
1 (Kinetic)
& $-3\,\delta Z_A$
& $0$
\; & $\dfrac{9}{2}\,\delta Z_A$
& $0$
\\
\midrule
2 (Kinetic)
& $0$
& $-\dfrac{1}{4}\,\delta Z_A$
\; & $0$
& $\dfrac{1}{12}\,\delta Z_A$
\\
\midrule
3 (Kinetic)
& $0$
& $-\dfrac{1}{4}\,\delta Z_A$
\; & $0$
& $\dfrac{1}{12}\,\delta Z_A$
\\
\midrule
1 (Mass)
& $0$
& $0$
\; & $-\dfrac{4}{3}\,\delta Z_A^{\rm mass}$
& $0$
\\
\midrule
2 (Mass)
& $0$
& $0$
\; & $\dfrac{1}{3}\,\delta Z_A^{\rm mass}$
& $-\dfrac{1}{12}\,\delta Z_A^{\rm mass}$
\\
\midrule
3 (Mass)
& $0$
& $0$
\; & $-\dfrac{1}{3}\,\delta Z_A^{\rm mass}$
& $-\dfrac{1}{12}\,\delta Z_A^{\rm mass}$
\\
\bottomrule[0.7mm]
\end{tabular}
\caption{One-loop counterterm insertions for the $\SLL$ operator. }
\label{tab:CT_SLL}
\end{table}

\begin{table}[H]
\centering
\setlength{\tabcolsep}{7pt}
\begin{tabular}{@{}c c c c c c c@{}}
\toprule[0.7mm]
Diagram
& $\SLL/\epsilon$
& $\TLL/\epsilon$
& $E^{(4)}_{LL}/\epsilon$
\; & $\SLL$
& $\TLL$
& $E^{(4)}_{LL}$
\\
\midrule[0.7mm]
1 (Fermion)
& $0$
& $0$
& $0$
\; & $0$
& $0$
& $0$
\\
\midrule
2 (Fermion)
& $-12\,\delta Z_\psi$
& $3\,\delta Z_\psi$
& $-\dfrac{1}{4}\,\delta Z_\psi$
\; & $24\,\delta Z_\psi$
& $-\dfrac{5}{2}\,\delta Z_\psi$
& $\dfrac{1}{12}\,\delta Z_\psi$
\\
\midrule
3 (Fermion)
& $-12\,\delta Z_\psi$
& $-3\,\delta Z_\psi$
& $-\dfrac{1}{4}\,\delta Z_\psi$
\; & $24\,\delta Z_\psi$
& $\dfrac{9}{2}\,\delta Z_\psi$
& $\dfrac{1}{12}\,\delta Z_\psi$
\\
\midrule
1 (Vertex)
& $0$
& $0$
& $0$
\; & $0$
& $0$
& $0$
\\
\midrule
2 (Vertex)
& $12\,\delta Z_V$
& $-3\,\delta Z_V$
& $\dfrac{1}{4}\,\delta Z_V$
\; & $-20\,\delta Z_V$
& $\dfrac{3}{2}\,\delta Z_V$
& $0$
\\
\midrule
3 (Vertex)
& $12\,\delta Z_V$
& $3\,\delta Z_V$
& $\dfrac{1}{4}\,\delta Z_V$
\; & $-20\,\delta Z_V$
& $-\dfrac{7}{2}\,\delta Z_V$
& $0$
\\
\midrule
1 (Kinetic)
& $0$
& $\delta Z_A$
& $0$
\; & $0$
& $-\dfrac{5}{6}\,\delta Z_A$
& $0$
\\
\midrule
2 (Kinetic)
& $-12\,\delta Z_A$
& $2\,\delta Z_A$
& $-\dfrac{1}{4}\,\delta Z_A$
\; & $24\,\delta Z_A$
& $-\dfrac{5}{3}\,\delta Z_A$
& $\dfrac{1}{12}\,\delta Z_A$
\\
\midrule
3 (Kinetic)
& $-12\,\delta Z_A$
& $-2\,\delta Z_A$
& $-\dfrac{1}{4}\,\delta Z_A$
\; & $24\,\delta Z_A$
& $\dfrac{11}{3}\,\delta Z_A$
& $\dfrac{1}{12}\,\delta Z_A$
\\
\midrule
1 (Mass)
& $0$
& $0$
& $0$
\; & $0$
& $0$
& $0$
\\
\midrule
2 (Mass)
& $0$
& $0$
& $0$
\; & $-4\,\delta Z_A^{\rm mass}$
& $\delta Z_A^{\rm mass}$
& $-\dfrac{1}{12}\,\delta Z_A^{\rm mass}$
\\
\midrule
3 (Mass)
& $0$
& $0$
& $0$
\; & $-4\,\delta Z_A^{\rm mass}$
& $-\delta Z_A^{\rm mass}$
& $-\dfrac{1}{12}\,\delta Z_A^{\rm mass}$
\\
\bottomrule[0.7mm]
\end{tabular}
\caption{One-loop counterterm insertions for the $\TLL$ operator.}
\label{tab:CT_TLL}
\end{table}

\begin{table}[H]
\centering
\setlength{\tabcolsep}{10pt}
\begin{tabular}{@{}c c c c c@{}}
\toprule[0.7mm]
Diagram
& $\VLL/\epsilon$
& $E^{(3)}_{LL}/\epsilon$
\; & $\VLL$
& $E^{(3)}_{LL}$
\\
\midrule[0.7mm]
1 (Fermion)
& $-\delta Z_\psi$
& $0$
\; & $\dfrac{7}{3}\,\delta Z_\psi$
& $0$
\\
\midrule
2 (Fermion)
& $4\,\delta Z_\psi$
& $\dfrac{1}{4}\,\delta Z_\psi$
\; & $-\dfrac{7}{3}\,\delta Z_\psi$
& $-\dfrac{1}{12}\,\delta Z_\psi$
\\
\midrule
3 (Fermion)
& $-\delta Z_\psi$
& $\dfrac{1}{4}\,\delta Z_\psi$
\; & $\dfrac{7}{3}\,\delta Z_\psi$
& $-\dfrac{1}{12}\,\delta Z_\psi$
\\
\midrule
1 (Vertex)
& $\delta Z_V$
& $0$
\; & $-2\,\delta Z_V$
& $0$
\\
\midrule
2 (Vertex)
& $-4\,\delta Z_V$
& $-\dfrac{1}{4}\,\delta Z_V$
\; & $\delta Z_V$
& $0$
\\
\midrule
3 (Vertex)
& $\delta Z_V$
& $-\dfrac{1}{4}\,\delta Z_V$
\; & $-2\,\delta Z_V$
& $0$
\\
\midrule
1 (Kinetic)
& $0$
& $0$
\; & $\dfrac{3}{2}\,\delta Z_A$
& $0$
\\
\midrule
2 (Kinetic)
& $3\,\delta Z_A$
& $\dfrac{1}{4}\,\delta Z_A$
\; & $-\dfrac{3}{2}\,\delta Z_A$
& $-\dfrac{1}{12}\,\delta Z_A$
\\
\midrule
3 (Kinetic)
& $0$
& $\dfrac{1}{4}\,\delta Z_A$
\; & $\dfrac{3}{2}\,\delta Z_A$
& $-\dfrac{1}{12}\,\delta Z_A$
\\
\midrule
1 (Mass)
& $0$
& $0$
\; & $-\dfrac{1}{3}\,\delta Z_A^{\rm mass}$
& $0$
\\
\midrule
2 (Mass)
& $0$
& $0$
\; & $\dfrac{4}{3}\,\delta Z_A^{\rm mass}$
& $\dfrac{1}{12}\,\delta Z_A^{\rm mass}$
\\
\midrule
3 (Mass)
& $0$
& $0$
\; & $-\dfrac{1}{3}\,\delta Z_A^{\rm mass}$
& $\dfrac{1}{12}\,\delta Z_A^{\rm mass}$
\\
\bottomrule[0.7mm]
\end{tabular}
\caption{One-loop counterterm insertions for the $\VLL$ operator.}
\label{tab:CT_VLL}
\end{table}

\begin{table}[H]
\centering
\setlength{\tabcolsep}{10pt}
\begin{tabular}{@{}c c c c c@{}}
\toprule[0.7mm]
Diagram
& $\VLR/\epsilon$
& $E^{(3)}_{LR}/\epsilon$
\; & $\VLR$
& $E^{(3)}_{LR}$
\\
\midrule[0.7mm]
1 (Fermion)
& $-\delta Z_\psi$
& $0$
\; & $\dfrac{7}{3}\,\delta Z_\psi$
& $0$
\\
\midrule
2 (Fermion)
& $\delta Z_\psi$
& $\dfrac{1}{4}\,\delta Z_\psi$
\; & $\dfrac{2}{3}\,\delta Z_\psi$
& $-\dfrac{1}{12}\,\delta Z_\psi$
\\
\midrule
3 (Fermion)
& $-4\,\delta Z_\psi$
& $\dfrac{1}{4}\,\delta Z_\psi$
\; & $\dfrac{16}{3}\,\delta Z_\psi$
& $-\dfrac{1}{12}\,\delta Z_\psi$
\\
\midrule
1 (Vertex)
& $\delta Z_V$
& $0$
\; & $-2\,\delta Z_V$
& $0$
\\
\midrule
2 (Vertex)
& $-\delta Z_V$
& $-\dfrac{1}{4}\,\delta Z_V$
\; & $-\delta Z_V$
& $0$
\\
\midrule
3 (Vertex)
& $4\,\delta Z_V$
& $-\dfrac{1}{4}\,\delta Z_V$
\; & $-4\,\delta Z_V$
& $0$
\\
\midrule
1 (Kinetic)
& $0$
& $0$
\; & $\dfrac{3}{2}\,\delta Z_A$
& $0$
\\
\midrule
2 (Kinetic)
& $0$
& $\dfrac{1}{4}\,\delta Z_A$
\; & $\dfrac{3}{2}\,\delta Z_A$
& $-\dfrac{1}{12}\,\delta Z_A$
\\
\midrule
3 (Kinetic)
& $-3\,\delta Z_A$
& $\dfrac{1}{4}\,\delta Z_A$
\; & $\dfrac{9}{2}\,\delta Z_A$
& $-\dfrac{1}{12}\,\delta Z_A$
\\
\midrule
1 (Mass)
& $0$
& $0$
\; & $-\dfrac{1}{3}\,\delta Z_A^{\rm mass}$
& $0$
\\
\midrule
2 (Mass)
& $0$
& $0$
\; & $\dfrac{1}{3}\,\delta Z_A^{\rm mass}$
& $\dfrac{1}{12}\,\delta Z_A^{\rm mass}$
\\
\midrule
3 (Mass)
& $0$
& $0$
\; & $-\dfrac{4}{3}\,\delta Z_A^{\rm mass}$
& $\dfrac{1}{12}\,\delta Z_A^{\rm mass}$
\\
\bottomrule[0.7mm]
\end{tabular}
\caption{One-loop counterterm insertions for the $\VLR$ operator.}
\label{tab:CT_VLR}
\end{table}

\begin{table}[H]
\centering
\setlength{\tabcolsep}{10pt}
\begin{tabular}{@{}c c c c c@{}}
\toprule[0.7mm]
Diagram
& $\SRL/\epsilon$
& $E^{(2)}_{RL}/\epsilon$
\; & $\SRL$
& $E^{(2)}_{RL}$
\\
\midrule[0.7mm]
1 (Fermion)
& $-4\,\delta Z_\psi$
& $0$
\; & $\dfrac{16}{3}\,\delta Z_\psi$
& $0$
\\
\midrule
2 (Fermion)
& $\delta Z_\psi$
& $\dfrac{1}{4}\,\delta Z_\psi$
\; & $\dfrac{2}{3}\,\delta Z_\psi$
& $-\dfrac{1}{12}\,\delta Z_\psi$
\\
\midrule
3 (Fermion)
& $-\delta Z_\psi$
& $\dfrac{1}{4}\,\delta Z_\psi$
\; & $\dfrac{7}{3}\,\delta Z_\psi$
& $-\dfrac{1}{12}\,\delta Z_\psi$
\\
\midrule
1 (Vertex)
& $4\,\delta Z_V$
& $0$
\; & $-4\,\delta Z_V$
& $0$
\\
\midrule
2 (Vertex)
& $-\delta Z_V$
& $-\dfrac{1}{4}\,\delta Z_V$
\; & $-\delta Z_V$
& $0$
\\
\midrule
3 (Vertex)
& $\delta Z_V$
& $-\dfrac{1}{4}\,\delta Z_V$
\; & $-2\,\delta Z_V$
& $0$
\\
\midrule
1 (Kinetic)
& $-3\,\delta Z_A$
& $0$
\; & $\dfrac{9}{2}\,\delta Z_A$
& $0$
\\
\midrule
2 (Kinetic)
& $0$
& $\dfrac{1}{4}\,\delta Z_A$
\; & $\dfrac{3}{2}\,\delta Z_A$
& $-\dfrac{1}{12}\,\delta Z_A$
\\
\midrule
3 (Kinetic)
& $0$
& $\dfrac{1}{4}\,\delta Z_A$
\; & $\dfrac{3}{2}\,\delta Z_A$
& $-\dfrac{1}{12}\,\delta Z_A$
\\
\midrule
1 (Mass)
& $0$
& $0$
\; & $-\dfrac{4}{3}\,\delta Z_A^{\rm mass}$
& $0$
\\
\midrule
2 (Mass)
& $0$
& $0$
\; & $\dfrac{1}{3}\,\delta Z_A^{\rm mass}$
& $\dfrac{1}{12}\,\delta Z_A^{\rm mass}$
\\
\midrule
3 (Mass)
& $0$
& $0$
\; & $-\dfrac{1}{3}\,\delta Z_A^{\rm mass}$
& $\dfrac{1}{12}\,\delta Z_A^{\rm mass}$
\\
\bottomrule[0.7mm]
\end{tabular}
\caption{One-loop counterterm insertions for the $\SRL$ operator.}
\label{tab:CT_SRL}
\end{table}

\subsection{Evanescent insertions}
\label{sec:1Lpolesev}
For the construction of the two-loop ADM, the finite parts of one-loop insertions of evanescent operators (EVs) are required. In this subsection we collect these contributions. We adopt the same definitions of EVs as in \cite{Dekens:2019ept}. For the $\VLL$, $\VLR$ and $\TLL$ insertions, we introduce additional EVs in order to reduce Dirac structures involving five and six gamma matrices:

\begin{equation}
\gamma^\mu\gamma^\nu\gamma^\rho\gamma^\sigma\gamma^\alpha P_L \otimes \gamma_\mu\gamma_\nu\gamma_\rho\gamma_\sigma\gamma_\alpha P_L = 32 (8-7 g_{\rm ev}\epsilon)\,\gamma^\mu P_L \otimes \gamma_\mu P_L + E^{(5)}_{LL}\,,
\end{equation}

\begin{equation}
\gamma^\mu\gamma^\nu\gamma^\rho\gamma^\sigma\gamma^\alpha P_L \otimes \gamma_\mu\gamma_\nu\gamma_\rho\gamma_\sigma\gamma_\alpha P_R = 16 (1+8 h_{\rm ev}\epsilon)\,\gamma^\mu P_L \otimes \gamma_\mu P_R + E^{(5)}_{LR}\,,
\end{equation}

\begin{equation}
\begin{aligned}
\gamma^\mu\gamma^\nu\gamma^\rho\gamma^\sigma\gamma^\alpha \gamma^\beta  P_L \otimes \gamma_\mu\gamma_\nu\gamma_\rho\gamma_\sigma\gamma_\alpha\gamma_\beta  P_L &= 128 (8-23 i_{\rm ev}\epsilon) P_L \otimes P_L \\
&- 32 (8-11 j_{\rm ev}\epsilon)\,\sigma^{\mu\nu} P_L \otimes \sigma_{\mu\nu} P_L + E^{(6)}_{LL}\,,
\end{aligned}
\end{equation}

where $g_{\rm ev}$, $h_{\rm ev}$, $i_{\rm ev}$ and $j_{\rm ev}$ denote scheme-dependent constants.

\begin{table}[H]
\centering
\setlength{\tabcolsep}{7pt}
\begin{tabular}{@{}c c c@{}}
\toprule[0.7mm]
Diagram & $\SLL$ & $\TLL$ \\
\midrule[0.7mm]
1  & $192$ & $-48 + 32\, e_{\rm ev}$ \\
\midrule
2  & $32 - 480\, d_{\rm ev} - 96\, e_{\rm ev} + 736\, i_{\rm ev}$ 
    & $-8 + 24\, d_{\rm ev} + 56\, e_{\rm ev} - 88\, j_{\rm ev}$ \\
\midrule
3  & $-160 - 480\, d_{\rm ev} - 96\, e_{\rm ev} + 736\, i_{\rm ev}$ 
    & $40 + 24\, d_{\rm ev} + 24\, e_{\rm ev} - 88\, j_{\rm ev}$ \\
\bottomrule[0.7mm]
\end{tabular}
\caption{Finite contribution to physical operators from the one-loop insertion of $E^{(4)}_{LL}$.}
\label{tab:E4}
\end{table}

\begin{table}[H]
\centering
\setlength{\tabcolsep}{7pt}
\begin{tabular}{@{}c c@{}}
\toprule[0.7mm]
Diagram & $\VLL$ \\
\midrule[0.7mm]
1  & $24$ \\
\midrule
2  & $-32\, b_{\rm ev} + 56\, g_{\rm ev}$ \\
\midrule
3  & $-24 - 32\, b_{\rm ev} + 56\, g_{\rm ev}$ \\
\bottomrule[0.7mm]
\end{tabular}
\caption{Finite contribution to physical operators from the one-loop insertion of $E^{(3)}_{LL}$.}
\label{tab:E3LL}
\end{table}

\begin{table}[H]
\centering
\setlength{\tabcolsep}{7pt}
\begin{tabular}{@{}c c@{}}
\toprule[0.7mm]
Diagram & $\VLR$ \\
\midrule[0.7mm]
1  & $-24$ \\
\midrule
2  & $8\, c_{\rm ev} - 32\, h_{\rm ev}$ \\
\midrule
3  & $24 + 8\, c_{\rm ev} - 32\, h_{\rm ev}$ \\
\bottomrule[0.7mm]
\end{tabular}
\caption{Finite contribution to physical operators from the one-loop insertion of $E^{(3)}_{LR}$.}
\label{tab:E4LR}
\end{table}

\begin{table}[H]
\centering
\setlength{\tabcolsep}{7pt}
\begin{tabular}{@{}c c@{}}
\toprule[0.7mm]
Diagram & $\SRL$ \\
\midrule[0.7mm]
1  & $-8 - 16\, a_{\rm ev}$ \\
\midrule
2  & $8\, a_{\rm ev} - 32\, f_{\rm ev}$ \\
\midrule
3  & $8 + 24\, a_{\rm ev} - 32\, f_{\rm ev}$ \\
\bottomrule[0.7mm]
\end{tabular}
\caption{Finite contribution to physical operators from the one-loop insertion of $E^{(2)}_{RL}$.}
\label{tab:E2RL}
\end{table}

\subsection{Two-loop operator insertions}\label{sec:2Lpoles}

\subsubsection{Abelian contributions}\label{sec:2LpolesU1SUNcommon}

\begin{table}[H]
\centering
\setlength{\tabcolsep}{8pt}
\renewcommand{\arraystretch}{1}
\begin{tabular}{@{}c c c c c c c@{}}
\toprule[0.7mm]
Diagram & 
$\SLL/\epsilon^2$ & $\TLL/\epsilon^2$ & $E^{(4)}_{LL}/\epsilon^2$ & $\SLL/\epsilon$ & $\TLL/\epsilon$  & $E^{(4)}_{LL}/\epsilon$\\
\midrule[0.7mm] 
4 & $8$ & $0$ & $0$ & $-8$ & $0$ & $0$ 
\\ 
\midrule 
5 & $2$ & $-1/2$ & $1/32$ & $-1$ & $-1/4$ & $1/64$ 
\\ 
\midrule 
6 & $2$ & $1/2$ & $1/32$ & $-1$ & $0$ & $1/64$ 
\\ 
\midrule 
7 & $0$ & $0$ & $0$ & $-2$ & $0$ & $0$ 
\\ 
\midrule 
8 & $0$ & $0$ & $0$ & $1$ & $1/4$ & $1/16$ 
\\ 
\midrule 
9 & $0$ & $0$ & $0$ & $1$ & $-1/4$ & $1/16$ 
\\ 
\midrule 
10 & $2$ & $0$ & $0$ & $-6$ & $0$ & $0$ 
\\
\end{tabular}
\caption{Pole structure of the two-loop $\SLL$ operator insertion.}
\label{tab:2L_SLL}
\end{table}
\newpage

\begin{table}[H]
\centering
\setlength{\tabcolsep}{7pt}
\begin{tabular}{@{}c c c c c c c c c@{}}
\toprule[0.7mm]
Diagram & $\TLL/\epsilon^2$ & $\SLL/\epsilon^2$ & $E^{(4)}_{LL}/\epsilon^2$ & $E^{(6)}_{LL}/\epsilon^2$ & $\TLL/\epsilon$ & $\SLL/\epsilon$ & $E^{(4)}_{LL}/\epsilon$ & $E^{(6)}_{LL}/\epsilon$ \\
\midrule[0.7mm] 
4  & $0$ & $0$ & $0$ & $0$ & $0$ & $0$ & $0$ & $0$
\\
\midrule 
5  & $6$ & $-24$ & $1/8$ & $-1/32$ & $-3$ & $52$ & $-1/4$ & $-1/64$
\\
\midrule 
6  & $6$ & $24$ & $9/8$ & $-1/32$ & $-7$ & $-40$ & $-1$ & $-1/64$
\\
\midrule 
7  & $0$ & $0$ & $0$ & $0$ & $2$ & $0$ & $0$ & $0$
\\
\midrule 
8  & $0$ & $0$ & $0$ & $0$ & $-1$ & $12$ & $3/2$ & $-1/16$
\\
\midrule 
9  & $0$ & $0$ & $0$ & $0$ & $-1$ & $-12$ & $1$ & $-1/16$
\\
\midrule 
10 & $0$ & $0$ & $0$ & $0$ & $1$ & $0$ & $0$ & $0$
\\
\end{tabular}
\caption{Pole structure of the two-loop $\TLL$ operator insertion.}
\label{tab:2L_TLL}
\end{table}
\newpage

\begin{table}[H]
\centering
\setlength{\tabcolsep}{8pt}
\renewcommand{\arraystretch}{1}
\begin{tabular}{@{}c c c c c c c@{}}
\toprule[0.7mm]
Diagram & 
$\VLL/\epsilon^2$ & $E^{(3)}_{LL}/\epsilon^2$ & $E^{(5)}_{LL}/\epsilon^2$ & $\VLL/\epsilon$ & $E^{(3)}_{LL}/\epsilon$  & $E^{(5)}_{LL}/\epsilon$\\
\midrule[0.7mm] 
4 & $1/2$ & $0$ & $0$ & $-3/4$ & $0$ & $0$
\\
\midrule 
5 & $8$ & $0$ & $1/32$ & $1$ & $1/4$ & $1/64$
\\
\midrule 
6 & $1/2$ & $-5/4$ & $1/32$ & $-3/4$ & $3/8$ & $1/64$
\\
\midrule 
7 & $0$ & $0$ & $0$ & $-2$ & $0$ & $0$
\\
\midrule 
8 & $0$ & $0$ & $0$ & $-2$ & $-5/4$ & $1/16$
\\
\midrule 
9 & $0$ & $0$ & $0$ & $-2$ & $-5/4$ & $1/16$
\\
\midrule 
10 & $1/2$ & $0$ & $0$ & $-5/4$ & $0$ & $0$
\\
\end{tabular}
\caption{Pole structure of the two-loop $\VLL$ operator insertion.}
\label{tab:2L_VLL}
\end{table}


\begin{table}[H]
\centering
\setlength{\tabcolsep}{8pt}
\renewcommand{\arraystretch}{1}
\begin{tabular}{@{}c c c c c c c@{}}
\toprule[0.7mm]
Diagram & 
$\VLR/\epsilon^2$ & $E^{(3)}_{LR}/\epsilon^2$ & $E^{(5)}_{LR}/\epsilon^2$ & $\VLR/\epsilon$ & $E^{(3)}_{LR}/\epsilon$  & $E^{(5)}_{LR}/\epsilon$\\
\midrule[0.7mm] 
4 & $1/2$ & $0$ & $0$ & $-3/4$ & $0$ & $0$
\\
\midrule 
5 & $1/2$ & $0$ & $1/32$ & $21/4$ & $1/4$ & $1/64$
\\
\midrule 
6 & $8$ & $-5/4$ & $1/32$ & $-8$ & $3/8$ & $1/64$
\\
\midrule 
7 & $0$ & $0$ & $0$ & $-2$ & $0$ & $0$
\\
\midrule 
8 & $0$ & $0$ & $0$ & $-2$ & $-5/4$ & $1/16$
\\
\midrule 
9 & $0$ & $0$ & $0$ & $-2$ & $-5/4$ & $1/16$
\\
\midrule 
10 & $1/2$ & $0$ & $0$ & $-5/4$ & $0$ & $0$
\\
\end{tabular}
\caption{Pole structure of the two-loop $\VLR$ operator insertion.}
\label{tab:2L_VLR}
\end{table}


\newpage
\begin{table}[H]
\centering
\setlength{\tabcolsep}{8pt}
\renewcommand{\arraystretch}{1}
\begin{tabular}{@{}c c c c c c c@{}}
\toprule[0.7mm]
Diagram & 
$\SRL/\epsilon^2$ & $E^{(2)}_{RL}/\epsilon^2$ & $E^{(4)}_{RL}/\epsilon^2$ & $\SRL/\epsilon$ & $E^{(2)}_{RL}/\epsilon$  & $E^{(4)}_{RL}/\epsilon$\\
\midrule[0.7mm] 
4 & $8$ & $0$ & $0$ & $-8$ & $0$ & $0$
\\
\midrule 
5 & $1/2$ & $0$ & $1/32$ & $21/4$ & $1/4$ & $1/64$
\\
\midrule 
6 & $1/2$ & $-1$ & $1/32$ & $-3/4$ & $0$ & $1/64$
\\
\midrule 
7 & $0$ & $0$ & $0$ & $-2$ & $0$ & $0$
\\
\midrule 
8 & $0$ & $0$ & $0$ & $-2$ & $-5/4$ & $1/16$
\\
\midrule 
9 & $0$ & $0$ & $0$ & $-2$ & $-3/4$ & $1/16$
\\
\midrule 
10 & $2$ & $0$ & $0$ & $-6$ & $0$ & $0$
\\
\end{tabular}
\caption{Pole structure of the two-loop $\SRL$ operator insertion.}
\label{tab:2L_SRL}
\end{table}


\subsubsection{$SU(N)$ tri-gauge-boson diagrams}
\label{subsec:trigluon}

\begin{table}[H]
\centering
\setlength{\tabcolsep}{10pt}
\begin{tabular}{@{}c c c c c@{}}
\toprule[0.7mm]
Diagram & $\SLL/\epsilon^2$ & $\TLL/\epsilon^2$ \; & $\SLL/\epsilon$ & $\TLL/\epsilon$ \\
\midrule[0.7mm]
25 & $6$ & $0$ \; & $-7$ & $0$ \\
\midrule
26 & $-3/2$ & $3/8$ \; & $1$ & $-1/16$ \\
\midrule
27 & $3/2$ & $3/8$ \; & $-1$ & $-1/16$ \\
\midrule
28 & $0$ & $0$ \; & $0$ & $-3/4$ \\
\bottomrule[0.7mm]
\end{tabular}
\caption{ Tri-gauge pole structure of the $\SLL$ insertion.}
\label{tab:trig_SLL}
\end{table}

\begin{table}[H]
\centering
\setlength{\tabcolsep}{10pt}
\begin{tabular}{@{}c c c c  c c c@{}}
\toprule[0.7mm]
Diagram & $\SLL/\epsilon^2$ & $\TLL/\epsilon^2$ & $E^{(4)}_{LL}/\epsilon^2$ \; & $\SLL/\epsilon$ & $\TLL/\epsilon$ & $E^{(4)}_{LL}/\epsilon$ \\
\midrule[0.7mm]
25 & $0$ & $0$ & $0$ \; & $0$ & $0$ & $0$ \\
\midrule
26 & $18$ & $-9/2$ & $3/8$ \; & $-33$ & $3$ & $-1/16$ \\
\midrule
27 & $18$ & $9/2$ & $3/8$ \; & $-33$ & $-6$ & $-1/16$ \\
\midrule
28 & $0$ & $0$ & $0$ \; & $36$ & $0$ & $3/4$ \\ 
\bottomrule[0.7mm]
\end{tabular}
\caption{ Tri-gauge pole structure of the $\TLL$ insertion.}
\label{tab:trig_TLL}
\end{table}

\begin{table}[H]
\centering
\setlength{\tabcolsep}{10pt}
\begin{tabular}{@{}c c c c c@{}}
\toprule[0.7mm]
Diagram & $\VLL/\epsilon^2$ & $E^{(3)}_{LL}/\epsilon^2$ \; & $\VLL/\epsilon$ & $E^{(3)}_{LL}/\epsilon$ \\
\midrule[0.7mm]
25 & $3/2$ & $0$ \; & $-13/4$ & $0$ \\
\midrule
26 & $-6$ & $-3/8$ \; & $5/2$ & $1/16$ \\
\midrule
27 & $3/2$ & $-3/8$ \; & $-13/4$ & $1/16$ \\
\midrule
28 & $0$ & $0$ \; & $0$ & $0$ \\
\bottomrule[0.7mm]
\end{tabular}
\caption{ Tri-gauge pole structure of the $\VLL$ insertion.}
\label{tab:trig_VLL}
\end{table}

\begin{table}[H]
\centering
\setlength{\tabcolsep}{10pt}
\begin{tabular}{@{}c c c  c c@{}}
\toprule[0.7mm]
Diagram & $\VLR/\epsilon^2$ & $E^{(3)}_{LR}/\epsilon^2$ \; & $\VLR/\epsilon$ & $E^{(3)}_{LR}/\epsilon$ \\
\midrule[0.7mm]
25 & $3/2$ & $0$ \; & $-13/4$ & $0$ \\
\midrule
26 & $-3/2$ & $-3/8$ \; & $-5/4$ & $1/16$ \\
\midrule
27 & $6$ & $-3/8$ \; & $-7$ & $1/16$ \\
\midrule
28 & $0$ & $0$ \; & $0$ & $0$ \\
\bottomrule[0.7mm]
\end{tabular}
\caption{ Tri-gauge pole structure of the $\VLR$ insertion.}
\label{tab:trig_VLR}
\end{table}

\begin{table}[H]
\centering
\setlength{\tabcolsep}{10pt}
\begin{tabular}{@{}c c c  c c@{}}
\toprule[0.7mm]
Diagram & $\SRL/\epsilon^2$ & $E^{(2)}_{RL}/\epsilon^2$ \; & $\SRL/\epsilon$ & $E^{(2)}_{RL}/\epsilon$ \\
\midrule[0.7mm]
25 & $6$ & $0$ \; & $-7$ & $0$ \\
\midrule
26 & $-3/2$ & $-3/8$ \; & $-5/4$ & $1/16$ \\
\midrule
27 & $3/2$ & $-3/8$ \; & $-13/4$ & $1/16$ \\
\midrule
28 & $0$ & $0$ \; & $0$ & $3/4$ \\
\bottomrule[0.7mm]
\end{tabular}
\caption{ Tri-gauge pole structure of the $\SRL$ insertion.}
\label{tab:trig_SRL}
\end{table}

\subsubsection{$SU(N)$ gauge boson self-energy diagrams}
\label{subsec:gluonSE}

\begin{table}[H]
\centering
\setlength{\tabcolsep}{12pt}
\begin{tabular}{@{}c c c @{}}
\toprule[0.7mm]
Diagram & $\SLL/\epsilon^2$ & $\SLL/\epsilon$ \\
\midrule[0.7mm]
29a & $-2N_T$ & $-2N_T$ \\
\midrule
29b & $9/4$ & $-9$ \\
\midrule
29c & $0$ & $4$ \\
\midrule
29d & $-1/ 4$ & $-1 / 2$ \\
\bottomrule[0.7mm]
\end{tabular}
\caption{ Diagram 29: pole structure of the $\SLL$ insertion.}
\label{tab:se29_SLL}
\end{table}

\begin{table}[H]
\centering
\setlength{\tabcolsep}{12pt}
\begin{tabular}{@{}c c c @{}}
\toprule[0.7mm]
Diagram & $\TLL/\epsilon^2$ & $\TLL/\epsilon$ \\
\midrule[0.7mm]
29a & $2 N_T / 3$ & $- 2 N_T / 3$ \\
\midrule
29b & $- 11 / 12$ & $7 / 12$ \\
\midrule
29c & $0$ & $0$ \\
\midrule
29d & $- 1 / 12$ & $1 / 12$ \\
\bottomrule[0.7mm]
\end{tabular}
\caption{ Diagram 29: pole structure of the $\TLL$ insertion.}
\label{tab:se29_TLL}
\end{table}

\begin{table}[H]
\centering
\setlength{\tabcolsep}{12pt}
\begin{tabular}{@{}c c c @{}}
\toprule[0.7mm]
Diagram & $\VLL/\epsilon^2$ & $\VLL/\epsilon$ \\
\midrule[0.7mm]
29a & 0 & $-N_T / 3$ \\
\midrule
29b & $-1/8$ & $-125 / 48$ \\
\midrule
29c & 0 & 1 \\
\midrule
29d & $-1/8$ & $-1 / 48$ \\
\bottomrule[0.7mm]
\end{tabular}
\caption{ Diagram 29: pole structure of the $\VLL$ insertion.}
\label{tab:se29_VLL}
\end{table}

\begin{table}[H]
\centering
\setlength{\tabcolsep}{12pt}
\begin{tabular}{@{}c c c @{}}
\toprule[0.7mm]
Diagram & $\VLR/\epsilon^2$ & $\VLR/\epsilon$ \\
\midrule[0.7mm]
29a & $0$ & $- N_T / 3$ \\
\midrule
29b & $- 1 / 8$ & $- 125 / 48$ \\
\midrule
29c & $0$ & $1$ \\
\midrule
29d & $- 1 / 8$ & $- 1 / 48$ \\
\bottomrule[0.7mm]
\end{tabular}
\caption{ Diagram 29: pole structure of the $\VLR$ insertion.}
\label{tab:se29_VLR}
\end{table}

\begin{table}[H]
\centering
\setlength{\tabcolsep}{12pt}
\begin{tabular}{@{}c c c @{}}
\toprule[0.7mm]
Diagram & $\SRL/\epsilon^2$ & $\SRL/\epsilon$ \\
\midrule[0.7mm]
29a & $- 2 N_T$ & $- 2 N_T$ \\
\midrule
29b & $9 / 4$ & $- 9$ \\
\midrule
29c & $0$ & $4$ \\
\midrule
29d & $- 1 / 4$ & $- 1 / 2$ \\
\bottomrule[0.7mm]
\end{tabular}
\caption{ Diagram 29: pole structure of the $\SRL$ insertion.}
\label{tab:se29_SRL}
\end{table}

\begin{table}[H]
\centering
\setlength{\tabcolsep}{12pt}
\begin{tabular}{@{}c c c c c@{}}
\toprule[0.7mm]
Diagram & $\SLL/\epsilon^2$ & $\TLL/\epsilon^2$ & $\SLL/\epsilon$ & $\TLL/\epsilon$ \\
\midrule[0.7mm]
30a & $0$ & $- N_T /6$ & $4 N_T /3$ & $- N_T /4$ \\
\midrule
30b & $1/8$ & $19/96$ & $17/12$ & $-77/192$ \\
\midrule
30c & $0$ & $0$ & $-1$ & $1/4$ \\
\midrule
30d & $1/8$ & $-1/96$ & $1/12$ & $-3/64$ \\
\bottomrule[0.7mm]
\end{tabular}
\caption{ Diagram 30: pole structure of the $\SLL$ insertion.}
\label{tab:se30_SLL}
\end{table}

\begin{table}[H]
\centering
\setlength{\tabcolsep}{12pt}
\begin{tabular}{@{}c c c c c c c@{}}
\toprule[0.7mm]
Diagram & $\SLL/\epsilon^2$ & $\TLL/\epsilon^2$ & $E^{(4)}_{LL}/\epsilon^2$
        & $\SLL/\epsilon$ & $\TLL/\epsilon$ & $E^{(4)}_{LL}/\epsilon$ \\
\midrule[0.7mm]
30a & $-8 N_T$ & $4 N_T /3$ & $- N_T /6$ & $4 N_T /3$ & $8 N_T /3$ & $- N_T /4$ \\
\midrule
30b & $19/2$ & $-35/24$ & $19/96$ & $-421/12$ & $65/12$ & $-77/192$ \\
\midrule
30c & $0$ & $0$ & $0$ & $12$ & $-3$ & $1/4$ \\
\midrule
30d & $-1/2$ & $5/24$ & $-1/96$ & $-17/12$ & $5/12$ & $-3/64$ \\
\bottomrule[0.7mm]
\end{tabular}
\caption{ Diagram 30: pole structure of the $\TLL$ insertion.}
\label{tab:se30_TLL}
\end{table}

\begin{table}[H]
\centering
\setlength{\tabcolsep}{12pt}
\begin{tabular}{@{}c c c c c@{}}
\toprule[0.7mm]
Diagram & $\VLL/\epsilon^2$ & $E^{(3)}_{LL}/\epsilon^2$ & $\VLL/\epsilon$ & $E^{(3)}_{LL}/\epsilon$ \\
\midrule[0.7mm]
30a & $2 N_T$ & $N_T /6$ & $4 N_T$ & $N_T /4$ \\
\midrule
30b & $-9/4$ & $-19/96$ & $53/8$ & $77/192$ \\
\midrule
30c & $0$ & $0$ & $-4$ & $-1/4$ \\
\midrule
30d & $1/4$ & $1/96$ & $5/8$ & $3/64$ \\
\bottomrule[0.7mm]
\end{tabular}
\caption{ Diagram 30: pole structure of the $\VLL$ insertion.}
\label{tab:se30_VLL}
\end{table}

\begin{table}[H]
\centering
\setlength{\tabcolsep}{12pt}
\begin{tabular}{@{}c c c c c@{}}
\toprule[0.7mm]
Diagram & $\VLR/\epsilon^2$ & $E^{(3)}_{LR}/\epsilon^2$ & $\VLR/\epsilon$ & $E^{(3)}_{LR}/\epsilon$ \\
\midrule[0.7mm]
30a & $0$ & $N_T /6$ & $7 N_T /3$ & $N_T /4$ \\
\midrule
30b & $1/8$ & $-19/96$ & $11/48$ & $77/192$ \\
\midrule
30c & $0$ & $0$ & $-1$ & $-1/4$ \\
\midrule
30d & $1/8$ & $1/96$ & $7/48$ & $3/64$ \\
\bottomrule[0.7mm]
\end{tabular}
\caption{ Diagram 30: pole structure of the $\VLR$ insertion.}
\label{tab:se30_VLR}
\end{table}

\begin{table}[H]
\centering
\setlength{\tabcolsep}{12pt}
\begin{tabular}{@{}c c c c c@{}}
\toprule[0.7mm]
Diagram & $\SRL/\epsilon^2$ & $E^{(2)}_{RL}/\epsilon^2$ & $\SRL/\epsilon$ & $E^{(2)}_{RL}/\epsilon$ \\
\midrule[0.7mm]
30a & $0$ & $N_T /6$ & $7 N_T /3$ & $N_T /4$ \\
\midrule
30b & $1/8$ & $-19/96$ & $11/48$ & $77/192$ \\
\midrule
30c & $0$ & $0$ & $-1$ & $-1/4$ \\
\midrule
30d & $1/8$ & $1/96$ & $7/48$ & $3/64$ \\
\bottomrule[0.7mm]
\end{tabular}
\caption{ Diagram 30: pole structure of the $\SRL$ insertion.}
\label{tab:se30_SRL}
\end{table}

\begin{table}[H]
\centering
\setlength{\tabcolsep}{12pt}
\begin{tabular}{@{}c c c c c@{}}
\toprule[0.7mm]
Diagram & $\SLL/\epsilon^2$ & $\TLL/\epsilon^2$ & $\SLL/\epsilon$ & $\TLL/\epsilon$ \\
\midrule[0.7mm]
31a & $0$ & $- N_T /6$ & $- 4 N_T /3$ & $- N_T /4$ \\
\midrule
31b & $-1/8$ & $19/96$ & $-17/12$ & $-77/192$ \\
\midrule
31c & $0$ & $0$ & $1$ & $1/4$ \\
\midrule
31d & $-1/8$ & $-1/96$ & $-1/12$ & $-3/64$ \\
\bottomrule[0.7mm]
\end{tabular}
\caption{ Diagram 31: pole structure of the $\SLL$ insertion.}
\label{tab:se31_SLL}
\end{table}

\begin{table}[H]
\centering
\setlength{\tabcolsep}{12pt}
\begin{tabular}{@{}c c c c c c c@{}}
\toprule[0.7mm]
Diagram & $\SLL/\epsilon^2$ & $\TLL/\epsilon^2$ & $E^{(4)}_{LL}/\epsilon^2$
        & $\SLL/\epsilon$ & $\TLL/\epsilon$ & $E^{(4)}_{LL}/\epsilon$ \\
\midrule[0.7mm]
31a & $- 8 N_T$ & $- 4 N_T /3$ & $- N_T /6$
     & $4 N_T /3$ & $- 4 N_T /3$ & $- N_T /4$ \\
\midrule
31b & $19/2$ & $35/24$ & $19/96$
     & $-421/12$ & $-7$ & $-77/192$ \\
\midrule
31c & $0$ & $0$ & $0$
     & $12$ & $3$ & $1/4$ \\
\midrule
31d & $-1/2$ & $-5/24$ & $-1/96$
     & $-17/12$ & $-1/3$ & $-3/64$ \\
\bottomrule[0.7mm]
\end{tabular}
\caption{ Diagram 31: pole structure of the $\TLL$ insertion.}
\label{tab:se31_TLL}
\end{table}

\begin{table}[H]
\centering
\setlength{\tabcolsep}{12pt}
\begin{tabular}{@{}c c c c c@{}}
\toprule[0.7mm]
Diagram & $\VLL/\epsilon^2$ & $E^{(3)}_{LL}/\epsilon^2$ & $\VLL/\epsilon$ & $E^{(3)}_{LL}/\epsilon$ \\
\midrule[0.7mm]
31a & $0$ & $N_T /6$ & $- N_T /3$ & $N_T /4$ \\
\midrule
31b & $-1/8$ & $-19/96$ & $-125/48$ & $77/192$ \\
\midrule
31c & $0$ & $0$ & $1$ & $-1/4$ \\
\midrule
31d & $-1/8$ & $1/96$ & $-1/48$ & $3/64$ \\
\bottomrule[0.7mm]
\end{tabular}
\caption{ Diagram 31: pole structure of the $\VLL$ insertion.}
\label{tab:se31_VLL}
\end{table}

\begin{table}[H]
\centering
\setlength{\tabcolsep}{12pt}
\begin{tabular}{@{}c c c c c@{}}
\toprule[0.7mm]
Diagram & $\VLR/\epsilon^2$ & $E^{(3)}_{LR}/\epsilon^2$ & $\VLR/\epsilon$ & $E^{(3)}_{LR}/\epsilon$ \\
\midrule[0.7mm]
31a & $- 2 N_T$ & $N_T /6$ & $- 2 N_T$ & $N_T /4$ \\
\midrule
31b & $9/4$ & $-19/96$ & $-9$ & $77/192$ \\
\midrule
31c & $0$ & $0$ & $4$ & $-1/4$ \\
\midrule
31d & $-1/4$ & $1/96$ & $-1/2$ & $3/64$ \\
\bottomrule[0.7mm]
\end{tabular}
\caption{ Diagram 31: pole structure of the $\VLR$ insertion.}
\label{tab:se31_VLR}
\end{table}

\begin{table}[H]
\centering
\setlength{\tabcolsep}{12pt}
\begin{tabular}{@{}c c c c c@{}}
\toprule[0.7mm]
Diagram & $\SRL/\epsilon^2$ & $E^{(2)}_{RL}/\epsilon^2$ & $\SRL/\epsilon$ & $E^{(2)}_{RL}/\epsilon$ \\
\midrule[0.7mm]
31a & $0$ & $N_T /6$ & $- N_T /3$ & $N_T /4$ \\
\midrule
31b & $-1/8$ & $-19/96$ & $-125/48$ & $77/192$ \\
\midrule
31c & $0$ & $0$ & $1$ & $-1/4$ \\
\midrule
31d & $-1/8$ & $1/96$ & $-1/48$ & $3/64$ \\
\bottomrule[0.7mm]
\end{tabular}
\caption{ Diagram 31: pole structure of the $\SRL$ insertion.}
\label{tab:se31_SRL}
\end{table}

\bibliographystyle{JHEP}
\bibliography{refs}

@article{Schnetz:2022nsc,
    author = "Schnetz, Oliver",
    title = "{{\ensuremath{\phi}}4 theory at seven loops}",
    eprint = "2212.03663",
    archivePrefix = "arXiv",
    primaryClass = "hep-th",
    doi = "10.1103/PhysRevD.107.036002",
    journal = "Phys. Rev. D",
    volume = "107",
    number = "3",
    pages = "036002",
    year = "2023"
}

@article{Fuentes-Martin:2024agf,
    author = "Fuentes-Mart\'\i{}n, Javier and Moreno-S\'anchez, Adri\'an and Palavri\'c, Ajdin and Thomsen, Anders Eller",
    title = "{A Guide to Functional Methods Beyond One-Loop Order}",
    eprint = "2412.12270",
    archivePrefix = "arXiv",
    primaryClass = "hep-ph",
    month = "12",
    year = "2024"
}

@article{Jenkins:2023rtg,
    author = "Jenkins, Elizabeth E. and Manohar, Aneesh V. and Naterop, Luca and Pag\`es, Julie",
    title = "{An algebraic formula for two loop renormalization of scalar quantum field theory}",
    eprint = "2308.06315",
    archivePrefix = "arXiv",
    primaryClass = "hep-ph",
    reportNumber = "ZU-TH 45/23, PSI-PR-23-29",
    doi = "10.1007/JHEP12(2023)165",
    journal = "JHEP",
    volume = "12",
    pages = "165",
    year = "2023"
}

@article{Jenkins:2023bls,
    author = "Jenkins, Elizabeth E. and Manohar, Aneesh V. and Naterop, Luca and Pag\`es, Julie",
    title = "{Two loop renormalization of scalar theories using a geometric approach}",
    eprint = "2310.19883",
    archivePrefix = "arXiv",
    primaryClass = "hep-ph",
    reportNumber = "ZU-TH 69/23, PSI-PR-23-39",
    doi = "10.1007/JHEP02(2024)131",
    journal = "JHEP",
    volume = "02",
    pages = "131",
    year = "2024"
}

@article{Aebischer:2025hsx,
    author = "Aebischer, Jason and Morell, Pol and Pesut, Marko and Virto, Javier",
    title = "{Two-loop anomalous dimensions in the LEFT: dimension-six four-fermion operators in NDR}",
    eprint = "2501.08384",
    archivePrefix = "arXiv",
    primaryClass = "hep-ph",
    reportNumber = "CERN-TH-2025-007",
    doi = "10.1007/JHEP05(2026)043",
    journal = "JHEP",
    volume = "05",
    pages = "043",
    year = "2026"
}

@article{DasBakshi:2026ief,
    author = "Das Bakshi, Supratim and Chala, Mikael and Ren, Zhe",
    title = "{Renormalization of the SMEFT to Dimension Eight: Fermionic Interactions II}",
    eprint = "2606.19202",
    archivePrefix = "arXiv",
    primaryClass = "hep-ph",
    month = "6",
    year = "2026"
}

@article{Zhang:2024clp,
    author = "Zhang, Di",
    title = "{Renormalization Group Equations for the Dimension-7 SMEFT Operators}",
    eprint = "2409.02622",
    archivePrefix = "arXiv",
    primaryClass = "hep-ph",
    doi = "10.22323/1.476.0776",
    journal = "PoS",
    volume = "ICHEP2024",
    pages = "776",
    year = "2025"
}

@article{Zhang:2023ndw,
    author = "Zhang, Di",
    title = "{Revisiting renormalization group equations of the SMEFT dimension-seven operators}",
    eprint = "2310.11055",
    archivePrefix = "arXiv",
    primaryClass = "hep-ph",
    reportNumber = "TUM-HEP 1475/23",
    doi = "10.1007/JHEP02(2024)133",
    journal = "JHEP",
    volume = "02",
    pages = "133",
    year = "2024"
}

@article{Antusch:2001ck,
    author = {Antusch, Stefan and Drees, Manuel and Kersten, J{\"o}rn and Lindner, Manfred and Ratz, Michael},
    title = "{Neutrino mass operator renormalization revisited}",
    eprint = "hep-ph/0108005",
    archivePrefix = "arXiv",
    reportNumber = "TUM-HEP-424-01",
    doi = "10.1016/S0370-2693(01)01127-3",
    journal = "Phys. Lett. B",
    volume = "519",
    pages = "238--242",
    year = "2001"
}

@article{Babu:1993qv,
    author = "Babu, K. S. and Leung, Chung Ngoc and Pantaleone, James T.",
    title = "{Renormalization of the neutrino mass operator}",
    eprint = "hep-ph/9309223",
    archivePrefix = "arXiv",
    reportNumber = "IUHET-252, UDHEP-93-03, BA-93-44",
    doi = "10.1016/0370-2693(93)90801-N",
    journal = "Phys. Lett. B",
    volume = "319",
    pages = "191--198",
    year = "1993"
}

@article{Chankowski:1993tx,
    author = "Chankowski, Piotr H. and Pluciennik, Zbigniew",
    title = "{Renormalization group equations for seesaw neutrino masses}",
    eprint = "hep-ph/9306333",
    archivePrefix = "arXiv",
    reportNumber = "ZU-TH-20-93, DFPD-93-TH-44",
    doi = "10.1016/0370-2693(93)90330-K",
    journal = "Phys. Lett. B",
    volume = "316",
    pages = "312--317",
    year = "1993"
}

@article{Aebischer:2022anv,
    author = "Aebischer, Jason and Buras, Andrzej J. and Kumar, Jacky",
    title = "{NLO QCD renormalization group evolution for nonleptonic \ensuremath{\Delta}F=2 transitions in the SMEFT}",
    eprint = "2203.11224",
    archivePrefix = "arXiv",
    primaryClass = "hep-ph",
    doi = "10.1103/PhysRevD.106.035003",
    journal = "Phys. Rev. D",
    volume = "106",
    number = "3",
    pages = "035003",
    year = "2022"
}

@article{Assi:2023zid,
    author = "Assi, Beno\^\i{}t and Helset, Andreas and Manohar, Aneesh V. and Pag\`es, Julie and Shen, Chia-Hsien",
    title = "{Fermion geometry and the renormalization of the Standard Model Effective Field Theory}",
    eprint = "2307.03187",
    archivePrefix = "arXiv",
    primaryClass = "hep-ph",
    reportNumber = "CALT-TH-2023-024, FERMILAB-PUB-23-362-T",
    doi = "10.1007/JHEP11(2023)201",
    journal = "JHEP",
    volume = "11",
    pages = "201",
    year = "2023"
}

@article{Helset:2022pde,
    author = "Helset, Andreas and Jenkins, Elizabeth E. and Manohar, Aneesh V.",
    title = "{Renormalization of the Standard Model Effective Field Theory from geometry}",
    eprint = "2212.03253",
    archivePrefix = "arXiv",
    primaryClass = "hep-ph",
    reportNumber = "CALT-TH-2022-041",
    doi = "10.1007/JHEP02(2023)063",
    journal = "JHEP",
    volume = "02",
    pages = "063",
    year = "2023"
}

@article{Jiang:2020mhe,
    author = "Jiang, Minyuan and Ma, Teng and Shu, Jing",
    title = "{Renormalization Group Evolution from On-shell SMEFT}",
    eprint = "2005.10261",
    archivePrefix = "arXiv",
    primaryClass = "hep-ph",
    doi = "10.1007/JHEP01(2021)101",
    journal = "JHEP",
    volume = "01",
    pages = "101",
    year = "2021"
}

@article{Chala:2023xjy,
    author = "Chala, Mikael and Li, Xu",
    title = "{Positivity restrictions on the mixing of dimension-eight SMEFT operators}",
    eprint = "2309.16611",
    archivePrefix = "arXiv",
    primaryClass = "hep-ph",
    doi = "10.1103/PhysRevD.109.065015",
    journal = "Phys. Rev. D",
    volume = "109",
    number = "6",
    pages = "065015",
    year = "2024"
}

@article{Bakshi:2024wzz,
    author = "Bakshi, S. D. and Chala, M. and D{\'\i}az-Carmona, {\'A}. and Ren, Z. and Vilches, F.",
    title = "{Renormalization of the SMEFT to dimension eight: Fermionic interactions I}",
    eprint = "2409.15408",
    archivePrefix = "arXiv",
    primaryClass = "hep-ph",
    doi = "10.1007/JHEP12(2024)214",
    journal = "JHEP",
    volume = "12",
    pages = "214",
    year = "2025"
}

@article{Jenkins:2017jig,
    author = "Jenkins, Elizabeth E. and Manohar, Aneesh V. and Stoffer, Peter",
    title = "{Low-Energy Effective Field Theory below the Electroweak Scale: Operators and Matching}",
    eprint = "1709.04486",
    archivePrefix = "arXiv",
    primaryClass = "hep-ph",
    doi = "10.1007/JHEP03(2018)016",
    journal = "JHEP",
    volume = "03",
    pages = "016",
    year = "2018",
    note = "[Erratum: JHEP 12, 043 (2023)]"
}

@article{Naterop:2023dek,
    author = "Naterop, Luca and Stoffer, Peter",
    title = "{Low-energy effective field theory below the electroweak scale: one-loop renormalization in the {\textquoteright}t Hooft-Veltman scheme}",
    eprint = "2310.13051",
    archivePrefix = "arXiv",
    primaryClass = "hep-ph",
    reportNumber = "PSI-PR-23-24, ZU-TH 66/23",
    doi = "10.1007/JHEP02(2024)068",
    journal = "JHEP",
    volume = "02",
    pages = "068",
    year = "2024"
}

@article{Zhang:2025ywe,
    author = "Zhang, Di",
    title = "{Two-loop renormalization group equations in the {\ensuremath{\nu}}SMEFT}",
    eprint = "2504.00792",
    archivePrefix = "arXiv",
    primaryClass = "hep-ph",
    doi = "10.1007/JHEP06(2025)106",
    journal = "JHEP",
    volume = "06",
    pages = "106",
    year = "2025"
}

@article{Naterop:2025cwg,
    author = "Naterop, Luca and Stoffer, Peter",
    title = "{Renormalization-group equations of the LEFT at two loops: dimension-six operators}",
    eprint = "2507.08926",
    archivePrefix = "arXiv",
    primaryClass = "hep-ph",
    reportNumber = "ZU-TH 48/25",
    doi = "10.1007/JHEP02(2026)016",
    journal = "JHEP",
    volume = "02",
    pages = "016",
    year = "2026"
}

@article{Naterop:2024cfx,
    author = "Naterop, Luca and Stoffer, Peter",
    title = "{Renormalization-group equations of the LEFT at two loops: dimension-five effects}",
    eprint = "2412.13251",
    archivePrefix = "arXiv",
    primaryClass = "hep-ph",
    reportNumber = "PSI-PR-24-30, ZU-TH 66/24",
    doi = "10.1007/JHEP06(2025)007",
    journal = "JHEP",
    volume = "06",
    pages = "007",
    year = "2025"
}

@article{DasBakshi:2023htx,
    author = "Das Bakshi, Supratim and D\'\i{}az-Carmona, \'Alvaro",
    title = "{Renormalisation of SMEFT bosonic interactions up to dimension eight by LNV operators}",
    eprint = "2301.07151",
    archivePrefix = "arXiv",
    primaryClass = "hep-ph",
    doi = "10.1007/JHEP06(2023)123",
    journal = "JHEP",
    volume = "06",
    pages = "123",
    year = "2023"
}

@article{DasBakshi:2022mwk,
    author = "Das Bakshi, Supratim and Chala, Mikael and D\'\i{}az-Carmona, \'Alvaro and Guedes, Guilherme",
    title = "{Towards the renormalisation of the Standard Model effective field theory to dimension eight: bosonic interactions II}",
    eprint = "2205.03301",
    archivePrefix = "arXiv",
    primaryClass = "hep-ph",
    doi = "10.1140/epjp/s13360-022-03194-5",
    journal = "Eur. Phys. J. Plus",
    volume = "137",
    number = "8",
    pages = "973",
    year = "2022"
}

@article{Chala:2021pll,
    author = "Chala, Mikael and Guedes, Guilherme and Ramos, Maria and Santiago, Jose",
    title = "{Towards the renormalisation of the Standard Model effective field theory to dimension eight: Bosonic interactions I}",
    eprint = "2106.05291",
    archivePrefix = "arXiv",
    primaryClass = "hep-ph",
    doi = "10.21468/SciPostPhys.11.3.065",
    journal = "SciPost Phys.",
    volume = "11",
    pages = "065",
    year = "2021"
}

@article{Ibarra:2024tpt,
    author = "Ibarra, Alejandro and Leister, Nicholas and Zhang, Di",
    title = "{Complete Two-loop Renormalization Group Equation of the Weinberg Operator}",
    eprint = "2411.08011",
    archivePrefix = "arXiv",
    primaryClass = "hep-ph",
    reportNumber = "MITP-24-081",
    month = "11",
    year = "2024"
}

@article{Aebischer:2025qhh,
    author = "Aebischer, Jason and Buras, Andrzej J. and Kumar, Jacky",
    title = "{SMEFT ATLAS: The landscape beyond the Standard Model}",
    eprint = "2507.05926",
    archivePrefix = "arXiv",
    primaryClass = "hep-ph",
    reportNumber = "AJB-25-1, CERN-TH-2025-129, LA-UR-24-24665",
    doi = "10.1016/j.physrep.2026.07.003",
    journal = "Phys. Rept.",
    volume = "1198",
    pages = "1--181",
    year = "2026"
}

@article{Born:2024mgz,
    author = "Born, Lukas and Fuentes-Mart{\'\i}n, Javier and Kvedarait{\.{e}}, Sandra and Thomsen, Anders Eller",
    title = "{Two-loop running in the bosonic SMEFT using functional methods}",
    eprint = "2410.07320",
    archivePrefix = "arXiv",
    primaryClass = "hep-ph",
    doi = "10.1007/JHEP05(2025)121",
    journal = "JHEP",
    volume = "05",
    pages = "121",
    year = "2025"
}

@article{Naterop:2025lzc,
    author = "Naterop, Luca and Stoffer, Peter",
    title = "{Renormalization-group equations of the LEFT at two loops: dimension-six baryon-number-violating operators}",
    eprint = "2505.03871",
    archivePrefix = "arXiv",
    primaryClass = "hep-ph",
    reportNumber = "PSI-PR-25-09, ZU-TH 31/25, INT-PUB-25-013",
    doi = "10.1007/JHEP07(2025)237",
    journal = "JHEP",
    volume = "07",
    pages = "237",
    year = "2025"
}

@article{Jenkins:2017dyc,
    author = "Jenkins, Elizabeth E. and Manohar, Aneesh V. and Stoffer, Peter",
    title = "{Low-Energy Effective Field Theory below the Electroweak Scale: Anomalous Dimensions}",
    eprint = "1711.05270",
    archivePrefix = "arXiv",
    primaryClass = "hep-ph",
    doi = "10.1007/JHEP01(2018)084",
    journal = "JHEP",
    volume = "01",
    pages = "084",
    year = "2018",
    note = "[Erratum: JHEP 12, 042 (2023)]"
}

@article{Alonso:2014zka,
    author = "Alonso, Rodrigo and Chang, Hsi-Ming and Jenkins, Elizabeth E. and Manohar, Aneesh V. and Shotwell, Brian",
    title = "{Renormalization group evolution of dimension-six baryon number violating operators}",
    eprint = "1405.0486",
    archivePrefix = "arXiv",
    primaryClass = "hep-ph",
    doi = "10.1016/j.physletb.2014.05.065",
    journal = "Phys. Lett. B",
    volume = "734",
    pages = "302--307",
    year = "2014"
}

@article{Machacek:1983tz,
    author = "Machacek, Marie E. and Vaughn, Michael T.",
    title = "{Two Loop Renormalization Group Equations in a General Quantum Field Theory. 1. Wave Function Renormalization}",
    reportNumber = "NUB-2590, HUTP-83/A003",
    doi = "10.1016/0550-3213(83)90610-7",
    journal = "Nucl. Phys. B",
    volume = "222",
    pages = "83--103",
    year = "1983"
}

@article{Davies:2021mnc,
    author = "Davies, Joshua and Herren, Florian and Thomsen, Anders Eller",
    title = "{General gauge-Yukawa-quartic $\beta$-functions at 4-3-2-loop order}",
    eprint = "2110.05496",
    archivePrefix = "arXiv",
    primaryClass = "hep-ph",
    reportNumber = "FERMILAB-PUB-21-471-T",
    doi = "10.1007/JHEP01(2022)051",
    journal = "JHEP",
    volume = "01",
    pages = "051",
    year = "2022"
}

@article{Bednyakov:2021qxa,
    author = "Bednyakov, Alexander and Pikelner, Andrey",
    title = "{Four-Loop Gauge and Three-Loop Yukawa Beta Functions in a General Renormalizable Theory}",
    eprint = "2105.09918",
    archivePrefix = "arXiv",
    primaryClass = "hep-ph",
    doi = "10.1103/PhysRevLett.127.041801",
    journal = "Phys. Rev. Lett.",
    volume = "127",
    number = "4",
    pages = "041801",
    year = "2021"
}

@article{Schienbein:2018fsw,
    author = "Schienbein, Ingo and Staub, Florian and Steudtner, Tom and Svirina, Kseniia",
    title = "{Revisiting RGEs for general gauge theories}",
    eprint = "1809.06797",
    archivePrefix = "arXiv",
    primaryClass = "hep-ph",
    reportNumber = "KA-TP-27-2018",
    doi = "10.1016/j.nuclphysb.2018.12.001",
    journal = "Nucl. Phys. B",
    volume = "939",
    pages = "1--48",
    year = "2019",
    note = "[Erratum: Nucl.Phys.B 966, 115339 (2021)]"
}

@article{Steudtner:2024teg,
    author = "Steudtner, Tom and Thomsen, Anders Eller",
    title = "{General quartic \ensuremath{\beta}-function at three loops}",
    eprint = "2408.05267",
    archivePrefix = "arXiv",
    primaryClass = "hep-ph",
    doi = "10.1007/JHEP10(2024)163",
    journal = "JHEP",
    volume = "10",
    pages = "163",
    year = "2024"
}

@article{Steudtner:2021fzs,
    author = "Steudtner, Tom",
    title = "{Towards general scalar-Yukawa renormalisation group equations at three-loop order}",
    eprint = "2101.05823",
    archivePrefix = "arXiv",
    primaryClass = "hep-th",
    reportNumber = "DO-TH 21/02",
    doi = "10.1007/JHEP05(2021)060",
    journal = "JHEP",
    volume = "05",
    pages = "060",
    year = "2021"
}

@article{Jack:2023zjt,
    author = "Jack, Ian and Osborn, Hugh and Steudtner, Tom",
    title = "{Explorations in scalar fermion theories: \ensuremath{\beta}-functions, supersymmetry and fixed points}",
    eprint = "2301.10903",
    archivePrefix = "arXiv",
    primaryClass = "hep-th",
    reportNumber = "DO-TH 22/06",
    doi = "10.1007/JHEP02(2024)038",
    journal = "JHEP",
    volume = "02",
    pages = "038",
    year = "2024"
}

@article{Poole:2019txl,
    author = "Poole, C. and Thomsen, A. E.",
    title = "{Weyl Consistency Conditions and $\gamma_5$}",
    eprint = "1901.02749",
    archivePrefix = "arXiv",
    primaryClass = "hep-th",
    doi = "10.1103/PhysRevLett.123.041602",
    journal = "Phys. Rev. Lett.",
    volume = "123",
    number = "4",
    pages = "041602",
    year = "2019"
}

@article{Sperling:2013xqa,
    author = {Sperling, Marcus and St\"ockinger, Dominik and Voigt, Alexander},
    title = "{Renormalization of vacuum expectation values in spontaneously broken gauge theories: Two-loop results}",
    eprint = "1310.7629",
    archivePrefix = "arXiv",
    primaryClass = "hep-ph",
    doi = "10.1007/JHEP01(2014)068",
    journal = "JHEP",
    volume = "01",
    pages = "068",
    year = "2014"
}

@article{Sperling:2013eva,
    author = {Sperling, Marcus and St\"ockinger, Dominik and Voigt, Alexander},
    title = "{Renormalization of vacuum expectation values in spontaneously broken gauge theories}",
    eprint = "1305.1548",
    archivePrefix = "arXiv",
    primaryClass = "hep-ph",
    doi = "10.1007/JHEP07(2013)132",
    journal = "JHEP",
    volume = "07",
    pages = "132",
    year = "2013"
}

@article{Mihaila:2012pz,
    author = "Mihaila, Luminita N. and Salomon, Jens and Steinhauser, Matthias",
    title = "{Renormalization constants and beta functions for the gauge couplings of the Standard Model to three-loop order}",
    eprint = "1208.3357",
    archivePrefix = "arXiv",
    primaryClass = "hep-ph",
    reportNumber = "SFB-CPP-12-61, TTP12-30",
    doi = "10.1103/PhysRevD.86.096008",
    journal = "Phys. Rev. D",
    volume = "86",
    pages = "096008",
    year = "2012"
}

@article{Jack:1984vj,
    author = "Jack, I. and Osborn, H.",
    title = "{General Background Field Calculations With Fermion Fields}",
    reportNumber = "DAMTP-84-2",
    doi = "10.1016/0550-3213(85)90088-4",
    journal = "Nucl. Phys. B",
    volume = "249",
    pages = "472--506",
    year = "1985"
}

@article{Poole:2019kcm,
    author = "Poole, Colin and Thomsen, Anders Eller",
    title = "{Constraints on 3- and 4-loop $\beta$-functions in a general four-dimensional Quantum Field Theory}",
    eprint = "1906.04625",
    archivePrefix = "arXiv",
    primaryClass = "hep-th",
    doi = "10.1007/JHEP09(2019)055",
    journal = "JHEP",
    volume = "09",
    pages = "055",
    year = "2019"
}

@article{Pickering:2001aq,
    author = "Pickering, A. G. M. and Gracey, J. A. and Jones, D. R. T.",
    title = "{Three loop gauge beta function for the most general single gauge coupling theory}",
    eprint = "hep-ph/0104247",
    archivePrefix = "arXiv",
    reportNumber = "IUHET-434, LTH-498",
    doi = "10.1016/S0370-2693(01)00624-4",
    journal = "Phys. Lett. B",
    volume = "510",
    pages = "347--354",
    year = "2001",
    note = "[Erratum: Phys.Lett.B 535, 377 (2002)]"
}

@article{Bednyakov:2021ojn,
    author = "Bednyakov, A. and Pikelner, A.",
    title = "{Six-loop beta functions in general scalar theory}",
    eprint = "2102.12832",
    archivePrefix = "arXiv",
    primaryClass = "hep-ph",
    doi = "10.1007/JHEP04(2021)233",
    journal = "JHEP",
    volume = "04",
    pages = "233",
    year = "2021"
}

@article{Bednyakov:2025sri,
    author = "Bednyakov, A. V.",
    title = "{On the scalar sector of 2HDM: ring of basis invariants, syzygies, and six-loop renormalization-group equations}",
    eprint = "2501.14087",
    archivePrefix = "arXiv",
    primaryClass = "hep-ph",
    month = "1",
    year = "2025"
}

@article{Luo:2002ti,
    author = "Luo, Ming-xing and Wang, Hua-wen and Xiao, Yong",
    title = "{Two loop renormalization group equations in general gauge field theories}",
    eprint = "hep-ph/0211440",
    archivePrefix = "arXiv",
    doi = "10.1103/PhysRevD.67.065019",
    journal = "Phys. Rev. D",
    volume = "67",
    pages = "065019",
    year = "2003"
}

@article{Machacek:1984zw,
    author = "Machacek, Marie E. and Vaughn, Michael T.",
    title = "{Two Loop Renormalization Group Equations in a General Quantum Field Theory. 3. Scalar Quartic Couplings}",
    reportNumber = "NUB-2653-REV, NUB-2653",
    doi = "10.1016/0550-3213(85)90040-9",
    journal = "Nucl. Phys. B",
    volume = "249",
    pages = "70--92",
    year = "1985"
}

@article{Henning:2014wua,
    author = "Henning, Brian and Lu, Xiaochuan and Murayama, Hitoshi",
    title = "{How to use the Standard Model effective field theory}",
    eprint = "1412.1837",
    archivePrefix = "arXiv",
    primaryClass = "hep-ph",
    reportNumber = "UCB-PTH-14-40, IPMU14-0353",
    doi = "10.1007/JHEP01(2016)023",
    journal = "JHEP",
    volume = "01",
    pages = "023",
    year = "2016"
}

@article{Drozd:2015rsp,
    author = "Drozd, Aleksandra and Ellis, John and Quevillon, J{\'e}r{\'e}mie and You, Tevong",
    title = "{The Universal One-Loop Effective Action}",
    eprint = "1512.03003",
    archivePrefix = "arXiv",
    primaryClass = "hep-ph",
    reportNumber = "KCL-PH-TH-2015-54, LCTS-2015-42, CERN-PH-TH-2015-284, DAMTP-2015-88, CAVENDISH-HEP-15-12",
    doi = "10.1007/JHEP03(2016)180",
    journal = "JHEP",
    volume = "03",
    pages = "180",
    year = "2016"
}

@article{delAguila:2016zcb,
    author = "del Aguila, Francisco and Kunszt, Zoltan and Santiago, Jose",
    title = "{One-loop effective lagrangians after matching}",
    eprint = "1602.00126",
    archivePrefix = "arXiv",
    primaryClass = "hep-ph",
    reportNumber = "UG-FT-319-16, CAFPE-189-16",
    doi = "10.1140/epjc/s10052-016-4081-1",
    journal = "Eur. Phys. J. C",
    volume = "76",
    number = "5",
    pages = "244",
    year = "2016"
}

@article{Henning:2016lyp,
    author = "Henning, Brian and Lu, Xiaochuan and Murayama, Hitoshi",
    title = "{One-loop Matching and Running with Covariant Derivative Expansion}",
    eprint = "1604.01019",
    archivePrefix = "arXiv",
    primaryClass = "hep-ph",
    reportNumber = "IPMU16-0042",
    doi = "10.1007/JHEP01(2018)123",
    journal = "JHEP",
    volume = "01",
    pages = "123",
    year = "2018"
}

@article{Fuentes-Martin:2016uol,
    author = "Fuentes-Martin, Javier and Portoles, Jorge and Ruiz-Femenia, Pedro",
    title = "{Integrating out heavy particles with functional methods: a simplified framework}",
    eprint = "1607.02142",
    archivePrefix = "arXiv",
    primaryClass = "hep-ph",
    reportNumber = "IFIC-16-28, TUM-HEP-1047-16",
    doi = "10.1007/JHEP09(2016)156",
    journal = "JHEP",
    volume = "09",
    pages = "156",
    year = "2016"
}

@article{Cohen:2020fcu,
    author = "Cohen, Timothy and Lu, Xiaochuan and Zhang, Zhengkang",
    title = "{Functional Prescription for EFT Matching}",
    eprint = "2011.02484",
    archivePrefix = "arXiv",
    primaryClass = "hep-ph",
    reportNumber = "CALT-TH-2020-047",
    doi = "10.1007/JHEP02(2021)228",
    journal = "JHEP",
    volume = "02",
    pages = "228",
    year = "2021"
}

@article{Cohen:2020qvb,
    author = "Cohen, Timothy and Lu, Xiaochuan and Zhang, Zhengkang",
    title = "{STrEAMlining EFT Matching}",
    eprint = "2012.07851",
    archivePrefix = "arXiv",
    primaryClass = "hep-ph",
    reportNumber = "CALT-TH-2020-056",
    doi = "10.21468/SciPostPhys.10.5.098",
    journal = "SciPost Phys.",
    volume = "10",
    number = "5",
    pages = "098",
    year = "2021"
}

@article{Fuentes-Martin:2020udw,
    author = {Fuentes-Martin, Javier and K{\"o}nig, Matthias and Pag{\`e}s, Julie and Thomsen, Anders Eller and Wilsch, Felix},
    title = "{SuperTracer: A Calculator of Functional Supertraces for One-Loop EFT Matching}",
    eprint = "2012.08506",
    archivePrefix = "arXiv",
    primaryClass = "hep-ph",
    reportNumber = "MITP-20-076, TUM-HEP-1302/20, ZU-TH-54/20",
    doi = "10.1007/JHEP04(2021)281",
    journal = "JHEP",
    volume = "04",
    pages = "281",
    year = "2021"
}

@article{Zhang:2016pja,
    author = "Zhang, Zhengkang",
    title = "{Covariant diagrams for one-loop matching}",
    eprint = "1610.00710",
    archivePrefix = "arXiv",
    primaryClass = "hep-ph",
    reportNumber = "MCTP-16-23, DESY-16-188",
    doi = "10.1007/JHEP05(2017)152",
    journal = "JHEP",
    volume = "05",
    pages = "152",
    year = "2017"
}

@article{Machacek:1983fi,
    author = "Machacek, Marie E. and Vaughn, Michael T.",
    title = "{Two Loop Renormalization Group Equations in a General Quantum Field Theory. 2. Yukawa Couplings}",
    reportNumber = "NUB 2611",
    doi = "10.1016/0550-3213(84)90533-9",
    journal = "Nucl. Phys. B",
    volume = "236",
    pages = "221--232",
    year = "1984"
}

@article{Fonseca:2025zjb,
    author = "Fonseca, Renato M. and Olgoso, Pablo and Santiago, Jos\'e",
    title = "{Renormalization of general Effective Field Theories: Formalism and renormalization of bosonic operators}",
    eprint = "2501.13185",
    archivePrefix = "arXiv",
    primaryClass = "hep-ph",
    month = "1",
    year = "2025"
}

@article{Bresciani:2024shu,
    author = "Bresciani, Luigi C. and Brunello, Giacomo and Levati, Gabriele and Mastrolia, Pierpaolo and Paradisi, Paride",
    title = "{Renormalization of effective field theories via on-shell methods: the case of axion-like particles}",
    eprint = "2412.04160",
    archivePrefix = "arXiv",
    primaryClass = "hep-ph",
    month = "12",
    year = "2024"
}

@article{Jenkins:2013zja,
    author = "Jenkins, Elizabeth E. and Manohar, Aneesh V. and Trott, Michael",
    title = "{Renormalization Group Evolution of the Standard Model Dimension Six Operators I: Formalism and lambda Dependence}",
    eprint = "1308.2627",
    archivePrefix = "arXiv",
    primaryClass = "hep-ph",
    doi = "10.1007/JHEP10(2013)087",
    journal = "JHEP",
    volume = "10",
    pages = "087",
    year = "2013"
}

@article{Alonso:2013hga,
    author = "Alonso, Rodrigo and Jenkins, Elizabeth E. and Manohar, Aneesh V. and Trott, Michael",
    title = "{Renormalization Group Evolution of the Standard Model Dimension Six Operators III: Gauge Coupling Dependence and Phenomenology}",
    eprint = "1312.2014",
    archivePrefix = "arXiv",
    primaryClass = "hep-ph",
    reportNumber = "CERN-PH-TH-2013-305, CERN-PH-TH/2013-305",
    doi = "10.1007/JHEP04(2014)159",
    journal = "JHEP",
    volume = "04",
    pages = "159",
    year = "2014"
}

@article{Jenkins:2013wua,
    author = "Jenkins, Elizabeth E. and Manohar, Aneesh V. and Trott, Michael",
    title = "{Renormalization Group Evolution of the Standard Model Dimension Six Operators II: Yukawa Dependence}",
    eprint = "1310.4838",
    archivePrefix = "arXiv",
    primaryClass = "hep-ph",
    reportNumber = "CERN-PH-TH/2015-247",
    doi = "10.1007/JHEP01(2014)035",
    journal = "JHEP",
    volume = "01",
    pages = "035",
    year = "2014"
}

@article{Fuentes-Martin:2023ljp,
    author = "Fuentes-Mart{\'\i}n, Javier and Palavri{\'c}, Ajdin and Thomsen, Anders Eller",
    title = "{Functional matching and renormalization group equations at two-loop order}",
    eprint = "2311.13630",
    archivePrefix = "arXiv",
    primaryClass = "hep-ph",
    doi = "10.1016/j.physletb.2024.138557",
    journal = "Phys. Lett. B",
    volume = "851",
    pages = "138557",
    year = "2024"
}

@article{Dekens:2019ept,
    author = "Dekens, Wouter and Stoffer, Peter",
    title = "{Low-energy effective field theory below the electroweak scale: matching at one loop}",
    eprint = "1908.05295",
    archivePrefix = "arXiv",
    primaryClass = "hep-ph",
    doi = "10.1007/JHEP10(2019)197",
    journal = "JHEP",
    volume = "10",
    pages = "197",
    year = "2019",
    note = "[Erratum: JHEP 11, 148 (2022)]"
}

@article{Carmona:2021xtq,
    author = "Carmona, Adrian and Lazopoulos, Achilleas and Olgoso, Pablo and Santiago, Jose",
    title = "{Matchmakereft: automated tree-level and one-loop matching}",
    eprint = "2112.10787",
    archivePrefix = "arXiv",
    primaryClass = "hep-ph",
    doi = "10.21468/SciPostPhys.12.6.198",
    journal = "SciPost Phys.",
    volume = "12",
    number = "6",
    pages = "198",
    year = "2022"
}

@article{Fuentes-Martin:2022jrf,
    author = {Fuentes-Mart{\'\i}n, Javier and K{\"o}nig, Matthias and Pag{\`e}s, Julie and Thomsen, Anders Eller and Wilsch, Felix},
    title = "{A proof of concept for matchete: an automated tool for matching effective theories}",
    eprint = "2212.04510",
    archivePrefix = "arXiv",
    primaryClass = "hep-ph",
    reportNumber = "MITP-22-105, TUM-HEP-1443/22, ZU-TH-58/22",
    doi = "10.1140/epjc/s10052-023-11726-1",
    journal = "Eur. Phys. J. C",
    volume = "83",
    number = "7",
    pages = "662",
    year = "2023"
}

@article{Buras:1989xd,
    author = "Buras, Andrzej J. and Weisz, Peter H.",
    title = "{QCD Nonleading Corrections to Weak Decays in Dimensional Regularization and 't Hooft-Veltman Schemes}",
    reportNumber = "MPI-PAE/PTh-42/89, TUM-T32-189",
    doi = "10.1016/0550-3213(90)90223-Z",
    journal = "Nucl. Phys. B",
    volume = "333",
    pages = "66--99",
    year = "1990"
}

@article{Chetyrkin:1997fm,
    author = "Chetyrkin, Konstantin G. and Misiak, Mikolaj and Munz, Manfred",
    title = "{Beta functions and anomalous dimensions up to three loops}",
    eprint = "hep-ph/9711266",
    archivePrefix = "arXiv",
    reportNumber = "MPI-PHT-97-45, TTP-97-43, ZU-TH-16-97, TUM-HEP-284-97, IFT-11-97",
    doi = "10.1016/S0550-3213(98)00122-9",
    journal = "Nucl. Phys. B",
    volume = "518",
    pages = "473--494",
    year = "1998"
}

@article{Herrlich:1994kh,
    author = "Herrlich, Stefan and Nierste, Ulrich",
    title = "{Evanescent operators, scheme dependences and double insertions}",
    eprint = "hep-ph/9412375",
    archivePrefix = "arXiv",
    reportNumber = "TUM-T31-66-94, PSI-PR-94-37",
    doi = "10.1016/0550-3213(95)00474-7",
    journal = "Nucl. Phys. B",
    volume = "455",
    pages = "39--58",
    year = "1995"
}

@article{Aebischer:2017gaw,
    author = "Aebischer, Jason and Fael, Matteo and Greub, Christoph and Virto, Javier",
    title = "{B physics Beyond the Standard Model at One Loop: Complete Renormalization Group Evolution below the Electroweak Scale}",
    eprint = "1704.06639",
    archivePrefix = "arXiv",
    primaryClass = "hep-ph",
    doi = "10.1007/JHEP09(2017)158",
    journal = "JHEP",
    volume = "09",
    pages = "158",
    year = "2017"
}

@article{Buras:2000if,
    author = "Buras, Andrzej J. and Misiak, Mikolaj and Urban, Joerg",
    title = "{Two loop QCD anomalous dimensions of flavor changing four quark operators within and beyond the standard model}",
    eprint = "hep-ph/0005183",
    archivePrefix = "arXiv",
    reportNumber = "TUM-HEP-371-00, CERN-TH-2000-124, IFT-13-2000, TUM-HEP-371/00, CERN-TH/2000-124, IFT-13/2000",
    doi = "10.1016/S0550-3213(00)00437-5",
    journal = "Nucl. Phys. B",
    volume = "586",
    pages = "397--426",
    year = "2000",
    note = "[Erratum: Nucl.Phys.B 1002, 116529 (2024)]"
}

@article{Morell:2024aml,
    author = "Morell, Pol and Virto, Javier",
    title = "{On the two-loop penguin contributions to the Anomalous Dimensions of four-quark operators}",
    eprint = "2402.00249",
    archivePrefix = "arXiv",
    primaryClass = "hep-ph",
    doi = "10.1007/JHEP04(2024)105",
    journal = "JHEP",
    volume = "04",
    pages = "105",
    year = "2024"
}

@article{Aebischer:2021raf,
    author = "Aebischer, Jason and Bobeth, Christoph and Buras, Andrzej J. and Kumar, Jacky and Misiak, Miko{\l}aj",
    title = "{General non-leptonic {\ensuremath{\Delta}}F = 1 WET at the NLO in QCD}",
    eprint = "2107.10262",
    archivePrefix = "arXiv",
    primaryClass = "hep-ph",
    reportNumber = "AJB-21-5",
    doi = "10.1007/JHEP11(2021)227",
    journal = "JHEP",
    volume = "11",
    pages = "227",
    year = "2021"
}

@article{Buras:1992tc,
    author = "Buras, Andrzej J. and Jamin, Matthias and Lautenbacher, Markus E. and Weisz, Peter H.",
    title = "{Two loop anomalous dimension matrix for $\Delta S = 1$ weak nonleptonic decays I: $\mathcal{O}(\alpha_s^2)$}",
    eprint = "hep-ph/9211304",
    archivePrefix = "arXiv",
    reportNumber = "MPI-PAE-PTH-106-92, TUM-T31-18-92",
    doi = "10.1016/0550-3213(93)90397-8",
    journal = "Nucl. Phys. B",
    volume = "400",
    pages = "37--74",
    year = "1993"
}

@article{Ciuchini:1993vr,
    author = "Ciuchini, Marco and Franco, E. and Martinelli, G. and Reina, L.",
    title = "{The Delta S = 1 effective Hamiltonian including next-to-leading order QCD and QED corrections}",
    eprint = "hep-ph/9304257",
    archivePrefix = "arXiv",
    reportNumber = "LPTENS-93-11, ROME-913-1992, ULB-TH-93-03",
    doi = "10.1016/0550-3213(94)90118-X",
    journal = "Nucl. Phys. B",
    volume = "415",
    pages = "403--462",
    year = "1994"
}

@article{Ciuchini:1993ks,
    author = "Ciuchini, Marco and Franco, E. and Martinelli, G. and Reina, L. and Silvestrini, L.",
    title = "{Scheme independence of the effective Hamiltonian for b ---{\ensuremath{>}} s gamma and b ---{\ensuremath{>}} s g decays}",
    eprint = "hep-ph/9307364",
    archivePrefix = "arXiv",
    reportNumber = "LPTENS-93-28, ROME-93-958, ULB-TH-93-09",
    doi = "10.1016/0370-2693(93)90668-8",
    journal = "Phys. Lett. B",
    volume = "316",
    pages = "127--136",
    year = "1993"
}

@article{Buras:1991jm,
    author = "Buras, Andrzej J. and Jamin, Matthias and Lautenbacher, M. E. and Weisz, Peter H.",
    title = "{Effective Hamiltonians for $\Delta S = 1$ and $\Delta B = 1$ nonleptonic decays beyond the leading logarithmic approximation}",
    reportNumber = "MPI-PAE-PTH-56-91, TUM-T31-16-91",
    doi = "10.1016/0550-3213(92)90345-C",
    journal = "Nucl. Phys. B",
    volume = "370",
    pages = "69--104",
    year = "1992",
    note = "[Addendum: Nucl.Phys.B 375, 501 (1992)]"
}

@article{Chetyrkin:1997gb,
    author = "Chetyrkin, Konstantin G. and Misiak, Mikolaj and Munz, Manfred",
    title = "{$|\Delta F| = 1$ nonleptonic effective Hamiltonian in a simpler scheme}",
    eprint = "hep-ph/9711280",
    archivePrefix = "arXiv",
    reportNumber = "MPI-PHT-97-51, TTP-97-44, ZU-TH-17-97, TUM-HEP-285-97, IFT-12-97",
    doi = "10.1016/S0550-3213(98)00131-X",
    journal = "Nucl. Phys. B",
    volume = "520",
    pages = "279--297",
    year = "1998"
}

@article{Gorbahn:2004my,
    author = "Gorbahn, Martin and Haisch, Ulrich",
    title = "{Effective Hamiltonian for non-leptonic $|\Delta F| = 1$ decays at NNLO in QCD}",
    eprint = "hep-ph/0411071",
    archivePrefix = "arXiv",
    reportNumber = "IPPP-04-66, DCPT-04-132, FERMILAB-PUB-04-281-T",
    doi = "10.1016/j.nuclphysb.2005.01.047",
    journal = "Nucl. Phys. B",
    volume = "713",
    pages = "291--332",
    year = "2005"
}

@article{Buras:1992zv,
    author = "Buras, Andrzej J. and Jamin, Matthias and Lautenbacher, Markus E.",
    title = "{Two loop anomalous dimension matrix for Delta S = 1 weak nonleptonic decays. 2. O(alpha-alpha-s)}",
    eprint = "hep-ph/9211321",
    archivePrefix = "arXiv",
    reportNumber = "MPI-PAE-TH-107-92, TUM-T31-30-92",
    doi = "10.1016/0550-3213(93)90398-9",
    journal = "Nucl. Phys. B",
    volume = "400",
    pages = "75--102",
    year = "1993"
}

@article{Buchalla:2019wsc,
    author = "Buchalla, Gerhard and Celis, Alejandro and Krause, Claudius and Toelstede, Jan-Niklas",
    title = "{Master Formula for One-Loop Renormalization of Bosonic SMEFT Operators}",
    eprint = "1904.07840",
    archivePrefix = "arXiv",
    primaryClass = "hep-ph",
    reportNumber = "LMU-ASC\textasciitilde{}15/19, FERMILAB-PUB-19-003-T",
    month = "4",
    year = "2019"
}

@article{Isidori:2023pyp,
    author = "Isidori, Gino and Wilsch, Felix and Wyler, Daniel",
    title = "{The standard model effective field theory at work}",
    eprint = "2303.16922",
    archivePrefix = "arXiv",
    primaryClass = "hep-ph",
    reportNumber = "ZU-TH 14/23",
    doi = "10.1103/RevModPhys.96.015006",
    journal = "Rev. Mod. Phys.",
    volume = "96",
    number = "1",
    pages = "015006",
    year = "2024"
}

@article{Brivio:2017vri,
    author = "Brivio, Ilaria and Trott, Michael",
    title = "{The Standard Model as an Effective Field Theory}",
    eprint = "1706.08945",
    archivePrefix = "arXiv",
    primaryClass = "hep-ph",
    doi = "10.1016/j.physrep.2018.11.002",
    journal = "Phys. Rept.",
    volume = "793",
    pages = "1--98",
    year = "2019"
}

@article{Grzadkowski:2010es,
    author = "Grzadkowski, B. and Iskrzynski, M. and Misiak, M. and Rosiek, J.",
    title = "{Dimension-Six Terms in the Standard Model Lagrangian}",
    eprint = "1008.4884",
    archivePrefix = "arXiv",
    primaryClass = "hep-ph",
    reportNumber = "IFT-9-2010, TTP10-35",
    doi = "10.1007/JHEP10(2010)085",
    journal = "JHEP",
    volume = "10",
    pages = "085",
    year = "2010"
}

@article{Buchmuller:1985jz,
    author = "Buchmuller, W. and Wyler, D.",
    title = "{Effective Lagrangian Analysis of New Interactions and Flavor Conservation}",
    reportNumber = "CERN-TH-4254/85",
    doi = "10.1016/0550-3213(86)90262-2",
    journal = "Nucl. Phys. B",
    volume = "268",
    pages = "621--653",
    year = "1986"
}

@article{EliasMiro:2020tdv,
    author = "Elias Mir\'o, Joan and Ingoldby, James and Riembau, Marc",
    title = "{EFT anomalous dimensions from the S-matrix}",
    eprint = "2005.06983",
    archivePrefix = "arXiv",
    primaryClass = "hep-ph",
    doi = "10.1007/JHEP09(2020)163",
    journal = "JHEP",
    volume = "09",
    pages = "163",
    year = "2020"
}

@article{Bern:2020ikv,
    author = "Bern, Zvi and Parra-Martinez, Julio and Sawyer, Eric",
    title = "{Structure of two-loop SMEFT anomalous dimensions via on-shell methods}",
    eprint = "2005.12917",
    archivePrefix = "arXiv",
    primaryClass = "hep-ph",
    doi = "10.1007/JHEP10(2020)211",
    journal = "JHEP",
    volume = "10",
    pages = "211",
    year = "2020"
}

@article{Panico:2018hal,
    author = "Panico, Giuliano and Pomarol, Alex and Riembau, Marc",
    title = "{EFT approach to the electron Electric Dipole Moment at the two-loop level}",
    eprint = "1810.09413",
    archivePrefix = "arXiv",
    primaryClass = "hep-ph",
    reportNumber = "DESY-18-185",
    doi = "10.1007/JHEP04(2019)090",
    journal = "JHEP",
    volume = "04",
    pages = "090",
    year = "2019"
}

@article{Baratella:2020lzz,
    author = "Baratella, Pietro and Fernandez, Clara and Pomarol, Alex",
    title = "{Renormalization of Higher-Dimensional Operators from On-shell Amplitudes}",
    eprint = "2005.07129",
    archivePrefix = "arXiv",
    primaryClass = "hep-ph",
    doi = "10.1016/j.nuclphysb.2020.115155",
    journal = "Nucl. Phys. B",
    volume = "959",
    pages = "115155",
    year = "2020"
}

@article{Caron-Huot:2016cwu,
    author = "Caron-Huot, Simon and Wilhelm, Matthias",
    title = "{Renormalization group coefficients and the S-matrix}",
    eprint = "1607.06448",
    archivePrefix = "arXiv",
    primaryClass = "hep-th",
    doi = "10.1007/JHEP12(2016)010",
    journal = "JHEP",
    volume = "12",
    pages = "010",
    year = "2016"
}

@article{Cheung:2015aba,
    author = "Cheung, Clifford and Shen, Chia-Hsien",
    title = "{Nonrenormalization Theorems without Supersymmetry}",
    eprint = "1505.01844",
    archivePrefix = "arXiv",
    primaryClass = "hep-ph",
    reportNumber = "CALT-2015-024",
    doi = "10.1103/PhysRevLett.115.071601",
    journal = "Phys. Rev. Lett.",
    volume = "115",
    number = "7",
    pages = "071601",
    year = "2015"
}

@article{Bresciani:2023jsu,
    author = "Bresciani, L. C. and Levati, G. and Mastrolia, P. and Paradisi, P.",
    title = "{Anomalous dimensions via on-shell methods: Operator mixing and leading mass effects}",
    eprint = "2312.05206",
    archivePrefix = "arXiv",
    primaryClass = "hep-ph",
    doi = "10.1103/PhysRevD.110.056041",
    journal = "Phys. Rev. D",
    volume = "110",
    number = "5",
    pages = "056041",
    year = "2024"
}

@article{Misiak:2025xzq,
    author = "Misiak, Miko{\l}aj and Nalec, Ignacy",
    title = "{One-loop renormalization group equations in generic effective field theories. Part I. Bosonic operators}",
    eprint = "2501.17134",
    archivePrefix = "arXiv",
    primaryClass = "hep-ph",
    doi = "10.1007/JHEP06(2025)210",
    journal = "JHEP",
    volume = "06",
    pages = "210",
    year = "2025"
}

@article{AccettulliHuber:2021uoa,
    author = "Accettulli Huber, Manuel and De Angelis, Stefano",
    title = "{Standard Model EFTs via on-shell methods}",
    eprint = "2108.03669",
    archivePrefix = "arXiv",
    primaryClass = "hep-th",
    reportNumber = "QMUL-PH-21-32, SAGEX-21-17-E",
    doi = "10.1007/JHEP11(2021)221",
    journal = "JHEP",
    volume = "11",
    pages = "221",
    year = "2021"
}

@article{Bern:2019wie,
    author = "Bern, Zvi and Parra-Martinez, Julio and Sawyer, Eric",
    title = "{Nonrenormalization and Operator Mixing via On-Shell Methods}",
    eprint = "1910.05831",
    archivePrefix = "arXiv",
    primaryClass = "hep-ph",
    reportNumber = "UCLA/TEP/2019/105, CERN-TH-2019-160",
    doi = "10.1103/PhysRevLett.124.051601",
    journal = "Phys. Rev. Lett.",
    volume = "124",
    number = "5",
    pages = "051601",
    year = "2020"
}

@article{Aebischer:2025zxg,
    author = "Aebischer, Jason and Bresciani, Luigi C. and Selimovic, Nudzeim",
    title = "{Anomalous dimension of a general effective gauge theory. Part I. Bosonic sector}",
    eprint = "2502.14030",
    archivePrefix = "arXiv",
    primaryClass = "hep-ph",
    reportNumber = "CERN-TH-2025-032",
    doi = "10.1007/JHEP08(2025)209",
    journal = "JHEP",
    volume = "08",
    pages = "209",
    year = "2025"
}

@article{Liao:2024xel,
    author = "Liao, Yi and Ma, Xiao-Dong and Wang, Hao-Lin",
    title = "{Probing dimension-8 SMEFT operators through neutral meson mixing}",
    eprint = "2409.10305",
    archivePrefix = "arXiv",
    primaryClass = "hep-ph",
    doi = "10.1007/JHEP03(2025)133",
    journal = "JHEP",
    volume = "03",
    pages = "133",
    year = "2025"
}

@article{Wu:2025qto,
    author = "Wu, Chao and Xiao, Ming-Lei and Yu, Jiang-Hao and Zheng, Yu-Hui",
    title = "{On-Shell Renormalization of Dim-8 SMEFT from Complete Amplitude Basis: I. Four-Fermion Operators}",
    eprint = "2512.21724",
    archivePrefix = "arXiv",
    primaryClass = "hep-ph",
    month = "12",
    year = "2025"
}

@article{Born:2026xkr,
    author = "Born, Lukas and Fuentes-Mart{\'\i}n, Javier and Thomsen, Anders Eller",
    title = "{Next-to-Leading Order Running in the SMEFT}",
    eprint = "2601.19974",
    archivePrefix = "arXiv",
    primaryClass = "hep-ph",
    month = "1",
    year = "2026"
}

@article{Machado:2022ozb,
    author = "Machado, Camila S. and Renner, Sophie and Sutherland, Dave",
    title = "{Building blocks of the flavourful SMEFT RG}",
    eprint = "2210.09316",
    archivePrefix = "arXiv",
    primaryClass = "hep-ph",
    reportNumber = "DESY-22-161",
    doi = "10.1007/JHEP03(2023)226",
    journal = "JHEP",
    volume = "03",
    pages = "226",
    year = "2023"
}

@article{Chala:2024llp,
    author = "Chala, Mikael and L{\'o}pez Miras, Javier and Santiago, Jos{\'e} and Vilches, Fuensanta",
    title = "{Efficient on-shell matching}",
    eprint = "2411.12798",
    archivePrefix = "arXiv",
    primaryClass = "hep-ph",
    doi = "10.21468/SciPostPhys.18.6.185",
    journal = "SciPost Phys.",
    volume = "18",
    number = "6",
    pages = "185",
    year = "2025"
}

@article{LopezMiras:2025gar,
    author = "L{\'o}pez Miras, Javier and Vilches, Fuensanta",
    title = "{Automation of a matching on-shell calculator}",
    eprint = "2505.21353",
    archivePrefix = "arXiv",
    primaryClass = "hep-ph",
    doi = "10.1016/j.cpc.2025.109935",
    journal = "Comput. Phys. Commun.",
    volume = "320",
    pages = "109935",
    year = "2026"
}

@article{Aebischer:2023djt,
    author = "Aebischer, Jason and Pesut, Marko and Polonsky, Zachary",
    title = "{Renormalization scheme factorization of one-loop Fierz identities}",
    eprint = "2306.16449",
    archivePrefix = "arXiv",
    primaryClass = "hep-ph",
    doi = "10.1007/JHEP01(2024)060",
    journal = "JHEP",
    volume = "01",
    pages = "060",
    year = "2024"
}

@article{Aebischer:2024xnf,
    author = "Aebischer, Jason and Pesut, Marko and Polonsky, Zachary",
    title = "{A simple dirac prescription for two-loop anomalous dimension matrices}",
    eprint = "2401.16904",
    archivePrefix = "arXiv",
    primaryClass = "hep-ph",
    doi = "10.1140/epjc/s10052-024-13101-0",
    journal = "Eur. Phys. J. C",
    volume = "84",
    number = "7",
    pages = "750",
    year = "2024"
}

@article{Aebischer:2022rxf,
    author = "Aebischer, Jason and Pesut, Marko and Polonsky, Zachary",
    title = "{Dipole operators in Fierz identities}",
    eprint = "2211.01379",
    archivePrefix = "arXiv",
    primaryClass = "hep-ph",
    doi = "10.1016/j.physletb.2023.137968",
    journal = "Phys. Lett. B",
    volume = "842",
    pages = "137968",
    year = "2023"
}

@article{Aebischer:2022aze,
    author = "Aebischer, Jason and Pesut, Marko",
    title = "{One-loop Fierz transformations}",
    eprint = "2208.10513",
    archivePrefix = "arXiv",
    primaryClass = "hep-ph",
    doi = "10.1007/JHEP10(2022)090",
    journal = "JHEP",
    volume = "10",
    pages = "090",
    year = "2022"
}

@article{Guedes:2025sax,
    author = "Guedes, Guilherme and Roosmale Nepveu, Jasper",
    title = "{Two-loop renormalization of general bosonic effective field theories}",
    eprint = "2512.08827",
    archivePrefix = "arXiv",
    primaryClass = "hep-ph",
    reportNumber = "CERN-TH-2025-250",
    month = "12",
    year = "2025"
}

@article{Duhr:2025zqw,
    author = "Duhr, Claude and Vasquez, Andres and Ventura, Giuseppe and Vryonidou, Eleni",
    title = "{Two-loop renormalisation of quark and gluon fields in the SMEFT}",
    eprint = "2503.01954",
    archivePrefix = "arXiv",
    primaryClass = "hep-ph",
    reportNumber = "BONN-TH-2025-07",
    doi = "10.1007/JHEP07(2025)160",
    journal = "JHEP",
    volume = "07",
    pages = "160",
    year = "2025"
}

@article{DiNoi:2025arz,
    author = {Di Noi, Stefano and Gr{\"o}ber, Ramona},
    title = "{Two loops, four tops and two $\gamma_5$ schemes: A renormalization story}",
    eprint = "2507.10295",
    archivePrefix = "arXiv",
    primaryClass = "hep-ph",
    reportNumber = "KA-TP-18-2025",
    doi = "10.1016/j.physletb.2025.139878",
    journal = "Phys. Lett. B",
    volume = "869",
    pages = "139878",
    year = "2025"
}

@article{Haisch:2025vqj,
    author = "Haisch, Ulrich and Niggetiedt, Marco",
    title = "{Precision tests of third-generation four-quark operators: $gg \to h$ and $h \to \gamma\gamma$}",
    eprint = "2507.20803",
    archivePrefix = "arXiv",
    primaryClass = "hep-ph",
    reportNumber = "MPP-2025-147",
    month = "7",
    year = "2025"
}

@article{Aebischer:2025ddl,
    author = "Aebischer, Jason and Bresciani, Luigi C. and Selimovic, Nudzeim",
    title = "{Anomalous Dimension of a General Effective Gauge Theory II: Fermionic Sector}",
    eprint = "2512.16890",
    archivePrefix = "arXiv",
    primaryClass = "hep-ph",
    month = "12",
    year = "2025"
}

@article{Fonseca:2025cls,
    author = "Fonseca, Renato M. and Olgoso, Pablo and Santiago, Jos{\'e}",
    title = "{Renormalization of general Effective Field Theories: Renormalization of fermionic operators}",
    eprint = "2512.15866",
    archivePrefix = "arXiv",
    primaryClass = "hep-ph",
    month = "12",
    year = "2025"
}

@article{Duhr:2025yor,
    author = "Duhr, Claude and Ventura, Giuseppe and Vryonidou, Eleni",
    title = "{Two-loop renormalisation of quark and gluon fields in the SMEFT in the on-shell scheme}",
    eprint = "2508.04500",
    archivePrefix = "arXiv",
    primaryClass = "hep-ph",
    reportNumber = "BONN-TH-2025-27",
    doi = "10.1007/JHEP11(2025)046",
    journal = "JHEP",
    volume = "11",
    pages = "046",
    year = "2025"
}

@article{DiNoi:2025tka,
    author = {Di Noi, Stefano and Erdelyi, Barbara Anna and Gr{\"o}ber, Ramona},
    title = "{Complete two-loop Yukawa-induced running of the Higgs-gluon coupling in SMEFT}",
    eprint = "2510.14680",
    archivePrefix = "arXiv",
    primaryClass = "hep-ph",
    reportNumber = "KA-TP-31-2025, COMETA-2025-47",
    month = "10",
    year = "2025"
}

@article{Banik:2025wpi,
    author = "Banik, Sumit and Crivellin, Andreas and Naterop, Luca and Stoffer, Peter",
    title = "{Two-loop anomalous dimensions for baryon-number-violating operators in SMEFT}",
    eprint = "2510.08682",
    archivePrefix = "arXiv",
    primaryClass = "hep-ph",
    reportNumber = "ZU-TH 61/25",
    doi = "10.1007/JHEP02(2026)017",
    journal = "JHEP",
    volume = "02",
    pages = "017",
    year = "2026"
}

@article{DiNoi:2024ajj,
    author = {Di Noi, Stefano and Gr\"ober, Ramona and Mandal, Manoj K.},
    title = "{Two-loop running effects in Higgs physics in Standard Model Effective Field Theory}",
    eprint = "2408.03252",
    archivePrefix = "arXiv",
    primaryClass = "hep-ph",
    reportNumber = "COMETA-2024-19",
    doi = "10.1007/JHEP12(2024)220",
    journal = "JHEP",
    volume = "12",
    pages = "220",
    year = "2025"
}

@article{Hahn:2000kx,
    author = "Hahn, Thomas",
    title = "{Generating Feynman diagrams and amplitudes with FeynArts 3}",
    eprint = "hep-ph/0012260",
    archivePrefix = "arXiv",
    reportNumber = "KA-TP-23-2000",
    doi = "10.1016/S0010-4655(01)00290-9",
    journal = "Comput. Phys. Commun.",
    volume = "140",
    pages = "418--431",
    year = "2001"
}

@article{Shtabovenko:2023xyz,
    author = "Shtabovenko, Vladyslav and Mertig, Rolf and Orellana, Federico",
    title = "{FeynCalc 10: Do multiloop integrals dream of computer codes?}",
    eprint = "2312.14089",
    archivePrefix = "arXiv",
    primaryClass = "hep-ph",
    doi = "10.1016/j.cpc.2024.109357",
    journal = "Comput. Phys. Commun.",
    volume = "306",
    pages = "109357",
    year = "2025"
}

@article{Shtabovenko:2020gxv,
    author = "Shtabovenko, Vladyslav and Mertig, Rolf and Orellana, Frederik",
    title = "{FeynCalc 9.3: New features and improvements}",
    eprint = "2001.04407",
    archivePrefix = "arXiv",
    primaryClass = "hep-ph",
    reportNumber = "P3H-20-002, TTP19-020, TUM-EFT 130/19",
    doi = "10.1016/j.cpc.2020.107478",
    journal = "Comput. Phys. Commun.",
    volume = "256",
    pages = "107478",
    year = "2020"
}

@article{Shtabovenko:2016whf,
    author = "Shtabovenko, Vladyslav",
    title = "{FeynHelpers: Connecting FeynCalc to FIRE and Package-X}",
    eprint = "1611.06793",
    archivePrefix = "arXiv",
    primaryClass = "physics.comp-ph",
    reportNumber = "TUM-EFT-75-15",
    doi = "10.1016/j.cpc.2017.04.014",
    journal = "Comput. Phys. Commun.",
    volume = "218",
    pages = "48--65",
    year = "2017"
}

@article{Ma:2019gtx,
    author = "Ma, Teng and Shu, Jing and Xiao, Ming-Lei",
    title = "{Standard model effective field theory from on-shell amplitudes*}",
    eprint = "1902.06752",
    archivePrefix = "arXiv",
    primaryClass = "hep-ph",
    doi = "10.1088/1674-1137/aca200",
    journal = "Chin. Phys. C",
    volume = "47",
    number = "2",
    pages = "023105",
    year = "2023"
}

@article{Shtabovenko:2025lxq,
    author = "Shtabovenko, Vladyslav",
    title = "{FeynCalc 10.2 and FeynHelpers 2: Multiloop calculations streamlined}",
    eprint = "2512.19858",
    archivePrefix = "arXiv",
    primaryClass = "hep-ph",
    reportNumber = "P3H-25-112, SI-HEP-2025-31",
    month = "12",
    year = "2025"
}

@article{Lange:2025fba,
    author = "Lange, Fabian and Usovitsch, Johann and Wu, Zihao",
    title = "{Kira 3: integral reduction with efficient seeding and optimized equation selection}",
    eprint = "2505.20197",
    archivePrefix = "arXiv",
    primaryClass = "hep-ph",
    reportNumber = "ZU-TH 39/25, HU-EP-25/17-RTG",
    doi = "10.1016/j.cpc.2025.109999",
    journal = "Comput. Phys. Commun.",
    volume = "322",
    pages = "109999",
    year = "2026"
}

@article{Dedes:2023zws,
    author = "Dedes, A. and Rosiek, J. and Ryczkowski, M. and Suxho, K. and Trifyllis, L.",
    title = "{SmeftFR v3 {\textendash} Feynman rules generator for the Standard Model Effective Field Theory}",
    eprint = "2302.01353",
    archivePrefix = "arXiv",
    primaryClass = "hep-ph",
    doi = "10.1016/j.cpc.2023.108943",
    journal = "Comput. Phys. Commun.",
    volume = "294",
    pages = "108943",
    year = "2024"
}

\end{document}